\documentclass[12pt]{article}
\usepackage[latin9]{inputenc}
\usepackage{color}
\usepackage{amsmath}
\usepackage{amssymb}
\usepackage{soul} 
\usepackage{lscape}
\usepackage{float}
\usepackage{accents}

\usepackage{bbold}
\usepackage[unicode=true,
 bookmarks=false,
 breaklinks=false,pdfborder={0 0 1},backref=section,colorlinks=false]
 {hyperref}
 
\makeatletter

\usepackage{slashed}

\usepackage{slashed}
\usepackage{color}
\usepackage{verbatim}\usepackage{latexsym}
\usepackage{graphicx}
\usepackage{arydshln}
\usepackage{cite}
\usepackage{graphicx}
\usepackage{bm}
\usepackage{adjustbox}
\usepackage{booktabs}
\usepackage{multirow}
\usepackage{rotating}
\usepackage{xcolor}
\usepackage{multicol}

\newcommand{\hd}[1]{\textcolor{black}{#1}}

\newcommand{\suf}{\ensuremath{su\phi}}
\newcommand{\sdf}{\ensuremath{sd\phi}}
\newcommand{\sef}{\ensuremath{se\phi}}
\newcommand{\sfsq}{\ensuremath{s\phi\square}}
\newcommand{\ssq}{\ensuremath{s\square}}
\newcommand{\Op}{\ensuremath{\mathcal{O}}}
\newcommand{\ROp}{\ensuremath{\mathcal{R}}}
\newcommand{\ii}{\ensuremath{\mathrm{i}}}
\newcommand*{\tran}{{\mkern-1.5mu\mathsf{T}}}

\graphicspath{ {./NewDiagrams/} }

\makeatother

\begin{document}
\thispagestyle{empty}
\begin{flushright}
\end{flushright}
\vspace{0.8cm}

\begin{center}
{\Large\sc Running in the ALPs}
\vspace{0.8cm}

\textbf{
Mikael~Chala$^{\,a}$, Guilherme Guedes$^{a,\,b}$, Maria
Ramos$^{a,\,b}$ and Jose Santiago$^{\,a}$}\\ 
\vspace{0.8cm} 
{\em {$^a$CAFPE and Departamento de F\'isica Te\'orica y del Cosmos,
Universidad de Granada, Campus de Fuentenueva, E--18071 Granada, Spain}}\\[0.5cm]
{\em {$^b$Laborat\'orio de Instrumenta\c cao e F\'isica Experimental de Part\'iculas, Departamento de
F\'isica da Universidade do Minho, Campus de Gualtar, 4710-057 Braga, Portugal}}\\[0.2cm]
\end{center}
\vspace{0.4cm}
\begin{abstract}
The couplings of axion-like particles are probed by
different experiments across a huge range of energy scales. Accordingly, a
consistent analysis of the corresponding constraints requires the use
of the renormalization group equations.   
We compute the full one-loop renormalization group
evolution of all --relevant and marginal-- parameters in the effective
field theory for axion-like particles up to dimension five, above and
below the electroweak scale, assuming only that new physics does not violate
CP.
We also include a detailed
discussion of the different bases used in the literature, the
relations among them and the interplay of the CP and shift symmetries.
\end{abstract}

\newpage

\tableofcontents

\section{Introduction}
Axion-like particles (ALPs), $s$, are pseudo-scalar singlets of the Standard Model (SM) gauge group,
whose interactions are typically assumed to respect an approximate shift symmetry $s\to
s+\sigma$ with constant $\sigma$. They arise naturally in theories with a spontaneously broken global symmetry such as the Peccei-Quinn solution to the strong CP problem~\cite{Peccei:1977hh,Peccei:1977ur,Weinberg:1977ma,Wilczek:1977pj}, composite Higgs models~\cite{Gripaios:2009pe,Gripaios:2016mmi,Chala:2017sjk} and others~\cite{Wilczek:1982rv,Chikashige:1980ui}, as well as in different explanations for dark matter~\cite{Preskill:1982cy,Abbott:1982af,Dine:1982ah} and for the flavour~\cite{Davidson:1981zd,Wilczek:1982rv,Ema:2016ops,Calibbi:2016hwq} and hierarchy~\cite{Graham:2015cka} problems.
The shift symmetry must be broken explicitly by at least the
ALP mass, and potentially also by its marginal couplings to the Higgs boson. The 
phenomenology of ALPs is mostly triggered by effective operators, the first of which arise at dimension
five. Their impact has been studied at photon regeneration experiments~\cite{Povey:2010hs,Wagner:2010mi,Essig:2013lka,Betz:2013dza}, beam dumps~\cite{Bjorken:2009mm,Andreas:2012mt} and high-energy colliders including LEP~\cite{Jaeckel:2015jla,Bauer:2017ris,Craig:2018kne} and more recently the LHC~\cite{Jaeckel:2015jla,Knapen:2016moh,Brivio:2017ije,Bauer:2017ris,Bauer:2017nlg,Alonso-Alvarez:2018irt,Craig:2018kne,Ebadi:2019gij,Gavela:2019cmq,Coelho:2020saz,Haghighat:2020nuh,Goncalves:2020bqi} and future facilities~\cite{Bauer:2018uxu,Yue:2019gbh,Inan:2020aal}; see also Refs.~\cite{Feng:2018pew,Ariga:2018uku,Kling:2020mch}. Searches for ALPs produced from the blackbody photons in the solar 
core~\cite{Arik:2011rx} and in other astrophysical events~\cite{Raffelt:2006cw,Lee:2018lcj,Chang:2018rso,Jaeckel:2019xpa,Carenza:2019pxu,Ertas:2020xcc,Lucente:2020whw,Carenza:2020cis} have been  perfomed too. ALP searches in flavour experiments have been studied recently in Refs.~\cite{Marciano:2016yhf,Gavela:2019wzg,Bauer:2019gfk,Cornella:2019uxs,Calibbi:2020jvd,MartinCamalich:2020dfe}; while CP violation signatures of ALPs have been considered in Ref.~\cite{DiLuzio:2020oah}. (Recent reviews on the physics of ALPs can be found for example in Refs.~\cite{DiLuzio:2020wdo,Choi:2020rgn}.)

These experiments span a huge range of energies, across which the Wilson coefficients of the ALP
effective field theory (EFT) run, and mix, following the corresponding renormalization group equations
(RGEs). Different computations of parts of the RGEs are spread in the literature~\cite{Bauer:2016lbe,Choi:2017gpf,Bauer:2017ris}. However, to 
the best of our knowledge, there is no systematic study of the entire ALP anomalous dimension matrix in any concrete basis of operators. 
We fill this gap in this work, extending also previous computations in
different ways. In particular 
we compute the gauge dependence of the ALP-fermion-fermion operators
running, as well as the RGE dependence on the ALP-Higgs marginal coupling. Moreover, we work in a basis closer in spirit 
to the Warsaw basis~\cite{Grzadkowski:2010es} of the SMEFT;
\textit{i.e.} involving operators with less derivatives, not limiting
to purely shift-invariant interactions. However, we cross check  
(and update where necessary) previous partial results performed with
different sets of operators. Most importantly, we match at tree level at the
electroweak (EW) scale the ALP EFT onto the low-energy version in
which the heavy top quark, the Higgs and the $Z$ and $W$ gauge bosons
are integrated out, and we compute the running within this ALP low-energy
EFT (ALP LEFT) too, including the mixing of higher-dimensional operators
into renormalizable ones, as well as the mixing between purely SM EFT
operators and others that do involve the ALP. For the sake of generality, we compute this running for an arbitrary ALP LEFT, \textit{i.e.} independently of whether the EFT above the EW scale is the ALP EFT or a more generic theory. As far as we are aware,
essentially all results in this latter EFT are completely new.

The article is organised as follows. In Section~\ref{sec:eft} we introduce the ALP Lagrangian, including
a Green basis of effective operators and their on-shell relations. We
compute the one-loop counterterms for effective operators in
Section~\ref{sec:divergences}. In Section \ref{sec:rges} we obtain the
complete anomalous dimension matrix for dimension-five operators at
one loop. The different layers of the EFT valid at energies below the
EW scale as well as their connection through renormalization and
matching are discussed in Section~\ref{sec:lalp}. In
Section~\ref{sec:pheno} we present some phenomenological implications
of the previous results, most importantly the possibility of probing
ALP interactions to the $Z$ boson or to the top quark through their
mixing into ALP-lepton operators. We conclude in
Section~\ref{sec:conclusions}.  In Appendix~\ref{app:HEdiags} we
provide the different Feynman diagrams computed for the
renormalization of the ALP EFT. In Appendix~\ref{app:bases} we report
our results in a different basis commonly used too in phenomenological
studies, while in Appendix~\ref{app:4dim} we collect the
renormalization group running of renormalizable parameters within this
EFT. Finally, in Appendix~\ref{diagrams:left} we report the Feynman
diagrams neccessary for the computation of the RGEs in the ALP LEFT. 

\section{Effective field theory for ALPs}
\label{sec:eft}

The renormalizable Lagrangian of the SM extended with a real pseudo-scalar
singlet, $s$, reads,
\begin{align}\nonumber
  \mathcal{L}_{SM+s}  =&
  -\frac{1}{4}G_{\mu\nu}^{A}G_{A}^{\mu\nu}
  -\frac{1}{4}W_{\mu\nu}^{a}W_{a}^{\mu\nu}
  -\frac{1}{4}B_{\mu\nu}B^{\mu\nu}\\\nonumber
&
+\overline{q_{L}^{\alpha}}\ii\slashed{D}q_{L}^{\alpha}
+\overline{l_{L}^{\alpha}}\ii\slashed{D}l_{L}^{\alpha}
+\overline{u_{R}^{\alpha}}\ii\slashed{D}u_{R}^{\alpha}
+\overline{d_{R}^{\alpha}}\ii\slashed{D}d_{R}^{\alpha}
+\overline{e_{R}^{\alpha}}\ii\slashed{D}e_{R}^{\alpha}
\\\nonumber
& +\left(D_{\mu}\phi\right)^{\dagger}\left(D^{\mu}\phi\right)
-\mu^{2}|\phi|^{2}-\lambda|\phi|^{4}
-\left(
y_{\alpha\beta}^{u}\overline{q_{L}^{\alpha}}\widetilde{\phi}u_{R}^{\beta}
+y_{\alpha\beta}^{d}\overline{q_{L}^{\alpha}}\phi d_{R}^{\beta}
+y_{\alpha\beta}^{e}\overline{l_{L}^{\alpha}}\phi e_{R}^{\beta}
+\text{h.c.}\right)\\
& +\frac{1}{2}\left(\partial_{\mu}s\right)\left(\partial^{\mu}s\right)-\frac{1}{2}m^{2}s^{2}
-\frac{\kappa_s}{3!}s^{3}-\frac{\lambda_s}{4!}s^{4} 
-\kappa_{s\phi}s|\phi|^{2}-\frac{\lambda_{s\phi}}{2}s^{2}|\phi|^{2}\,,
\end{align}
where $\alpha$ and $\beta$ are flavour indices, $q_L$ and $l_L$ denote
the left-handed (LH) quark and lepton doublets, respectively, and
$u_R$, $d_R$ and $e_R$ the right-handed (RH) up-type,
down-type quark and charged lepton singlets, respectively. 
The gluon and the EW gauge bosons are represented, as usual, by $G$
and by $W$ and $B$, respectively. The Higgs doublet is called $\phi$, while
its conjugate is given by
$\widetilde{\phi} =\epsilon \phi^\ast= \mathrm{i}\sigma_2 \phi^\ast $,
with $\sigma_2$ being the second Pauli matrix. A possible tadpole term has been
eliminated via a field redefinition of $s$ and
we use the minus-sign convention for the covariant derivative.
In the renormalizable Lagrangian above, all coefficients are real
except for the Yukawa couplings. Complex phases in these Yukawa
couplings, as well as the couplings $\kappa_s$ and $\kappa_{s\phi}$,
induce CP violation.
\begin{table}[t]
\centering{}
\begin{tabular}{|l|l|l|l|}
\hline
\multicolumn{1}{|c|}{Scalar}
& \multicolumn{1}{|c|}{Yukawa}
& \multicolumn{1}{|c|}{Derivative}
& \multicolumn{1}{|c|}{Gauge}
\\
\hline 
\rule{0pt}{16pt}
&
$\mathcal{O}_{\suf}^{\alpha\beta}=\ii s(
\overline{q_{L}^\alpha}
\widetilde{\phi}u_{R}^\beta-
\overline{u_{R}^\beta}
\widetilde{\phi}^\dagger q_L^\alpha)$
&
$\mathcal{R}_{\sfsq}=\ii s(\phi^{\dagger}D^{2}\phi- (D^2 \phi)^\dagger  \phi)$
&
$\mathcal{O}_{s\widetilde{G}}=sG_{\mu\nu}^{A}\widetilde{G}_{A}^{\mu\nu}$
\\
&
$\mathcal{O}_{\sdf}^{\alpha\beta}=\ii s
(\overline{q_{L}^\alpha}\phi d_{R}^\beta
- \overline{d_R^\beta} \phi^\dagger q_L^\alpha)$
&
$\mathcal{R}_{sq}^{\alpha\beta}=
s(\overline{q_{L}^\alpha}\slashed{D}q_{L}^\beta
+ \overline{q_L^\beta}  \overleftarrow{\slashed{D}}q_L^\alpha)$
&
$\mathcal{O}_{s\widetilde{W}}=sW_{\mu\nu}^{a}\widetilde{W}_{a}^{\mu\nu}$
\\
&
$\mathcal{O}_{\sef}^{\alpha\beta}=\ii s
(\overline{l_{L}^\alpha}\phi e_{R}^\beta
- \overline{e_R^\beta} \phi^\dagger l_L^\alpha)$
&
$\mathcal{R}_{sl}^{\alpha\beta}=
s(\overline{l_{L}^\alpha}\slashed{D}l_{L}^\beta
+ \overline{l_L^\beta}  \overleftarrow{\slashed{D}}l_L^\alpha)$
&
$\mathcal{O}_{s\widetilde{B}}=sB_{\mu\nu}\widetilde{B}^{\mu\nu}$
\\
&
&
$\mathcal{R}_{su}^{\alpha\beta}=
s(\overline{u_{R}^\alpha}\slashed{D}u_{R}^\beta
+ \overline{u_R^\beta}  \overleftarrow{\slashed{D}}u_R^\alpha)$
&\\
&
&
$\mathcal{R}_{sd}^{\alpha\beta}=
s(\overline{d_{R}^\alpha}\slashed{D}d_{R}^\beta
+ \overline{d_R^\beta}  \overleftarrow{\slashed{D}}d_R^\alpha)$
&
\\
&
&
$\mathcal{R}_{se}^{\alpha\beta}=
s(\overline{e_{R}^\alpha}\slashed{D}e_{R}^\beta
+ \overline{e_R^\beta}  \overleftarrow{\slashed{D}}e_R^\alpha)$
&
\rule[-6pt]{0pt}{17pt}
\\
\hline 
\rule{0pt}{16pt}
$\mathcal{O}_{s^5}=s^{5}$  &
$\mathcal{O}_{\widetilde{\suf}}^{\alpha\beta}=s(
\overline{q_{L}^\alpha}
\widetilde{\phi}u_{R}^\beta+
\overline{u_{R}^\beta}
\widetilde{\phi}^\dagger q_L^\alpha)$
&
$\mathcal{R}_{s\square}=s^{2}\partial_{\mu}\partial^{\mu}s$  &
$\mathcal{O}_{sG}=sG_{\mu\nu}^{A}G_{A}^{\mu\nu}$
\\[0.2cm]
$\mathcal{O}_{s^3}=s^{3}|\phi|^{2}$  &
$\mathcal{O}_{\widetilde{\sdf}}^{\alpha\beta}=s
(\overline{q_{L}^\alpha}\phi d_{R}^\beta + \overline{d_R^\beta} \phi^\dagger q_L^\alpha)$
&
$\mathcal{R}_{\phi s \square}=|\phi|^{2}\partial^{2}s$
& $\mathcal{O}_{sW}=sW_{\mu\nu}^{a}W_{a}^{\mu\nu}$\\[0.2cm]
$\mathcal{O}_{s}=s|\phi|^{4}$  &
$\mathcal{O}_{\widetilde{\sef}}^{\alpha\beta}=s
(\overline{l_{L}^\alpha}\phi e_{R}^\beta
+ \overline{e_R^\beta} \phi^\dagger l_L^\alpha)$
&
$\mathcal{R}_{\widetilde{\sfsq}}=s(\phi^{\dagger}D^{2}\phi+ (D^2 \phi)^\dagger  \phi)$
&
$\mathcal{O}_{sB}=sB_{\mu\nu}B^{\mu\nu}$ \\[0.2cm]
&  &
$\mathcal{R}_{\widetilde{sq}}^{\alpha\beta}=
s(\overline{q_{L}^\alpha}\ii\slashed{D}q_{L}^\beta
- \overline{q_L^\beta} \ii \overleftarrow{\slashed{D}}q_L^\alpha)
$
& 
\\[0.2cm]
&  & $\mathcal{R}_{\widetilde{sl}}^{\alpha\beta}=
s(\overline{l_{L}^\alpha}\ii\slashed{D}l_{L}^\beta
- \overline{l_L^\beta} \ii \overleftarrow{\slashed{D}}l_L^\alpha)
$
& 
\\[0.2cm]
&  & $\mathcal{R}_{\widetilde{su}}^{\alpha\beta}=
s(\overline{u_{R}^\alpha}\ii\slashed{D}u_{R}^\beta
- \overline{u_R^\beta} \ii \overleftarrow{\slashed{D}}u_R^\alpha)
$  &
\\[0.2cm]
&  & $\mathcal{R}_{\widetilde{sd}}^{\alpha\beta}=
s(\overline{d_{R}^\alpha}\ii\slashed{D}d_{R}^\beta
- \overline{d_R^\beta} \ii \overleftarrow{\slashed{D}}d_R^\alpha)
$
& \\[0.2cm]
&  & $\mathcal{R}_{\widetilde{se}}^{\alpha\beta}=
s(\overline{e_{R}^\alpha}\ii\slashed{D}e_{R}^\beta
- \overline{e_R^\beta} \ii \overleftarrow{\slashed{D}}e_R^\alpha)
$
& \\[0.2cm]
 \hline
\end{tabular}
\caption{\it Green basis of effective operators of dimension five. All
  operators are hermitian (operators with flavour indices are
  hermitian for each fixed value of $\alpha$ and $\beta$, $(\Op_{\alpha
    \beta})^\dagger = \Op_{\alpha\beta}$). The ones in the top
  (bottom) panel are CP 
  conserving (violating). The dual field strength tensor is defined by
  $\widetilde{B}^{\mu\nu}=\frac{1}{2}\epsilon_{\mu\nu\rho\sigma}
  B^{\rho \sigma}$ (with $\epsilon^{0123} = 1$) and likewise for $W$ and $G$.}\label{tab:eft}
\end{table}

The first tower of effective interactions arise at dimension five. 
We provide a Green basis (checked with
\texttt{BasisGen}~\cite{Criado:2019ugp})
of such set of operators in Table~\ref{tab:eft}. 
Any other dimension-five operator can be written in terms of these via
algebraic or integration by parts identities. We have collected the
operators in the table according to their CP properties, with CP-even
and CP-odd operators in the top and bottom panels of the table,
respectively. All the operators in the table are hermitian (for fixed
values of flavour indices if present) and the
corresponding Wilson coefficients are therefore real parameters
(real matrices for operators involving flavour). A minimal basis of
non-redundant operators, enough to describe all physical processes, is
given by the operators named with $\Op$. The ones denoted by $\ROp$ can
be written in terms of the ones in the minimal basis by performing
field redefinitions and are therefore equivalent to them in all
physical observables.
To $\mathcal{O}(1/\Lambda)$ accuracy, these field redefinitions
can be enforced through the equations of motion of the dimension-four
Lagrangian, namely,
\begin{align}
\partial^{2}s&=-m^{2}s-\frac{\kappa_s}{2}s^{2}-\frac{\lambda_s s^{3}}{3!}-\kappa_{s\phi}|\phi|^{2}-\lambda_{s\phi}s|\phi|^{2}\,,
\\
D^{2}\phi_{k}&=
-\mu^{2}\phi_{k}-2\lambda|\phi|^{2}\phi_{k}
-\kappa_{s\phi}s\phi_{k}-\frac{\lambda_{s\phi}}{2}s^{2}\phi_{k}
-y^{u}_{\alpha\beta}\overline{q_{Lj}^\alpha}\epsilon_{jk}u_{R}^\beta
-y^{d\,\ast}_{\alpha\beta} \overline{d_{R}^\beta}q_{Lk}^\alpha
-y^{e\,\ast}_{\alpha \beta} \overline{e_{R}^\beta}l_{Lk}^\alpha
\,,
\\ 
  i\slashed{D}q_{Lk}^\alpha&=
  y^{d}_{\alpha\beta}\phi_{k}d_{R}^\beta
  +y^{u}_{\alpha\beta}\widetilde{\phi}_{k}u_{R}^\beta\,,
  \qquad
  i\slashed{D}l_{Lk}^\alpha=
y^{e}_{\alpha\beta}\phi_{k}e_{R}^\beta\,,
\\[0.2cm]
i\slashed{D}u_{R}^\alpha&=
y^{u\,\ast}_{\beta\alpha}\widetilde{\phi}_{k}^{\dagger}q_{Lk}^\beta\,,
\qquad
\qquad
\qquad \,\,\, \, i\slashed{D}d_{R}^\alpha=
y^{d\,\ast}_{\beta \alpha}\phi_{k}^{\dagger}q_{Lk}^{\beta}\,,
\qquad i\slashed{D}e_{R}^\alpha=
y^{e\,\ast}_{\beta\alpha}\phi_{k}^{\dagger}l_{Lk}^{\beta}\,;
\end{align}
where we use latin indices for $SU(2)$. Using these equations
we can arrive at the following identities, valid for physical
observables:
\begin{align}
  r_{\sfsq}
  \ROp_{\sfsq} =&
-r_{\sfsq}\mathrm{Re}(y^u) \Op_{\suf}
+r_{\sfsq}\mathrm{Re}(y^d) \Op_{\sdf}
+r_{\sfsq}\mathrm{Re}(y^e) \Op_{\sef}
\nonumber \\
&
+r_{\sfsq}\mathrm{Im}(y^u) \Op_{\widetilde{\suf}}
-r_{\sfsq}\mathrm{Im}(y^d) \Op_{\widetilde{\sdf}}
-r_{\sfsq}\mathrm{Im}(y^e) \Op_{\widetilde{\sef}}
\,,
\\
r_{sq}\ROp_{sq} =&
-r_{sq}\mathrm{Re}(y^u) \Op_{\suf}
-r_{sq}\mathrm{Re}(y^d) \Op_{\sdf}
+r_{sq}\mathrm{Im}(y^u) \Op_{\widetilde{\suf}}
+r_{sq}\mathrm{Im}(y^d) \Op_{\widetilde{\sdf}}
\,,
\\
r_{sl}\ROp_{sl} =&
-r_{sl}\mathrm{Re}(y^e) \Op_{\sef}
+r_{sl}\mathrm{Im}(y^e) \Op_{\widetilde{\sef}}
\,,
\\
r_{su}\ROp_{su} =&\,
\mathrm{Re}(y^u)r^\tran_{su} \Op_{\suf}
-\mathrm{Im}(y^u) r^\tran_{su}\Op_{\widetilde{\suf}}
\,,
\\
r_{sd} \ROp_{sd} =&\,
\mathrm{Re}(y^d)r^\tran_{sd}\Op_{\sdf}
-\mathrm{Im}(y^d)r^\tran_{sd} \Op_{\widetilde{\sdf}}
\,,
\\
r_{se} \ROp_{se} =&\,
\mathrm{Re}(y^e)r^\tran_{se} \Op_{\sef}
-\mathrm{Im}(y^e) r^\tran_{se}\Op_{\widetilde{\sef}}
\,; \label{redundancies:cpeven}
\end{align}
where flavour indices are left implicit and we have always assumed
that each Wilson coefficient and its corresponding operator have the
flavour indices in the same order so that, for instance,
\begin{equation}
  \mathrm{Re}(y^e)r^\tran_{se} \Op_{\sef} \equiv
  \mathrm{Re}(y^e)_{\alpha\gamma} (r^\tran_{se})_{\gamma \beta}
  \Op_{\sef}^{\alpha \beta}\,,
\end{equation}
with repeated indices summed over.

For the CP-odd ones we have
\begin{align}
  r_{\ssq}
  \mathcal{R}_{\ssq} =&
  -r_{\ssq}m^2 s^3
  -r_{\ssq}\frac{\kappa_s}{2}s^4-r_{\ssq}\frac{\lambda_s}{3!}\mathcal{O}_{s^5}
  -r_{\ssq}\kappa_{s\phi} s^2 |\phi|^2 -r_{\ssq}\lambda_{s\phi}\mathcal{O}_{s^3}\,,
  \\
  r_{\widetilde{\sfsq}}
  \mathcal{R}_{\widetilde{\sfsq}} =&
-2r_{\widetilde{\sfsq}}\mu^2s|\phi|^2
- 4r_{\widetilde{\sfsq}}\lambda\mathcal{O}_s
-2 r_{\widetilde{\sfsq}}\kappa_{s\phi} s^2|\phi|^2
-r_{\widetilde{\sfsq}}\lambda_{s\phi}\mathcal{O}_{s^3}
\nonumber \\
&
-r_{\widetilde{\sfsq}}\mathrm{Re}(y^u) \Op_{\widetilde{\suf}}
-r_{\widetilde{\sfsq}}\mathrm{Re}(y^d) \Op_{\widetilde{\sdf}}
-r_{\widetilde{\sfsq}}\mathrm{Re}(y^e) \Op_{\widetilde{\sef}}
\nonumber \\
&
-r_{\widetilde{\sfsq}}\mathrm{Im}(y^u) \Op_{\suf}
-r_{\widetilde{\sfsq}}\mathrm{Im}(y^d) \Op_{\sdf}
-r_{\widetilde{\sfsq}}\mathrm{Im}(y^e) \Op_{\sef}\,,
\\
r_{\phi s \square}
\mathcal{R}_{\phi s \square} &=
-r_{\phi s \square}m^2 s |\phi|^2
-r_{\phi s \square}\frac{\kappa_s}{2} s^2 |\phi|^2
-r_{\phi s \square}\frac{\lambda_s}{3!}\mathcal{O}_{s^3}
-r_{\phi s \square}\kappa_{s\phi}|\phi|^4-\lambda_{s\phi}
\mathcal{O}_{s}
\,,
\\
r_{\widetilde{sq}}
\ROp_{\widetilde{sq}} =&\,
r_{\widetilde{sq}}\mathrm{Re}(y^u) \Op_{\widetilde{\suf}}
+r_{\widetilde{sq}}\mathrm{Re}(y^d) \Op_{\widetilde{\sdf}}
+r_{\widetilde{sq}}\mathrm{Im}(y^u) \Op_{\suf}
+r_{\widetilde{sq}}\mathrm{Im}(y^d) \Op_{\sdf}
\,,
\\
r_{\widetilde{sl}}
\ROp_{\widetilde{sl}} =&\,
r_{\widetilde{sl}}\mathrm{Re}(y^e) \Op_{\widetilde{\sef}}
+r_{\widetilde{sl}}\mathrm{Im}(y^e) \Op_{\sef}
\,,
\\
r_{\widetilde{su}}
\ROp_{\widetilde{su}} =&\,
\mathrm{Re}(y^u) r_{\widetilde{su}}^\tran
\Op_{\widetilde{\suf}}
+\mathrm{Im}(y^u)r_{\widetilde{su}}^\tran \Op_{\suf}
\,,
\\
r_{\widetilde{sd}}\ROp_{\widetilde{sd}} =&\,
\mathrm{Re}(y^d)r_{\widetilde{sd}}^\tran \Op_{\widetilde{\sdf}}
+\mathrm{Im}(y^d)r_{\widetilde{sd}}^\tran \Op_{\sdf}
\,,
\\
r_{\widetilde{se}}\ROp_{\widetilde{se}} =&\,
\mathrm{Re}(y^e) r_{\widetilde{se}}^\tran \Op_{\widetilde{\sef}}
+\mathrm{Im}(y^e)r_{\widetilde{sd}}^\tran \Op_{\sef}
\,.
\label{redundancies:cpodd}
\end{align}
For the sake of generality we have not made use
of the freedom to make $y^e$ and one of $y^u$ or $y^d$ diagonal,
with real and positive entries, in the equations above.

Note that, despite their field content,
the operators $\ROp_{\sfsq}$ and $\ROp_{\widetilde{\sfsq}}$
do not induce the process
$s\phi\to \phi Z$. 
Indeed, by using the field redefinitions above we note that these operators
are equivalent to Yukawa-like operators (or to operators with no gauge
bosons in the CP-odd case), which clearly do not trigger the aforementioned
process. An explicit calculation with these operators shows indeed
that the corresponding amplitude goes with $p_Z^2$, which vanishes
on-shell as the $Z$ boson is massless before EW symmetry breaking (EWSB)~\footnote{This can be trivially shown by considering the on-shell scattering amplitude $\mathcal{M}(1_s 2_\phi 3_\phi 4_Z)$. By dimensional arguments, $\mathcal{M}\sim \langle ij\rangle/\Lambda$ (or with squares). The only way to satisfy little-group covariance is therefore $i=j=4$, which makes $\mathcal{M}$ vanish due to the antisymmetry of brackets.}. 

In the following we will consider that CP is a good symmetry of the
EFT. This amounts to setting the coefficients of the CP-odd operators
to zero, including $\kappa_s=\kappa_{s\phi}=0$ in the renormalizable
Lagrangian and the ones in the bottom panel of Table~\ref{tab:eft}.
This is a radiatively stable choice up to the complex phase in the SM
Yukawa couplings. The main goal of the present paper is to obtain
the RGEs for the CP-even sector in isolation. However, we will provide our
results for arbitrary Yukawa couplings, so that the mixing of the
CP-even operators into the CP-odd sector via the imaginary part of
the SM Yukawa couplings can be
easily obtained.
Under these conditions, the relevant Lagrangian reads:
\begin{align}
\mathcal{L}_{\mathrm{CP-even}}&=
\sum_{\psi=u,d,e} a_{s\psi \phi} \Op_{s\psi
  \phi}
+\sum_{X=G,W,B} a_{s\widetilde{X}} \Op_{s\widetilde{X}}
+r_{\sfsq} \ROp_{\sfsq}
+\sum_{\Psi=q,l,u,d,e} r_{s\Psi}
\ROp_{s\Psi}
\,,\label{LCP-even}
\end{align}
where all the Wilson coefficients are real or real matrices in
flavour space.

For the sake of generality we are not enforcing shift
symmetry. However, in Appendix~\ref{app:bases} we provide conditions
on the Wilson coefficients that guarantee that this symmetry is
preserved. We also explain why, although customary, trading the
Yukawa-like operators 
$\mathcal{O}_{s\psi \phi}$, with $\psi=u,d,e$,
by the explicitly shift-invariant terms
$\partial_\mu s (\overline{\Psi}\gamma^\mu \Psi)$, with
$\Psi=q_L,l_L,u_R,d_R,e_R$~\cite{Georgi:1986df,Brivio:2017ije,Choi:2017gpf,Bauer:2017ris,Alonso-Alvarez:2018irt}, 
is not necessarily an optimal choice, as the set of operators thus 
constructed is overcomplete. Still, in the same appendix we will also
provide the RGEs of these operators under some simplifying
assumptions.

\section{Divergences at one loop}
\label{sec:divergences}
In order to obtain the RGEs of the ALP EFT, we have computed the divergences
generated by  
one-particle-irreducible (1PI) diagrams at one loop with off-shell momenta
and to order $\mathcal{O}(1/\Lambda)$. 
In doing so, we have employed the background field method in the Feynman
gauge in dimensional 
regularisation with space-time dimension $d = 4-2\epsilon$. The
$1/\epsilon$ poles obtained this 
way are gauge invariant. We have subsequently matched these onto the
Green basis of operators of Table~\ref{tab:eft}. 
We have implemented the model in
\texttt{FeynRules}~\cite{Alloul:2013bka} and used
\texttt{FeynArts}~\cite{Hahn:2000kx} and
\texttt{FormCalc}~\cite{Hahn:1998yk} for the calculations.
In a completely independent cross-check, we have evaluated by hand the
Yukawa and $\lambda_{s\phi}$ pieces of each 
of the Feynman diagrams as obtained with
\texttt{Qgraf}~\cite{Nogueira:1991ex}.  

In the remainder of this section we will go through the different
amplitudes that we need for matching the divergences in the ALP EFT. For each
amplitude we will provide the ultraviolet (UV) divergence, matched onto our Green
basis. We will denote the corresponding Wilson coefficients with a
prime in order to distinguish them from the ones of the operator
insertion in the one-loop calculation (which appear, without a prime,
on the right-hand side of our equations).
Recall that we assume the EFT to preserve CP and we are interested in
the RGEs of the CP-even operators among themselves. In particular we
will consider only insertions of CP-even operators in the one-loop
calculation. Within our assumptions the corresponding divergences can be
again parameterized in terms of the CP-even operators up to the
imaginary part of the SM Yukawa couplings. In this section we will
provide the matching in the full basis, including the contribution via
complex Yukawa couplings to the CP-odd operators so that the
interested reader can obtain the corresponding mixing.
In all these equations we leave flavour indices implicit.

\begin{figure}[H]
 \centering
 \includegraphics[width=0.19\columnwidth]{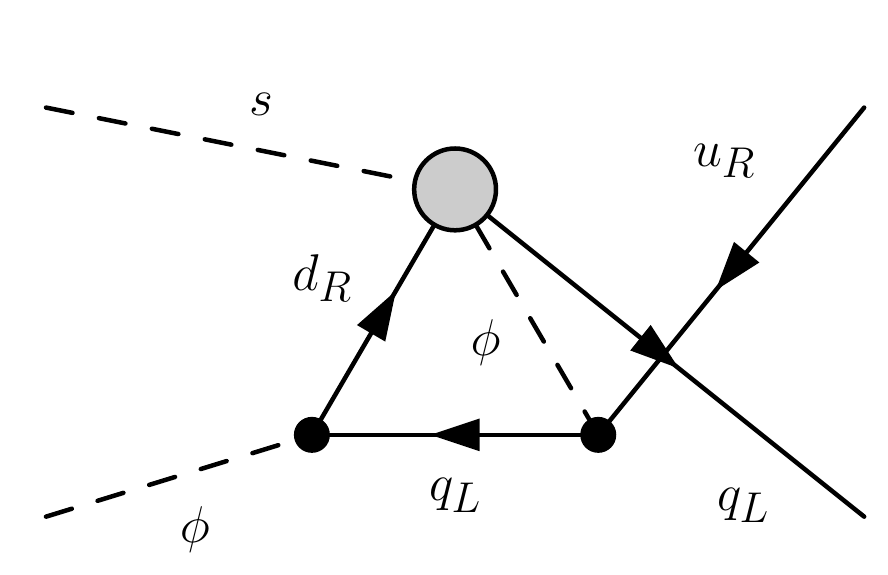}
 \includegraphics[width=0.19\columnwidth]{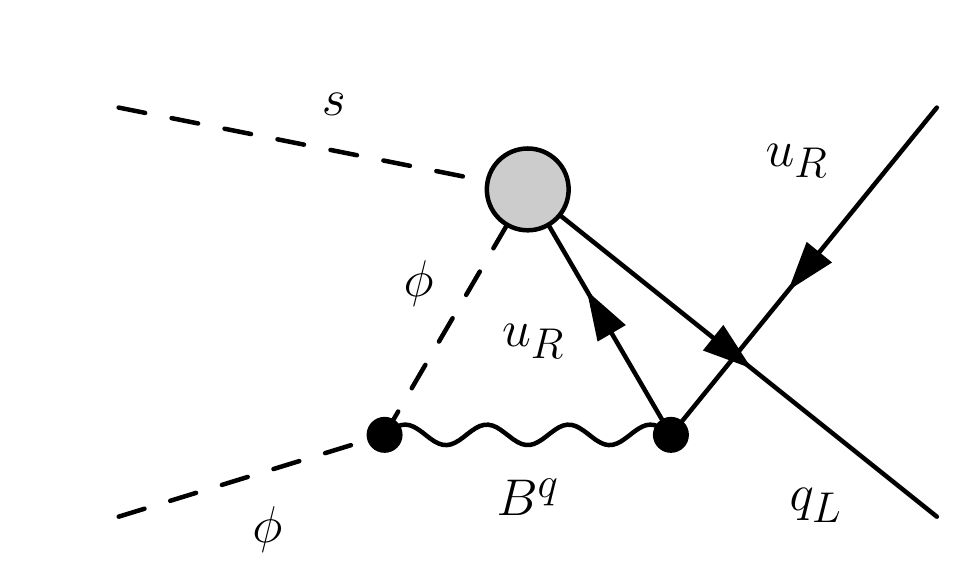}
 \includegraphics[width=0.19\columnwidth]{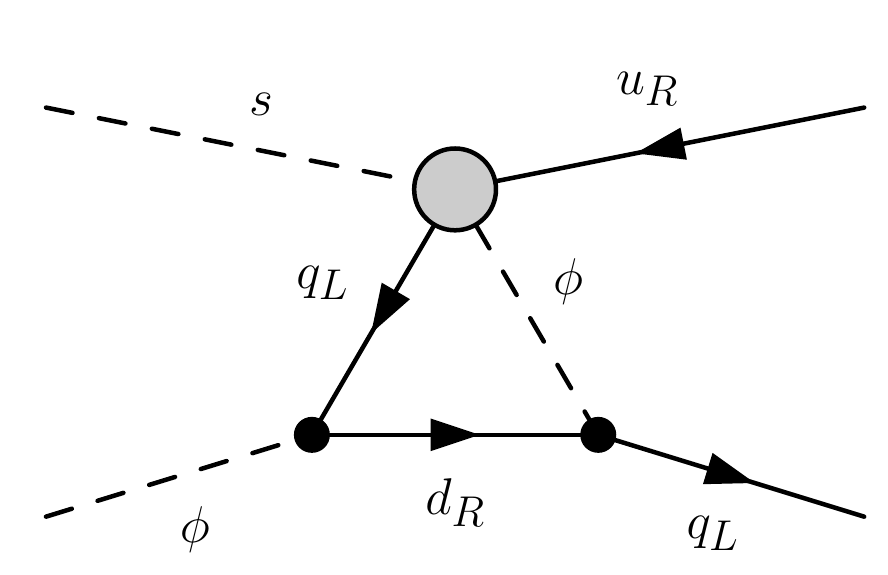}
 \includegraphics[width=0.19\columnwidth]{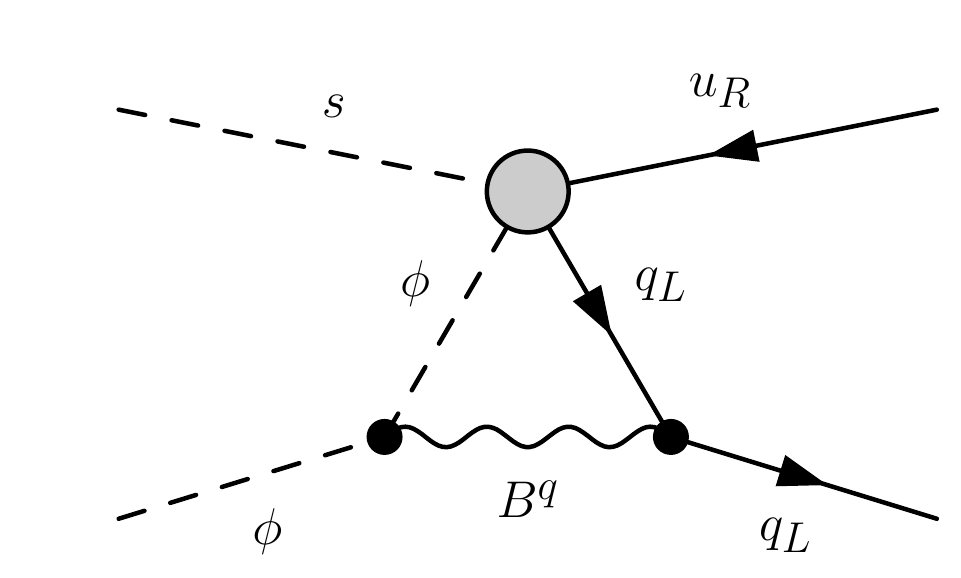}
 \includegraphics[width=0.19\columnwidth]{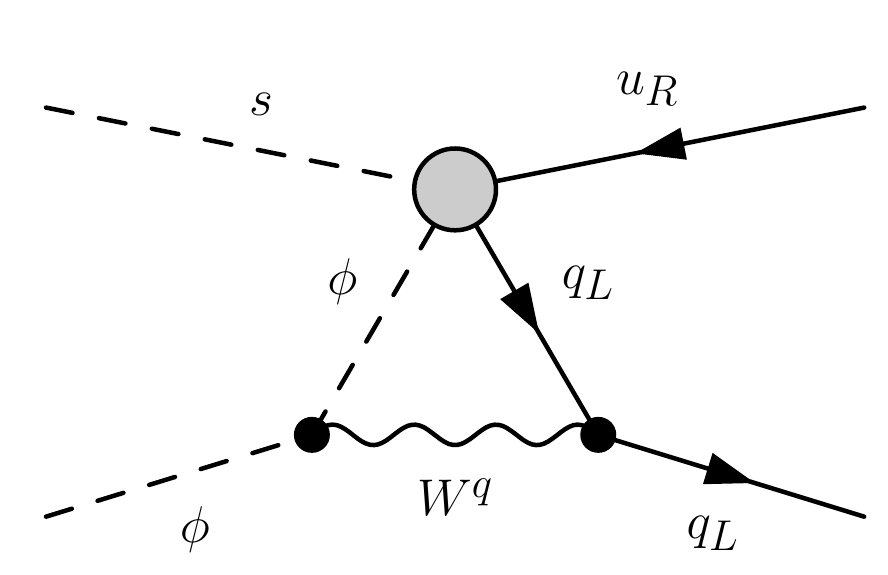}
 \includegraphics[width=0.19\columnwidth]{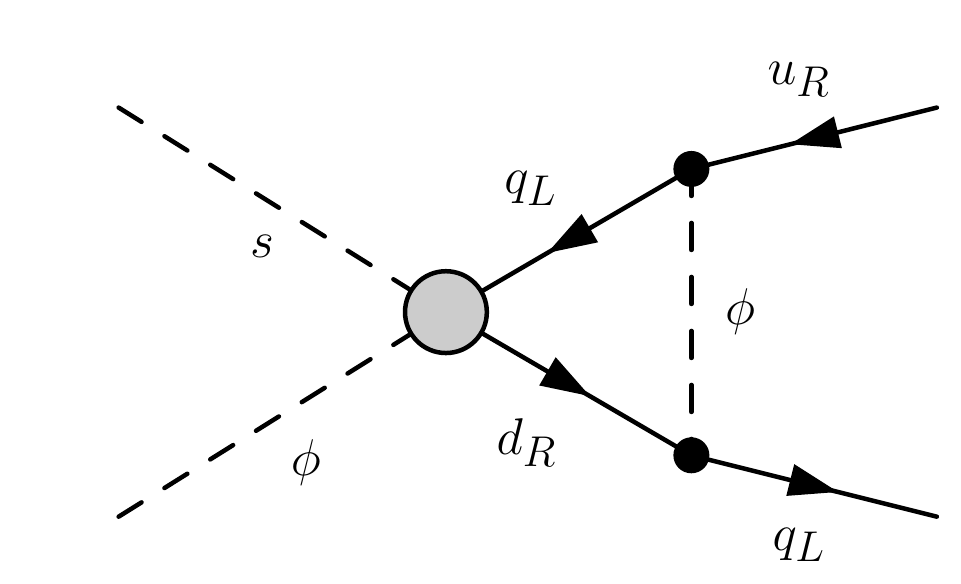}
 \includegraphics[width=0.19\columnwidth]{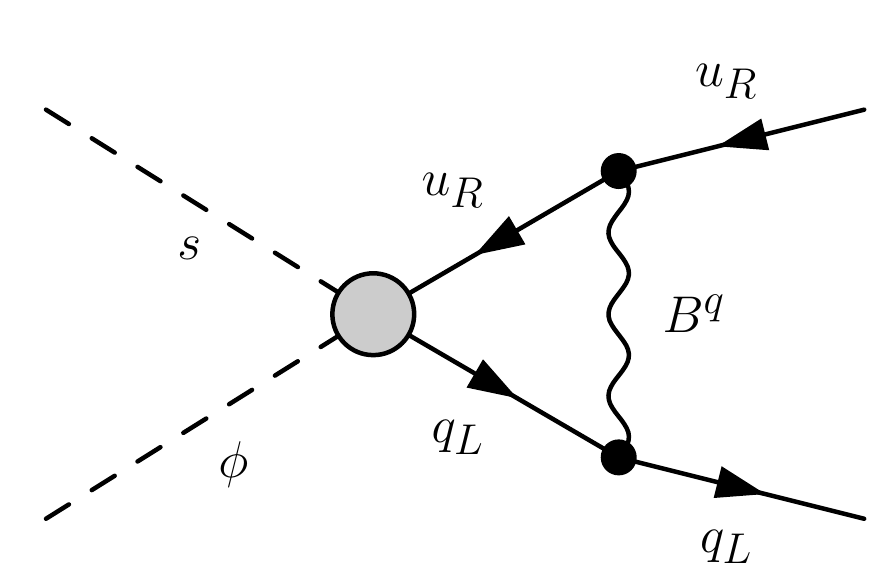}
 \includegraphics[width=0.19\columnwidth]{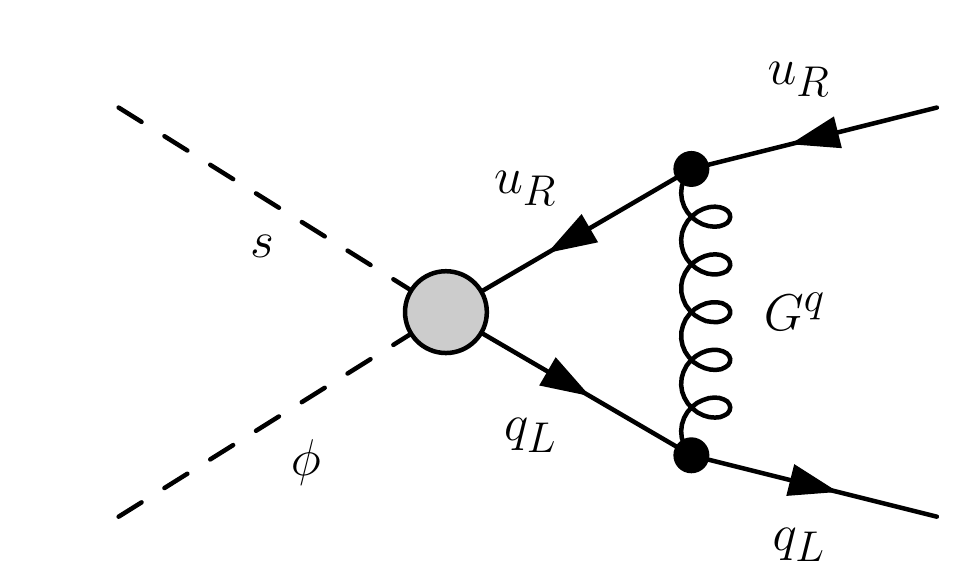}
 \includegraphics[width=0.19\columnwidth]{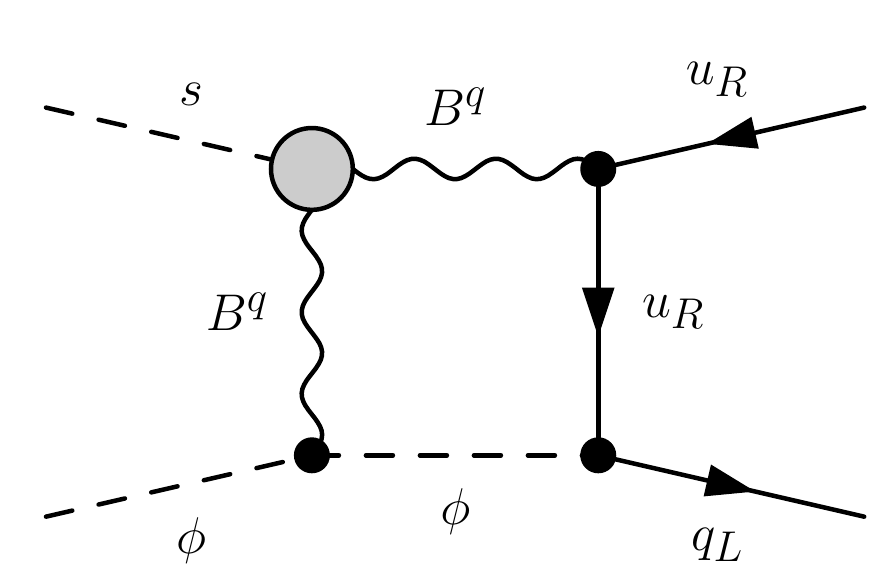}
 \includegraphics[width=0.19\columnwidth]{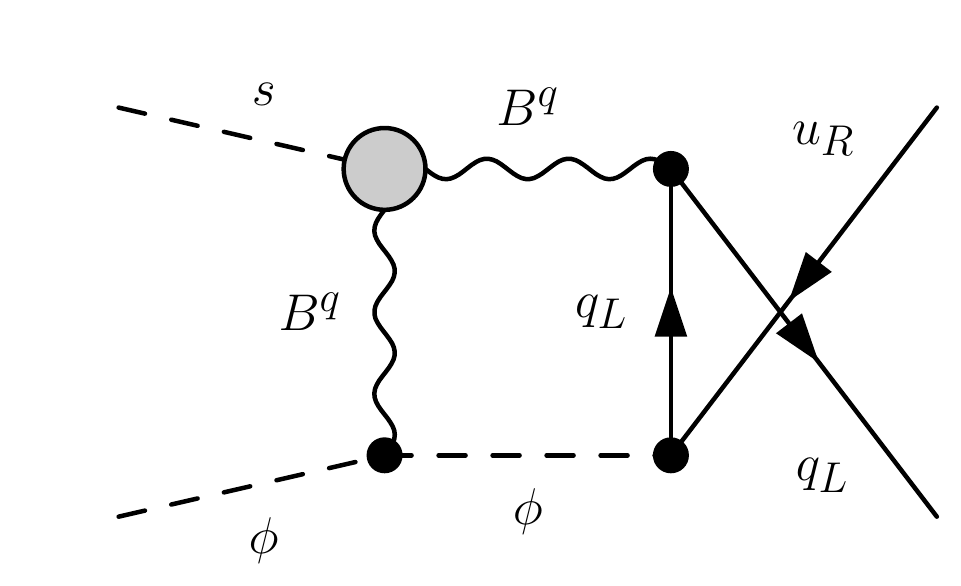}
 \includegraphics[width=0.19\columnwidth]{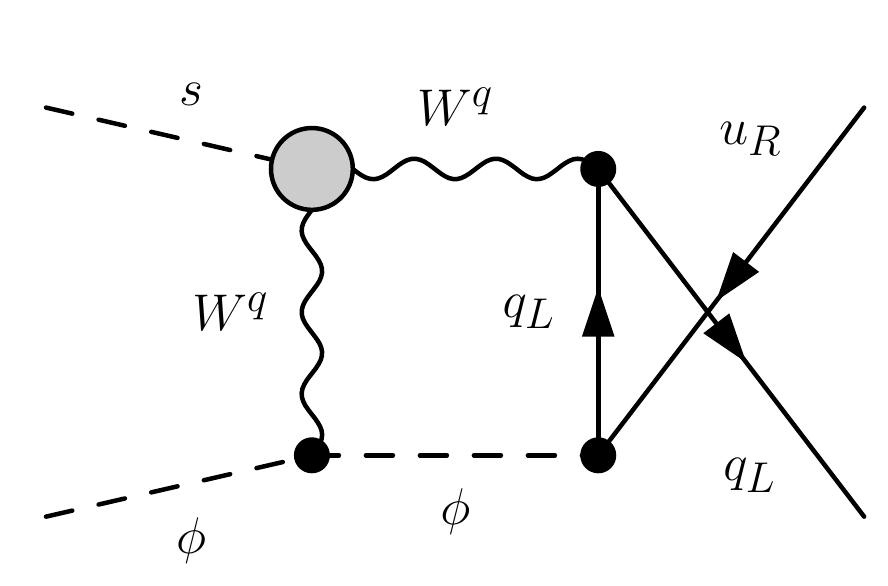}
 \includegraphics[width=0.19\columnwidth]{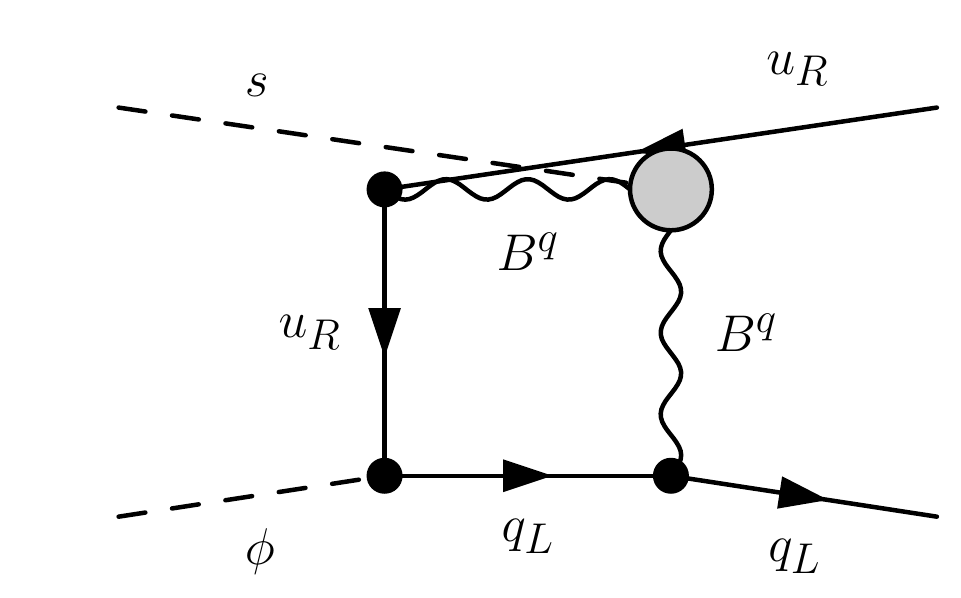}
 \includegraphics[width=0.19\columnwidth]{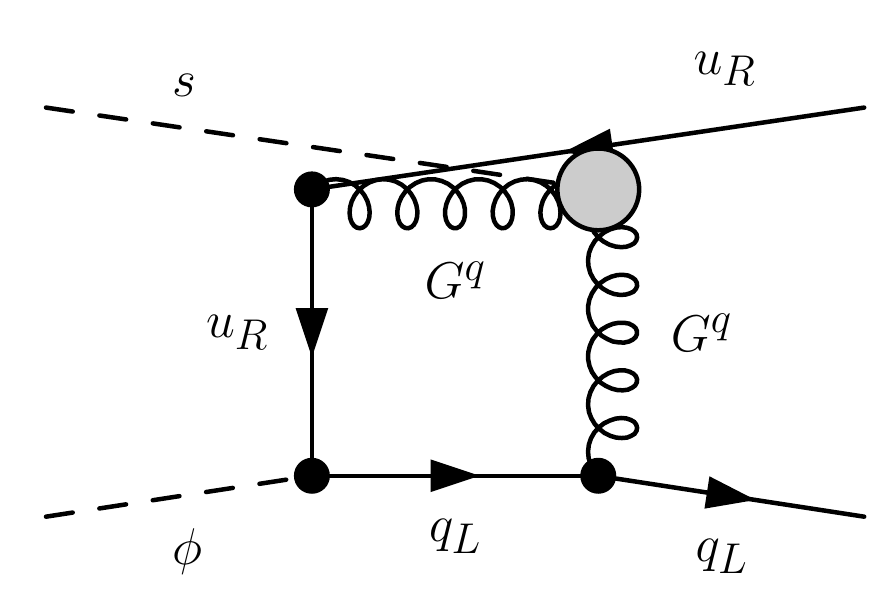}
 \includegraphics[width=0.19\columnwidth]{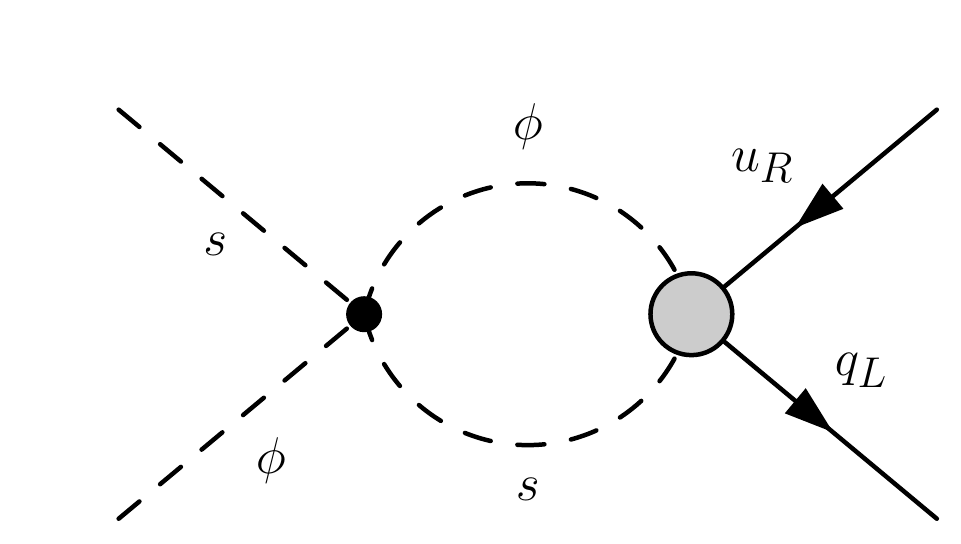}
 \caption{\it Feynman diagrams for $s(p_1)\phi^\dagger(p_2)\to q_L (p_3) \overline{u_R}(p_4)$. }\label{fig:shqu}
\end{figure}

\begin{figure}[H]
 \centering
 \includegraphics[width=0.19\columnwidth]{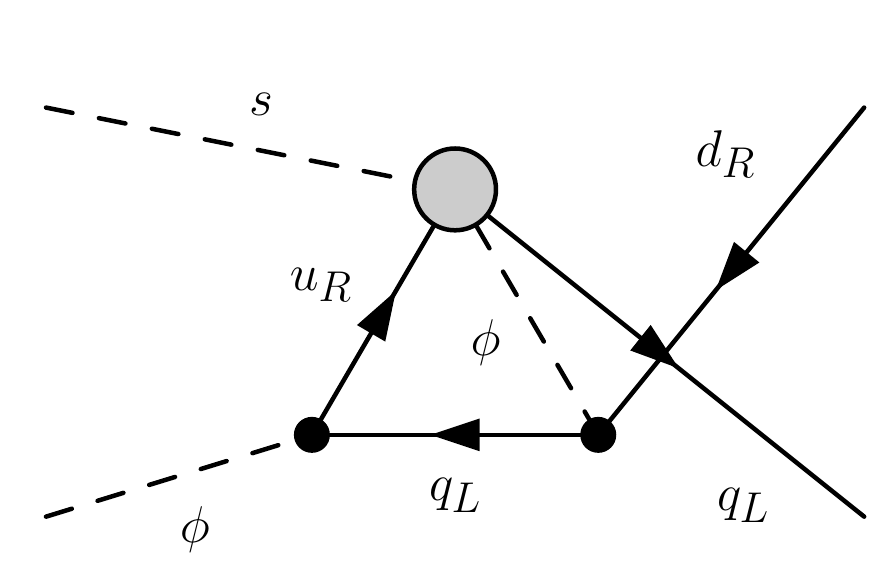}
 \includegraphics[width=0.19\columnwidth]{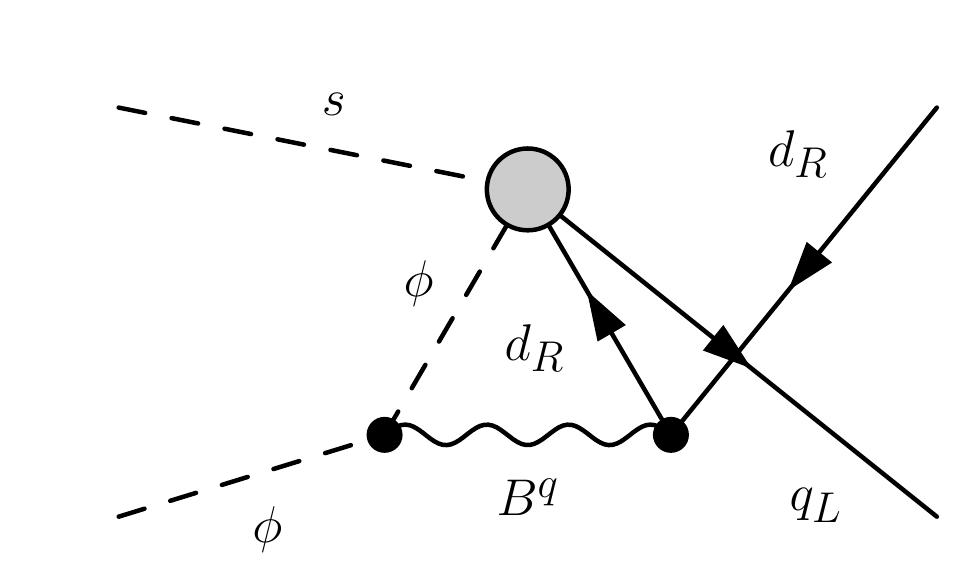}
 \includegraphics[width=0.19\columnwidth]{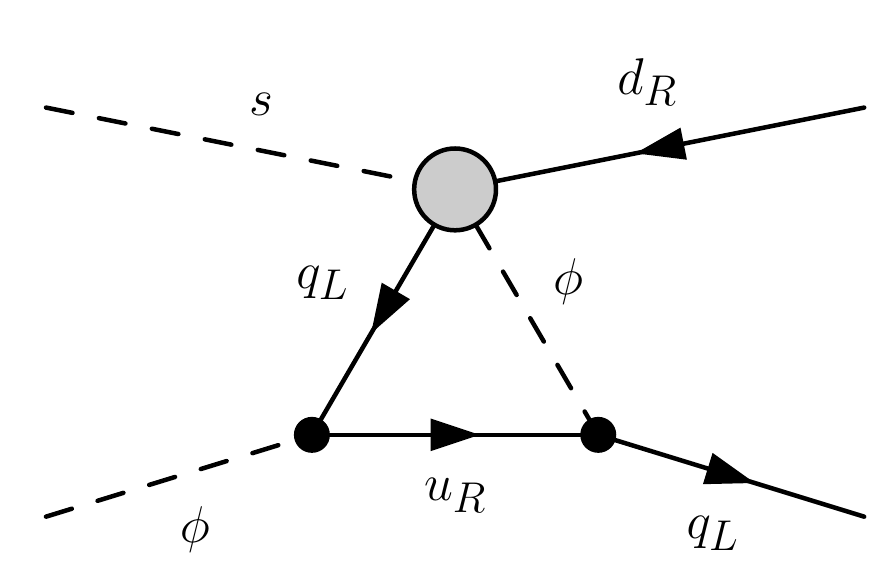}
 \includegraphics[width=0.19\columnwidth]{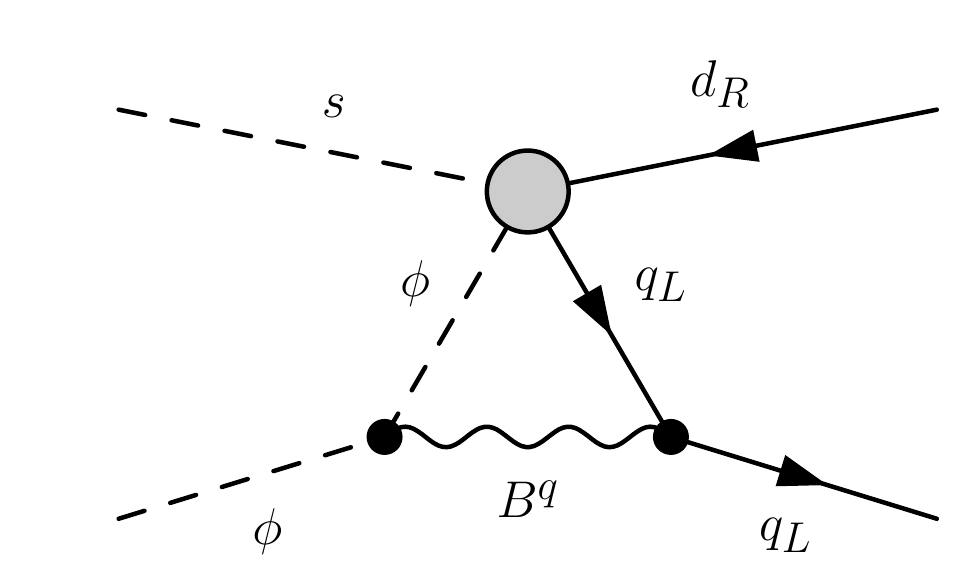}
 \includegraphics[width=0.19\columnwidth]{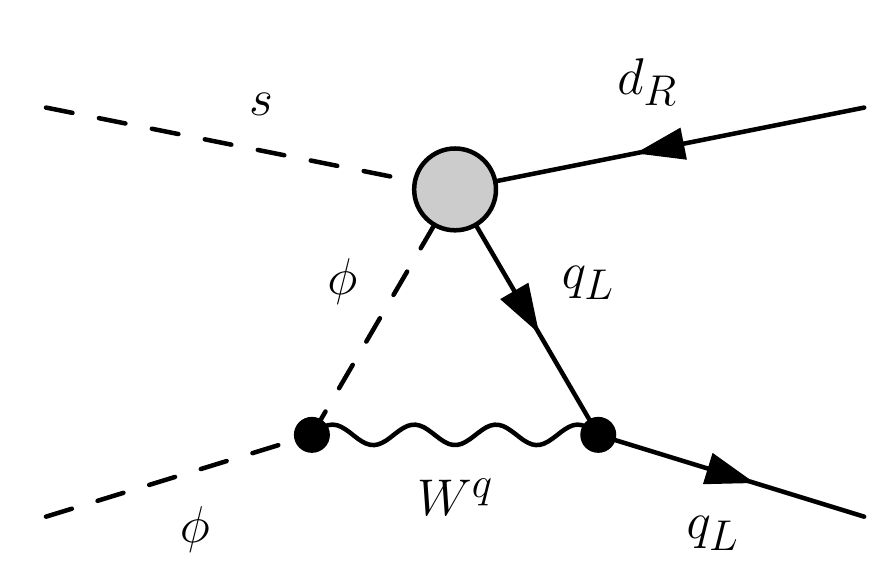}
 \includegraphics[width=0.19\columnwidth]{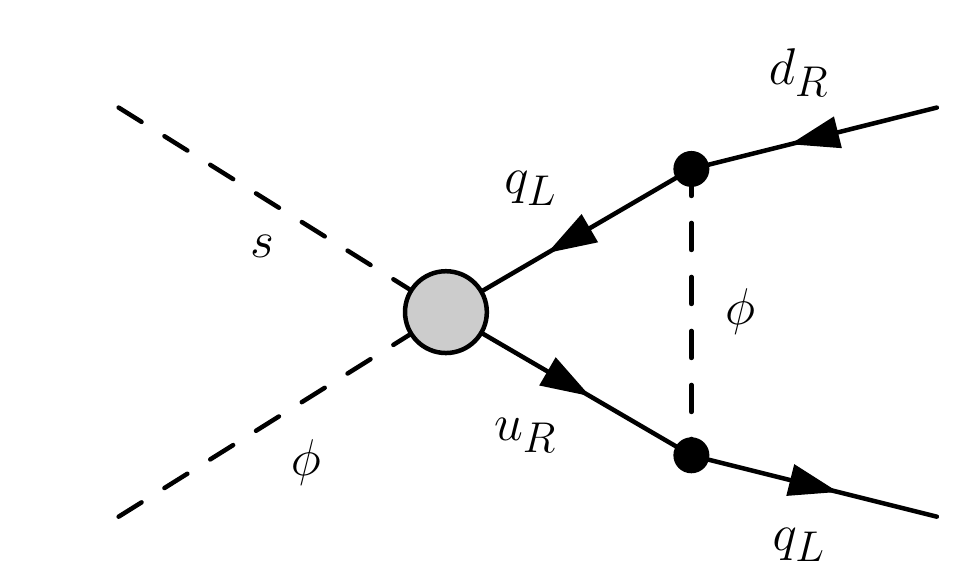}
 \includegraphics[width=0.19\columnwidth]{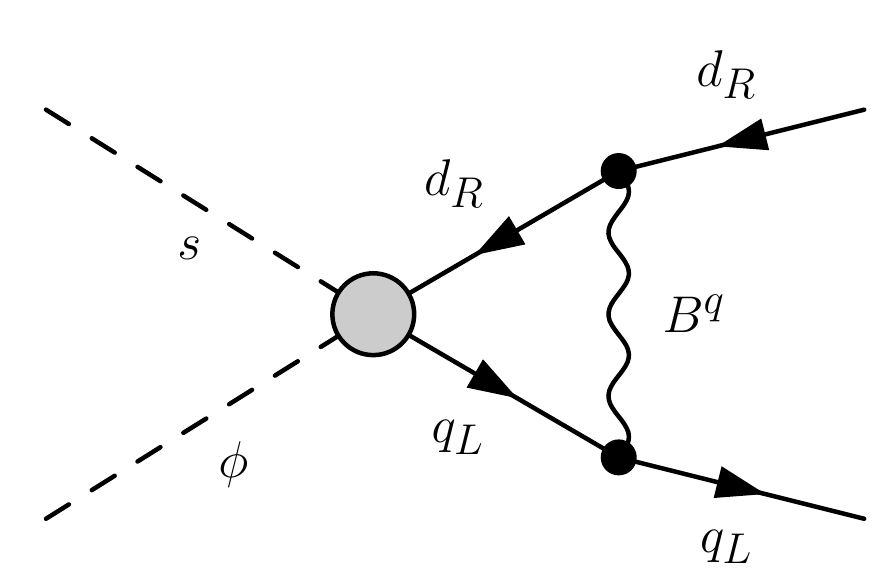}
 \includegraphics[width=0.19\columnwidth]{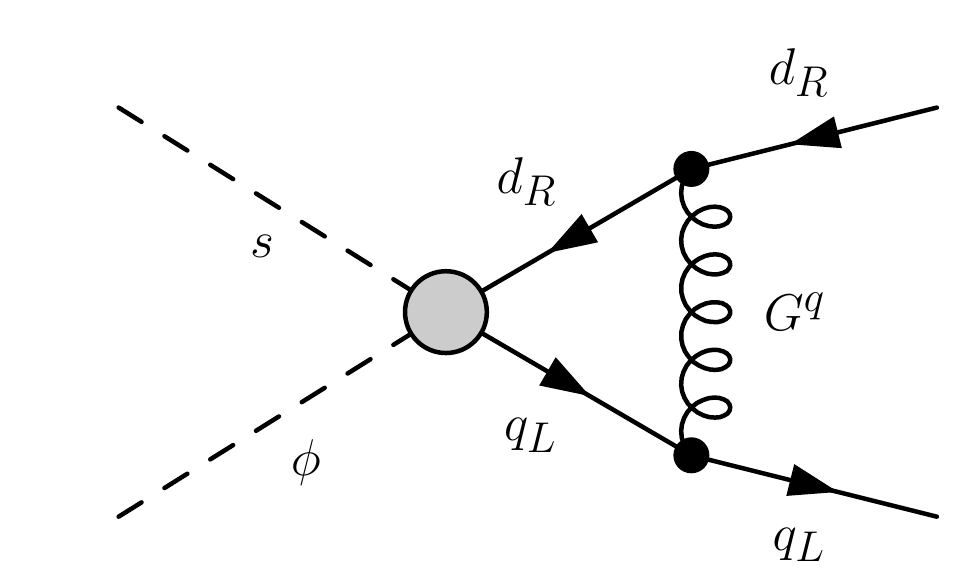}
 \includegraphics[width=0.19\columnwidth]{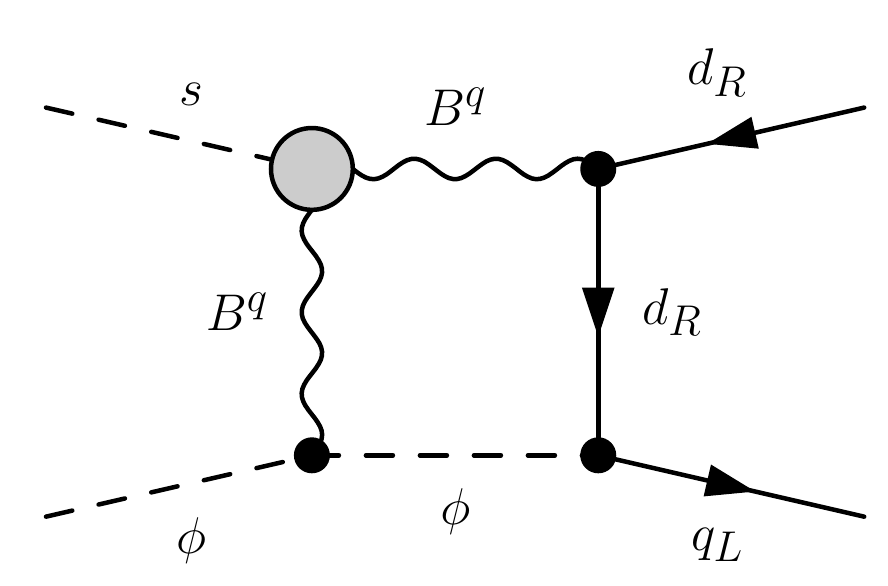}
 \includegraphics[width=0.19\columnwidth]{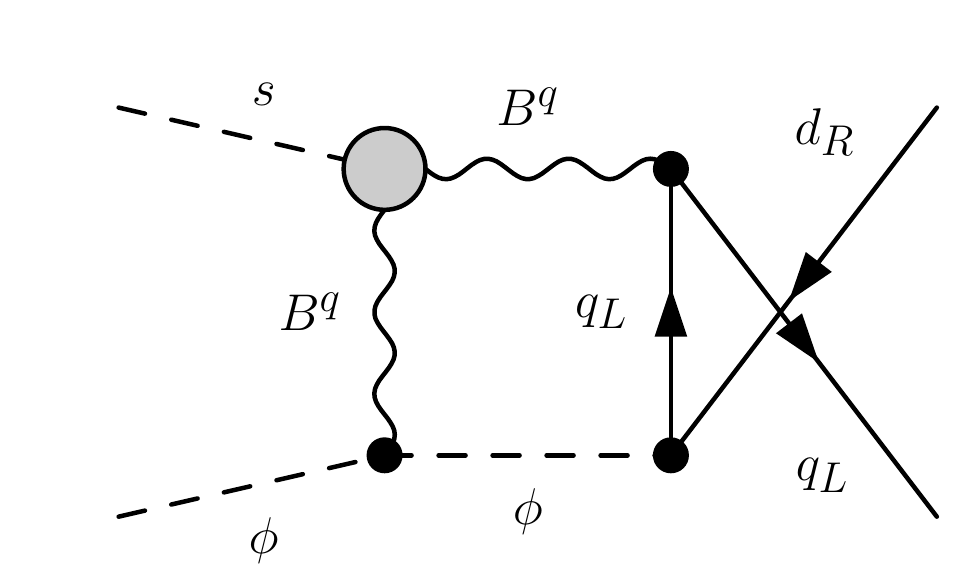}
 \includegraphics[width=0.19\columnwidth]{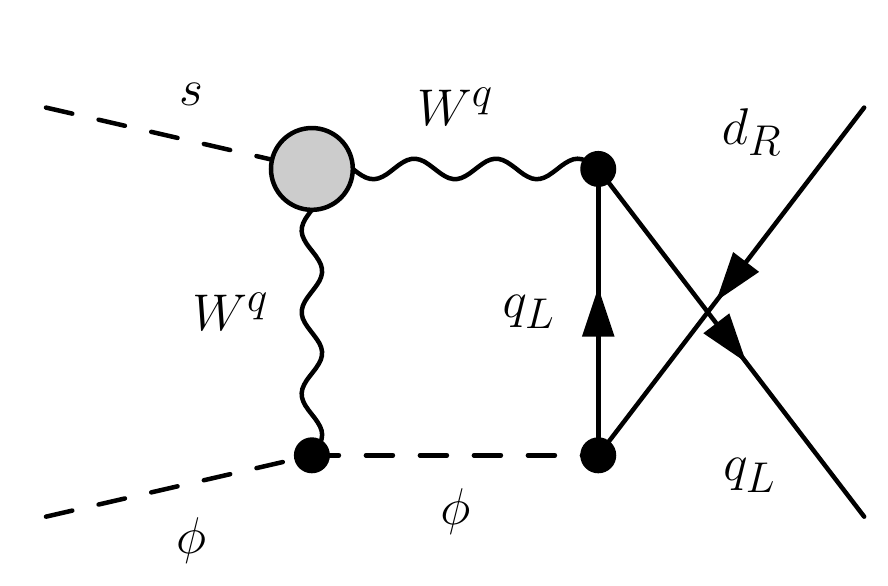}
 \includegraphics[width=0.19\columnwidth]{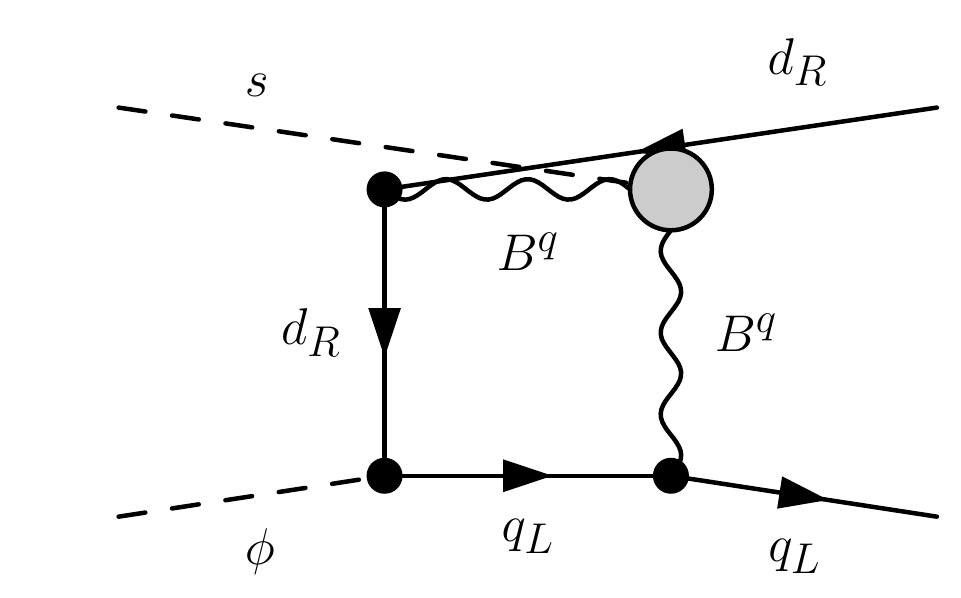}
 \includegraphics[width=0.19\columnwidth]{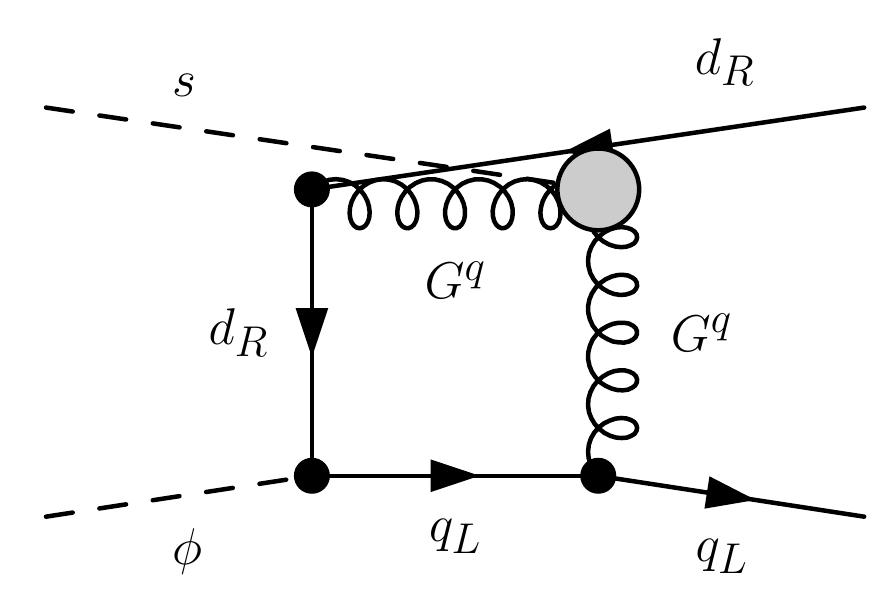}
 \includegraphics[width=0.19\columnwidth]{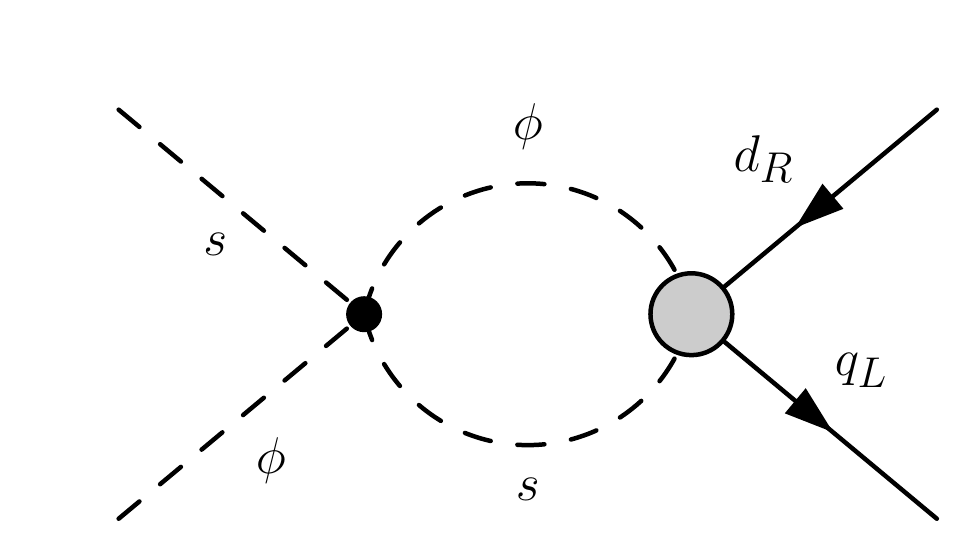}

 \caption{\it Feynman diagrams for $s(p_1)\phi(p_2)\to q_L (p_3) \overline{d_R}(p_4)$. }\label{fig:shqd}
\end{figure}

\begin{figure}[H]
 \centering
 \includegraphics[width=0.19\columnwidth]{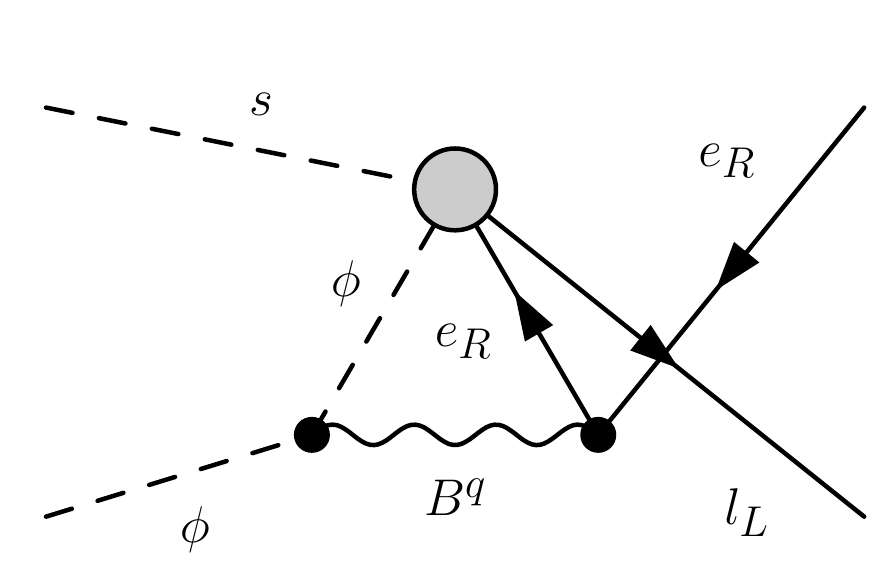}
 \includegraphics[width=0.19\columnwidth]{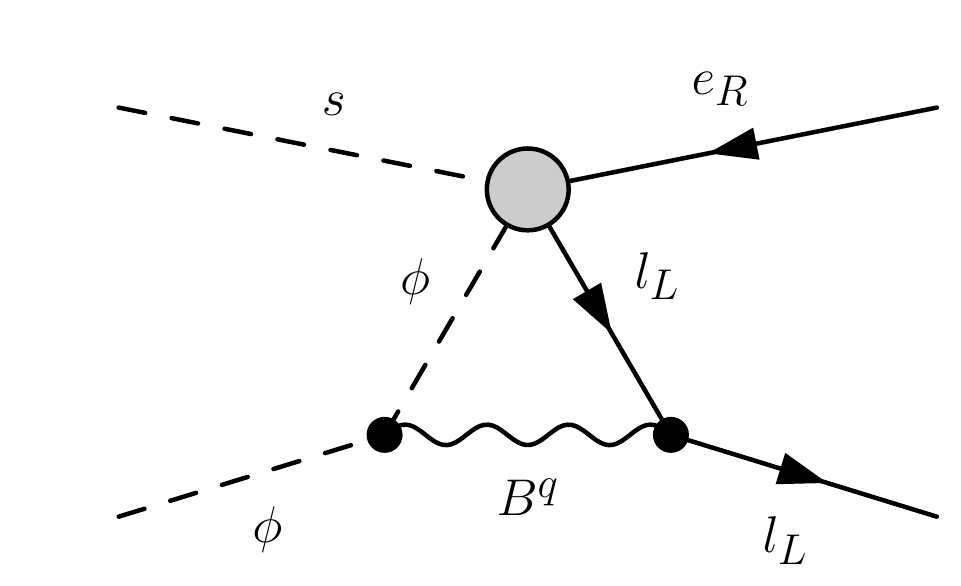}
 \includegraphics[width=0.19\columnwidth]{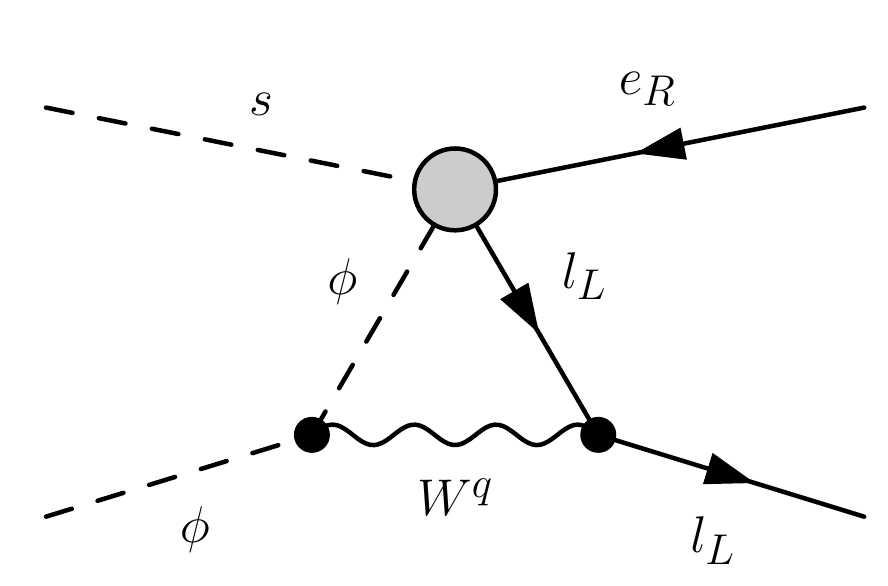}
 \includegraphics[width=0.19\columnwidth]{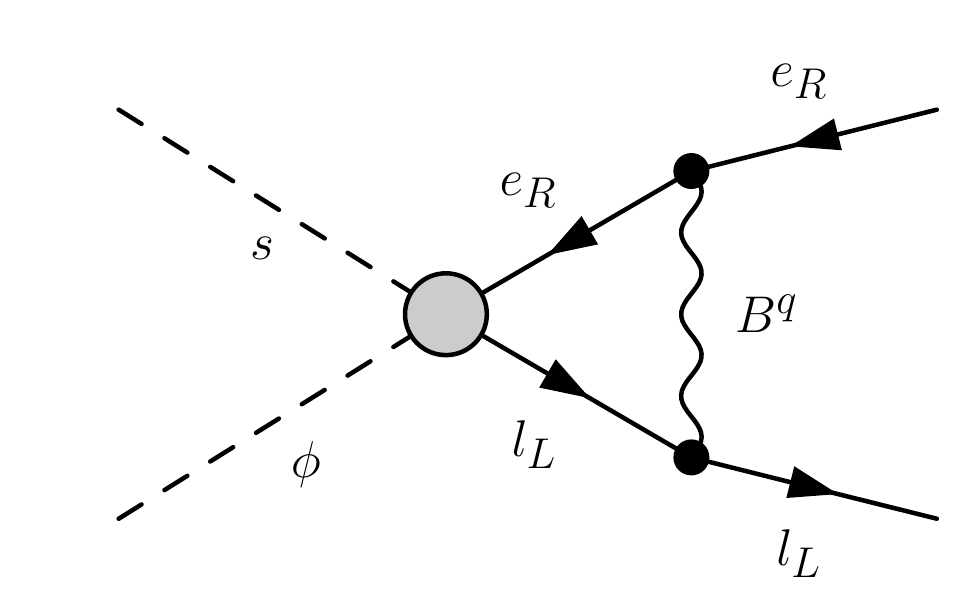}
 \includegraphics[width=0.19\columnwidth]{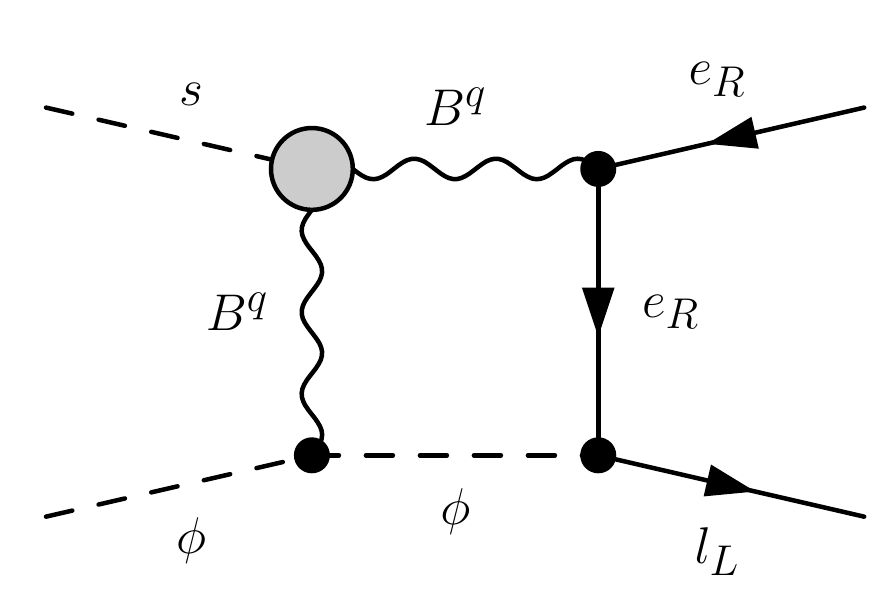}
 \includegraphics[width=0.19\columnwidth]{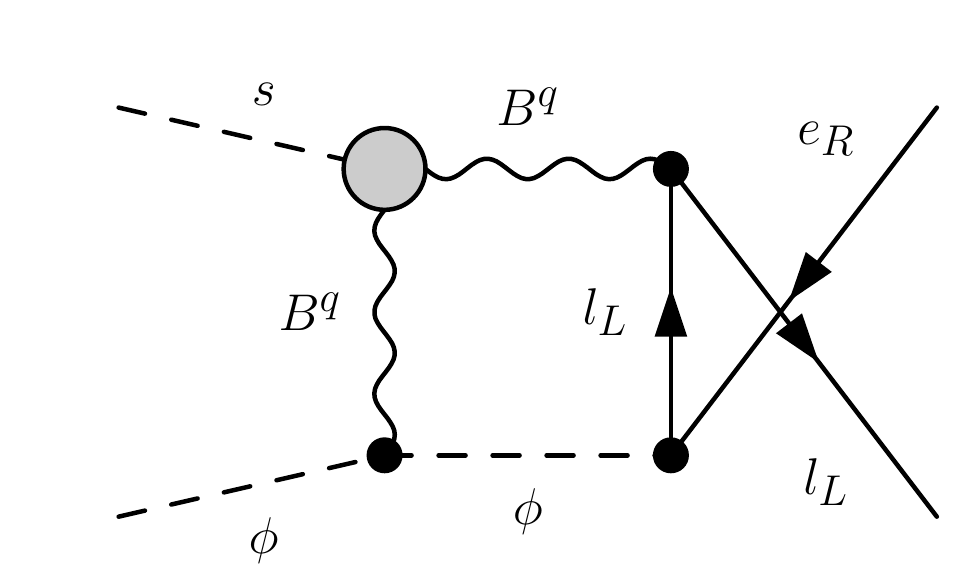}
 \includegraphics[width=0.19\columnwidth]{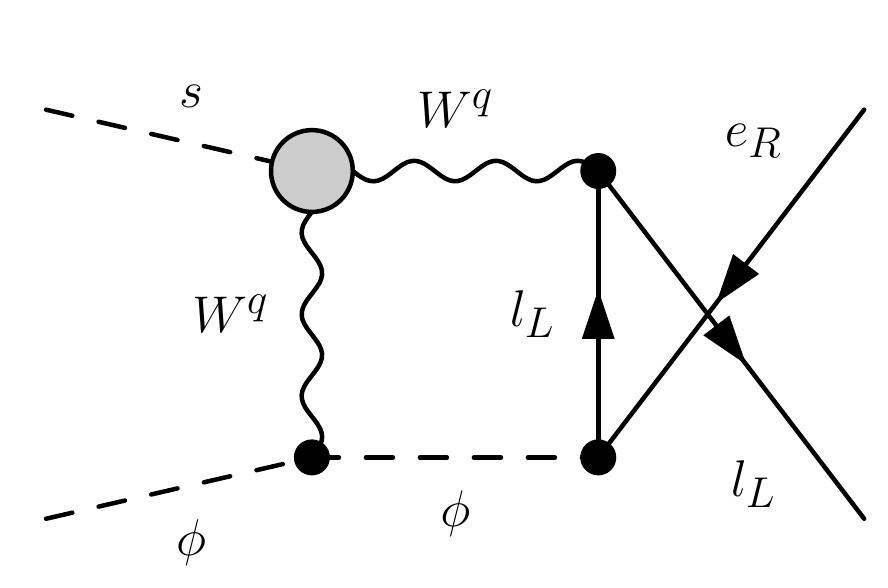}
 \includegraphics[width=0.19\columnwidth]{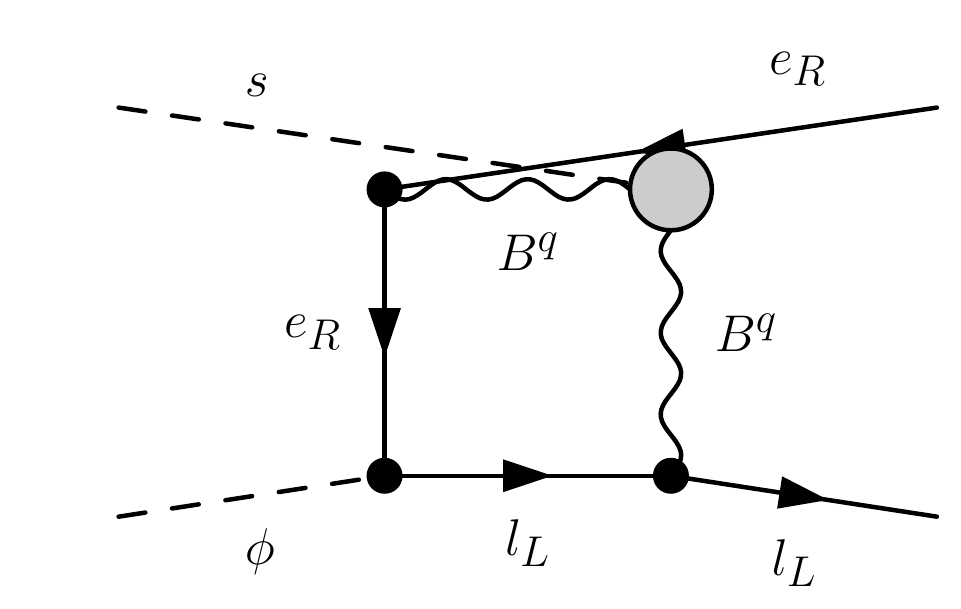}
 \includegraphics[width=0.19\columnwidth]{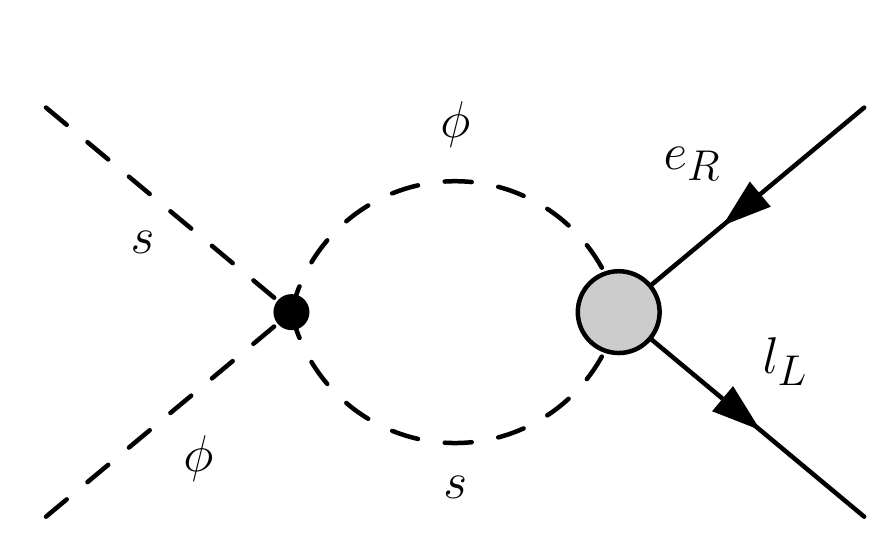}
 \caption{\it Feynman diagrams for $s(p_1)\phi(p_2)\to l_L (p_3) \overline{e_R}(p_4)$.}\label{fig:shle}
\end{figure}

\begin{figure}[H]
 \centering
 \includegraphics[width=0.2\columnwidth]{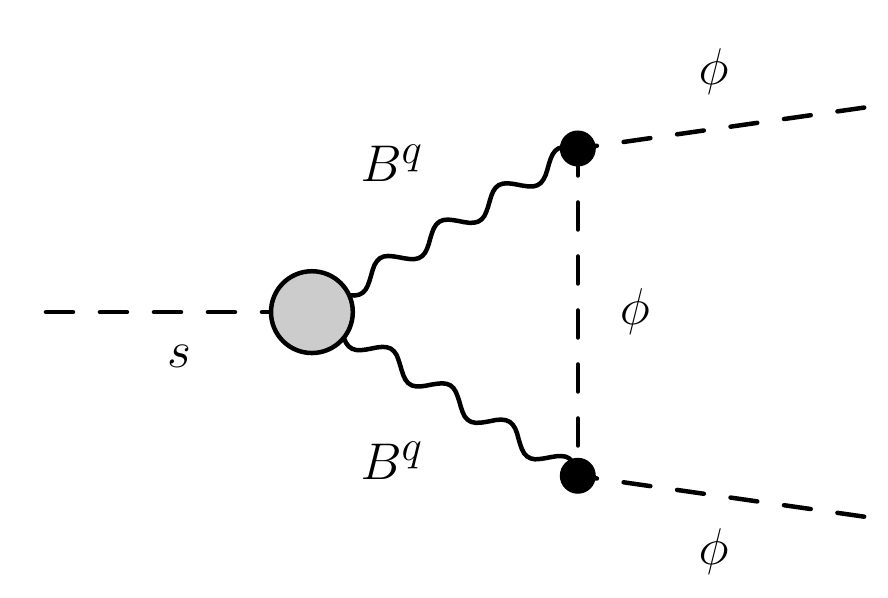}
 \includegraphics[width=0.2\columnwidth]{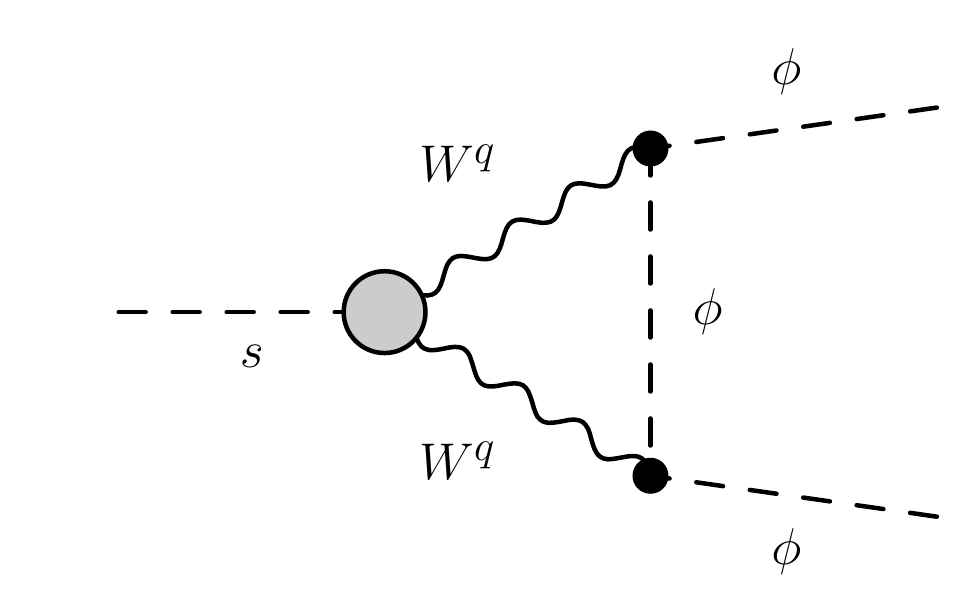}
 \includegraphics[width=0.2\columnwidth]{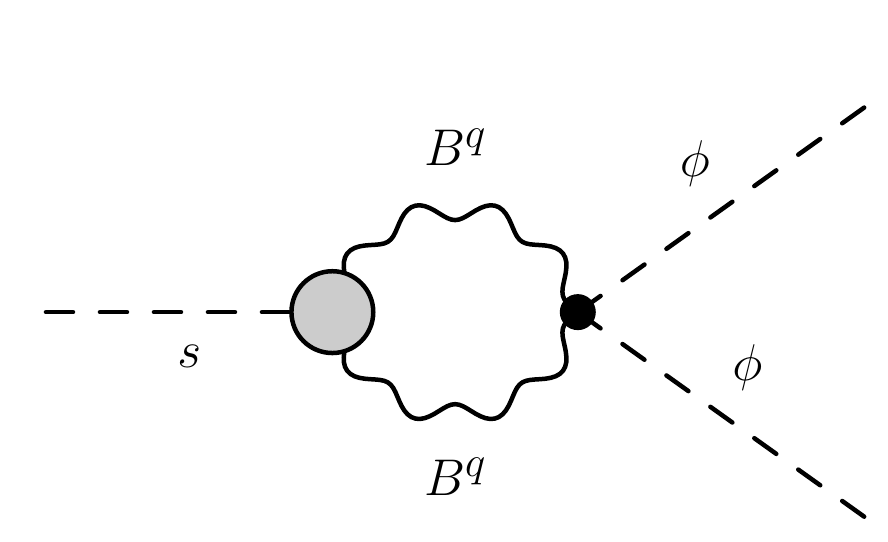}
 \includegraphics[width=0.2\columnwidth]{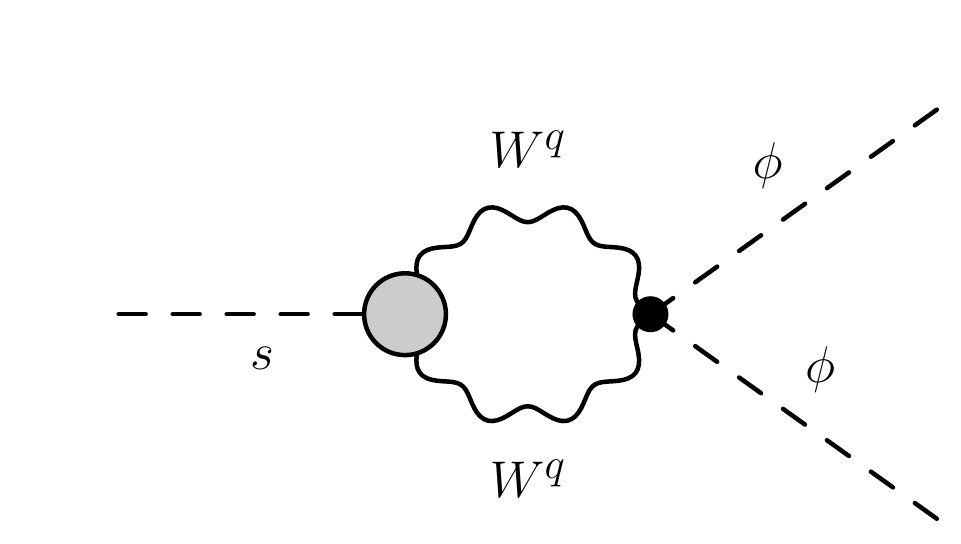}
 \includegraphics[width=0.2\columnwidth]{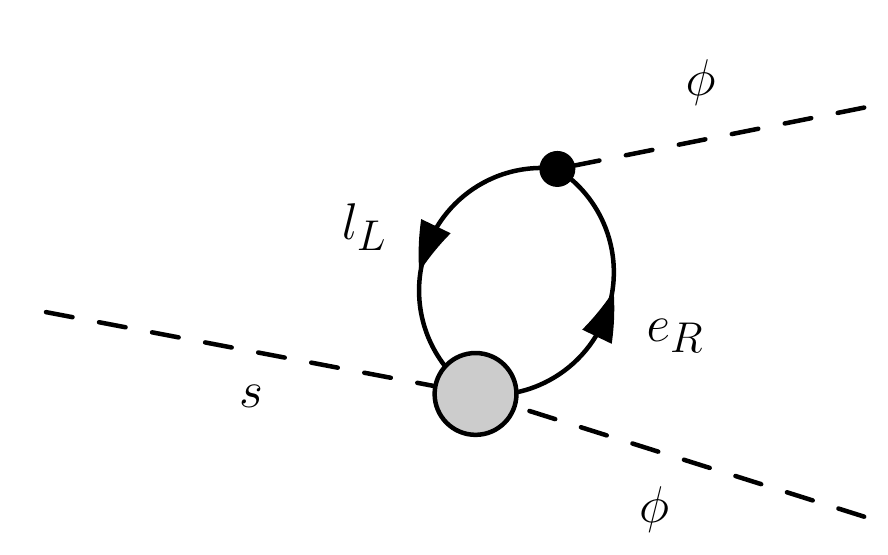}
 \includegraphics[width=0.2\columnwidth]{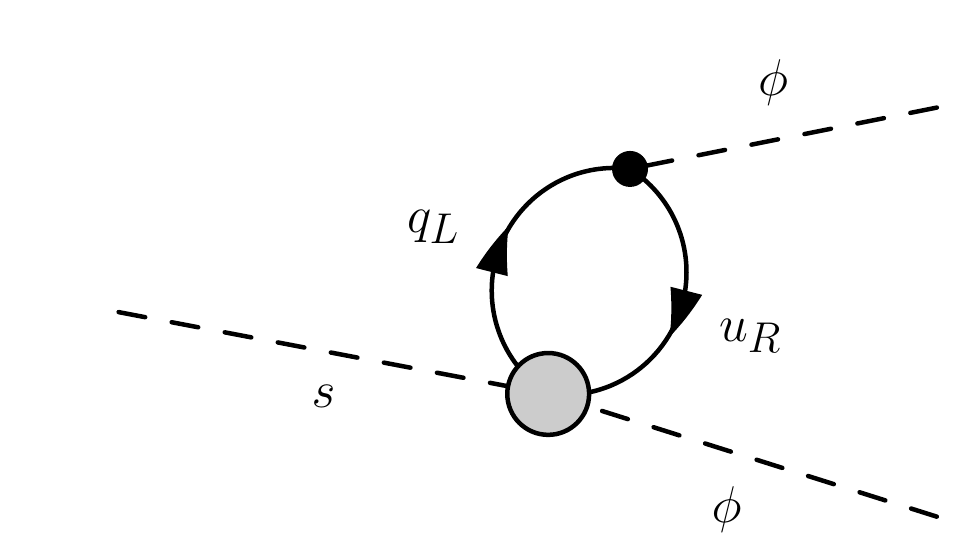}
 \includegraphics[width=0.2\columnwidth]{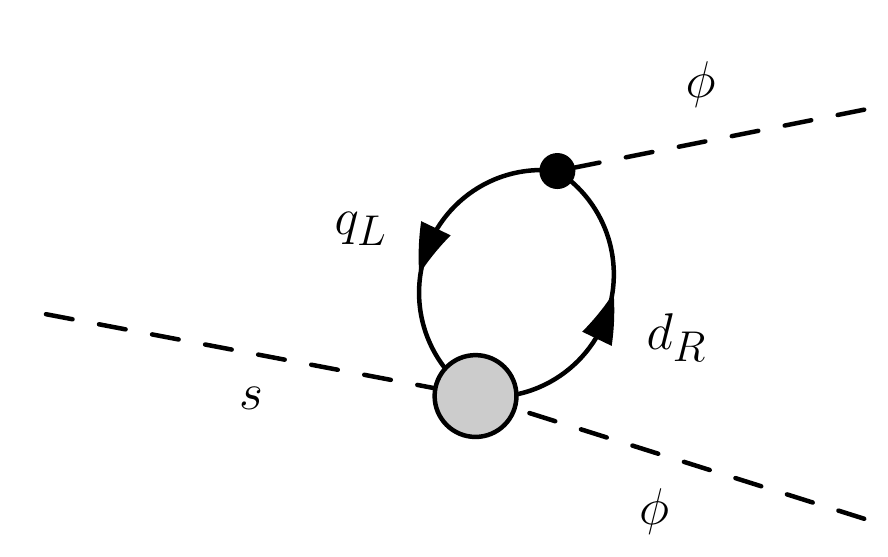}
 \includegraphics[width=0.2\columnwidth]{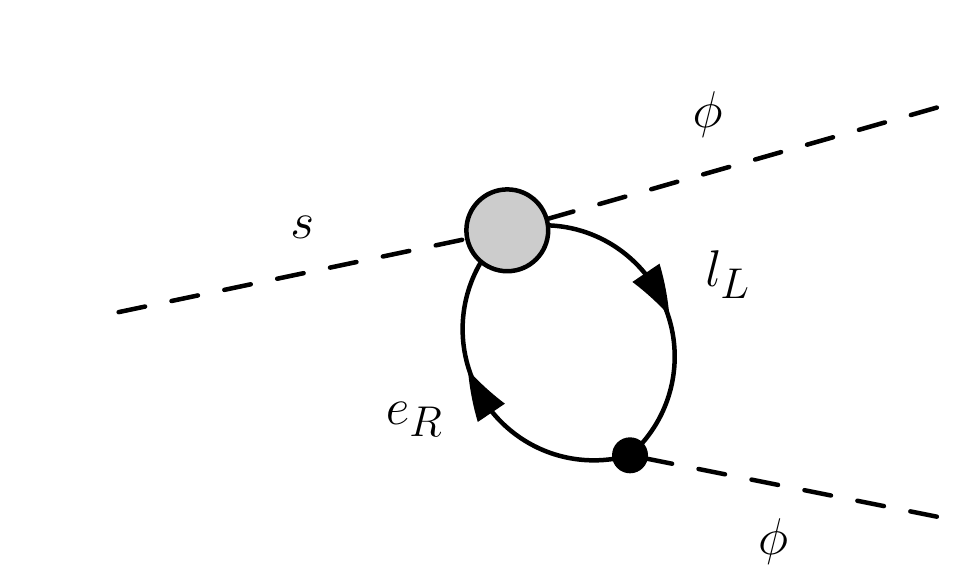}
 \includegraphics[width=0.2\columnwidth]{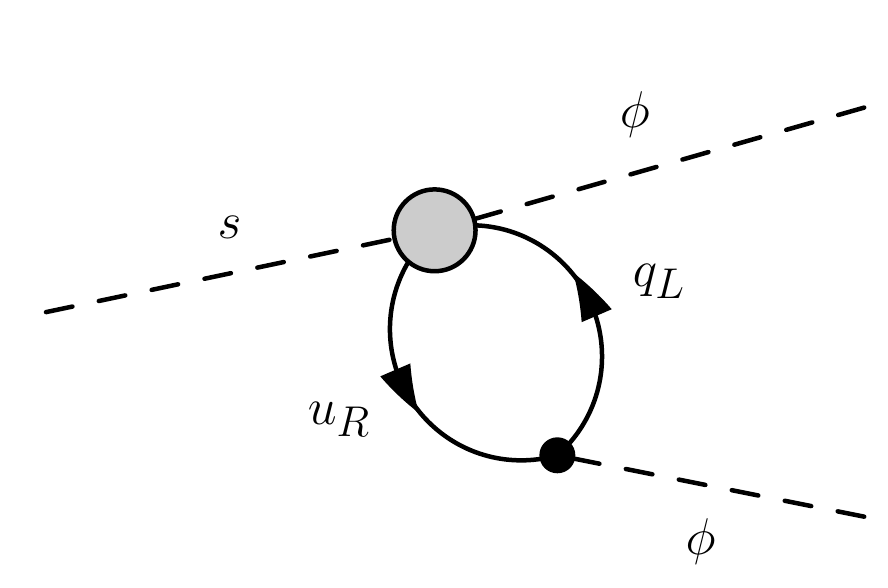}
 \includegraphics[width=0.2\columnwidth]{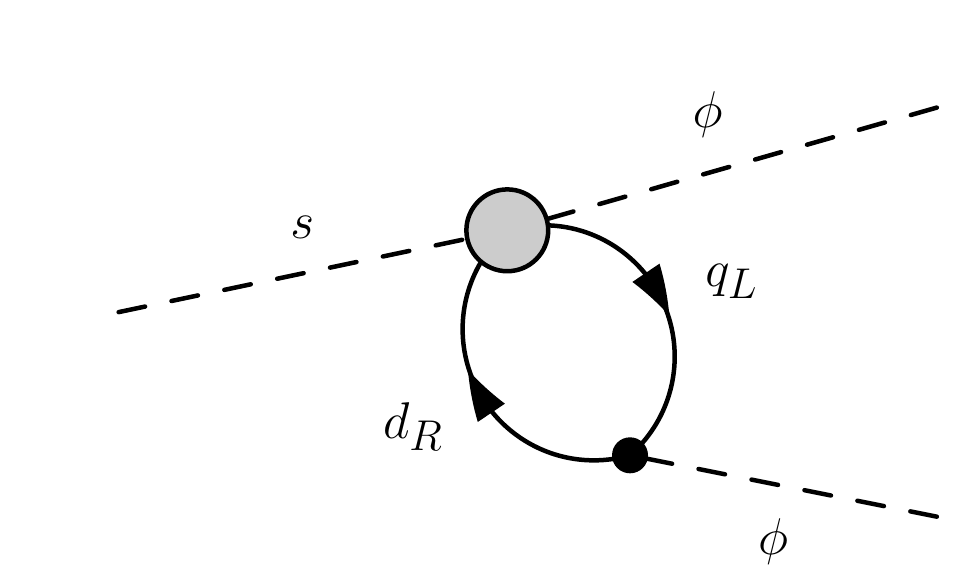}
 \caption{\it Feynman diagrams for $s(p_1)\to\phi(p_2)\phi^\dagger(p_3)$. }\label{fig:shh}
\end{figure}

\begin{figure}[H]
 \centering
 \includegraphics[width=0.2\columnwidth]{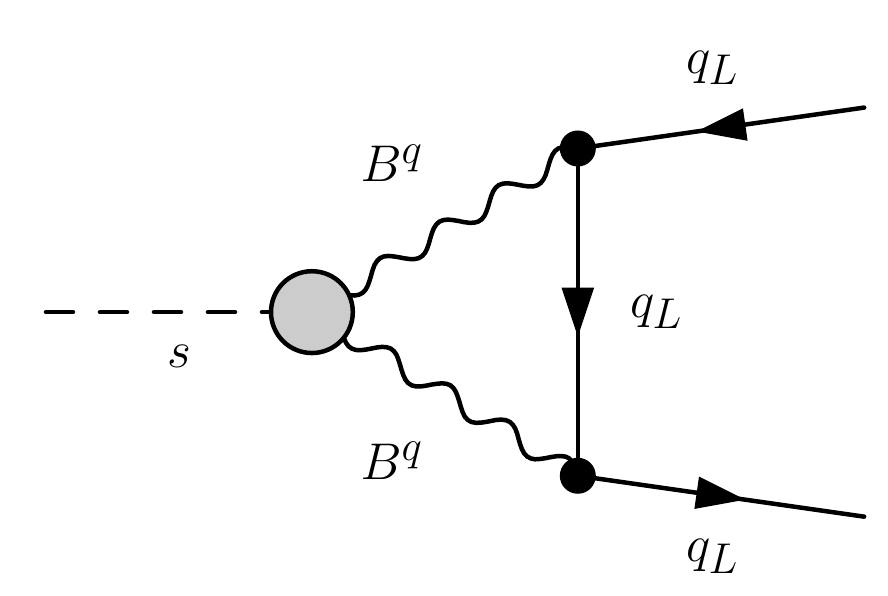}
 \includegraphics[width=0.2\columnwidth]{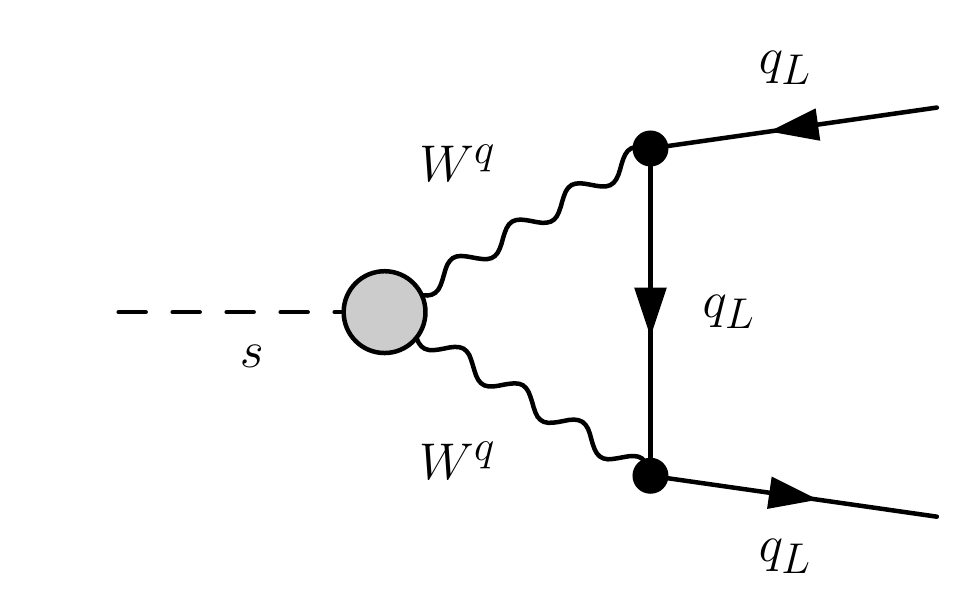}
 \includegraphics[width=0.2\columnwidth]{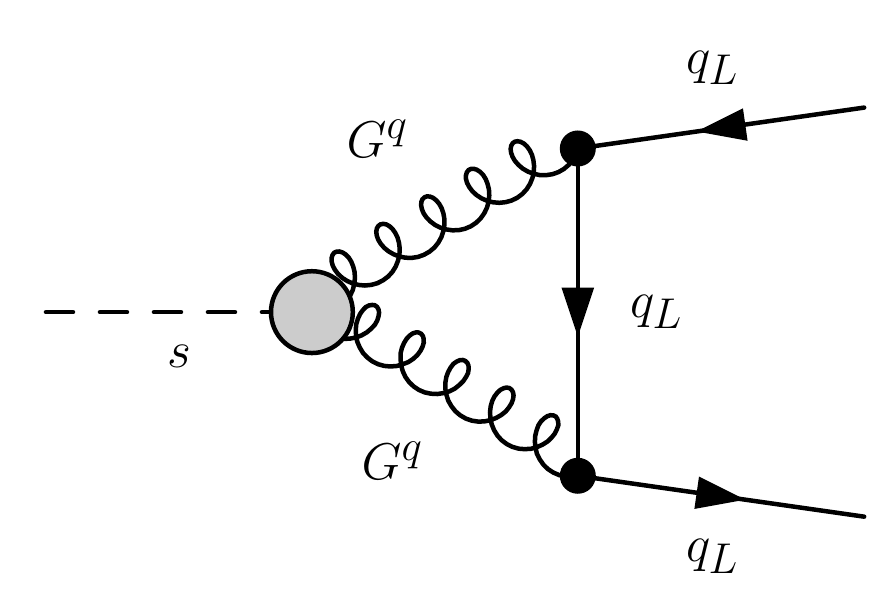}
 \includegraphics[width=0.2\columnwidth]{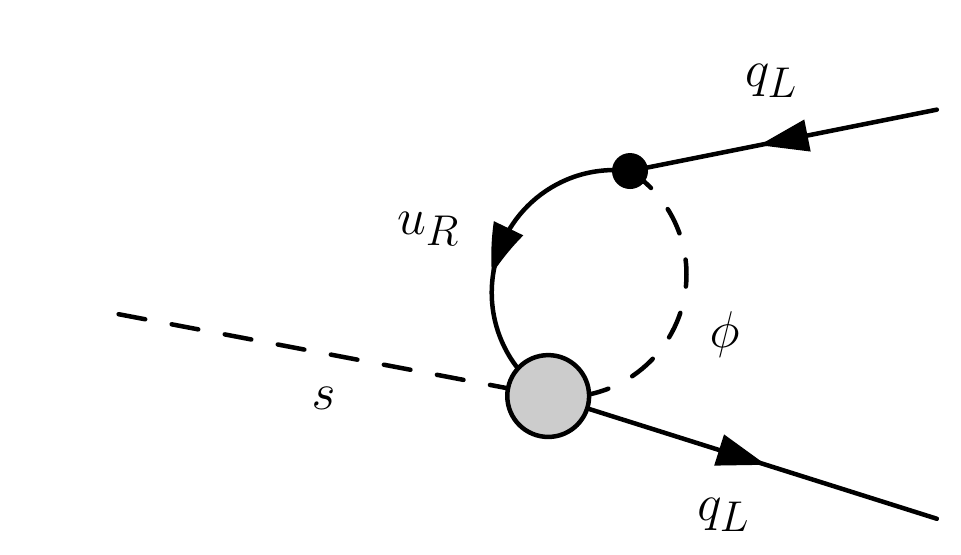}
 \includegraphics[width=0.2\columnwidth]{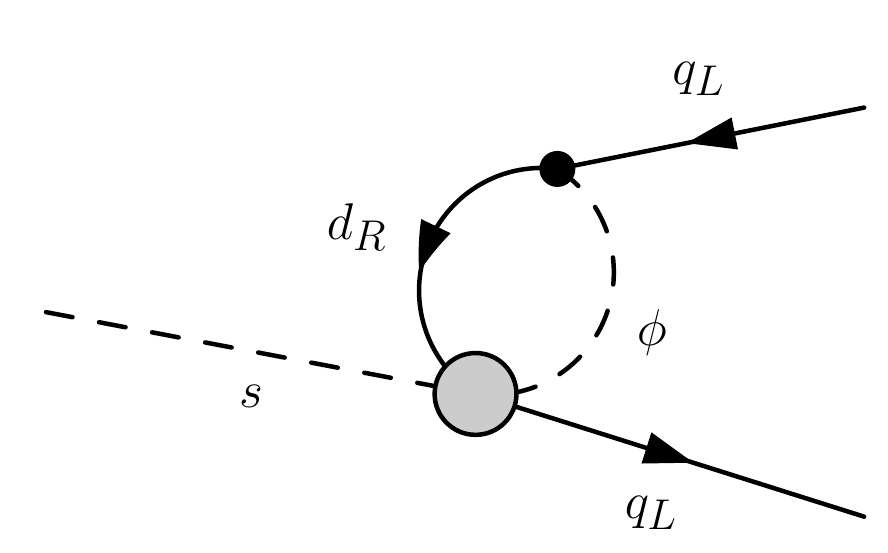}
 \includegraphics[width=0.2\columnwidth]{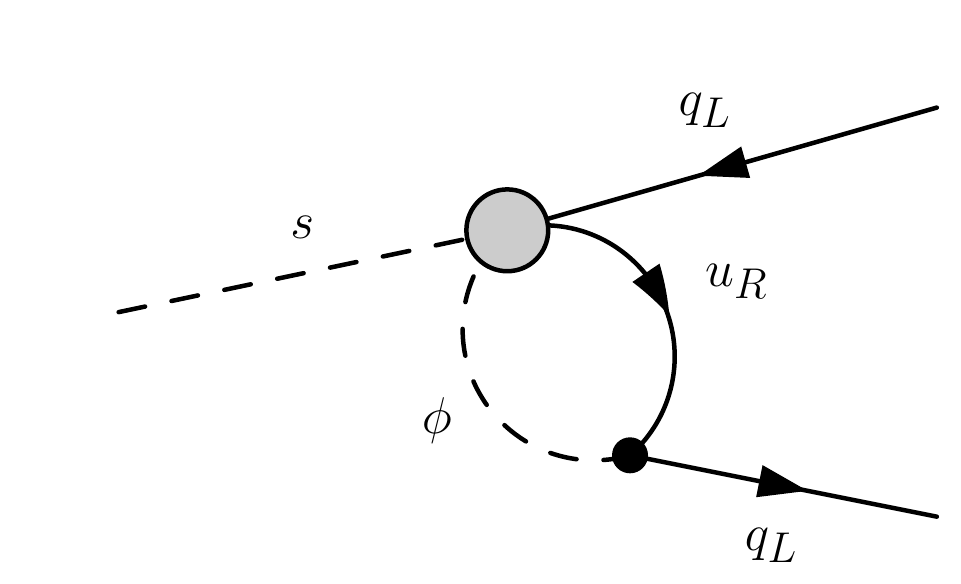}
 \includegraphics[width=0.2\columnwidth]{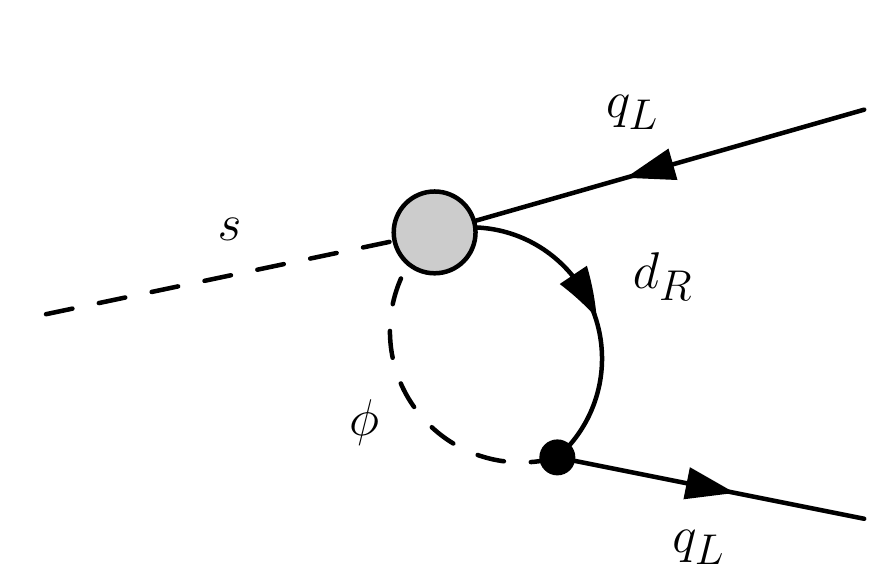}
 \caption{\it Feynman diagrams for $s(p_1)\to q_L(p_2) \overline{q_L}(p_3)$.}\label{fig:sqq}
\end{figure}

\begin{figure}[H]
 \centering
 \includegraphics[width=0.2\columnwidth]{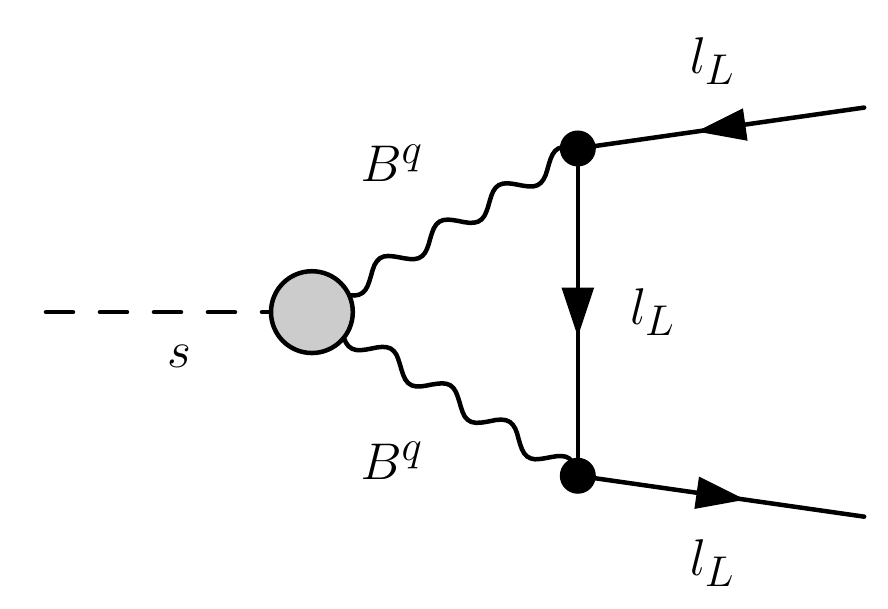}
 \includegraphics[width=0.2\columnwidth]{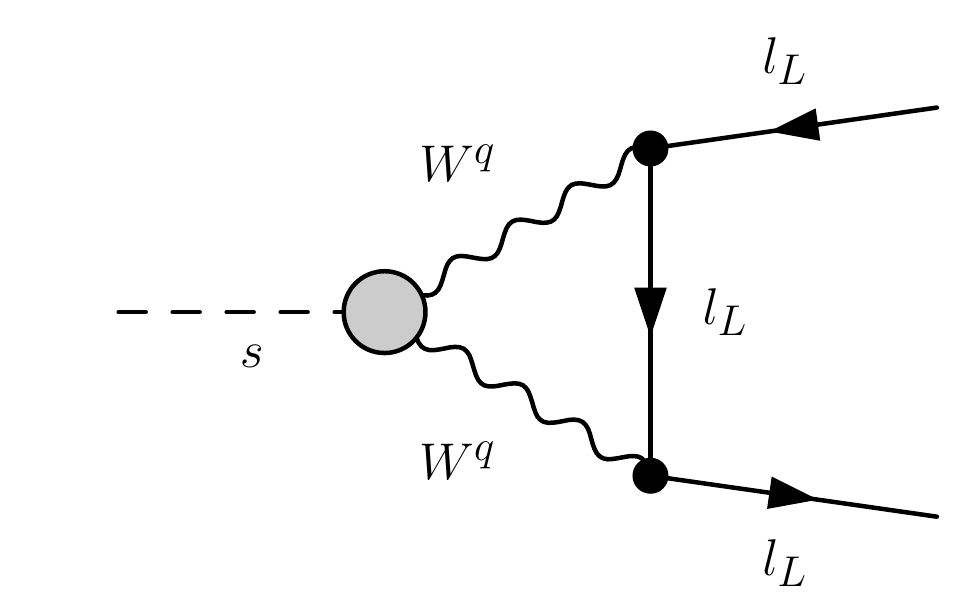}
 \includegraphics[width=0.2\columnwidth]{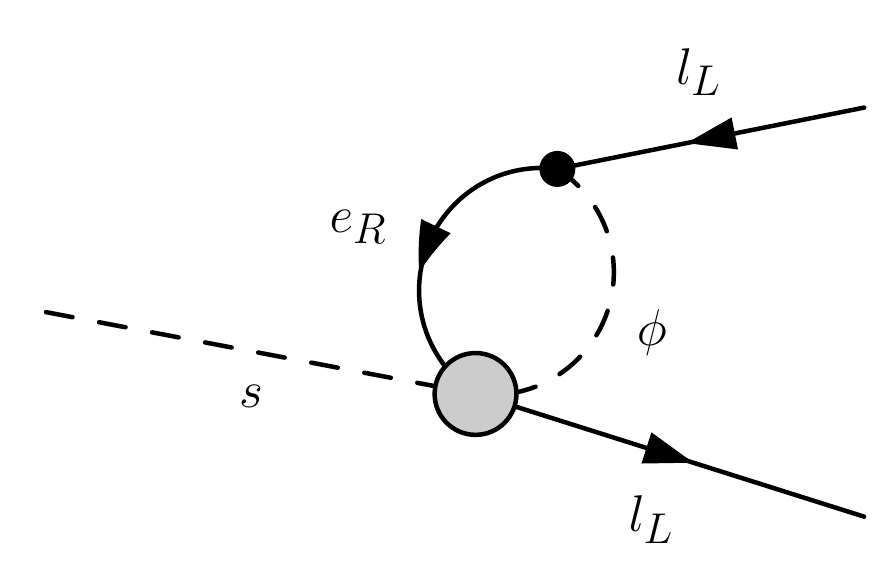}
 \includegraphics[width=0.2\columnwidth]{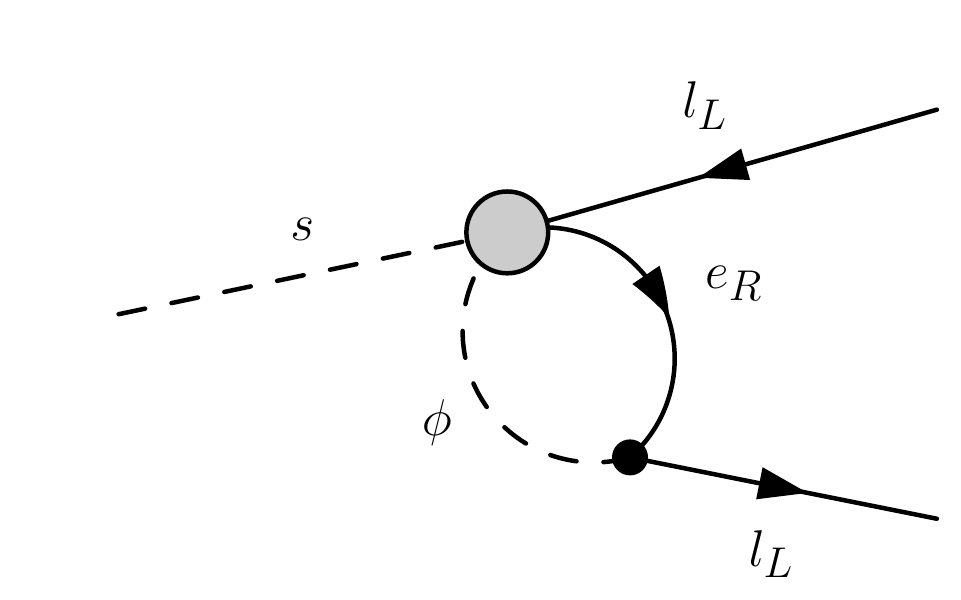}
 \caption{\it Feynman diagrams for $s(p_1)\to l_L(p_2) \overline{l_L}(p_3)$.}\label{fig:sll}
\end{figure}

\begin{figure}[H]
 \centering
 \includegraphics[width=0.2\columnwidth]{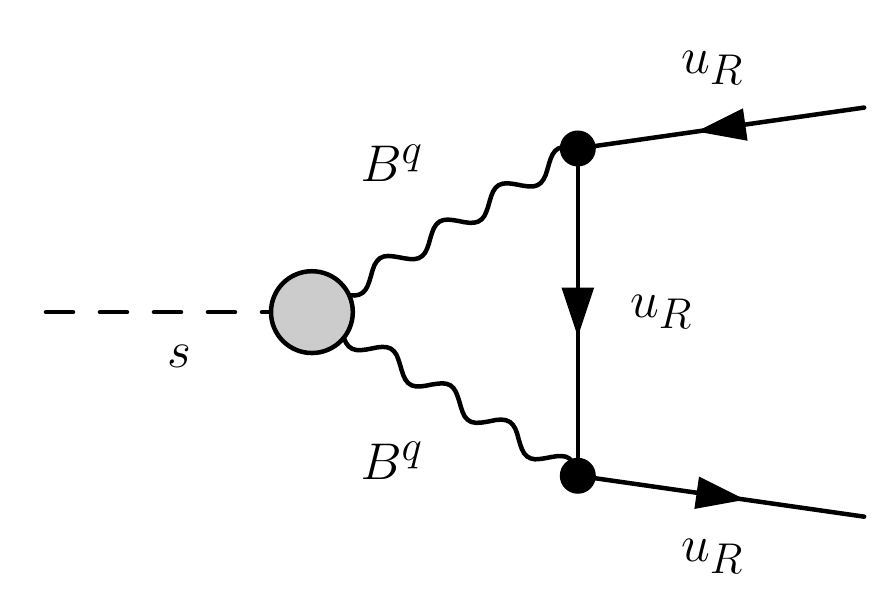}
 \includegraphics[width=0.2\columnwidth]{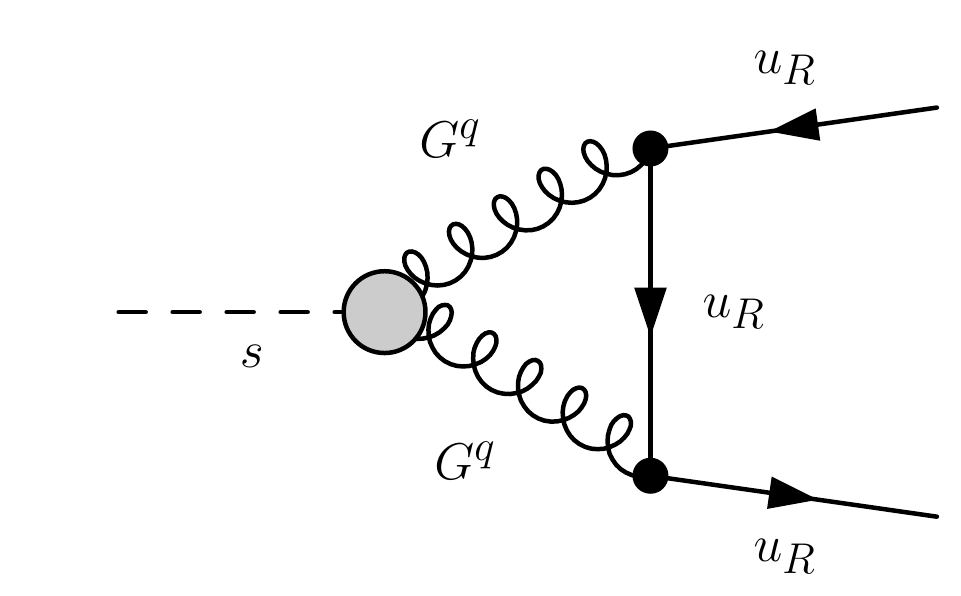}
 \includegraphics[width=0.2\columnwidth]{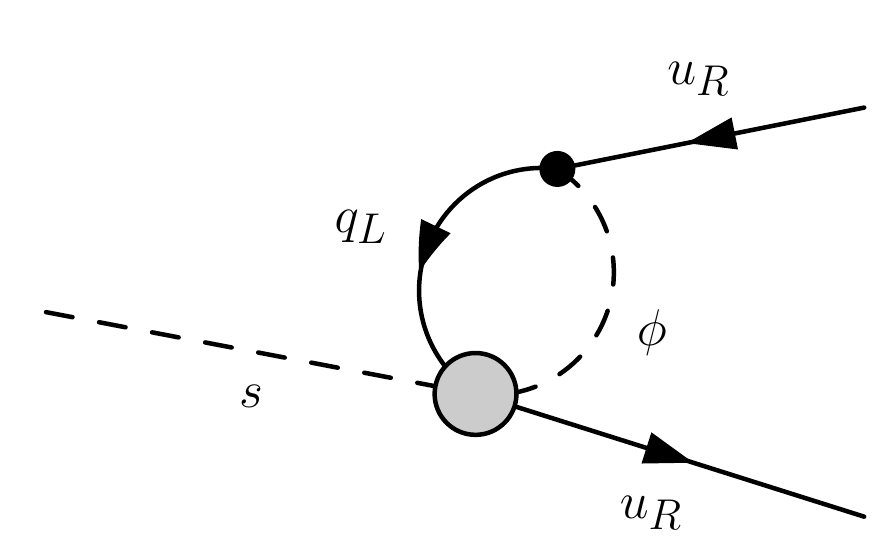}
 \includegraphics[width=0.2\columnwidth]{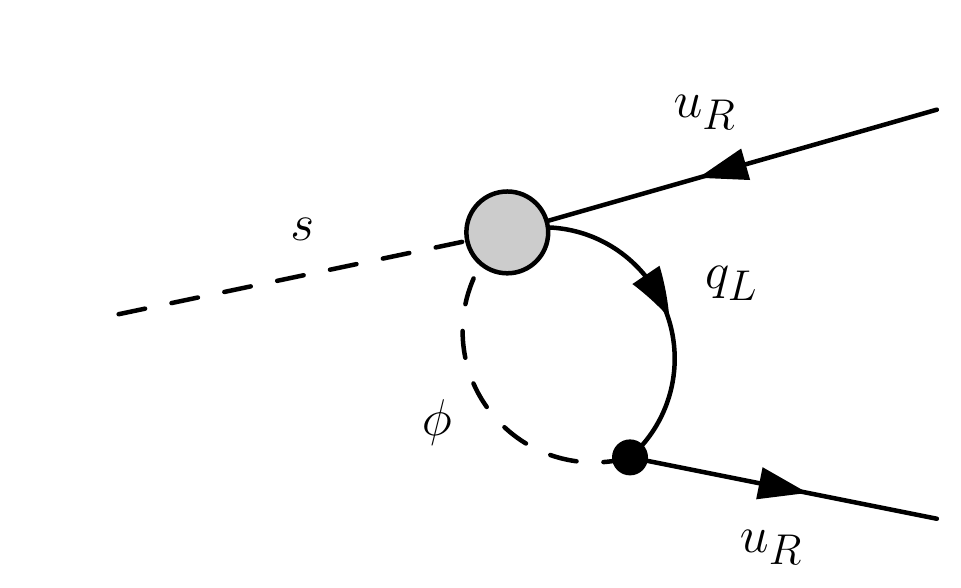}
 \caption{\it Feynman diagrams for $s(p_1)\to u_R(p_2) \overline{u_R}(p_3)$.}\label{fig:suu}
\end{figure}

\begin{figure}[H]
 \centering
 \includegraphics[width=0.2\columnwidth]{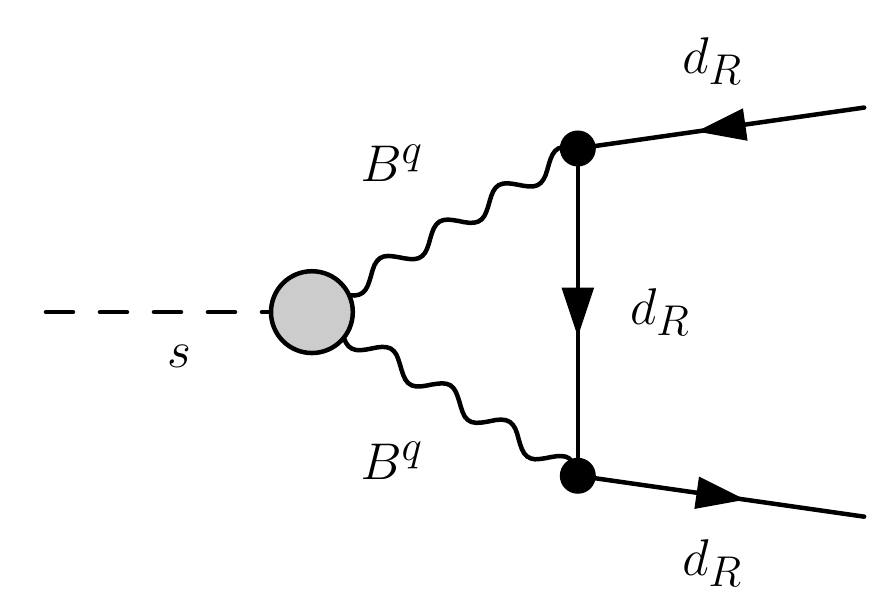}
 \includegraphics[width=0.2\columnwidth]{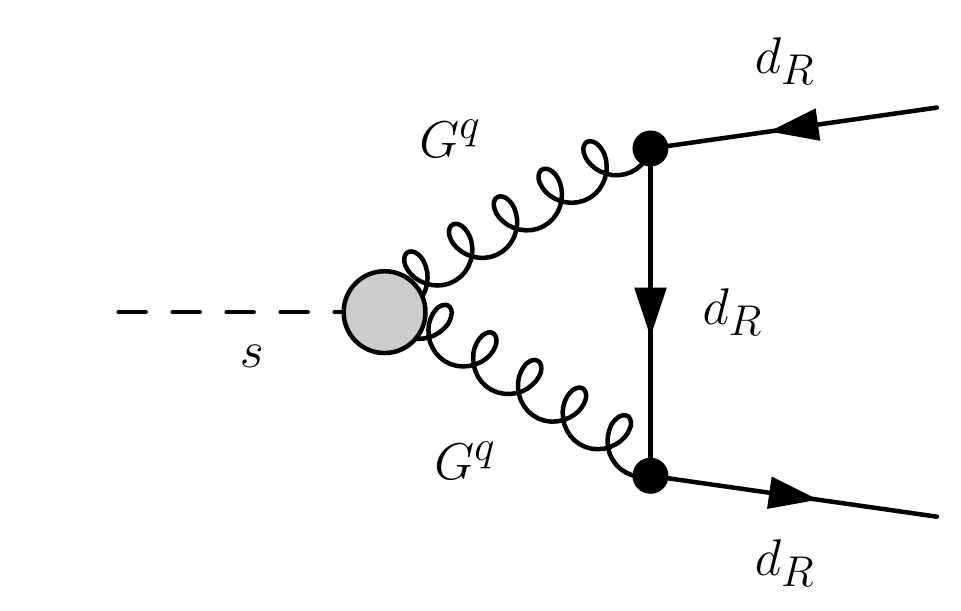}
 \includegraphics[width=0.2\columnwidth]{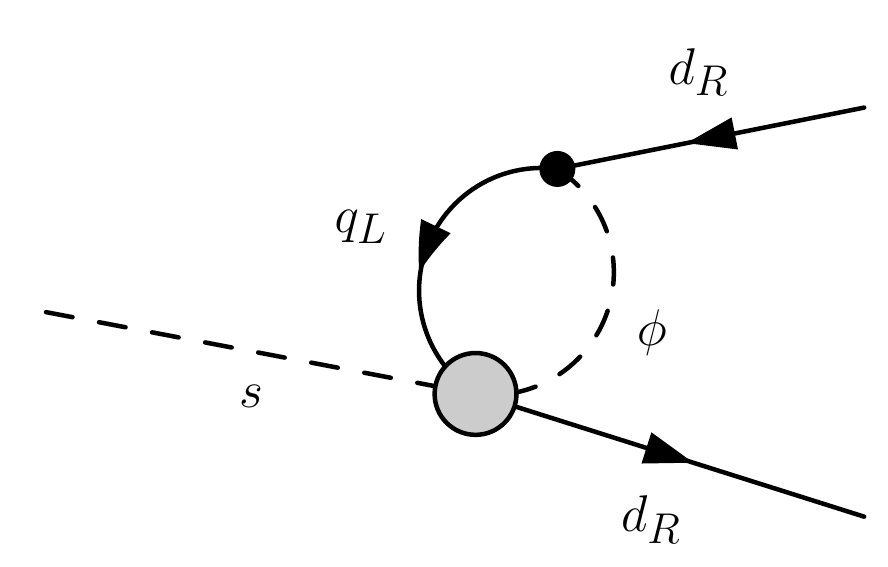}
 \includegraphics[width=0.2\columnwidth]{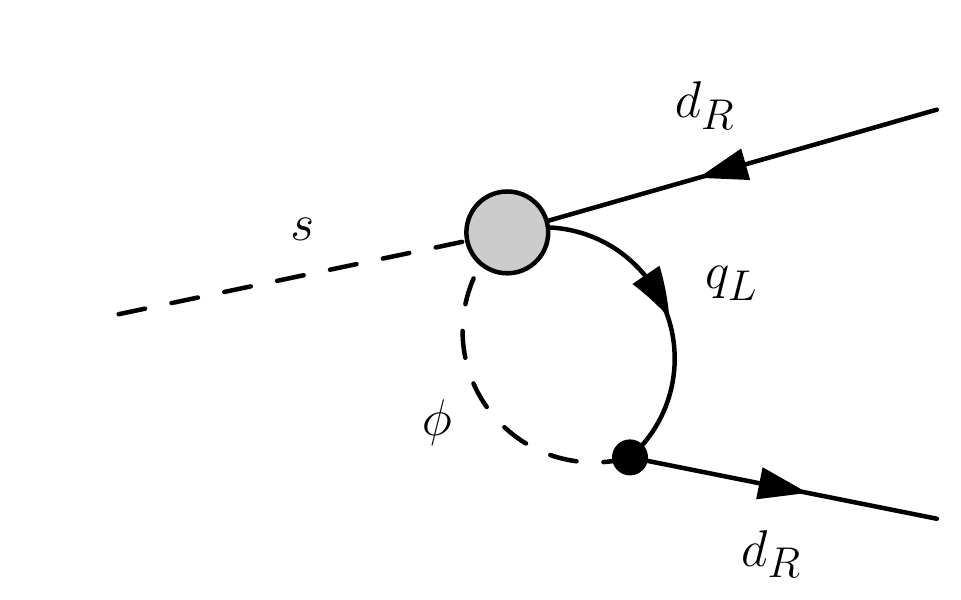}
 \caption{\it Feynman diagrams for $s(p_1)\to d_R(p_2) \overline{d_R}(p_3)$.}\label{fig:sdd}
\end{figure}

\begin{figure}[H]
 \centering
 \includegraphics[width=0.2\columnwidth]{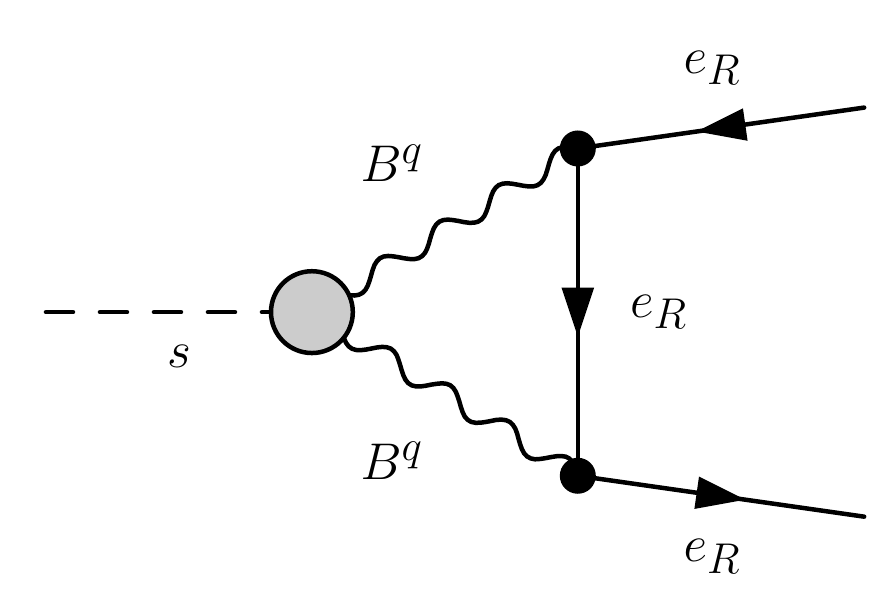}
 \includegraphics[width=0.2\columnwidth]{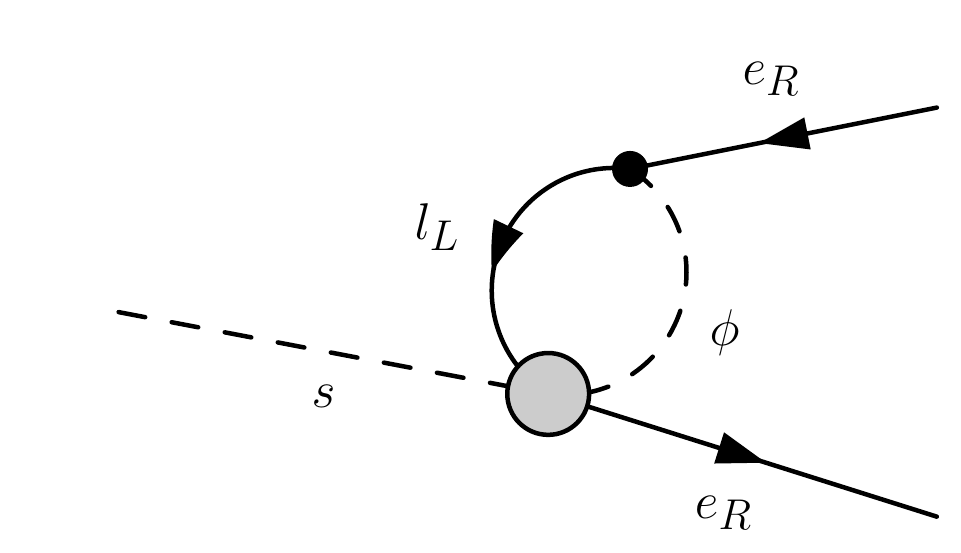}
 \includegraphics[width=0.2\columnwidth]{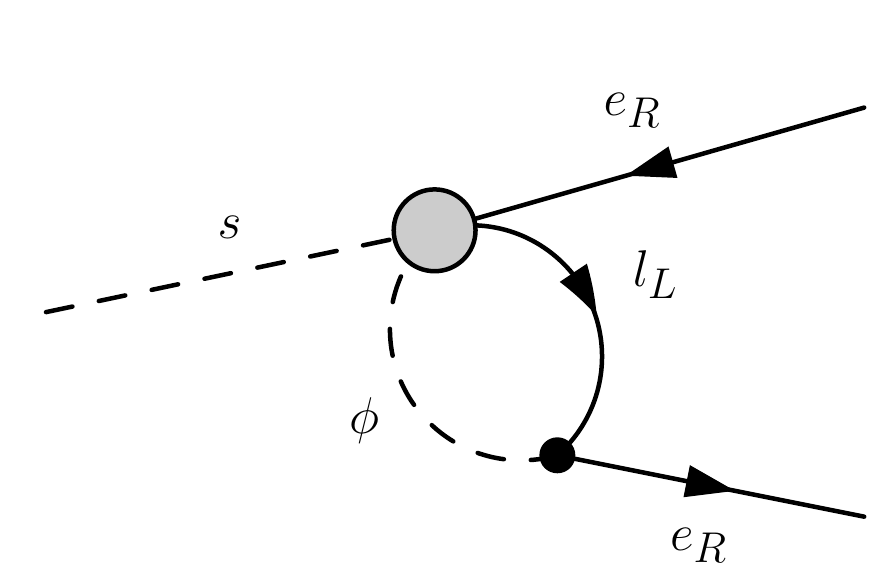}
 \caption{\it Feynman diagrams for $s(p_1)\to e_R(p_2) \overline{e_R}(p_3)$.}\label{fig:see}
\end{figure}

\begin{figure}[H]
 \centering
 \includegraphics[width=0.2\columnwidth]{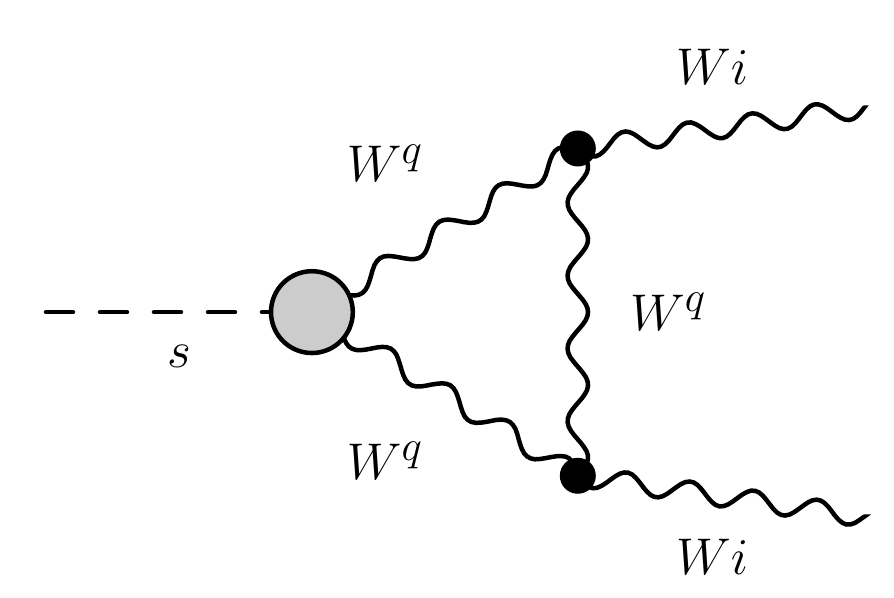}
 \includegraphics[width=0.2\columnwidth]{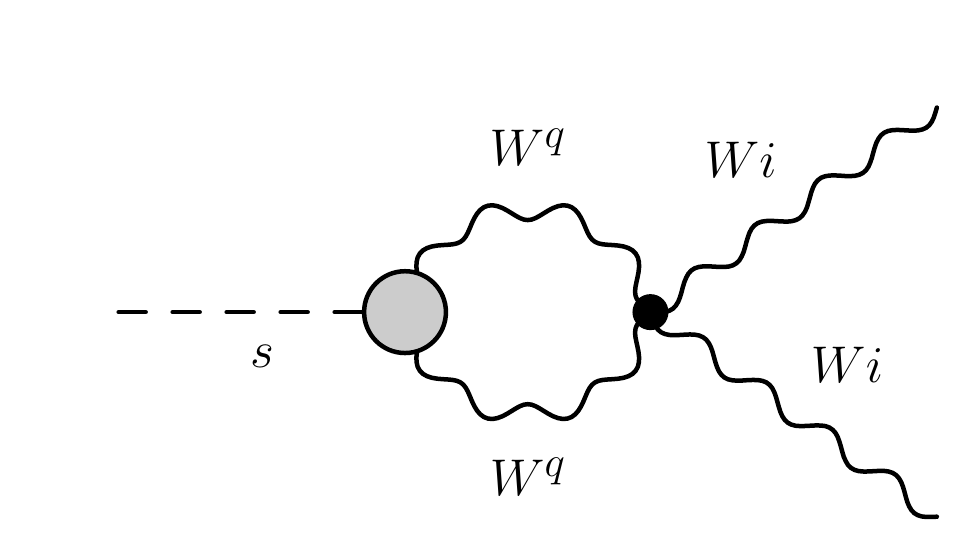}
 \includegraphics[width=0.2\columnwidth]{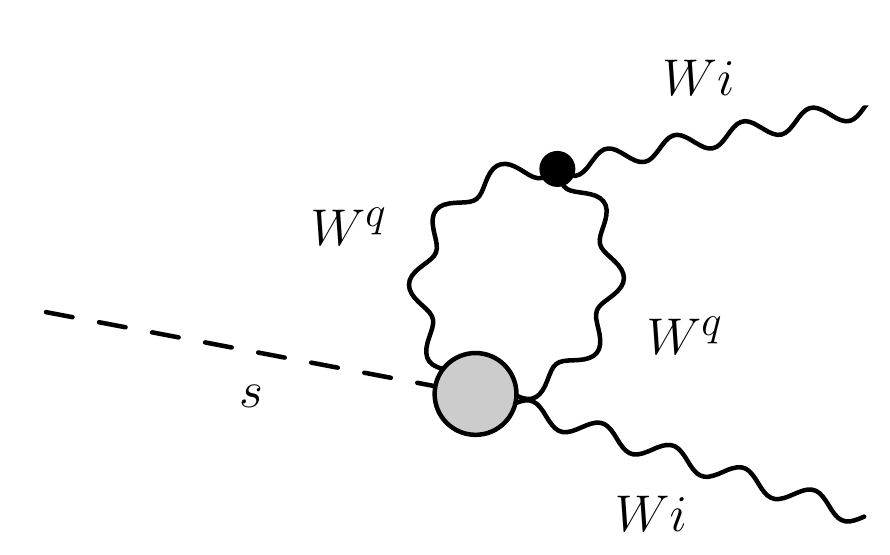}
 \includegraphics[width=0.2\columnwidth]{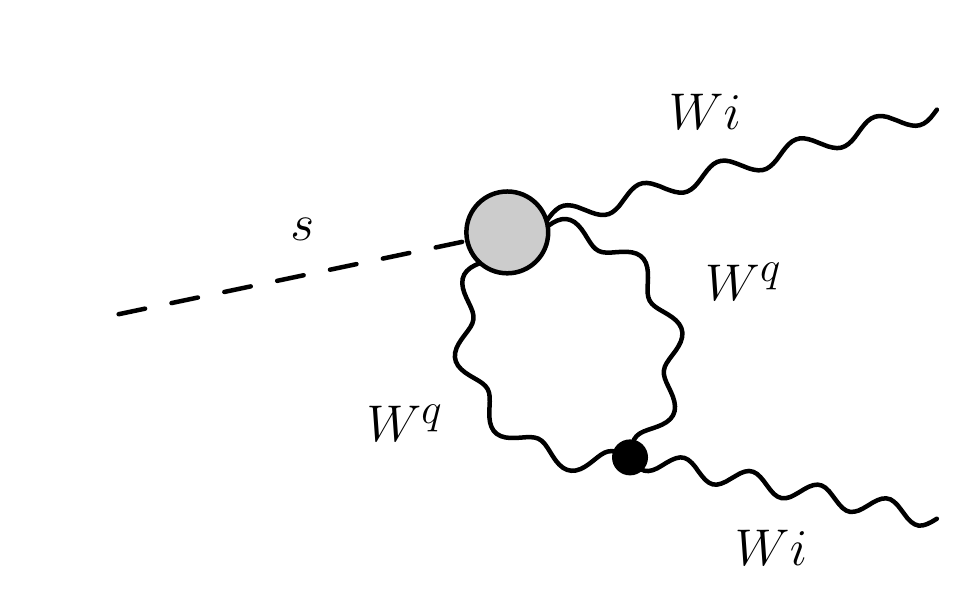}
 \caption{\it Feynman diagrams for $s(p_1)\to W^3 (p_2) W^3 (p_3)$.}\label{fig:sWW}
\end{figure}

\begin{figure}[H]
 \centering
 \includegraphics[width=0.2\columnwidth]{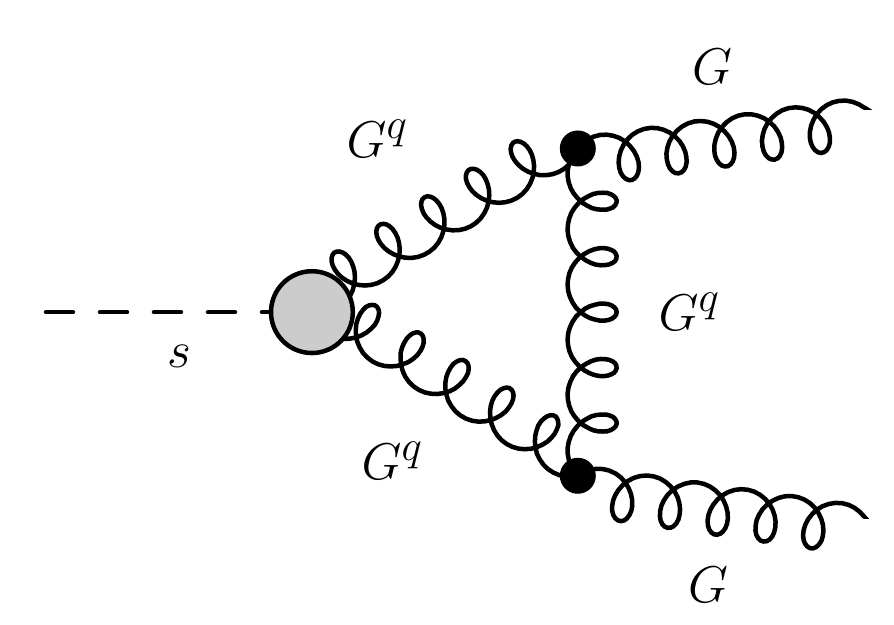}
 \includegraphics[width=0.2\columnwidth]{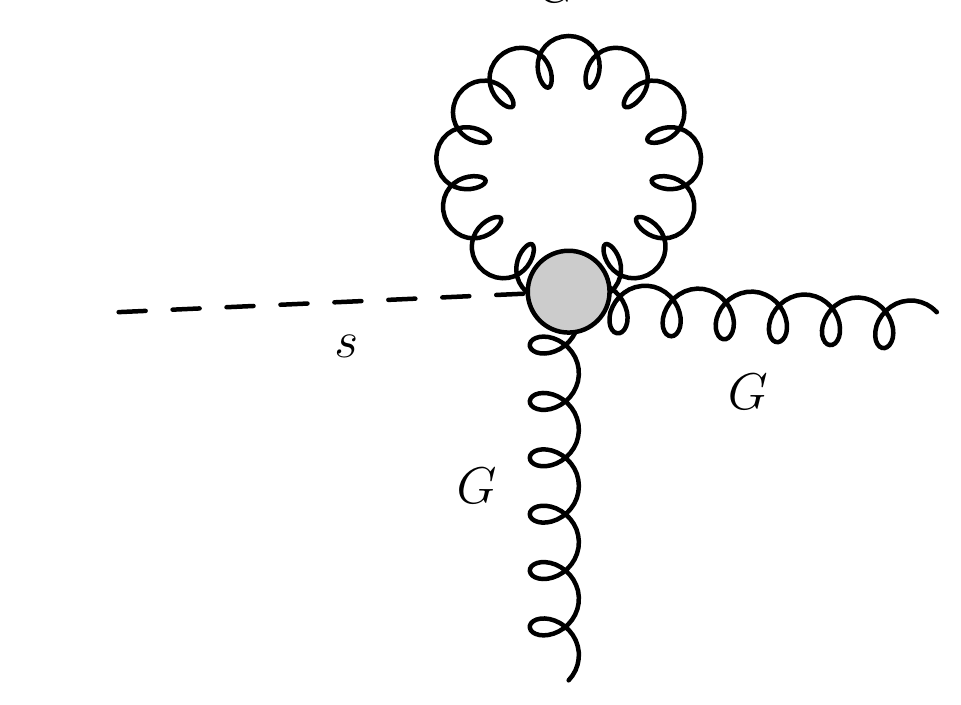}
 \includegraphics[width=0.2\columnwidth]{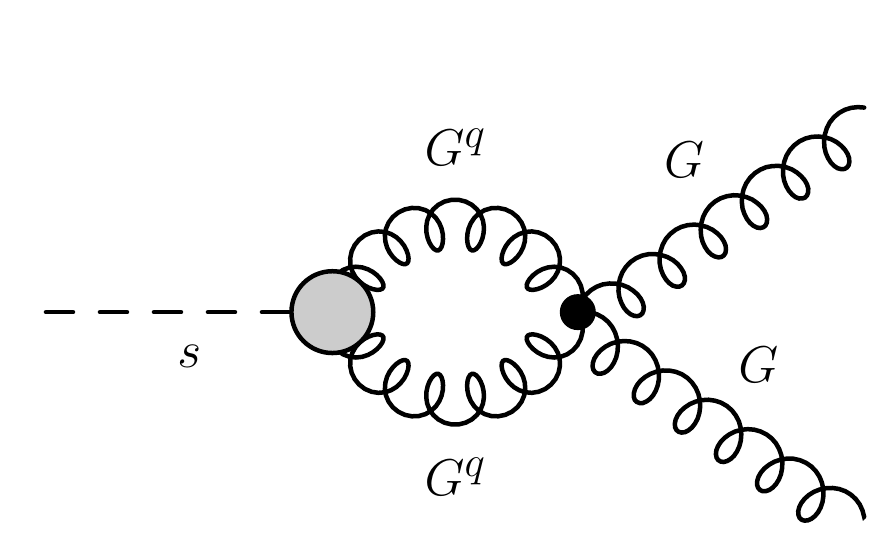}
 \includegraphics[width=0.2\columnwidth]{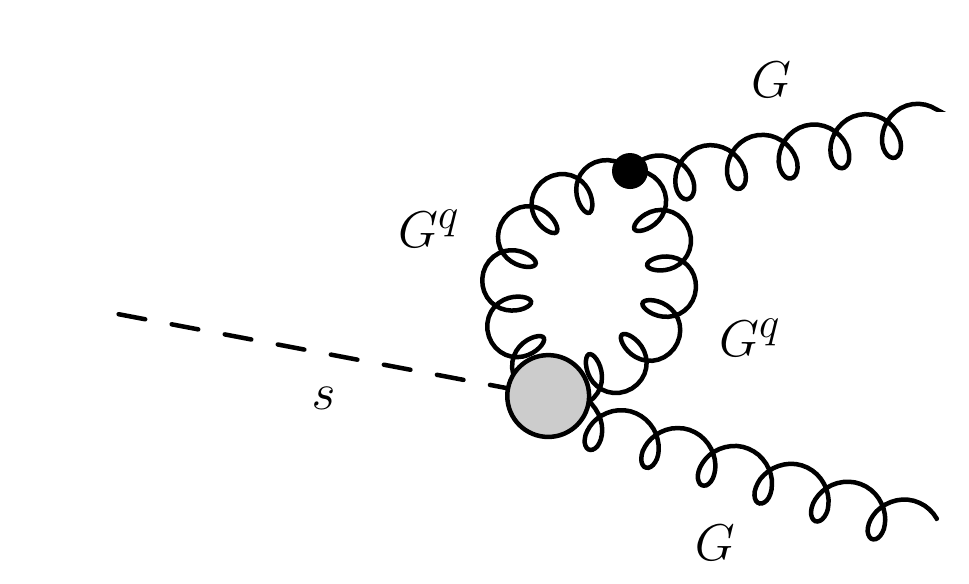}
 \includegraphics[width=0.2\columnwidth]{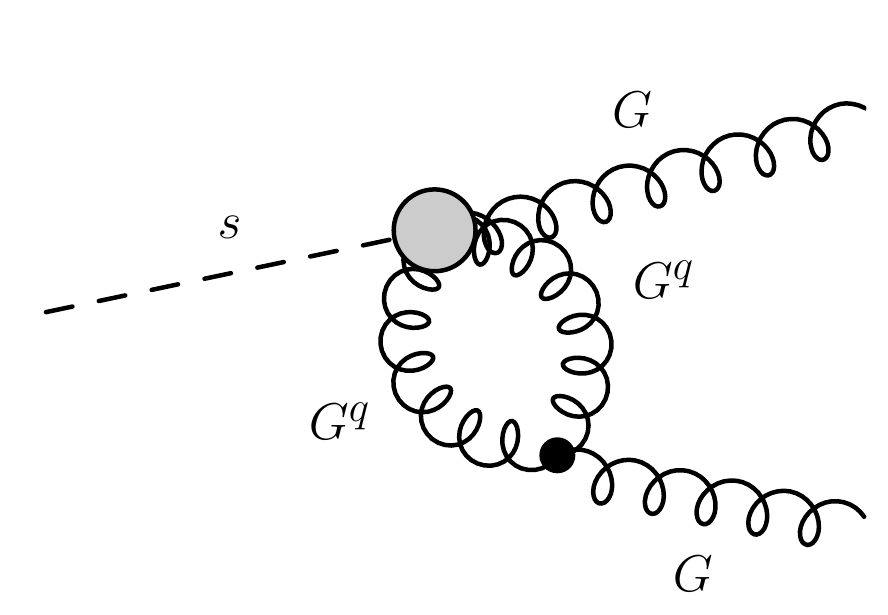}
 \caption{\it Feynman diagrams for $s(p_1)\to G (p_2) G(p_3)$.}\label{fig:sGG}
\end{figure}

\begin{itemize}
  \item{$s(p_1)\phi_i^\dagger(p_2)\rightarrow q_{Lj}^\alpha(p_3)
    \overline{u_R^\beta}(p_4)$}
    
The relevant diagrams, given in Fig.~\ref{fig:shqu}, produce the
following UV divergence,
\begin{align}
  a_{\suf}'-\ii a_{\widetilde{\suf}}'
  =
  -\frac{1}{(4\pi)^2 \epsilon} &
  \bigg\{
  \left[\lambda_{s\phi} 
  -  \left(\frac{25g_1^2}{36} + \frac{3 g_2^2}{4} + \frac{16g_3^2}{3}
  \right)\right]a_{\suf}
  \nonumber \\&
  -  y^d y^{d\dagger} a_{\suf}
  -  a_{\sdf} y^{d\dagger} y^u
  +  y^d a_{\sdf}^\tran y^u
  \bigg\} ~.
\end{align}
\newpage
\item{$s(p_1)\phi_i(p_2)\rightarrow q_{Lj}^\alpha(p_3)
    \overline{d_R^\beta}(p_4)$}
    
The relevant diagrams, given in Fig.~\ref{fig:shqd}, produce the
following UV divergence,
\begin{align}
  a_{\sdf}' -\ii a_{\widetilde{\sdf}}'= -\frac{1}{(4\pi)^2
    \epsilon}
  &\bigg\{
  \bigg[
    \lambda_{s\phi}
 - 
 \bigg( \frac{g_1^2}{36} + \frac{3 g_2^2}{4} + \frac{16 g_3^2
 }{3}\bigg)
\bigg]a_{\sdf}
\nonumber \\ & -  y^u y^{u\dagger} a_{\sdf} -  a_{\suf} y^{u\dagger} y^d
 +  y^u a_{\suf}^\tran y^d
  \bigg\} ~.
\end{align}

\item{$s(p_1)\phi_i(p_2)\rightarrow l_{Lj}^\alpha (p_3)
  \overline{e_R^\beta}(p_4)$}

  The diagrams in Fig.~\ref{fig:shle} give,
\begin{align}
  a_{\sef}'= -\frac{1}{(4\pi)^{2}\epsilon}  \left[\lambda_{s\phi}
    - \frac{9g_1^2}{4}  - \frac{3g_2^2}{4}	\right]
  a_{\sef}\,.
\end{align}


\item{$s(p_1)\rightarrow \phi_i (p_2) \phi_j^\dagger (p_3)$}

The diagrams are shown in Fig.~\ref{fig:shh}, and give
\begin{align}
  r^\prime_{\sfsq} -\ii r^\prime_{\widetilde{\sfsq}} &=
  - \frac{1}{16\pi^2 \epsilon} \left\{
  {\rm Tr}\left[ y^e a_{\sef}^\tran\right]
  + 3\, {\rm Tr} \left[ y^d a_{\sdf}^\tran
    - a_{\suf} y^{u\dagger}\right]
  \right\},\nonumber \\
  r_{\phi s \square}'&=0.
\end{align}


\item{$s(p_1)\rightarrow \Psi^\alpha(p_2) \overline{\Psi^\beta}(p_3)$}

For the process $s(p_1)\rightarrow \Psi^\alpha(p_2)
\overline{\Psi^\beta}(p_3)$, with $\Psi=q_L,l_L,u_R,d_R$ and $e_R$,
we collect the one-loop diagrams in
Figs.~\ref{fig:sqq}, \ref{fig:sll}, \ref{fig:suu}, \ref{fig:sdd},
\ref{fig:see}, respectively. 
The resulting divergences read:~\footnote{Note that only the
terms proportional to the
imaginary part of the Yukawa couplings contribute to the
corresponding CP-odd operators.}
\begin{align}
  r_{sq}^\prime
  +\ii r_{\widetilde{sq}}^\prime
  &
  = \frac{1}{32 \pi^2 \epsilon}
  \bigg[
    a_{\suf} y^{u\,\dagger }
    + a_{\sdf} y^{d\,\dagger}
    - \frac{g_1^2}{3} a_{s\widetilde{B}} - 9 g_2^2 a_{s\widetilde{W}}
    - 16 g_3^2 a_{s\widetilde{G}} 
    \bigg]\,,\\ 
  r_{sl}^\prime
  +\ii r_{\widetilde{sl}}^\prime
  &= \frac{1}{32\pi^2\epsilon}
  \bigg[
    a_{\sef} y^{e\,\dagger}
    - 3 g_1^2 a_{s\widetilde{B}} - 9 g_2^2a_{s\widetilde{W}}
    \bigg]\,,\\
  r_{su}^\prime
    +\ii r_{\widetilde{su}}^\prime
&=
  -\frac{1}{16\pi^2\epsilon}
  \bigg[
    a_{\suf}^\tran y^{u}
    - \frac{8}{3} g_1^2 a_{s\widetilde{B}} - 8 g_3^2 a_{s\widetilde{G}} 
    \bigg]\,,\\
  r_{sd}^\prime
    +\ii r_{\widetilde{sd}}^\prime
&=- \frac{1}{16\pi^2\epsilon}
  \bigg[
    a_{\sdf}^\tran y^{d}
    - \frac{2}{3} g_1^2 a_{s\widetilde{B}}
    - 8 g_3^2 a_{s\widetilde{G}}
    \bigg]\,,\\
  r_{se}^\prime
    +\ii r_{\widetilde{se}}^\prime
&=- \frac{1}{16\pi^2\epsilon}
  \bigg[a_{\sef}^\tran y^{e}- 6 g_1^2 a_{s\widetilde{B}}\bigg]\,.
\end{align}

We have partially cross-checked the results above by computing some
amplitudes related to the ones above by gauge invariance. As an
example the process $s B\to\overline{\Psi}\Psi$ allows us to cross check the
anti-symmetric combination $(r_{s\Psi})_{\alpha \beta}-(r_{s\Psi})_{\beta\alpha}$.

\item{$s(p_1) \to V(p_2) V(p_3)$}

No diagrams can be written at the order we are considering for the process $s\to
BB$. The Feynman diagrams for the amplitudes $s\to W^3 W^3$ and $s\to
GG$ are shown in Figs.~\ref{fig:sWW} and \ref{fig:sGG}, respectively.
The corresponding amplitudes are all non divergent. For $W$ bosons,
the second diagram is zero while the divergences of all others
together vanish. In the case of gluons, the second and third diagrams
vanish, while the divergences of the rest of the diagrams cancel each
other.  
It is evident from the diagrams that Yukawa-like operators do not
renormalize ALP-vector-vector ones (redundant
operators do not contribute to ALP-vector-vector couplings either).
This is in agreement with the non-renormalization results in
Refs.~\cite{Cheung:2015aba,Bern:2019wie}. 
\end{itemize}

CP-even renormalizable couplings do not receive any contribution from
dimension-five operators at one loop. This is easy to see from the fact
that CP-even renormalizable operators are even under $s\to -s$ whereas
all dimension-five operators are odd under such replacement and therefore
they cannot induce one-loop corrections to the renormalizable ones.

\subsection{Eliminating redundancy}

Once we have matched all the possible one-loop divergences onto
our Green basis, we can use the relations in
Eq.~\eqref{redundancies:cpeven} to obtain 
the divergences in the minimal basis. From this point on, even
  though we will continue writing Yukawa couplings in matrix form, we
  will neglect their complex phases. This amounts to the following
replacements:
\begin{align}
a'_{\suf} &\to a^\prime_{\suf} - r_{\sfsq}' y^u - r_{sq}' y^u +
y^{u} r_{su}^{\prime\,\tran}
\nonumber\\
&=\frac{-1}{(4\pi)^2 \epsilon}
\bigg[  
  \left(\lambda_{s\phi} -\frac{25g_1^2}{36}
  -\frac{3 g_2^2}{4} - \frac{16g_3^2}{3} \right)
  a_{\suf}- y^d y^{d\dagger} a_{\suf}
  - a_{\sdf} y^{d\dagger} y^u  \bigg.\nonumber \\
  & \phantom{=\frac{1}{(4\pi)^2}}\bigg.
  + y^d a_{\sdf}^\tran y^u
  + \frac{1}{2} a_{\suf} y^{u\,\dagger} y^u
  + \frac{1}{2} a_{\sdf} y^{d\,\dagger} y^u
  + y^u y^{u\,\dagger} a_{\suf} \bigg. 	\nonumber\\
  & \phantom{=\frac{1}{(4\pi)^2}}\bigg.
  - {\rm Tr}\left[y^e a_{\sef}^\tran
          + 3 y^d a_{\sdf}^\tran - 3 a_{\suf} y^{u\dagger} \right]
 y^u
  - 
  \left(\frac{17}{6} g_1^2 a_{s\widetilde{B}}
  +\frac{9}{2} g_2^2 a_{s\widetilde{W}}
  + 16 g_3^2 a_{s\widetilde{G}} \right)
  y^u
  \bigg]\,,
\end{align}
\begin{align}
a'_{\sdf} &\to a'_{\sdf}
+ r_{\sfsq}' y^d
- r_{sq}' y^{d} + y^d
  r_{sd}^{\prime\,\tran}
\nonumber \\
&=  \frac{-1}{(4\pi)^2 \epsilon} \left[ 
  \left(\lambda_{s\phi} - \frac{g_1^2}{36} - \frac{3 g_2^2}{4}
  - \frac{16 g_3^2 }{3}\right)
  a_{\sdf}
  - y^u y^{u\dagger} a_{\sdf}
  - a_{\suf} y^{u\dagger} y^d  \right.
  \nonumber\\
  & \phantom{=\frac{1}{(4\pi)^2}}\left.
  + y^u a_{\suf}^\tran y^d
  + \frac{1}{2} a_{\suf}
  y^{u\dagger} y^d
  + \frac{1}{2} a_{\sdf}
  y^{d\dagger} y^d
  + y^d y^{d\,\dagger}
      a_{\sdf}  \right.
      \nonumber\\
      & \phantom{=\frac{1}{(4\pi)^2}}\left.
  +{\rm Tr}\left[y^e a_{\sef}^\tran
          + 3 y^d a_{\sdf}^\tran - 3 a_{\suf} y^{u\dagger} \right]
 y^d
      -
      \left(\frac{5}{6}g_1^2 a_{s\widetilde{B}} 
      +\frac{9}{2} g_2^2 a_{s\widetilde{W}}
      + 16 g_3^2 a_{s\widetilde{G}}  \right)
      y^d\right]\,,
\end{align}
\begin{align}
a'_{\sef} &\to a'_{\sef} + r_{\sfsq}'
y^e - r_{sl}' y^{e} + y^e
  r_{se}^{\prime\,\tran}
\nonumber \\
&=\frac{-1}{(4\pi)^2 \epsilon} \left[ 
  \left(
  \lambda_{s\phi}
  - \frac{9g_1^2}{4}
  -  \frac{3g_2^2}{4}	\right)   a_{\sef}
  + \frac{1}{2} a_{\sef} y^{e\dagger} y^e    +  y^e y^{e\,\dagger} a_{\sef}
\right.
  \nonumber\\
  & \phantom{=\frac{1}{(4\pi)^2}}\left.
  +         {\rm Tr}\left[y^e a_{\sef}^\tran
          + 3 y^d a_{\sdf}^\tran - 3 a_{\suf} y^{u\dagger} \right]
  y^e
  - \left(\frac{15}{2} g_1^2 a_{s\widetilde{B}} +\frac{9}{2}g_2^2
  a_{s\widetilde{W}} \right)
  y^e
  \right]\,.
\end{align}

\section{Anomalous dimensions and comparison with the literature}
\label{sec:rges}

In the previous section we have determined completely the divergent
Lagrangian in the physical basis as  
\begin{equation}
  \mathcal{L}_{div} =\mathcal{O}_na_n' \equiv
  \mathcal{O}_n \frac{\mathcal{C}_{nm}}{32\pi^2 \epsilon}a_m\,,
\end{equation}
where $n,m$ run over all operators (including flavour indices when
present) and the coefficients
$\mathcal{C}_{nm}$ involve only dimension-four couplings. 
The $\beta$-function governing the RGEs is given by
\begin{equation}
\beta_{a_n} = 16\pi^2\mu \frac{d a_n}{d\mu} = \gamma_{nm} a_m\,,
\end{equation}
where $\gamma$ is the anomalous dimension matrix. It is completely
determined by the divergence matrix $\mathcal{C}$ up to the wave
function renormalization factor for the different operators:
\begin{equation}
 \gamma_{nm} = -(\mathcal{C}_{nm} + K^F_n \delta_{nm})\,,
\end{equation}
where $K^F$ parametrises the divergences
in the wave function renormalization factors of each operator:
\begin{equation}
  Z^F_n = 1 + \frac{K^F_n}{32\pi^2 \epsilon},
\end{equation}
with
\begin{align}
  &Z^F_{\Op_{\suf}}=\sqrt{Z_{q_L} Z_\phi Z_{u_R}},
&Z^F_{\Op_{s\widetilde{G}}}=\sqrt{Z_G}, \\
&Z^F_{\Op_{\sdf}}=\sqrt{Z_{q_L} Z_\phi Z_{d_R}}, 
&Z^F_{\Op_{s\widetilde{W}}}=\sqrt{Z_W}, \\
&Z^F_{\Op_{\sef}}=\sqrt{Z_{l_L} Z_\phi Z_{e_R}}, 
&Z^F_{\Op_{s\widetilde{B}}}=\sqrt{Z_B},
\end{align}
and the following wave function renormalization factors,
in agreement with Refs.~\cite{Buchalla:2019wsc,Chala:2020pbn},
\footnote{Note however that in Ref.~\cite{Buchalla:2019wsc} the Higgs
  is also split into background and quantum fields, therefore
  comparing $Z_{\phi}$ in this case is not straightforward. It can be
  also trivially seen that $Z_s$ vanishes.}
\begin{align}
  \label{Z:qL}
  Z_{q_L} &= 1  - \frac{1}{96 \pi^2 \epsilon}   \bigg[	\frac{1}{6}
  g_1^2 + \frac{9}{2} g_2^2 + 8 g_3^2
  +3y^uy^{u\,\dagger} +3 y^d y^{d\,\dagger}\bigg]\,,
  \\
\label{Z:lL}
 Z_{l_L}  &= 1  -\frac{1}{64 \pi^2 \epsilon} \bigg[g_1^2 + 3 g_2^2 + 2
    y^e y^{e\,\dagger}\bigg]\,,
  \end{align}\begin{align}
 Z_{u_R} &= 1  -\frac{1}{48\pi^2 \epsilon}   \bigg[ \frac{4}{3} g_1^2
    + 4 g_3^2	+ 3 y^{u\,\dagger} y^u\bigg]\,,   \\
 Z_{d_R} &= 1  -\frac{1}{48 \pi^2 \epsilon}\bigg[ \frac{1}{3} g_1^2 + 4 g_3^2 +   3 y^{d\,\dagger}y^d \bigg]\,,  \\
 Z_{e_R} &= 1  -\frac{1}{16 \pi^2 \epsilon}  \bigg[ g_1^2 +
  y^{e\,\dagger}y^e\bigg]\,,  
\\%
%
 Z_{\phi} &= 1 +  \frac{1}{32\pi^2 \epsilon} \bigg[ g_1^2 + 3 g_2^2
  -2 \gamma_\phi^{(Y)}
  \bigg]\,, \\ 
 Z_{B} &=
 1 - \frac{41 g_1^2}{96 \pi^2 \epsilon}\,, \\
 Z_{W} &=
 1 + \frac{19 g_2^2}{96 \pi^2 \epsilon}\,, \\
 Z_{G} &=
 1 + \frac{14 g_3^2}{32 \pi^2 \epsilon}\,,
\end{align}
where we have defined  
\begin{equation}
  \gamma_\phi^{(Y)}
  \equiv  {\rm Tr}\Big[y^{e\,\dagger} y^e
    + 3 y^{u\,\dagger} y^u + 3 y^{d\,\dagger} y^d\Big]\,.
\end{equation}

The final result for the $\beta$-functions, written as usual in matrix
form with flavour indices implicit, reads:
\begin{align}\label{eq:beta10}
  \beta_{a_{\suf}}   =  \, &2 \bigg[
    \bigg( \lambda_{s\phi}
    - \frac{17g_1^2}{24}
    - \frac{9 g_2^2}{8}
    -4 g_3^2
    + \frac{1}{2} \gamma_\phi^{(Y)}
    \bigg)
    a_{\suf}
    \nonumber\\ &
    -\frac{3}{4} y^d y^{d\dagger} a_{\suf}
    + \frac{5}{4}y^u y^{u\dagger} a_{\suf}
    +  a_{\suf} y^{u\dagger} y^u
    +y^d a_{\sdf}^\tran y^u
    - \frac{1}{2} a_{\sdf} y^{d\dagger} y^u
    \nonumber\\ &
    - \bigg(\frac{17 g_1^2}{6}a_{s\widetilde{B}}
    +\frac{9g_2^2}{2} a_{s\widetilde{W}}
    + 16 g_3^2 a_{s\widetilde{G}}
    +
 {\rm Tr}\left[y^e a_{\sef}^\tran
 + 3 y^d a_{\sdf}^\tran - 3 a_{\suf} y^{u\dagger}) \right]
\bigg) y^u
    \bigg]
  \,,
\end{align}
\begin{align}\label{eq:beta11}
  \beta_{a_{\sdf}} = \, &
  2 \bigg[ 
    \bigg(\lambda_{s\phi} - \frac{5 g_1^2}{24}
    - \frac{9 g_2^2}{8} - 4 g_3^2
    + \frac{1}{2} \gamma_\phi^{(Y)}
    \bigg)
    a_{\sdf}
    \nonumber\\&
    - \frac{3}{4}y^u y^{u\dagger} a_{\sdf}
    + \frac{5}{4}  y^d y^{d\dagger} a_{\sdf}
    +a_{\sdf} y^{d\dagger} y^d
    + y^u a_{\suf}^\tran y^d - \frac{1}{2} a_{\suf} y^{u\dagger} y^d
    \nonumber\\&
    - \bigg(\frac{5 g_1^2}{6}a_{s\widetilde{B}}
    +\frac{9 g_2^2}{2} a_{s\widetilde{W}}  + 16 g_3^2
    a_{s\widetilde{G}}
    -
     {\rm Tr}\left[y^e a_{\sef}^\tran
 + 3 y^d a_{\sdf}^\tran - 3 a_{\suf} y^{u\dagger}) \right]
\bigg) y^d\bigg]
\,,
\end{align}
\begin{align}\label{eq:beta12}
  \beta_{a_{\sef}}  = \,&
  2 \bigg[ a_{\sef} \bigg(\lambda_{s\phi}
    - \frac{15g_1^2}{8}
    -  \frac{9g_2^2}{8}
    + \frac{1}{2} \gamma_\phi^{(Y)} \bigg)
    +   \frac{5}{4} y^e y^{e\dagger} a_{\sef}
    +  a_{\sef} y^{e\dagger} y^e
    \nonumber\\ &
    - 
    \bigg(\frac{15 g_1^2}{2}a_{s\widetilde{B}}
    +\frac{9g_2^2}{2} a_{s\widetilde{W}}
    -
     {\rm Tr}\left[y^e a_{\sef}^\tran
 + 3 y^d a_{\sdf}^\tran - 3 a_{\suf} y^{u\dagger}) \right]
    \bigg) y^e \bigg]
  \,,
\end{align}
\begin{align}\label{eq:beta18}
\beta_{a_{s\widetilde{B}}} &= \frac{41}{3} g_1^2 a_{s\widetilde{B}}\,,
\\
\label{eq:beta16}
\beta_{a_{s\widetilde{W}}} &= -\frac{19}{3} g_2^2 a_{s\widetilde{W}}\,,
\\
\label{eq:beta14}
\beta_{a_{s\widetilde{G}}} &= -14g_3^2 a_{s\widetilde{G}}\,.
\end{align}
%
%
%

%
A more graphic picture of the operator mixing can be obtained for the case in which different
fermion families factorize, so that all dimension-five Wilson
coefficients are flavour diagonal, $a_{\alpha \beta} = \delta_{\alpha
  \beta} a_{\alpha}$ (and neglecting also off-diagonal Yukawa
  couplings). Writing
$\gamma_{nm}$,
where $n$ runs over $\Op_{\suf}^\alpha$, $\Op_{\sdf}^\alpha$,
$\Op_{\sef}^\alpha$, $\Op_{s\widetilde{G}}$, $\Op_{s\widetilde{W}}$
and $\Op_{s\widetilde{B}}$, and $m$ over the same operators but with
flavour index $\rho$, we can express the anomalous dimensions in the
following form:
\begin{equation}\label{eq:result1} \gamma =
\begin{pmatrix}
\gamma_{11}  + 6 y_u^\alpha y_u^\rho  \quad& 	y_d^\alpha y_u^\alpha  - 6y_u^\alpha y_d^\rho  \quad& -2 y_u^\alpha y_e^\rho \quad& -32 g_3^2 y_ u^\alpha \quad& -9 g_2^2 y_u^\alpha   \quad&  -\frac{17}{3} g_1^2 y_u^\alpha  \\[0.5cm]

y_u^\alpha y_d^\alpha - 6 y_d^\alpha y_u^\rho \quad& \gamma_{22} + 6 y_d^\alpha y_d^\rho & 2 y_d^\alpha y_e^\rho & -32 g_3^2 y_d^\alpha  & -9 g_2^2 y_d^\alpha&  -\frac{5}{3}g_1^2 y_d^\alpha   \\[0.5cm]

-6 y_e^\alpha y_u^\rho & 6 y_e^\alpha y_d^\rho & \gamma_{33} + 2y_e^\alpha y_e^\rho & 0 & -9 g_2^2 y_e^\alpha & -15g_1^2  y_e^\alpha  \\[0.5cm]

0 & 0 & 0 & -14 g_3^2 & 0 & 0 \\[0.5cm]

0 & 0 & 0 & 0 & -\frac{19}{3} g_2^2 & 0 \\[0.5cm]

0 & 0 & 0 & 0 & 0 & \frac{41}{3} g_1^2 \\
\end{pmatrix}\,,
\end{equation}
where a $\delta_{\alpha \rho}$ should be understood in every entry in
which the $\rho$-index does not explicitly appear, and we have defined
\begin{align}
\gamma_{11} &= 2 \lambda_{s\phi} - \frac{3}{2}  \left(y_d^\alpha\right)^2 +
\frac{9}{2} \left(y_u^\alpha\right)^2   -\frac{17}{12} g_1^2 -
\frac{9}{4} g_2^2 -8 g_3^2 +
     \gamma_\phi^{(Y)}\,,\\
\gamma_{22} &= 2 \lambda_{s\phi} - \frac{3}{2}  \left(y_u^\alpha\right)^2 +
\frac{9}{2} \left(y_d^\alpha\right)^2  -\frac{5}{12} g_1^2 -
\frac{9}{4} g_2^2 -8 g_3^2 + \gamma_\phi^{(Y)}\,,\\
\gamma_{33} &=  2 \lambda_{s\phi} + \frac{9}{2}  \left(y_e^\alpha\right)^2
-\frac{15}{4} g_1^2 - \frac{9}{4} g_2^2+  \gamma_\phi^{(Y)}\,.
\end{align}
Note that, due to the contribution to $r_{\sfsq}$, even in the flavour
diagonal case there is inter-generational mixing but the choice of
diagonal Wilson coefficients is radiatively stable (up to the small
non-diagonal terms in the SM Yukawa couplings).

Different pieces of the anomalous dimension matrix have been
previously computed in the literature. In particular, the mixing of
the operators $\mathcal{O}_{\suf}$, $\mathcal{O}_{\sdf}$ and
$\mathcal{O}_{\sef}$ driven by Yukawa interactions has been obtained in
Ref.~\cite{Choi:2017gpf}. %
Such work relies, however, on a different basis of effective interactions,
where the fermionic operators take the form $(\partial_\mu s)
\overline{\Psi} \gamma^\mu \Psi$. Using the relation between their
Wilson coefficients and $a_{\suf}$, $a_{\sdf}$ and $a_{\sef}$ we have
obtained the beta functions for the latter from the results
in~Ref.\cite{Choi:2017gpf} and compared them with our direct calculation
reported in
Eqs.~\eqref{eq:beta10}--\eqref{eq:beta12}.
The results completely agree
up to a sign difference in the terms proportional to
${\rm Tr}\left[y^e a_{\sef}^\tran
 + 3 y^d a_{\sdf}^\tran - 3 a_{\suf} y^{u\dagger}) \right] y^\psi$,
with $\psi=u,d,e$.
Unfortunately, we do not
find enough information to track the origin of this discrepancy. It is
also worth emphasizing that the set of effective operators used in
Ref.~\cite{Choi:2017gpf} is over-complete.
We provide in
Appendix~\ref{app:bases} the RGEs for a new basis
in which the redundant operators have been removed, under some
simplifying assumptions. 

In Ref.~\cite{Bauer:2016lbe} it was shown that
the Wilson coefficients of the 
operators 
$g_3^2 \mathcal{O}_{s\widetilde{G}}$, $g_2^2
\mathcal{O}_{s\widetilde{W}}$ and
$g_1^2 \mathcal{O}_{s\widetilde{B}}$,
are scale invariant, \textit{i.e.} they do not depend on $\mu$. This
is consistent with our results, for which all the running of
$a_{s\widetilde{G},s\widetilde{W},s\widetilde{B}}$ can be accounted
for by the running of the corresponding gauge couplings, which are
determined by the wave function renormalization of the gauge fields in
the background field method.
%
Similarly,  the 1-loop ALP couplings to fermions induced by the
$\mathcal{O}_{s\widetilde{G},s\widetilde{W},s\widetilde{B}}$ operators
in the effective Lagrangian were computed in
Ref.~\cite{Bauer:2017ris}, again in a different basis from ours. We
have compared the results, using
the RGEs derived in Appendix ~\ref{app:bases} and found exact agreement.

Finally, the RGEs for the dimension-four operators in the ALP EFT
Lagrangian must be provided in order to fully determine the theory. We
have obtained these with the help of \texttt{Pyr@te}~\cite{Lyonnet:2016xiz}; they are reported in
Appendix~\ref{app:4dim}. Let us remind once more that the RGEs of
renormalizable interactions are not perturbed by higher-dimensional
operators at order $\mathcal{O}(1/\Lambda)$. 

\section{Matching and running below the electroweak scale}
\label{sec:lalp}

At energies smaller than the EW scale, set by the Higgs vacuum
expectation value (VEV), $v\sim 246$ GeV, the ALP phenomenology must
be described by a different EFT, that we call ALP LEFT, organised in inverse
powers of $v$, in 
which the now massive 
top quark and the Higgs, $Z$ and $W$ bosons are not present.
Assuming still CP conservation, the corresponding ALP
LEFT Lagrangian, to dimension five, takes the following form: 
\begin{align}
 \mathcal{L}_{\text{LEFT}} &= \frac{1}{2}(\partial_\mu s)(\partial^\mu s)
 -\frac{1}{2}\tilde{m}^2s^2 - \frac{\tilde{\lambda}_s}{4!}
 s^4   -\frac{1}{4}  A_{\mu\nu} A^{\mu\nu}
  -\frac{1}{4}  G^A_{\mu\nu} G^{A\,\mu\nu}
\nonumber \\
 &+\sum_{\psi=u,d,e}\bigg\{
 \overline{\psi^\alpha}\ii\slashed{D}\psi^\alpha -
 \bigg[(\tilde{m}_\psi)_{\alpha\beta}\overline{\psi^\alpha_L}\psi_R^\beta
   -  s \,\ii (\tilde{c}_\psi)_{\alpha\beta}
   \overline{\psi^\alpha_L}\psi_R^\beta + \text{h.c.}\bigg]\bigg\}
 \nonumber \\[0.2cm]
&+ \tilde{a}_{s\widetilde{G}}s\,G_{\mu\nu}^A \widetilde{G}^{A\,\mu\nu}
+ \tilde{a}_{s\widetilde{A}} s
A_{\mu\nu} \widetilde{A}^{\mu\nu}
\nonumber \\[0.2cm]
  &+\sum_{\psi=u,d,e}\bigg\{(\tilde{a}_{\psi A})_{\alpha\beta}\overline{\psi_L^\alpha}\sigma^{\mu\nu} \psi_R^\beta A_{\mu\nu} + (\tilde{a}_{\psi G})_{\alpha\beta} \overline{\psi_L^\alpha}\sigma^{\mu\nu} T_A \psi_R^\beta G_{\mu\nu}^A + s^2 (\tilde{a}_\psi)_{\alpha\beta}\overline{\psi_L^\alpha}\psi_R^\beta+\text{h.c.}\bigg\}\,,
\label{lag:left}
\end{align}
where $\alpha,\beta$ are flavour indices that run over the three families for
$d$ and $e$ and over the lighter two for the case of $u$.

The assumed CP invariance forces all coefficients to be real
(matrices in case flavour is involved).
When no chiralities for the
fermions are explicitly written we assume $\psi=\psi_L+\psi_R$
and the covariant derivative in this regime reads $D_\mu =
\partial_\mu -\ii\tilde{e}QA_\mu -\ii\tilde{g}_3 T^A G_\mu^A$, with $Q$ being
the electric charge. We emphasise
that contrary to the ALP EFT above the EW scale, in this case there
are (lepton-number conserving) effective operators of the same dimension with
and without the ALP. 
Note that, as explicitly written above, we work in a flavour basis
in which mass matrices are not necessarily diagonal. At the 
level of computation, this is equivalent to promoting the masses to
Yukawa couplings of a spurion scalar field which is later set to its
VEV. Technically, every time we have to integrate out one of the SM
fermions (the top in this case, lighter fermions as we go to lower
energies, see below) we go to the physical basis in which the mass
matrix is diagonalised; off-diagonal mass terms being generated by
running to lower energies. 

The following redundant operators arise at dimension five:
\begin{equation}
  \mathcal{L}_R =  \sum_{\psi=u,d,e} \bigg[
    \left(\tilde{r}_{\psi \Box
    }\right)_{\alpha\beta} \overline{\psi_L^\alpha} D^2 \psi_R^\beta
    + \ii
  \left(\tilde{r}_{s \psi_L}\right)_{\alpha\beta} s \overline{\psi_L^\alpha} \ii
  \slashed{D} \psi_L^\beta + \ii \left(\tilde{r}_{s \psi_R}\right)_{\alpha\beta} s
  \overline{\psi_R^\alpha} \ii \slashed{D} \psi_R^\beta + \text{h.c.}\bigg]\,, 
\label{eq:LEFTred}
\end{equation}
where again the operators are CP-even for real Wilson coefficients.
The purely SMEFT redundant operator can be removed by making use of the relation
\begin{equation}
D^2 = \slashed{D}^2 + \frac{\sigma_{\mu\nu}}{2} \left( \tilde{e} Q A^{\mu\nu} + \tilde{g}_3 G^{\mu\nu}_A T_A\right)~.
\label{eq:Dslashed}
\end{equation}
The first term in the equation above can then be further reduced by
applying the equations of motion for fermions in the ALP LEFT,
\begin{equation}
\ii\slashed{D} \psi_\alpha = \left(\tilde{m}_\psi\right)_{\alpha\beta} \psi_R^\beta + (\tilde{m}_\psi^\dagger)_{\alpha\beta} \psi_L^\beta - \ii
\left( \tilde{c}_\psi \right)_{\alpha\beta} s \psi_R^\beta + \ii (
\tilde{c}_\psi^{\dagger}  )_{\alpha\beta} s \psi_L^\beta~. 
\end{equation}
There is an apparent ambiguity in this process for the $\slashed{D}^2$
term due to the possibility of performing integration by parts before
applying the equations of motion. However, this ambiguity simply
corresponds to a chiral rotation and therefore has 
no physical consequences (see~Ref.~\cite{Jenkins:2017dyc} for a related discussion).~\footnote{As an example of the mentioned apparent
ambiguity, we could choose to apply the equations of
motion without the splitting in Eq.~\eqref{eq:split}. In that case
we obtain the same contribution to the dimension-five operators but
a different contribution to the renormalizable ones. This difference
is however removed, after canonical normalization, by the following
chiral unitary rotation:
  \begin{equation}
    \psi_L \to \left(1+\frac{
      \tilde{r}_{\psi\Box}
      \tilde{m}_\psi^\dagger
      -\tilde{m}_\psi
      \tilde{r}_{\psi\Box}^\dagger
    }{4}\right)\psi_L\,,
    \qquad
    \psi_R \to \left(1+\frac{
      \tilde{m}_\psi^\dagger
      \tilde{r}_{\psi\Box}
      -\tilde{r}_{\psi\Box}^\dagger
      \tilde{m}_\psi
    }{4}\right)\psi_R\,.    
  \end{equation}
}
We
choose to split the covariant derivative 
symmetrically:
\begin{equation}
\slashed{D}^2=\frac{1}{2}(\slashed{D}^2+\overleftarrow{\slashed{D}}^2)\,.
\label{eq:split}
\end{equation}
In this case we obtain the following on-shell equivalence relations
(as usual we write our equations in matrix form in flavour space):
\begin{align}
  \overline{\psi_L}
  \tilde{r}_{\psi \Box}
  D^2 \psi_R
  + \text{h.c.}=&
  - \overline{\psi_L}
  \frac{\tilde{r}_{\psi \Box} \tilde{m}_\psi^\dagger
  +\tilde{m}_\psi\tilde{r}_{\psi \Box}^\dagger}{2}
  \ii \slashed{D}
  \psi_L
  - \overline{\psi_R}
  \frac{\tilde{r}_{\psi \Box}^\dagger \tilde{m}_\psi
  +\tilde{m}_\psi^\dagger\tilde{r}_{\psi \Box}}{2}
\ii \slashed{D}  \psi_R
  \nonumber \\
  +&\bigg[
    \ii s \overline{\psi_L}
    \frac{\tilde{m}_\psi \tilde{r}_{\psi \Box}^\dagger \tilde{c}_\psi
+\tilde{c}_\psi \tilde{r}_{\psi \Box}^\dagger \tilde{m}_\psi    }{2}
    \psi_R
    +s^2 \overline{\psi_L} \tilde{c}_\psi\tilde{r}_{\psi\Box}^\dagger\tilde{c}_\psi
    \psi_R 
  \nonumber \\
  &+
  \overline{\psi_L} \frac{e Q_\psi \tilde{r}_{\psi \Box}}{2}
  \sigma_{\mu\nu} \psi_R A^{\mu\nu}
+
  \overline{\psi_L} \frac{\tilde{g}_3  \tilde{r}_{\psi \Box}}{2} T_A
  \sigma_{\mu\nu} \psi_R G_A^{\mu\nu}
  +\text{h.c.}\bigg]
  \,,\label{red:psiBox}
  \\
  \ii s \overline{\psi_L} \tilde{r}_{s\psi_L} \ii \slashed{D} \psi_L + \text{h.c.}
  = &\,
  \ii s \overline{\psi_L} \tilde{r}_{s\psi_L} \tilde{m}_\psi \psi_R
  + s^2 \overline{\psi_L} \tilde{r}_{s\psi_L} \tilde{c}_\psi \psi_R
  +\text{h.c.}
  \,,\label{red:RspsiL}
  \\
  \ii s \overline{\psi_R} \tilde{r}_{s\psi_R} \ii \slashed{D} \psi_R + \text{h.c.}
  = &
  -\ii s \overline{\psi_L} \tilde{m}_\psi \tilde{r}_{s\psi_R}^\dagger \psi_R
  - s^2 \overline{\psi_L} \tilde{c}_\psi \tilde{r}_{s\psi_R}^\dagger \psi_R
  +\text{h.c.}
  \,.\label{red:RspsiR}
\end{align}

The parameters of the ALP LEFT can be fully fixed at the scale
$\mu =v$ by requiring that it describes exactly the same physics as
the EFT before EWSB at the scale $\mu$. Proceeding this way at tree
level, we obtain the following matching conditions for the
interactions in Eq.~\eqref{lag:left}:
\begin{align}\label{eq:matching}
  &\tilde{e} = g_2 s_w=g_1 c_w\,,&
&\tilde{m}^2 = m^2 + \frac{\lambda_{s\phi}}{2} v^2\,,&  \\
  & \tilde{g}_3 =g_3 \,,&
  & \tilde{\lambda}_s = \lambda_s - 3\frac{v^2}{m_h^2}
  \lambda_{s\phi}^2\,,&
\\
& (\tilde{m}_u)_{\alpha\beta} =\frac{v}{\sqrt{2}} (y^u)_{\alpha\beta}
\,,&
  & (\tilde{c}_u)_{\alpha\beta} = \frac{v}{\sqrt{2}} (a_{\suf})_{\alpha\beta}  \,,&\\ 
& (\tilde{m}_d)_{\alpha\beta} = \frac{v}{\sqrt{2}}
(y^d)_{\alpha\beta}
\,,&
& (\tilde{c}_d)_{\alpha\beta} = \frac{v}{\sqrt{2}} (a_{\sdf})_{\alpha\beta}\,,&\\
& (\tilde{m}_e)_{\alpha\beta} = \frac{v}{\sqrt{2}} (y^e)_{\alpha\beta}
\,,&
& (\tilde{c}_e)_{\alpha\beta} = \frac{v}{\sqrt{2}} (a_{\sef})_{\alpha\beta}\,,&\\
& \tilde{a}_{s\widetilde{G}} = a_{s\widetilde{G}}\,,&
& \tilde{a}_{s\widetilde{A}} = a_{s\widetilde{W}} s_w^2 + a_{s\widetilde{B}} c_w^2\,;&\label{eq:matching2}
\end{align}
where, as before, $\alpha$ and $\beta$ are flavour indices that run over the three
families for $d$ and $e$ and over the first two for $u$;   $c_\omega$ and $s_\omega$ are the cosine and sine of the Weinberg angle, respectively.
All the other Wilson coefficients vanish at the order we are
computing.
The fact that the three coefficients $\tilde{a}_{u,d,e}$
vanish might be surprising at first glance, as the
Higgs couples to both $s^2$ and to fermionic currents with overall
strength $\sim \lambda_{s\phi} y^\psi/v$. However, precisely because the Higgs boson
sets the scale of light masses~\cite{Jenkins:2017jig}, \textit{i.e.}
because $y^\psi\sim m_\psi/v$, the product $\lambda_{s\phi}
y^\psi/v \sim \lambda_{s\phi} m_\psi/v^2$ is
of higher order in the low-energy power counting and therefore
negligible. This is no longer true at dimension six (it is in the
pure SM EFT even at dimension six because an s-channel Higgs always
involves two powers of Yukawa couplings).
Similarly, a possible contribution proportional to two 
powers of $v a_{\suf}$ is also higher order in the $1/\Lambda$ expansion.
At energies below the bottom quark mass $m_b$, the effective
Lagrangian takes exactly the same form as in Eq.~\eqref{lag:left}
except that the flavour indices now run only over the
remaining fermions and the Wilson coefficients in the new EFT have to
be matched accordingly. The same logic applies as we cross new
fermionic thresholds.
At the order we are considering, however, the matching is straightforward and
the only thing we have to do is to remove the Wilson coefficients
involving the particle being integrated out.
The only exception arises
if $\tilde{c}_\psi$ is unsuppressed, in which case
integrating out a massive fermion would result in the following
matching condition:
\begin{equation}
  (\tilde{a}_\psi)_{\alpha \beta} = -\frac{(\tilde{c}_\psi)_{\alpha
      \gamma} (\tilde{c}_\psi)_{\gamma \beta}}{(\tilde{m}_\psi)_\gamma}
  \mbox{   (no sum over $\gamma$)}\,,
\end{equation}
where $\gamma$ corresponds to the flavour that is being integrated out
while $\alpha$ and $\beta$ run over lighter flavours of the same type of fermion.
This term is higher order if the ALP LEFT is obtained from the ALP
EFT. However, we prefer to keep this section completely general,
independently of which theory completes the ALP LEFT in the UV.

The running of the Wilson
coefficients between different thresholds is
very different from the running above the EW scale (the operators in
Table~\ref{tab:eft}). 
In particular, operators of different energy dimensions, as well
as operators with and without the ALP field, will now mix under renormalization. 

\subsection{Divergences at one loop}

\begin{figure}[H]
 \centering
 \includegraphics[width=0.20\columnwidth]{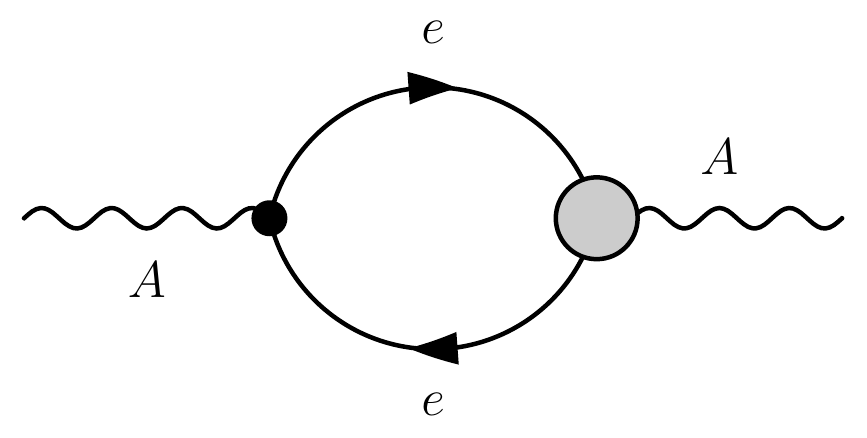}
 \includegraphics[width=0.20\columnwidth]{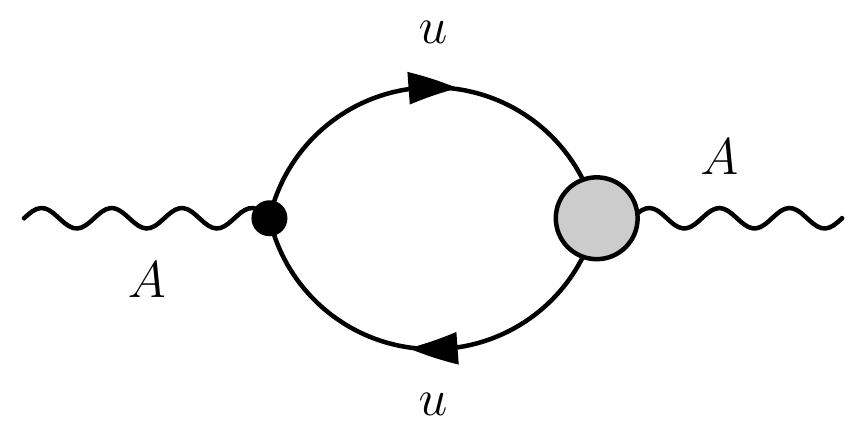}
 \includegraphics[width=0.20\columnwidth]{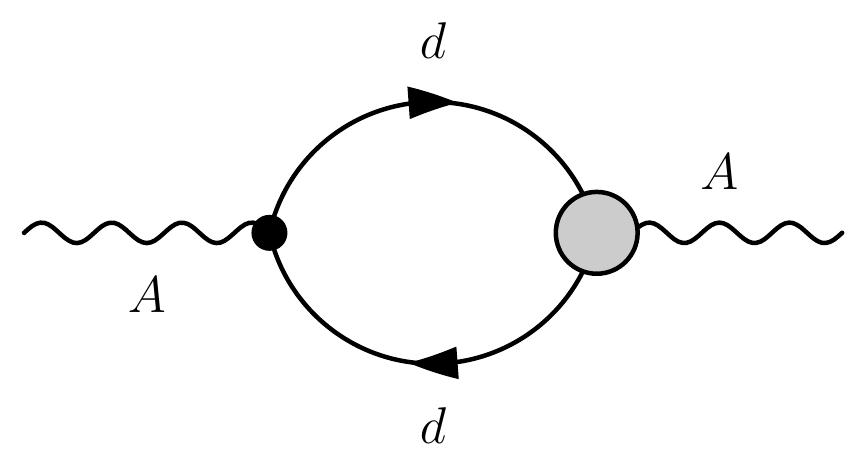}
 \includegraphics[width=0.20\columnwidth]{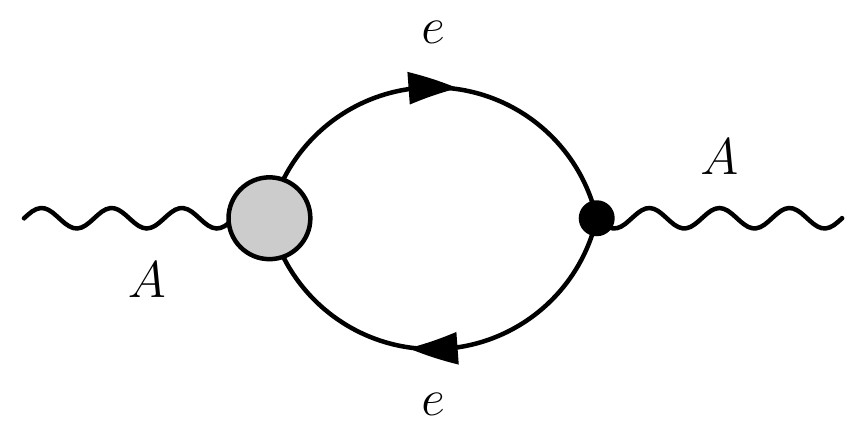}
 \includegraphics[width=0.20\columnwidth]{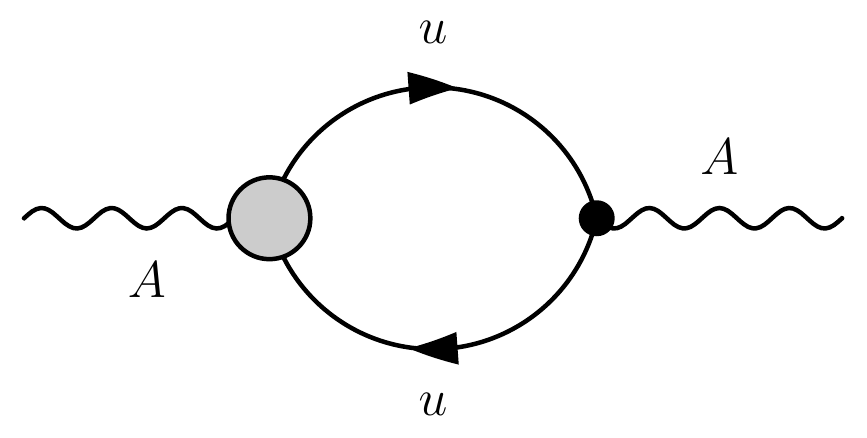}
 \includegraphics[width=0.20\columnwidth]{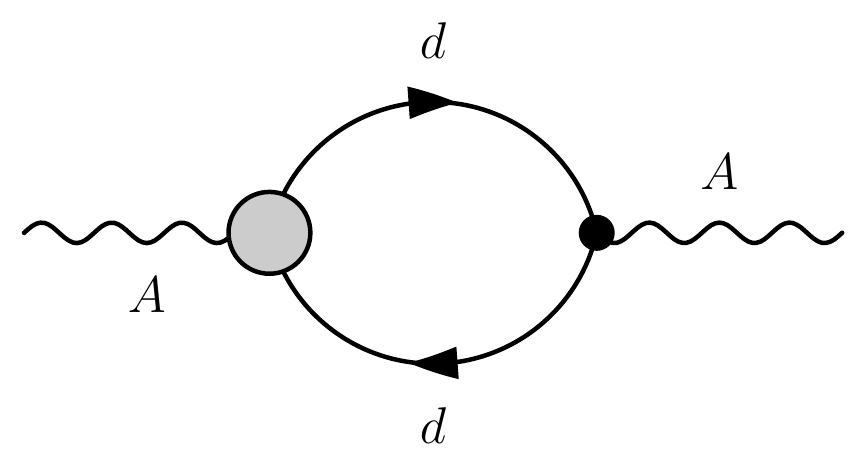}
 \caption{\it Feynman diagrams for $A(p_1)\to A(p_2)$.}\label{fig:A_A}
\end{figure}

\begin{figure}[H]
 \centering
 \includegraphics[width=0.20\columnwidth]{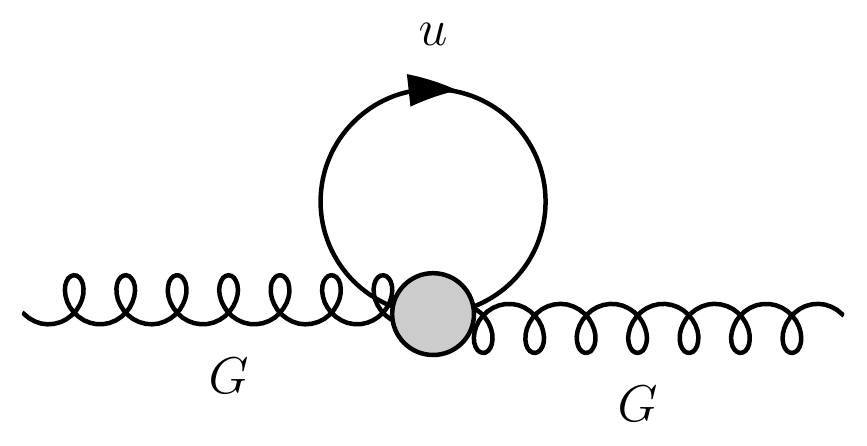}
 \includegraphics[width=0.20\columnwidth]{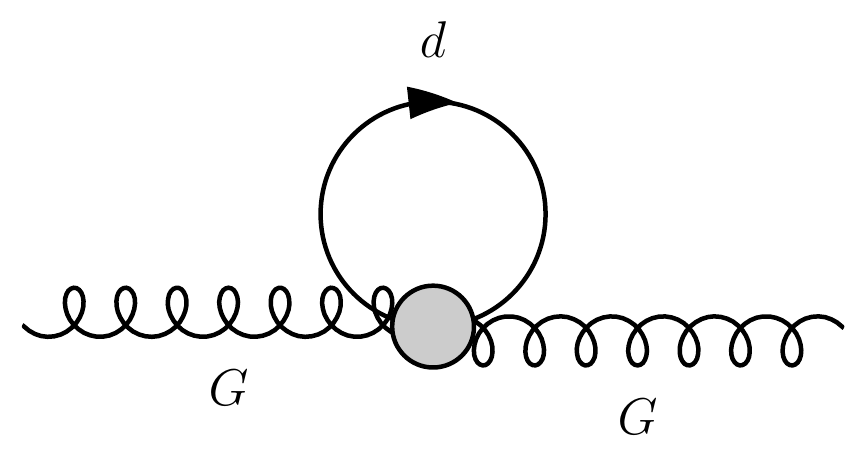}
 \includegraphics[width=0.20\columnwidth]{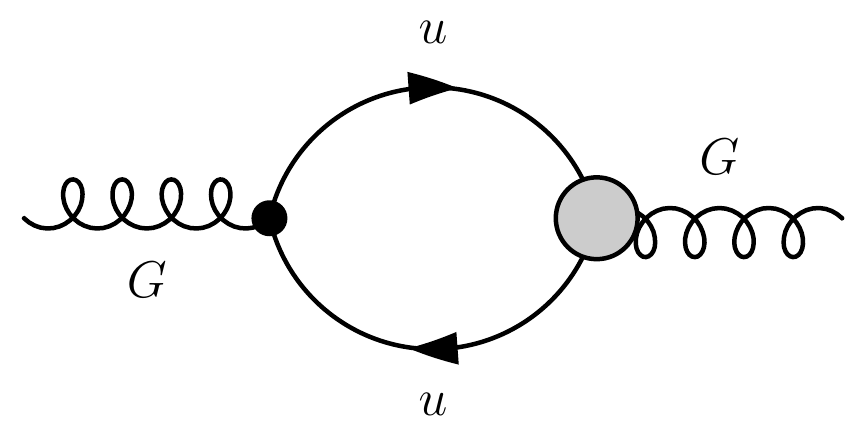}
 \includegraphics[width=0.20\columnwidth]{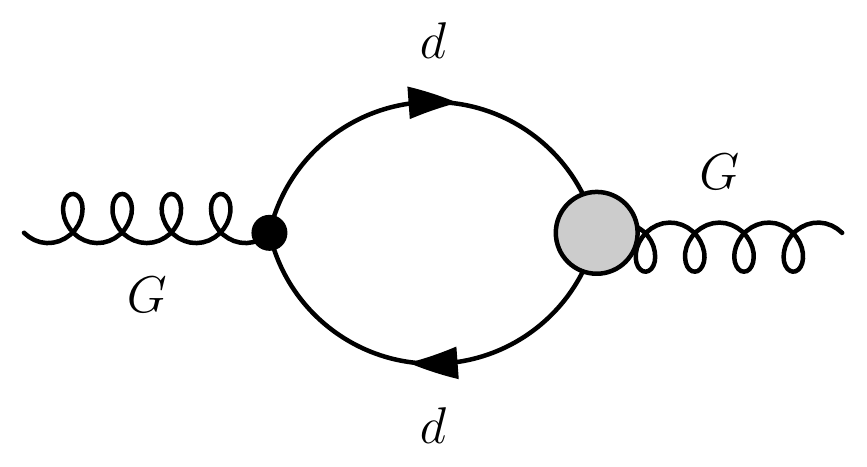}
 \includegraphics[width=0.20\columnwidth]{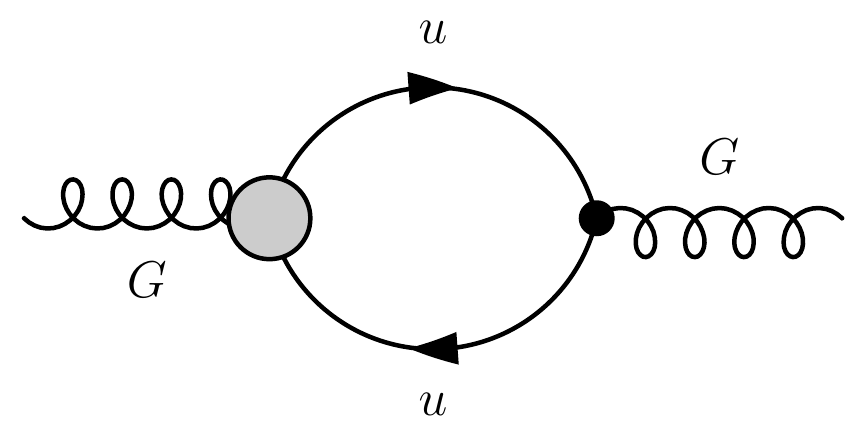}
 \includegraphics[width=0.20\columnwidth]{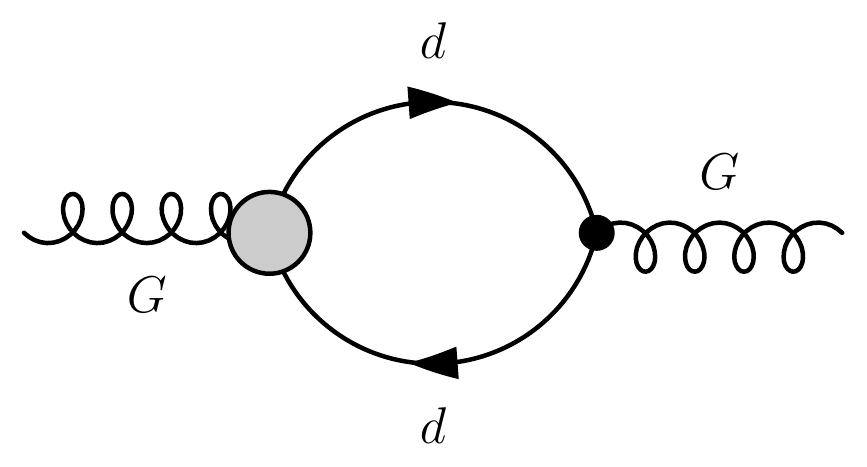}
 \caption{\it Feynman diagrams for $G(p_1)\to G(p_2)$.}\label{fig:G_G}
\end{figure}

\begin{figure}[H]
 \centering
 \includegraphics[width=0.20\columnwidth]{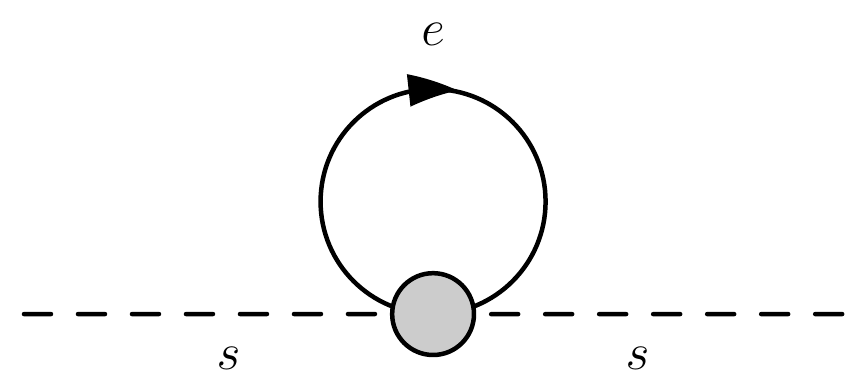}
 \includegraphics[width=0.20\columnwidth]{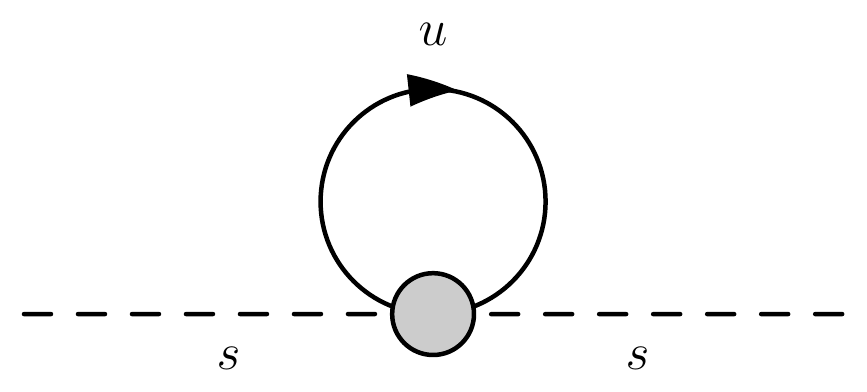}
 \includegraphics[width=0.20\columnwidth]{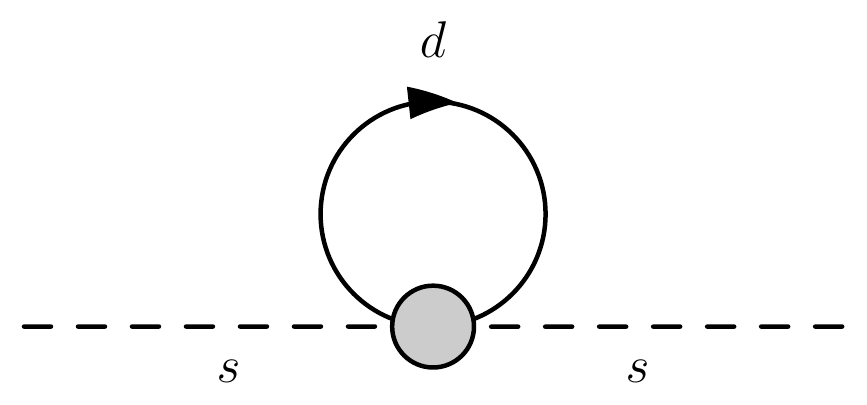}
 \caption{\it Feynman diagrams for $s(p_1)\to s(p_2)$.}\label{fig:s_s}
\end{figure}

\begin{figure}[H]
 \centering
 \includegraphics[width=0.20\columnwidth]{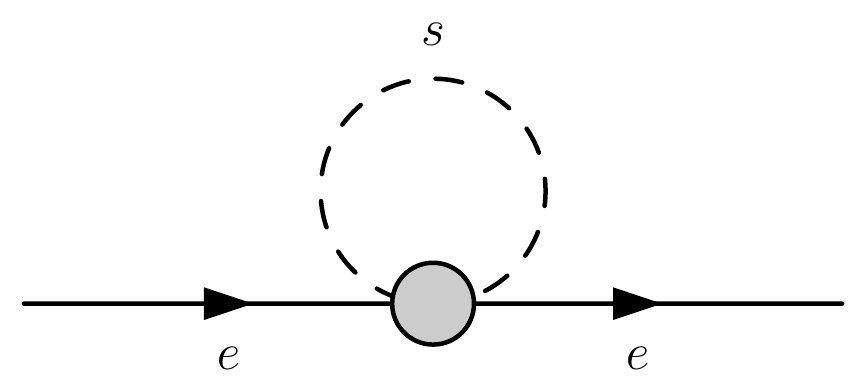}
 \includegraphics[width=0.20\columnwidth]{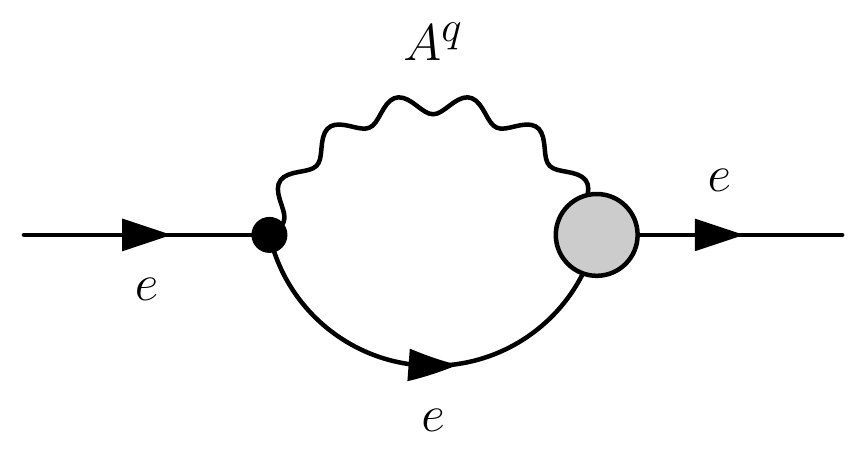}
 \includegraphics[width=0.20\columnwidth]{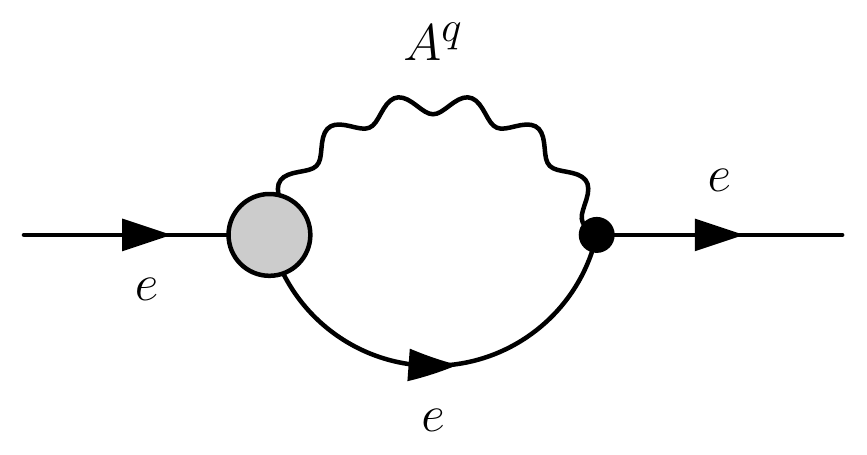}
 \caption{\it Feynman diagrams for $e(p_1)\to e(p_2)$.}\label{fig:e_e}
\end{figure}

\begin{figure}[H]
 \centering
 \includegraphics[width=0.20\columnwidth]{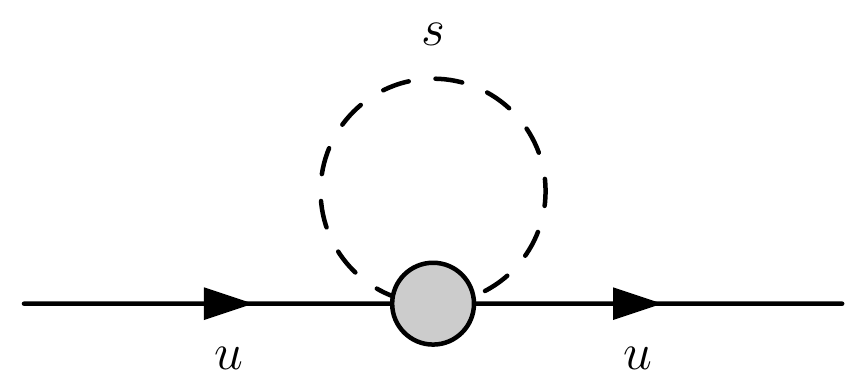}
 \includegraphics[width=0.20\columnwidth]{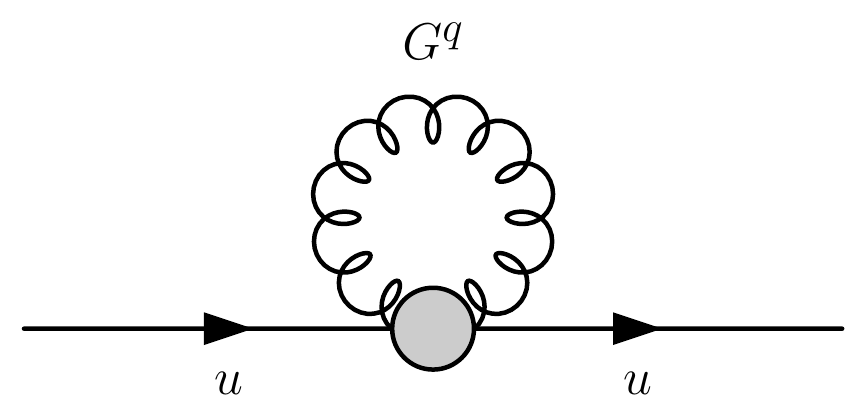}
 \includegraphics[width=0.20\columnwidth]{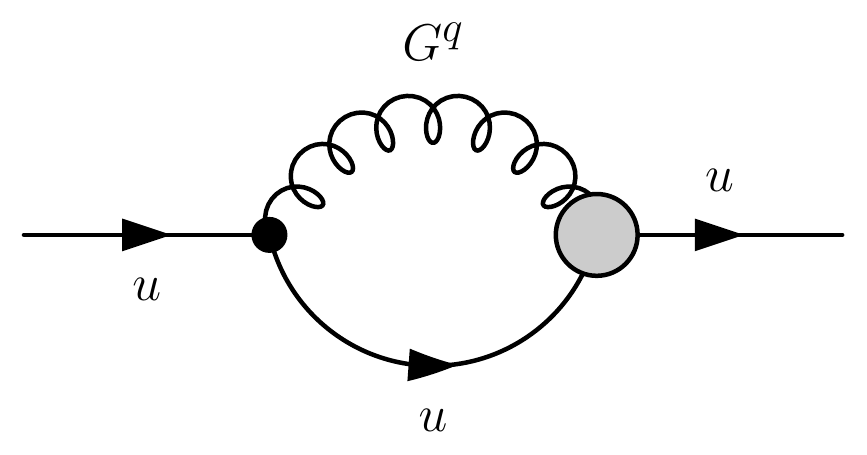}
 \includegraphics[width=0.20\columnwidth]{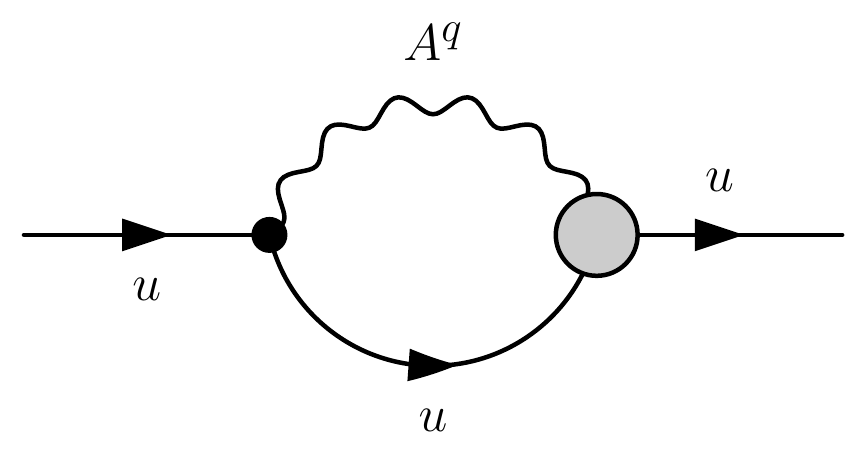}
 \includegraphics[width=0.20\columnwidth]{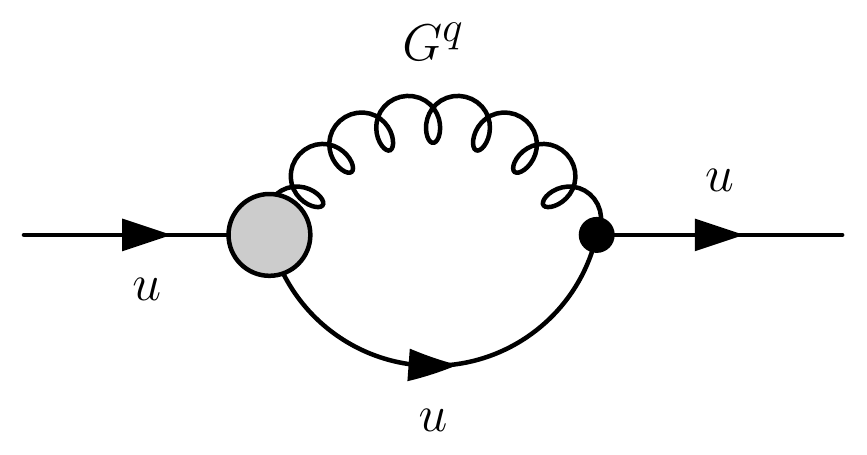}
 \includegraphics[width=0.20\columnwidth]{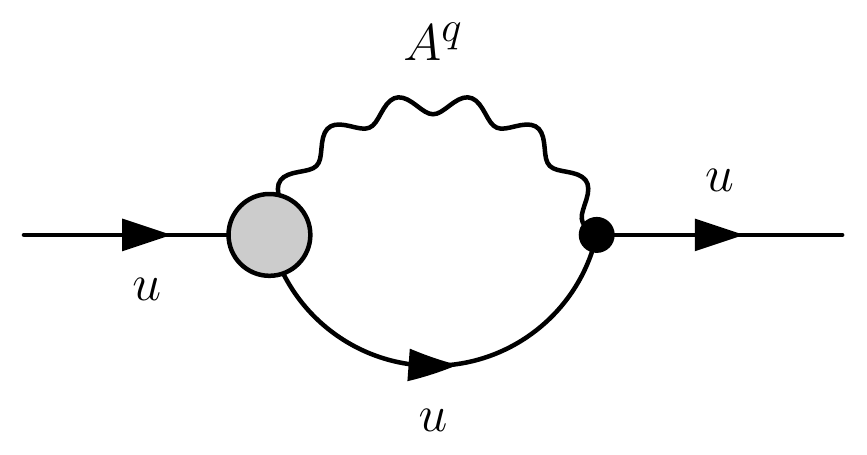}
 \caption{\it Feynman diagrams for $u(p_1)\to u(p_2)$.}\label{fig:u_u}
\end{figure}

\begin{figure}[H]
 \centering
 \includegraphics[width=0.20\columnwidth]{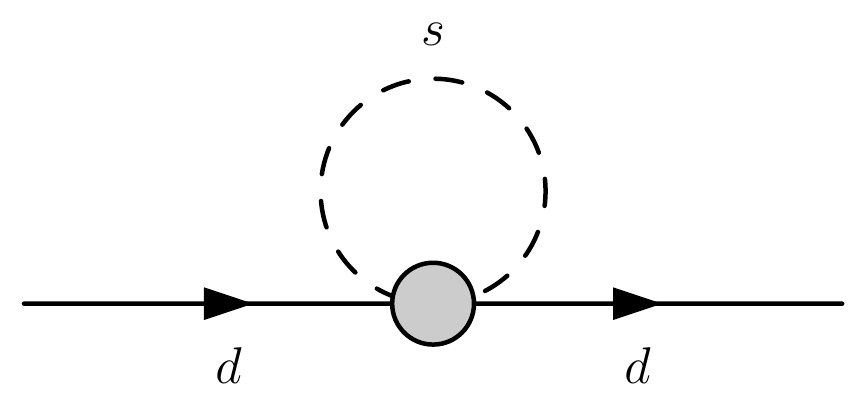}
 \includegraphics[width=0.20\columnwidth]{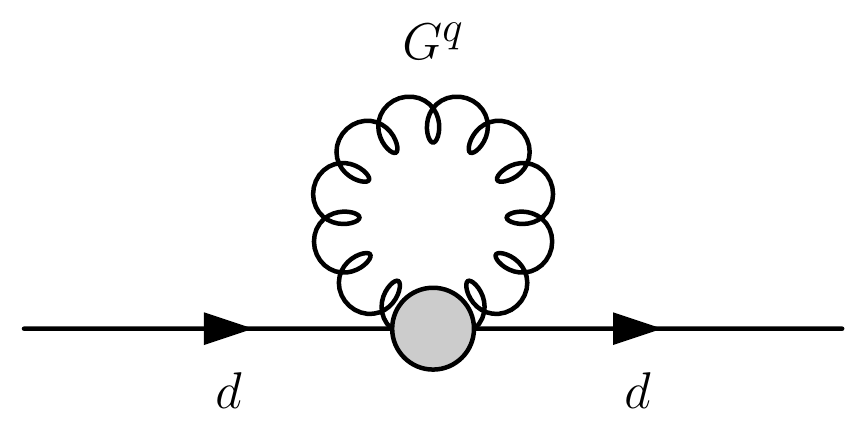}
 \includegraphics[width=0.20\columnwidth]{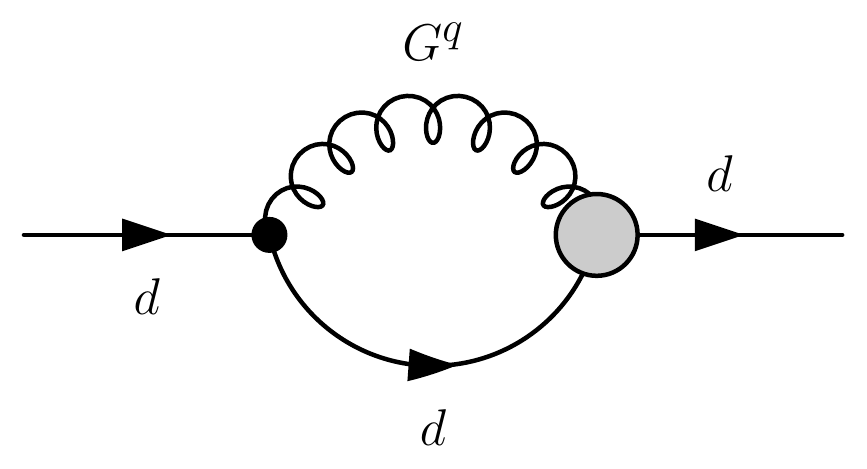}
 \includegraphics[width=0.20\columnwidth]{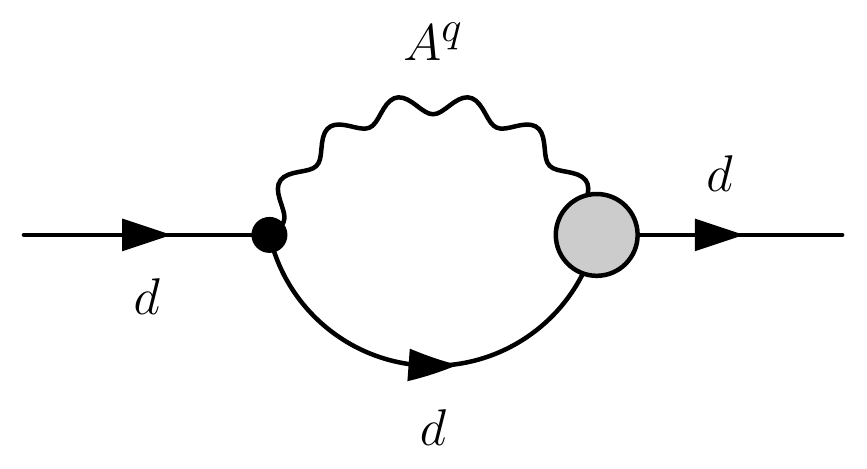}
 \includegraphics[width=0.20\columnwidth]{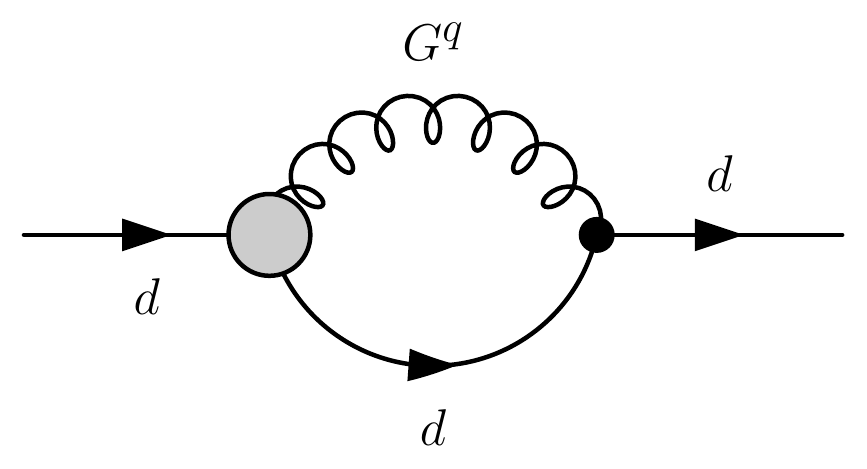}
 \includegraphics[width=0.20\columnwidth]{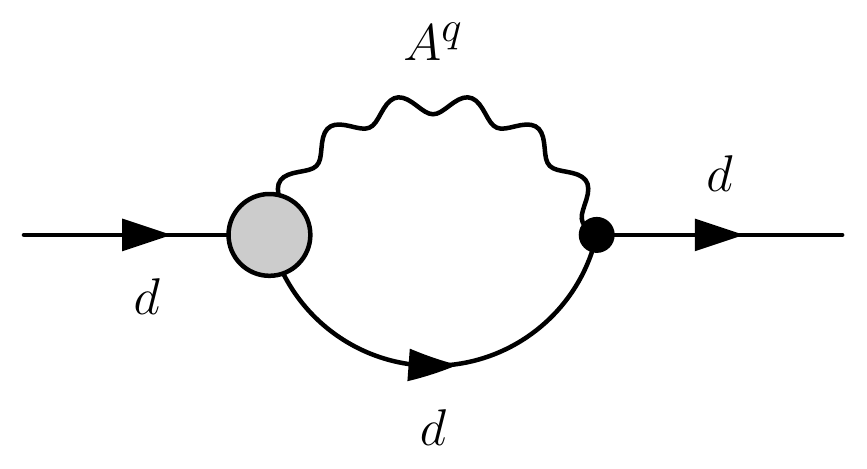}
 \caption{\it Feynman diagrams for $d(p_1)\to d(p_2)$.}\label{fig:d_d} 
\end{figure}

\begin{figure}[H]
 \centering
 \includegraphics[width=0.20\columnwidth]{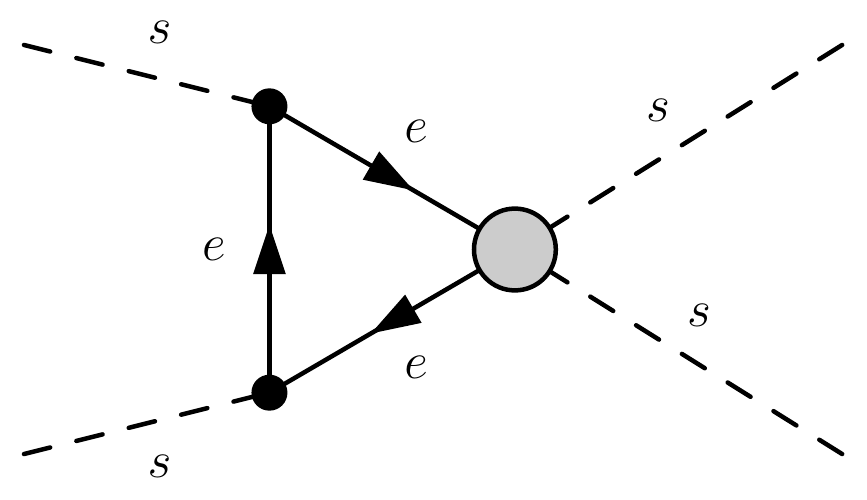}
 \includegraphics[width=0.20\columnwidth]{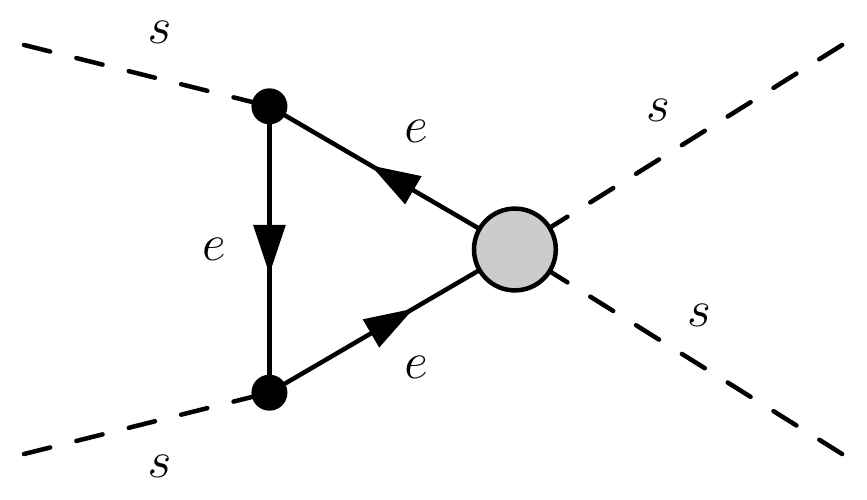}
 \includegraphics[width=0.20\columnwidth]{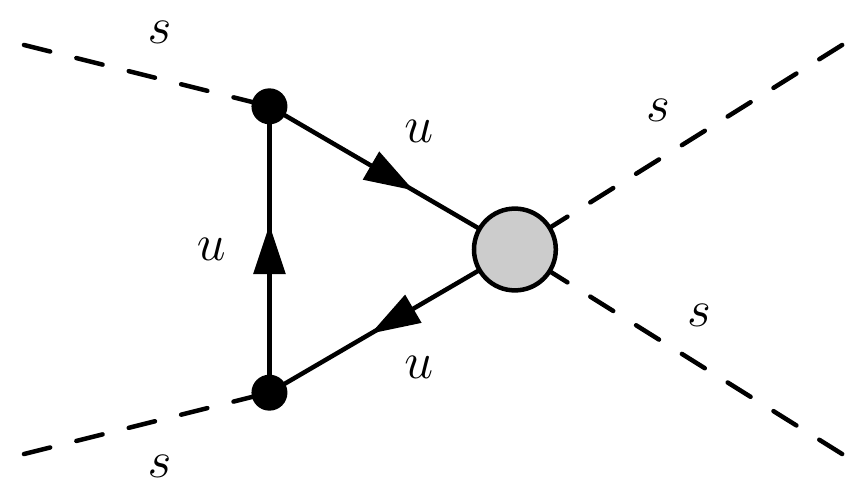}
 \includegraphics[width=0.20\columnwidth]{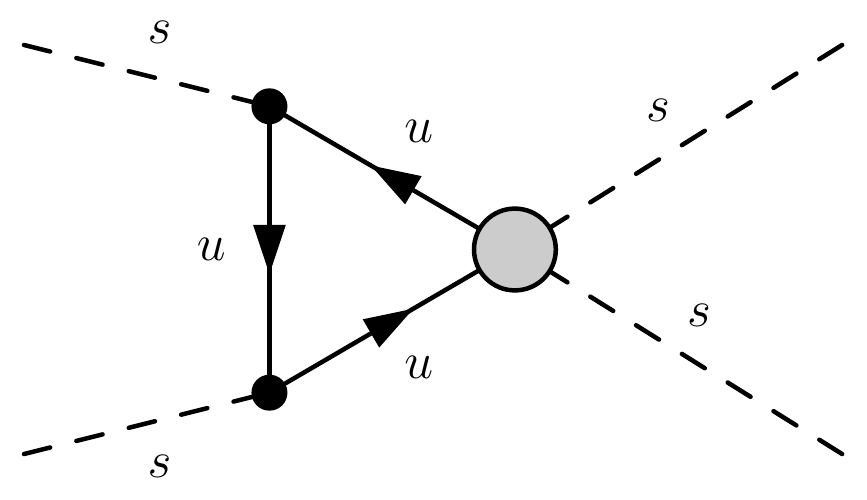}\\[0.5cm]
 \includegraphics[width=0.20\columnwidth]{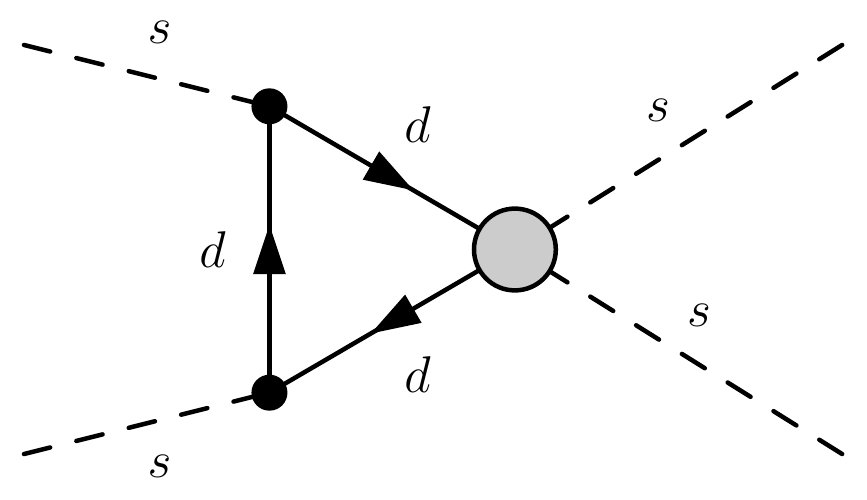}
 \includegraphics[width=0.20\columnwidth]{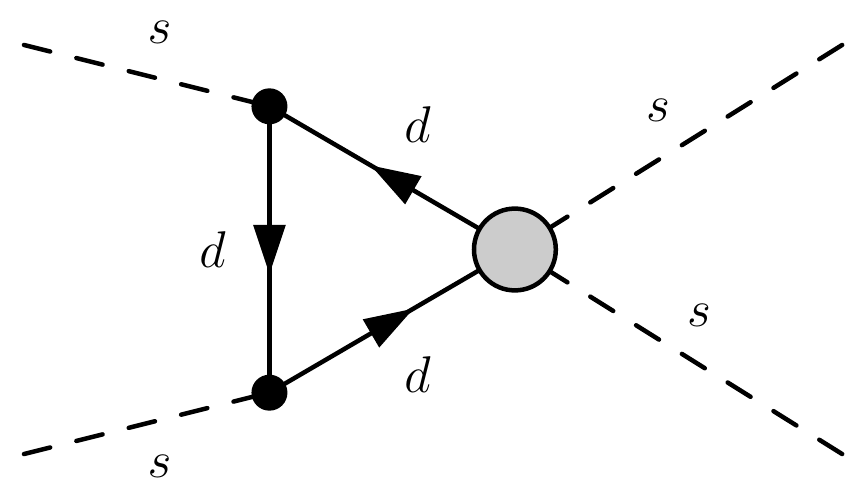}
 \includegraphics[width=0.20\columnwidth]{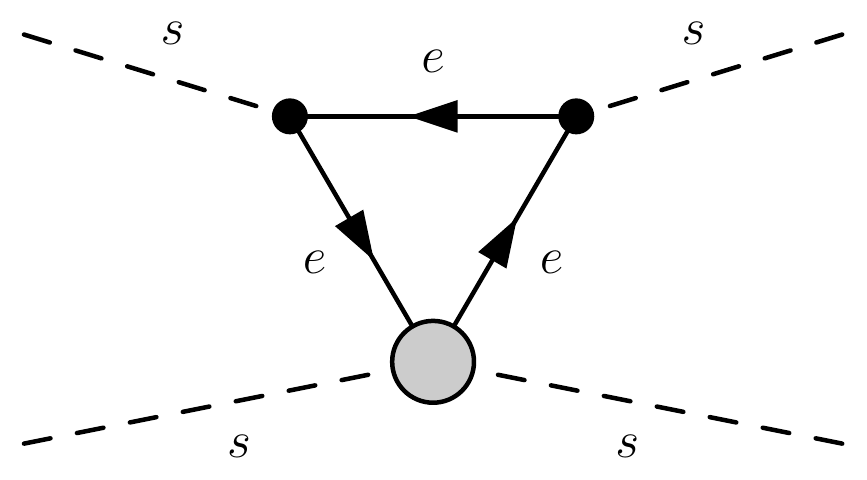}
 \includegraphics[width=0.20\columnwidth]{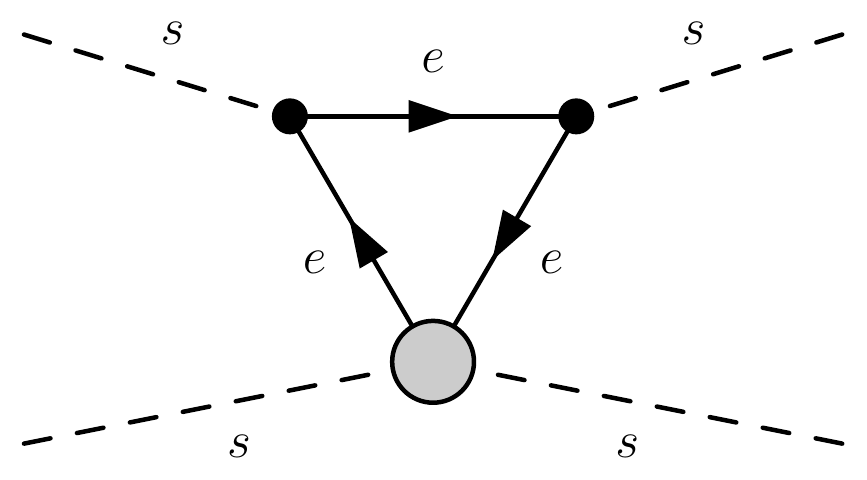}\\[0.5cm]
 \includegraphics[width=0.20\columnwidth]{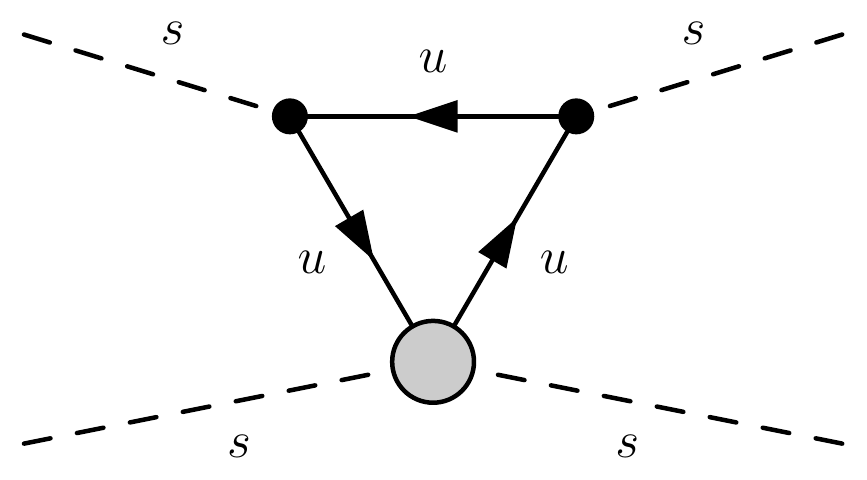}
 \includegraphics[width=0.20\columnwidth]{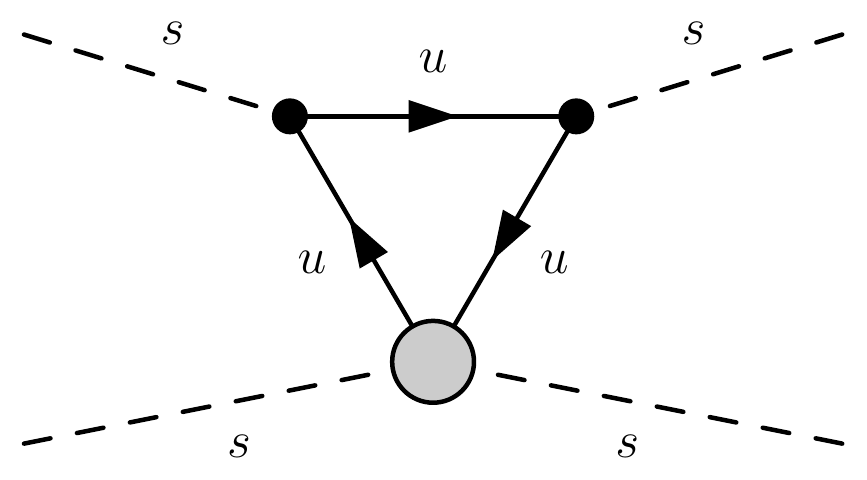}
 \includegraphics[width=0.20\columnwidth]{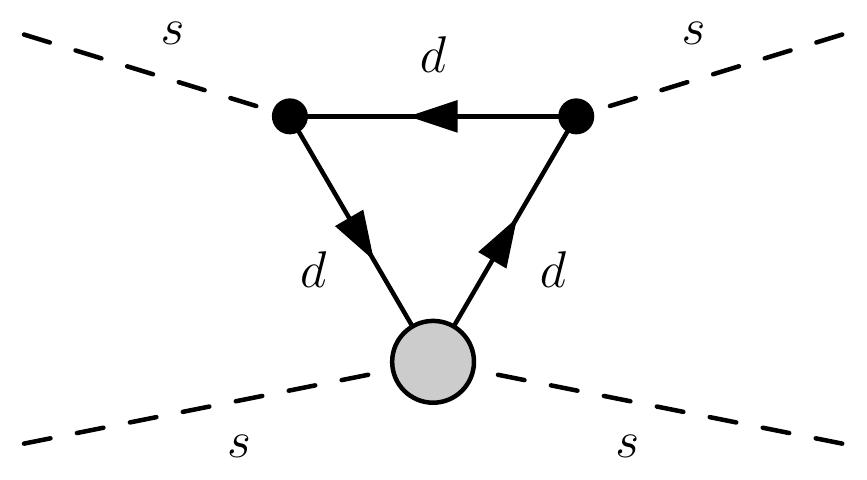}
 \includegraphics[width=0.20\columnwidth]{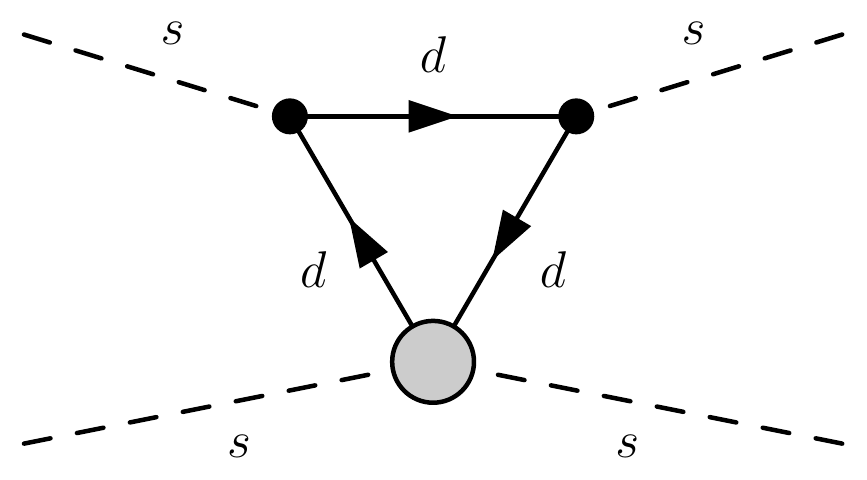}\\[0.5cm]
 \includegraphics[width=0.20\columnwidth]{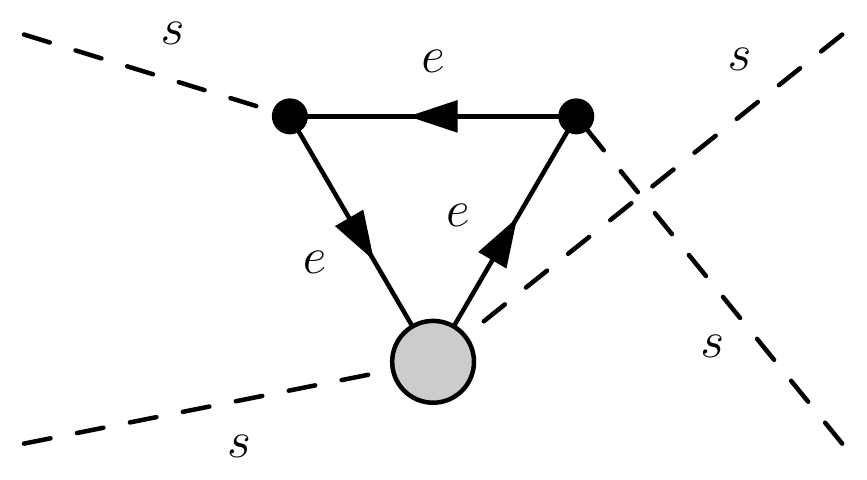}
 \includegraphics[width=0.20\columnwidth]{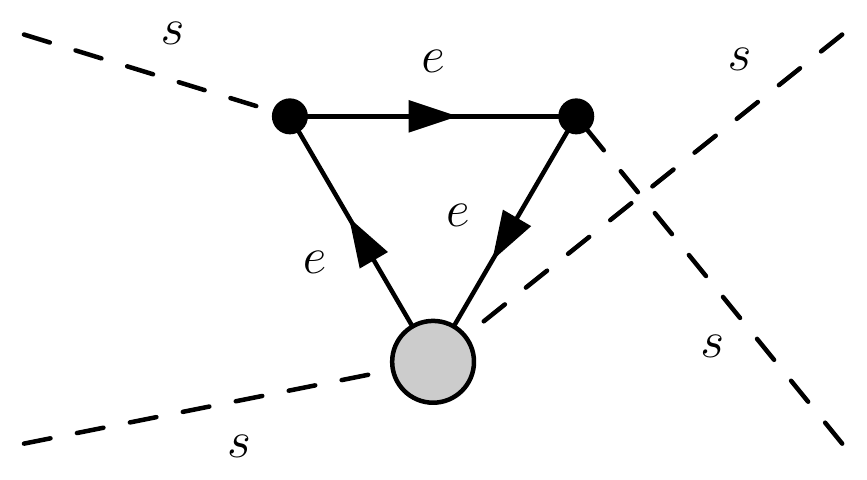}
 \includegraphics[width=0.20\columnwidth]{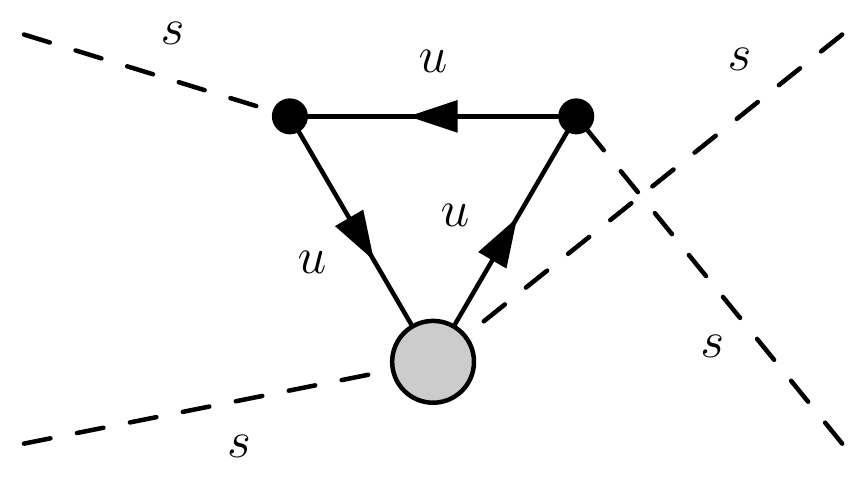}
 \includegraphics[width=0.20\columnwidth]{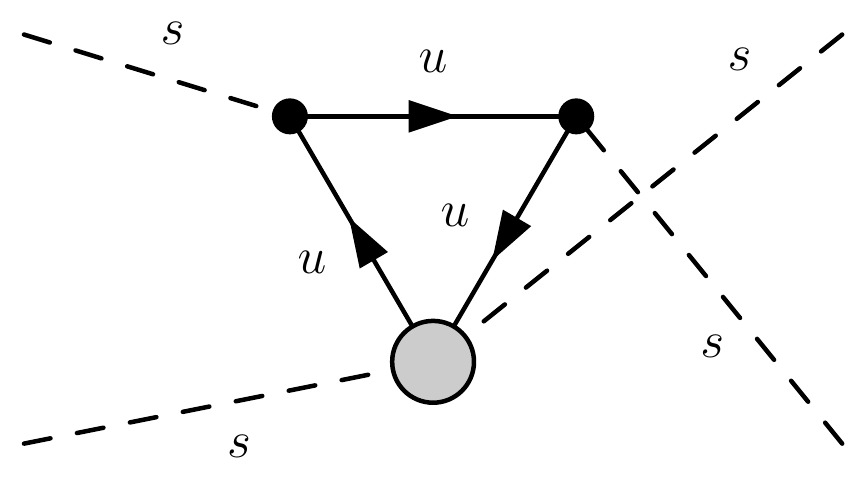}\\[0.5cm]
 \includegraphics[width=0.20\columnwidth]{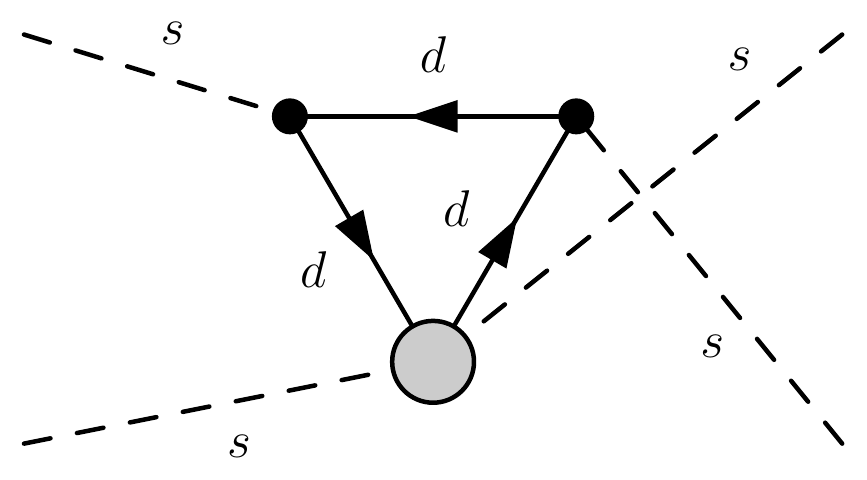}
 \includegraphics[width=0.20\columnwidth]{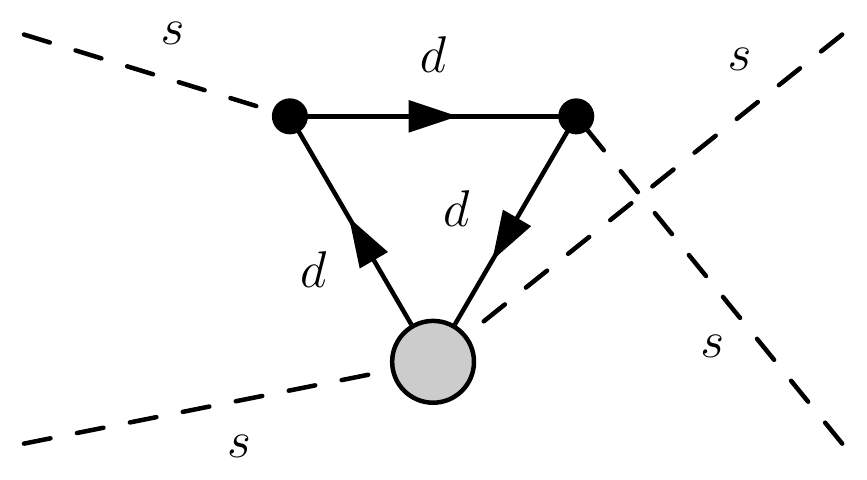}
 \includegraphics[width=0.20\columnwidth]{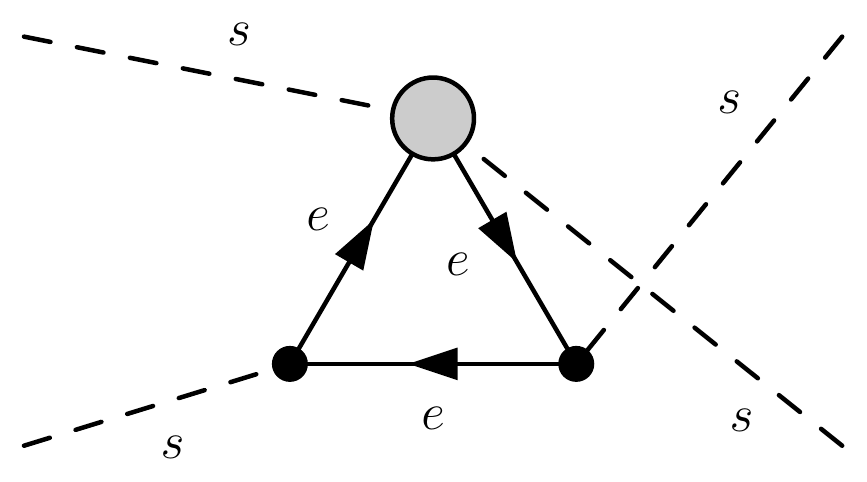}
 \includegraphics[width=0.20\columnwidth]{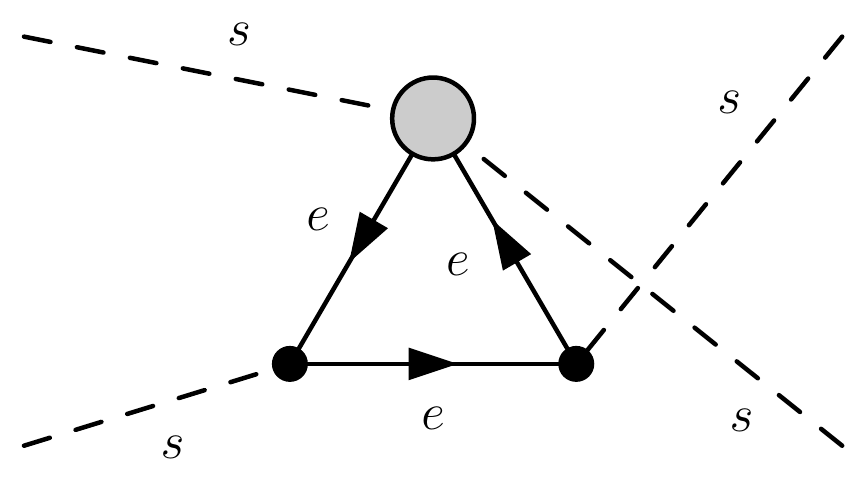}\\[0.5cm]
 \includegraphics[width=0.20\columnwidth]{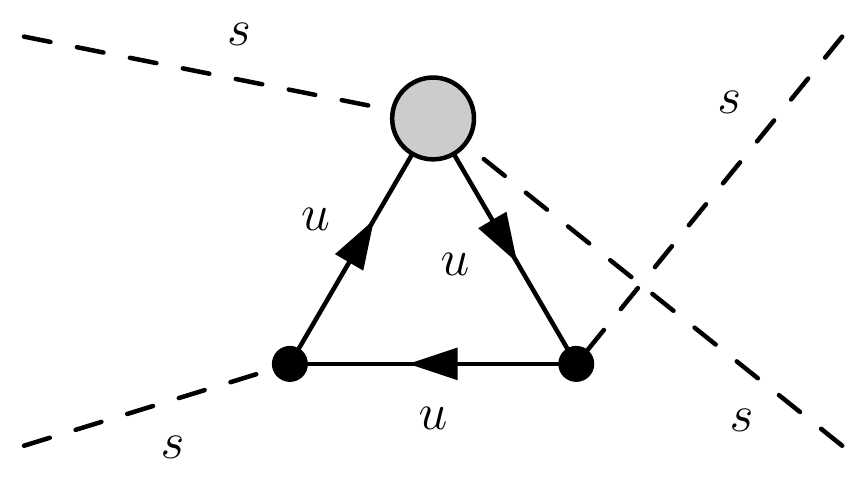}
 \includegraphics[width=0.20\columnwidth]{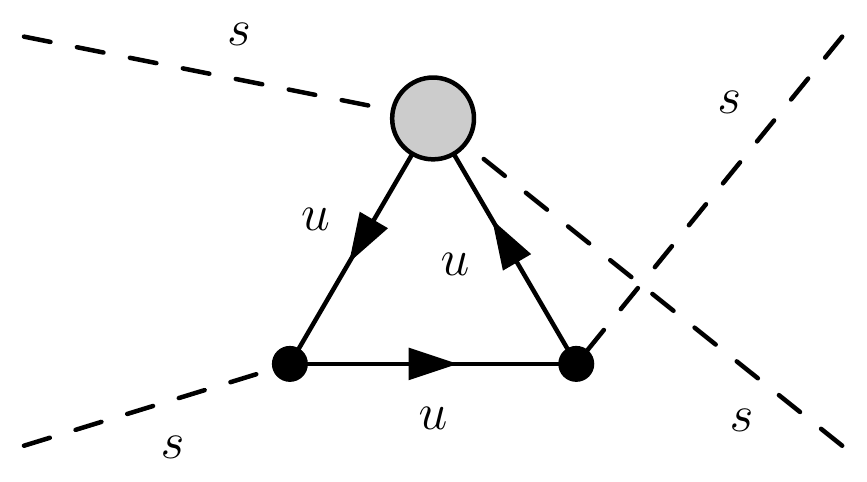}
 \includegraphics[width=0.20\columnwidth]{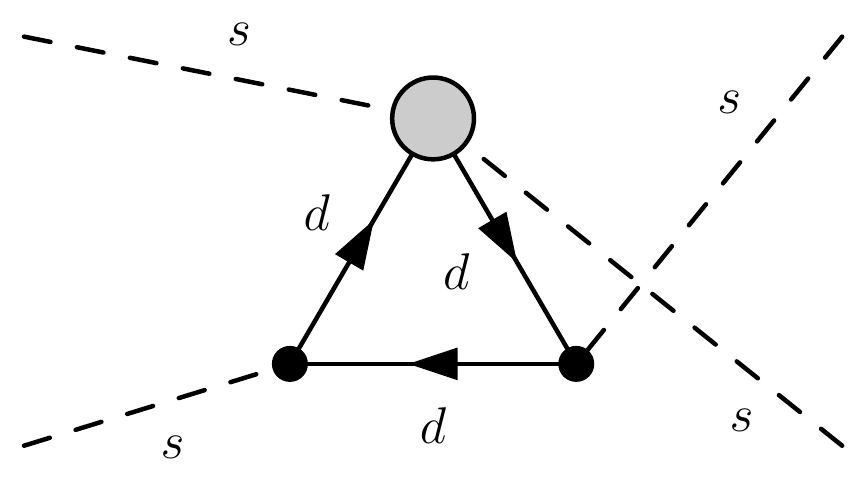}
 \includegraphics[width=0.20\columnwidth]{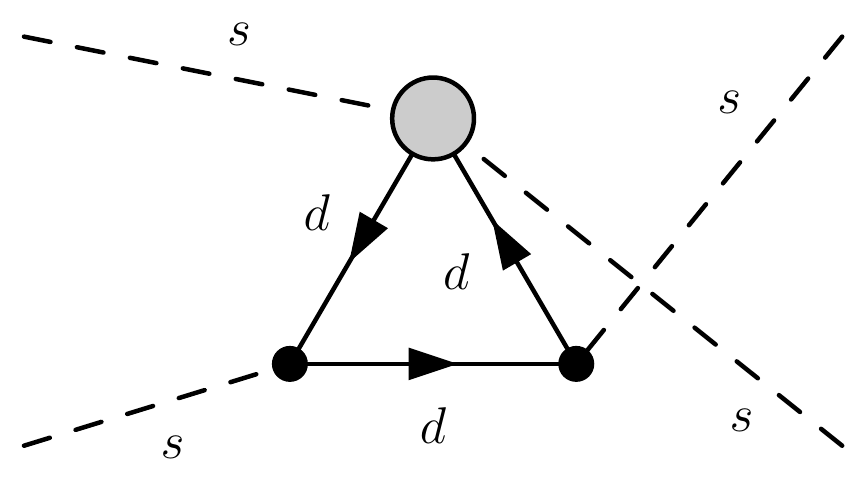}\\[0.5cm]
 \includegraphics[width=0.20\columnwidth]{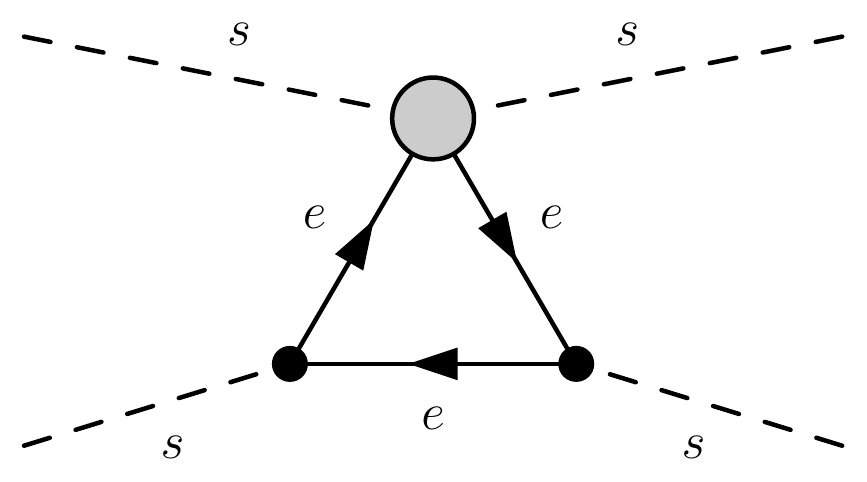}
 \includegraphics[width=0.20\columnwidth]{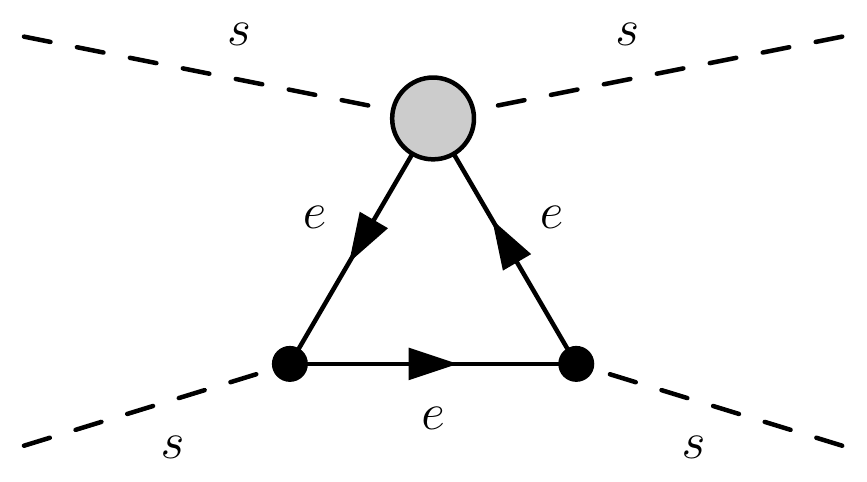}
 \includegraphics[width=0.20\columnwidth]{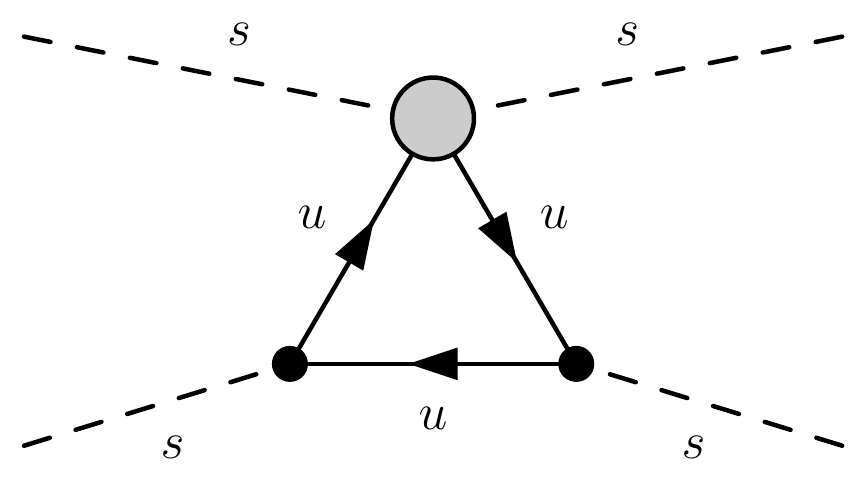}
 \includegraphics[width=0.20\columnwidth]{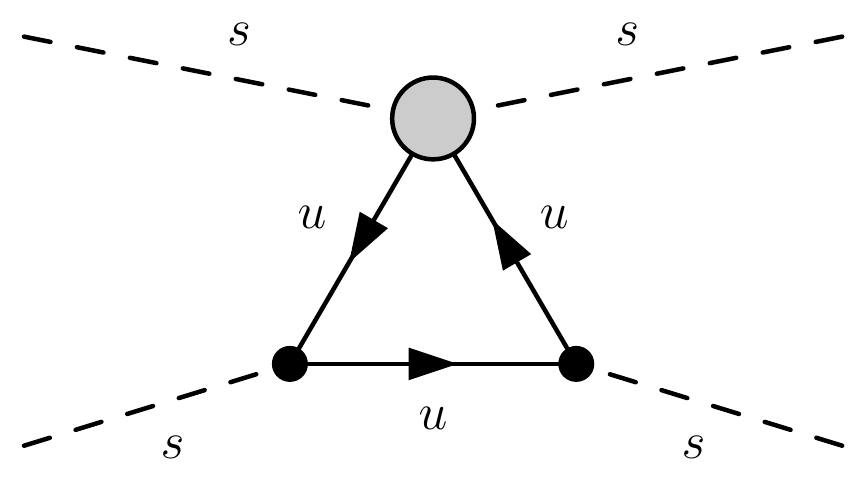}\\[0.5cm]
 \includegraphics[width=0.20\columnwidth]{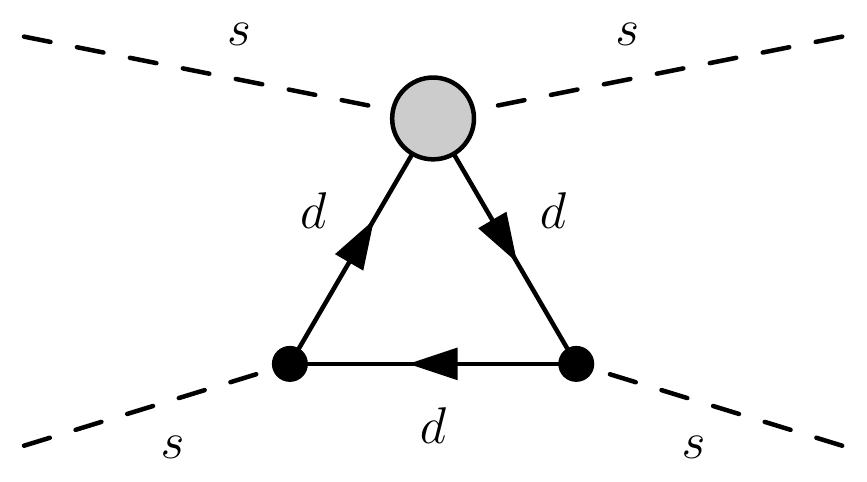}
 \includegraphics[width=0.20\columnwidth]{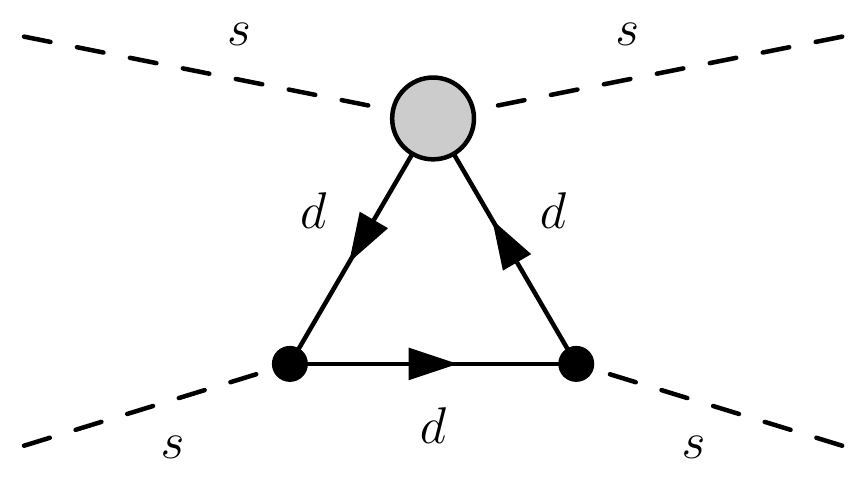}
 \includegraphics[width=0.20\columnwidth]{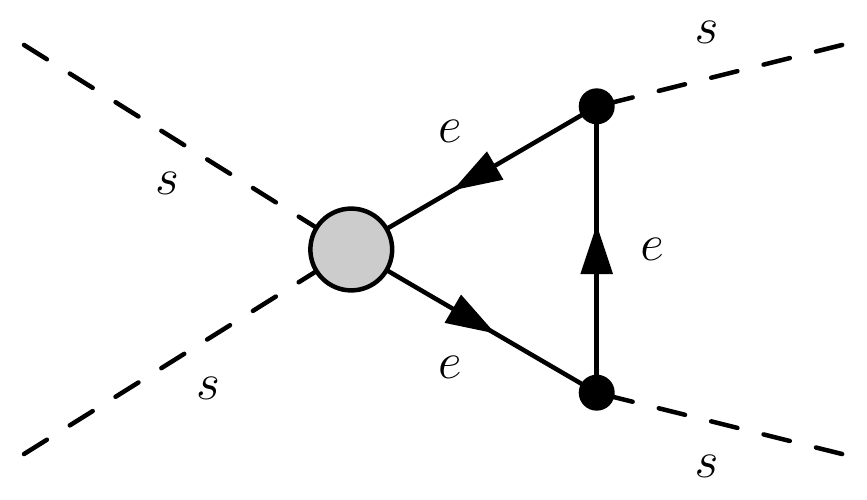}
 \includegraphics[width=0.20\columnwidth]{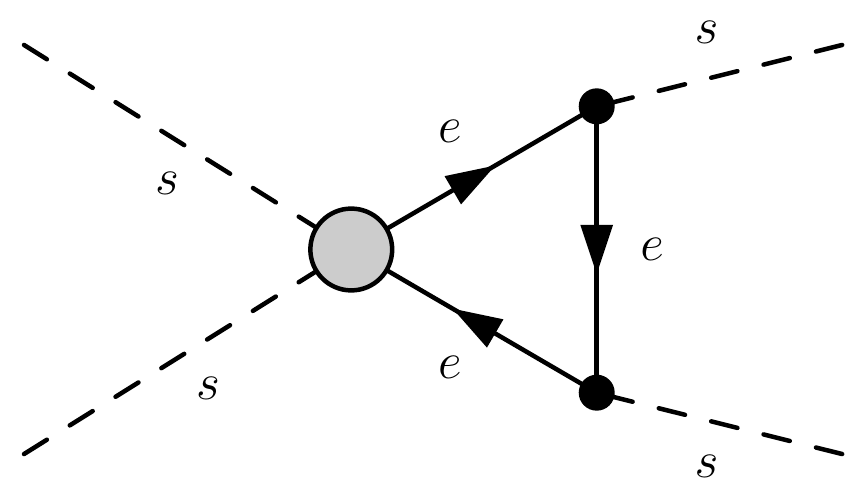}\\[0.5cm]
 \includegraphics[width=0.20\columnwidth]{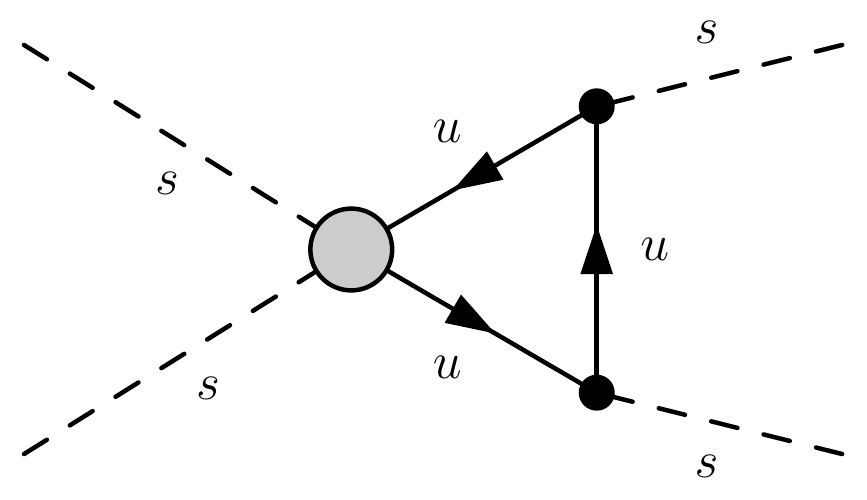}
 \includegraphics[width=0.20\columnwidth]{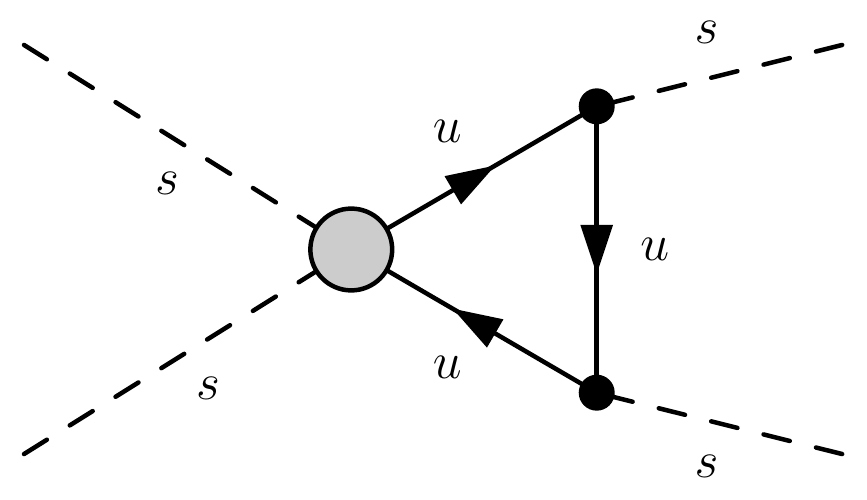}
 \includegraphics[width=0.20\columnwidth]{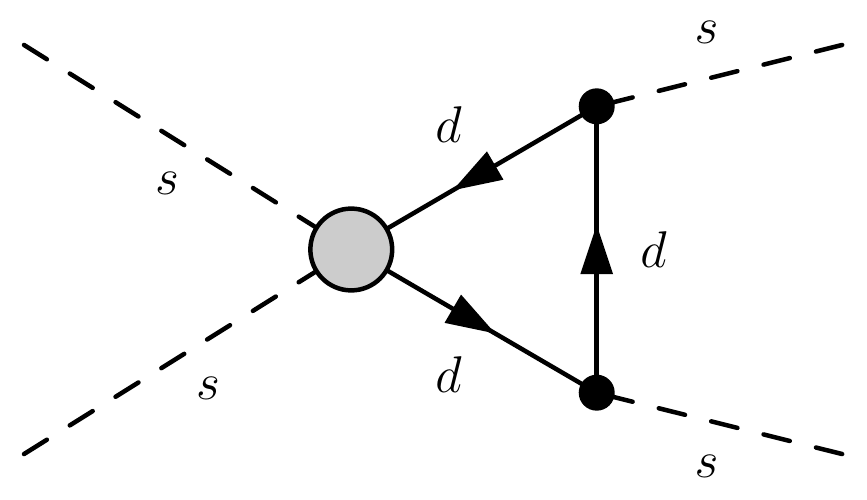}
 \includegraphics[width=0.20\columnwidth]{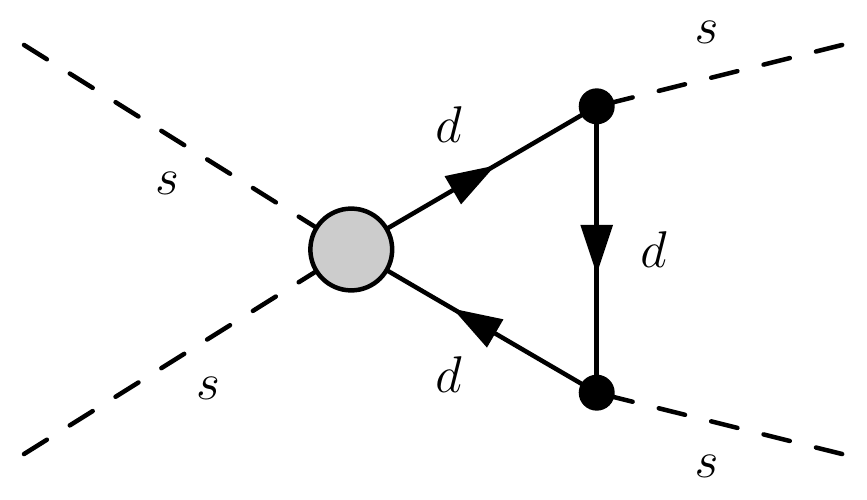}
 \caption{\it Feynman diagrams for $s(p_1)s(p_2)\to s(p_3)s(p_4)$.}\label{fig:ss_ss}
\end{figure}

\begin{figure}[H]
 \centering
 \includegraphics[width=0.20\columnwidth]{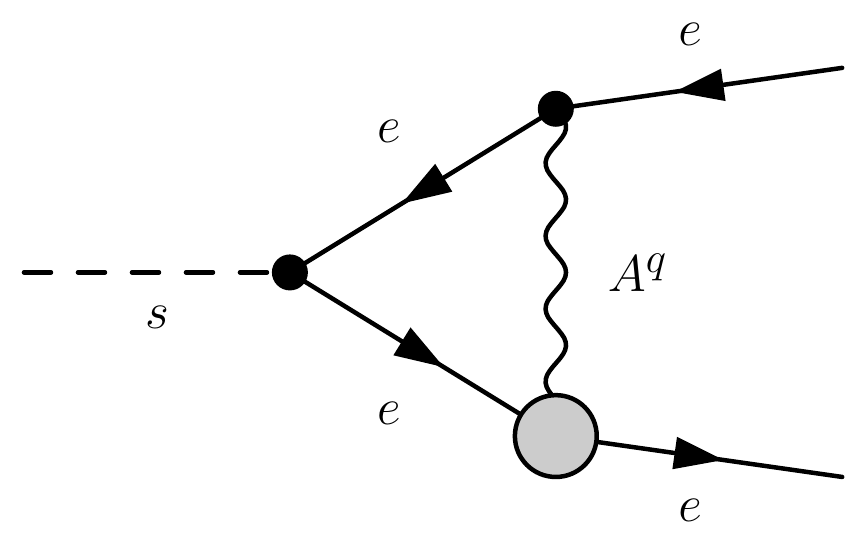}
 \includegraphics[width=0.20\columnwidth]{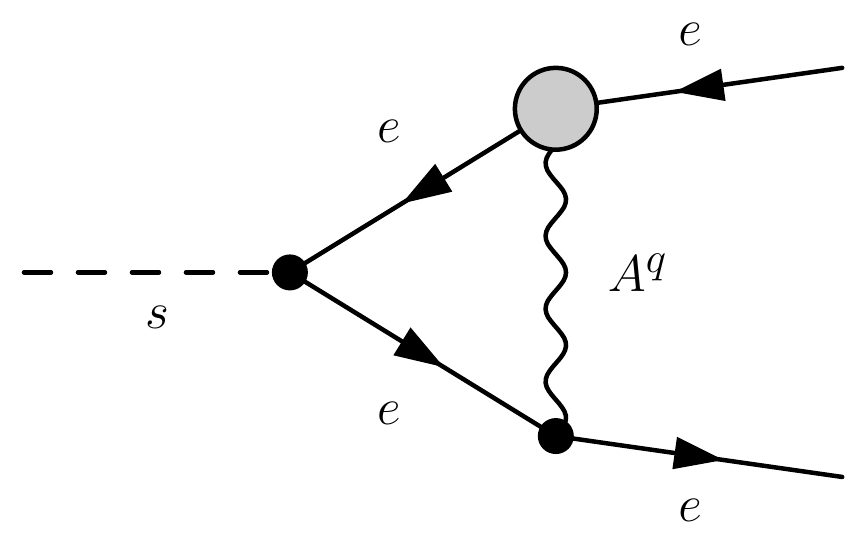}
 \includegraphics[width=0.20\columnwidth]{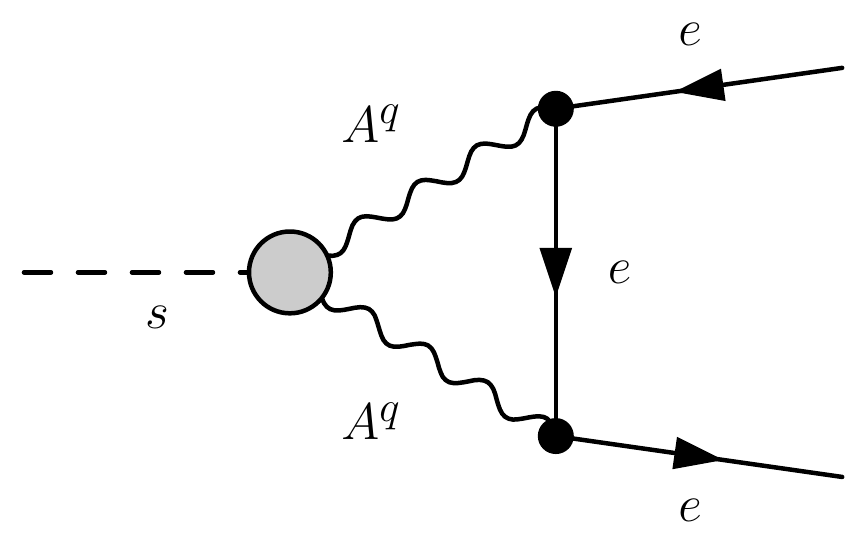}\\
 \includegraphics[width=0.20\columnwidth]{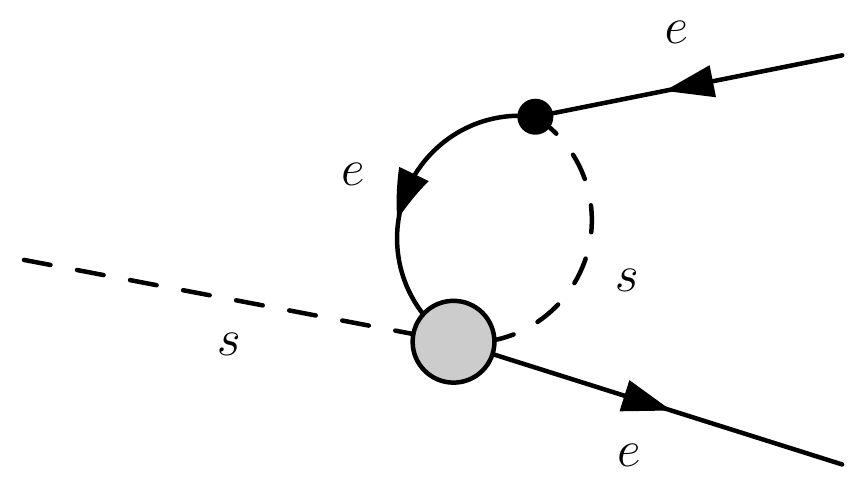}
 \includegraphics[width=0.20\columnwidth]{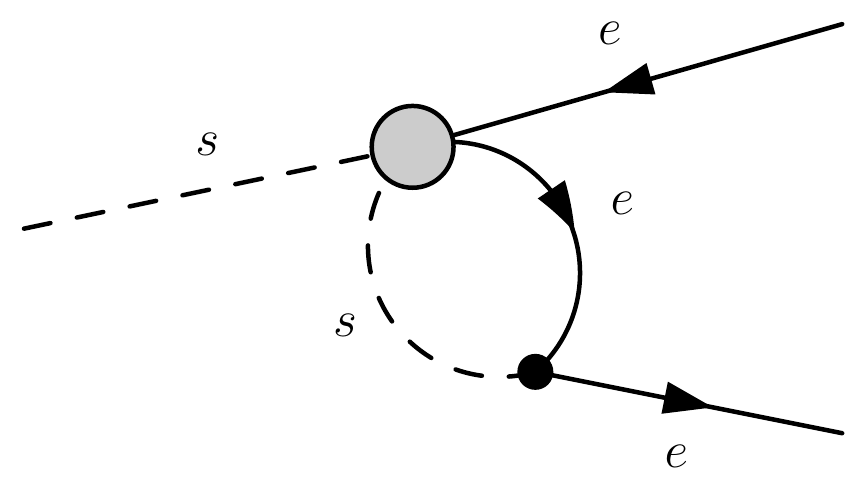}
 \caption{\it Feynman diagrams for $s(p_1)\to\overline{e}(p_2) e(p_3)$.}\label{fig:s_ee}
\end{figure}

\begin{figure}[H]
 \centering
 \includegraphics[width=0.20\columnwidth]{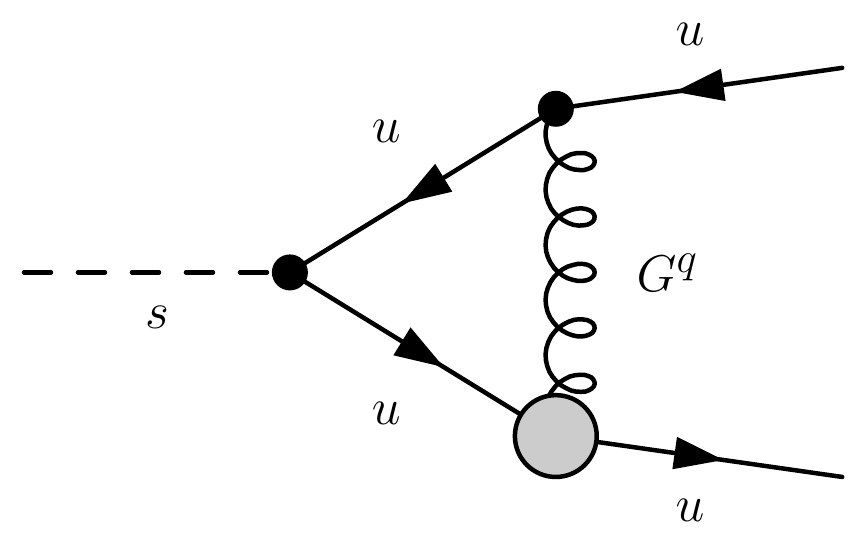}
 \includegraphics[width=0.20\columnwidth]{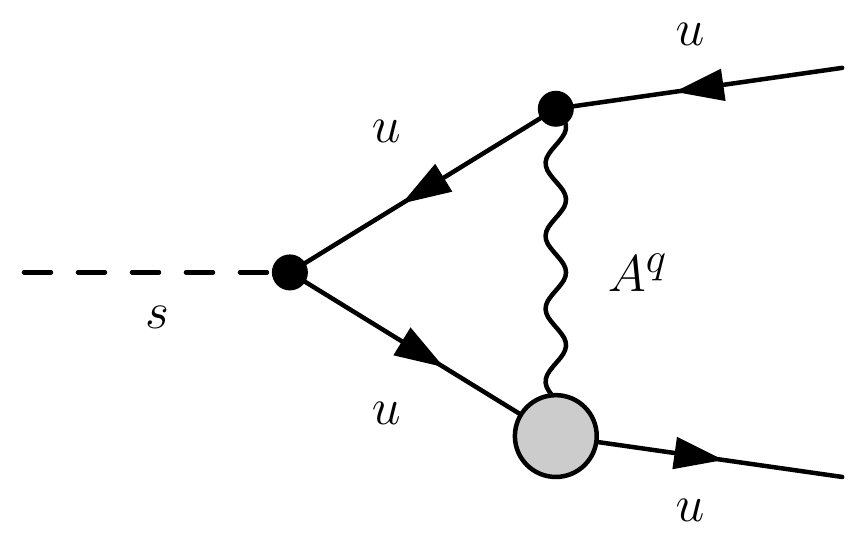}
 \includegraphics[width=0.20\columnwidth]{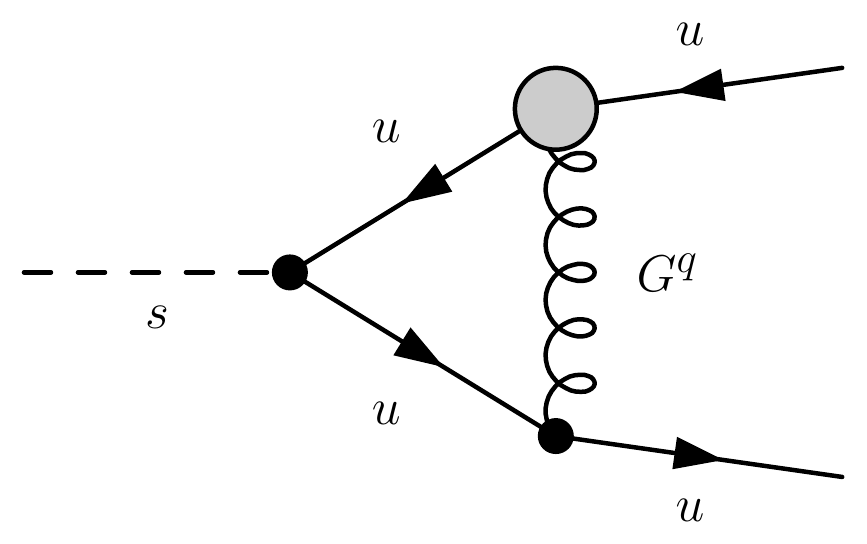}
 \includegraphics[width=0.20\columnwidth]{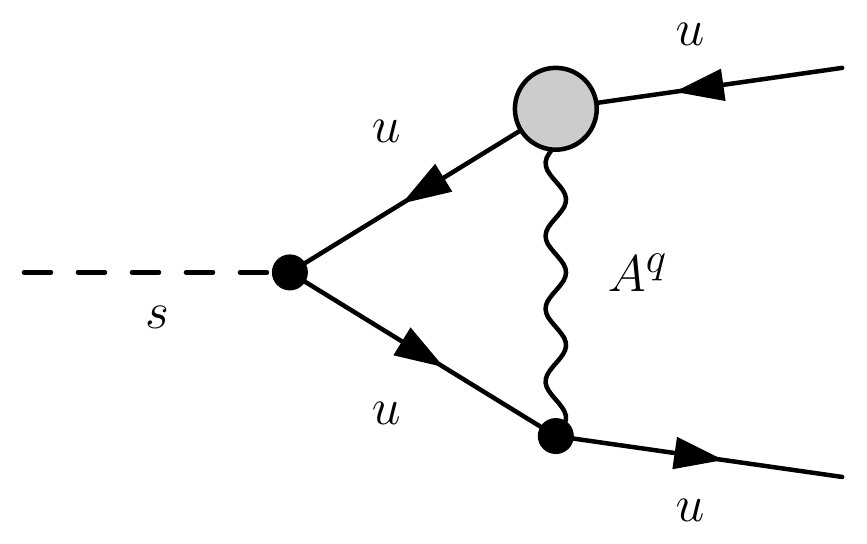}
 \includegraphics[width=0.20\columnwidth]{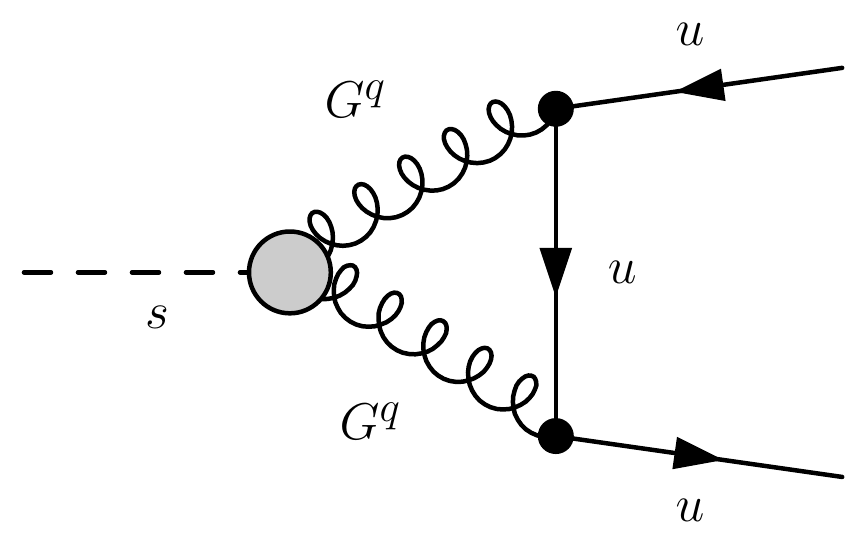}
 \includegraphics[width=0.20\columnwidth]{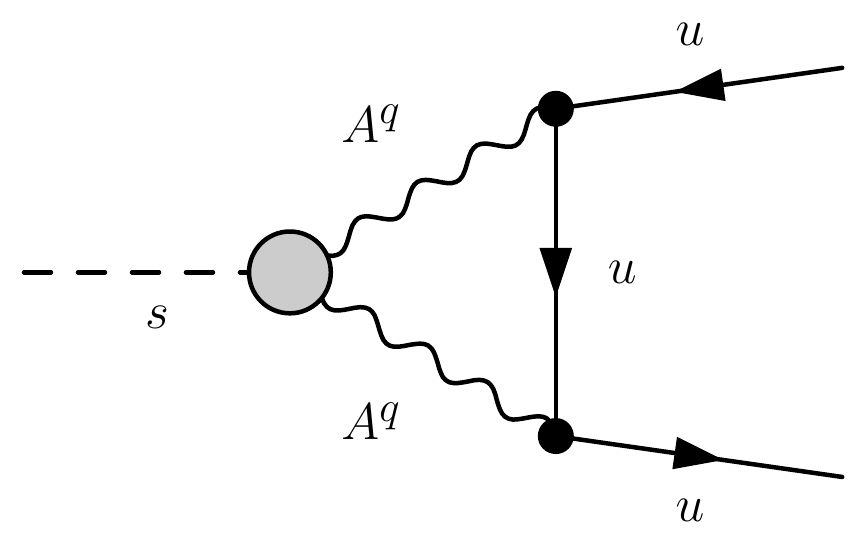}
 \includegraphics[width=0.20\columnwidth]{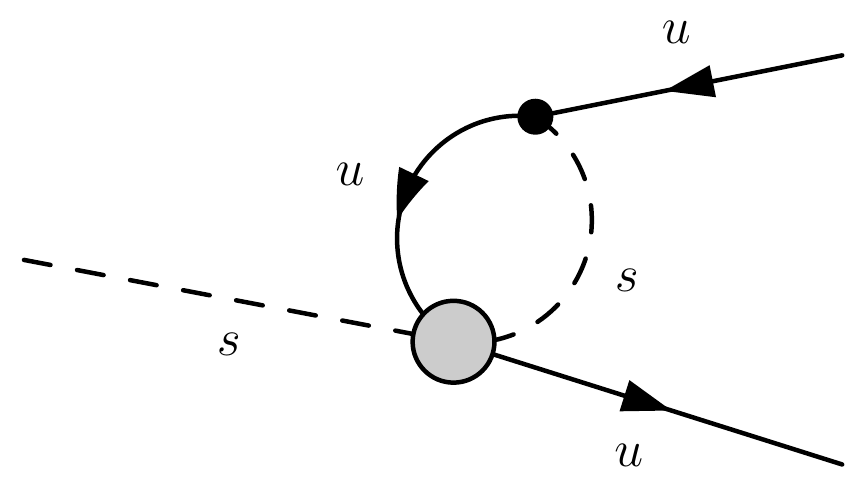}
 \includegraphics[width=0.20\columnwidth]{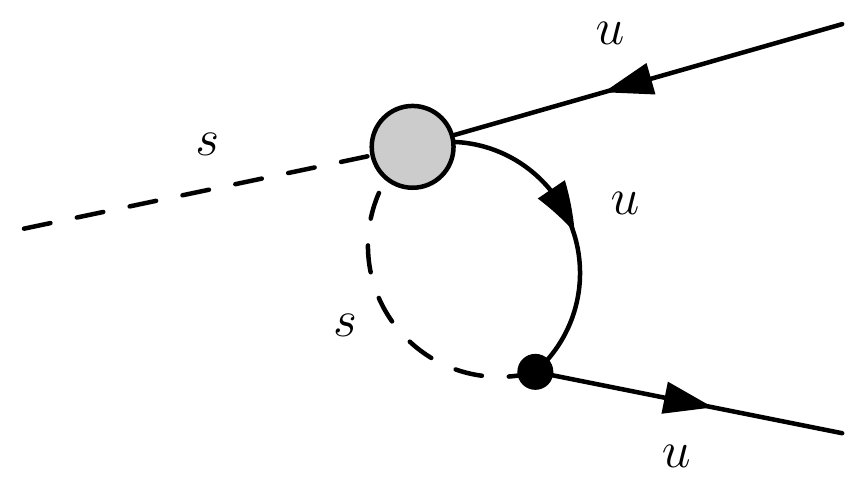}
 \caption{\it Feynman diagrams for $s(p_1)\to\overline{u}(p_2) u(p_3)$.}\label{fig:s_uu}
\end{figure}

\begin{figure}[H]
 \centering
 \includegraphics[width=0.20\columnwidth]{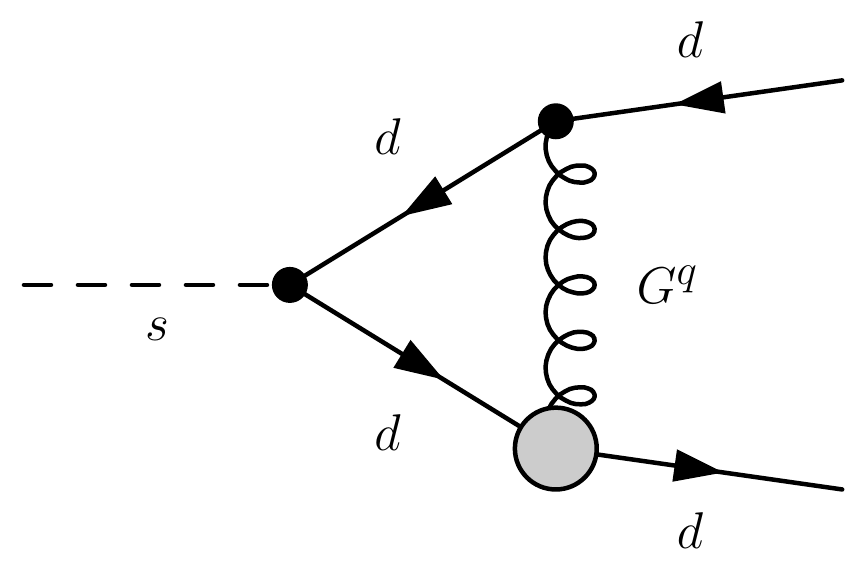}
 \includegraphics[width=0.20\columnwidth]{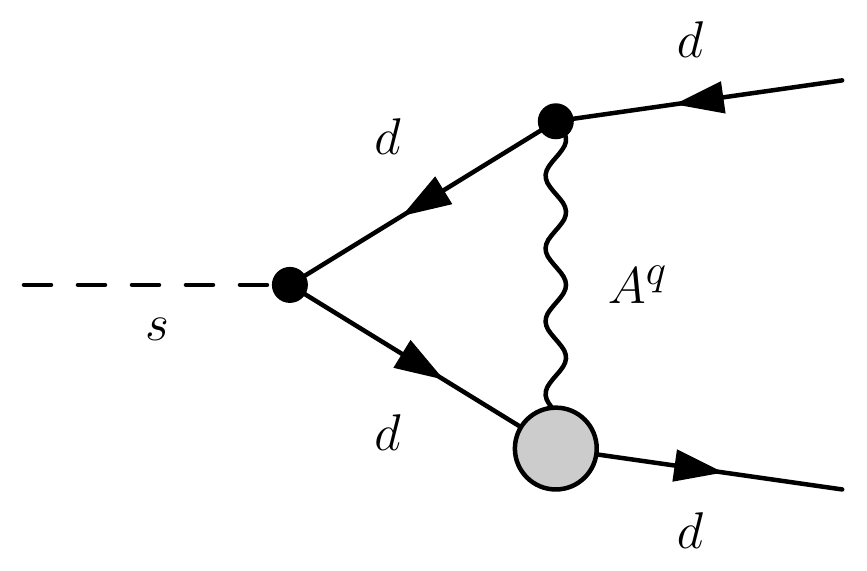}
 \includegraphics[width=0.20\columnwidth]{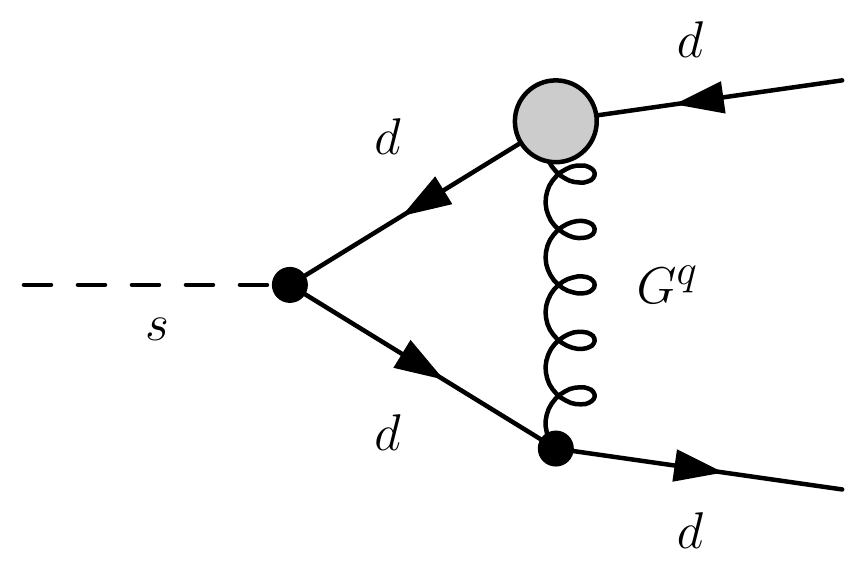}
 \includegraphics[width=0.20\columnwidth]{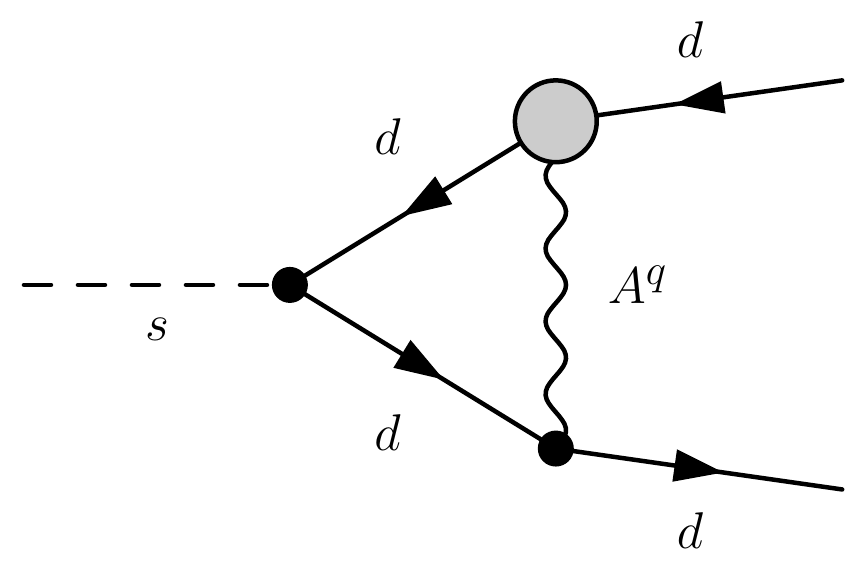}
 \includegraphics[width=0.20\columnwidth]{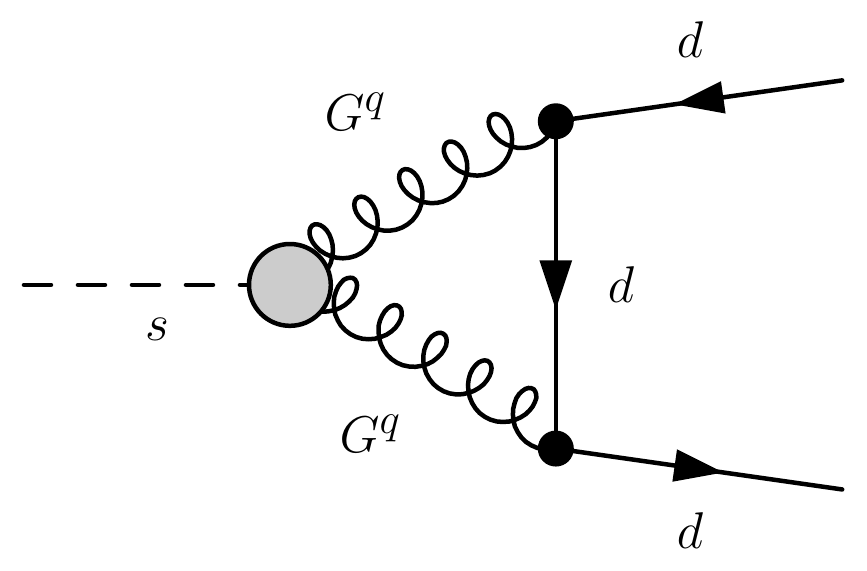}
 \includegraphics[width=0.20\columnwidth]{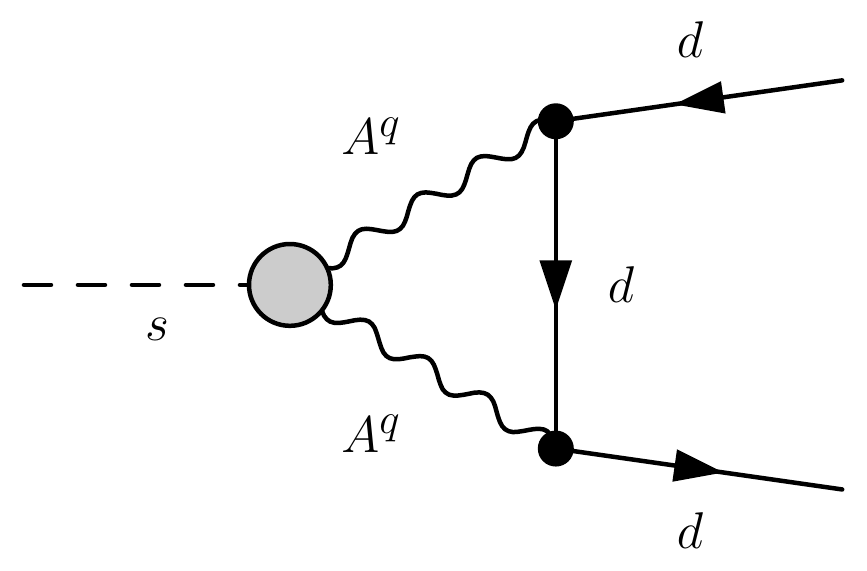}
 \includegraphics[width=0.20\columnwidth]{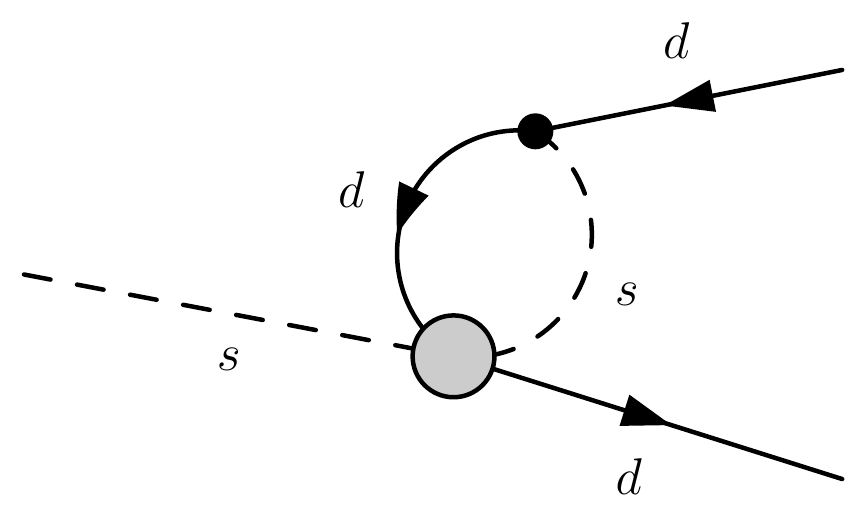}
 \includegraphics[width=0.20\columnwidth]{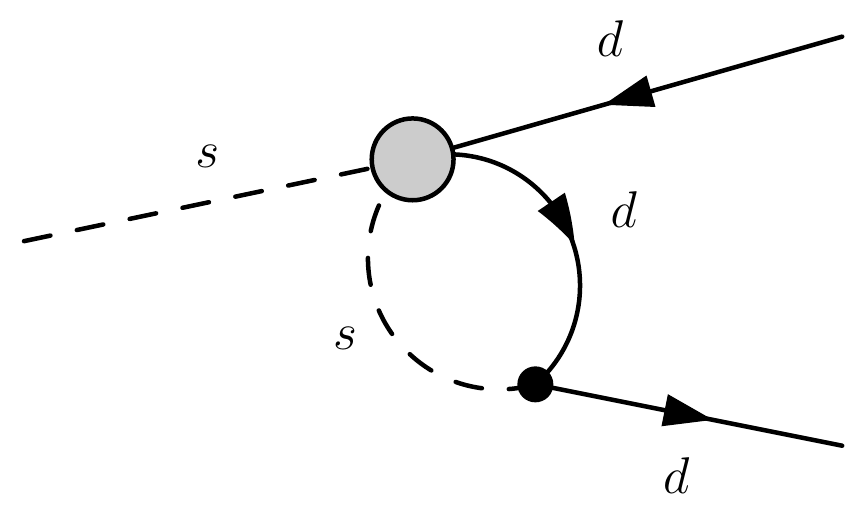}
 \caption{\it Feynman diagrams for $s(p_1)\to\overline{d}(p_2) d(p_3)$.}\label{fig:s_dd}
\end{figure}

Similar to how we proceeded in Section~\ref{sec:divergences}, we fix
the ALP LEFT divergences by computing a reduced set of 1PI
amplitudes, the $1/v$ term of which we reproduce below. Since the
dimension-five operators mix into renormalizable ones, we start with
the divergences that can be absorbed in the renormalizable
operators. In particular, the divergences associated to the kinetic
terms can be parametrised, at the one-loop order, in terms of the
wave function renormalization factors as follows:
\begin{align}
  \mathcal{L}_{\text{kin}}&=
  \overline{\psi_L} (1-\delta Z_L) \ii \slashed{D} \psi_L
  + \overline{\psi_R} (1-\delta Z_R) \ii \slashed{D} \psi_R
  +\frac{1}{2} (1-\delta  Z_s) (\partial_\mu s)(\partial^\mu s)
  \nonumber \\
  &-\frac{1}{4} (1-\delta Z_A) A_{\mu\nu} A^{\mu\nu}
  -\frac{1}{4} (1-\delta Z_G) G^A_{\mu\nu} G^{A\,\mu\nu}\,,
\end{align}
where the wave function renormalization factors are defined in general
by
\begin{equation}
  Z= 1+\delta Z\,,
\end{equation}
and the relative minus sign is due to the fact that the $Z$ factors
are conventionally defined to absorb, rather than parametrise, the
corresponding divergences.
As discussed above, the wave function renormalization factors have
contributions proportional to only renormalizable couplings (that
contribute to the running of the non-renormalizable ones; see Figs.~\ref{fig:A_A}--\ref{fig:d_d}) and to
dimension-five couplings (that contribute to the mixing into
renormalizable ones). We obtain the following result:
\begin{align}
& Z_{e_L}    =   1  - \frac{\alpha}{4 \pi \epsilon} -
  \frac{1}{32\pi^2\epsilon} \left(\tilde{c}_e \tilde{c}_e^{\dagger}
  \right) - \frac{3 \tilde{e} }{16\pi^2\epsilon} \left( \tilde{m}_e
  \tilde{a}_{eA}^\dagger + \tilde{a}_{eA} \tilde{m}_e^\dagger \right)\,,
  \label{ZeL:offshell}  \\ 
& Z_{e_R}  =  1  - \frac{\alpha}{4 \pi \epsilon} -
  \frac{1}{32\pi^2\epsilon} \left(\tilde{c}_e^{\dagger} \tilde{c}_e
  \right) - \frac{3 \tilde{e} }{16\pi^2\epsilon} \left(
  \tilde{a}_{eA}^\dagger \tilde{m}_e + \tilde{m}_e^\dagger  \tilde{a}_{eA} \right)\,,
  \label{ZeR:offshell}\\ 
& Z_{d_{L}} = 1  - \frac{1}{3 \pi \epsilon}\bigg[ \frac{1}{12} \alpha
    + \alpha_s \bigg]   - \frac{1}{32\pi^2\epsilon} \left(\tilde{c}_d
  \tilde{c}_d^{\dagger} \right)
  \nonumber
  \\
& ~~~~~~~ -\frac{ \tilde{e}  }{16\pi^2\epsilon} \left( \tilde{m}_d
  \tilde{a}_{dA}^\dagger + \tilde{a}_{dA} \tilde{m}_d^\dagger
  \right)+\frac{\tilde{g}_3}{4\pi^2\epsilon} \left( \tilde{m}_d \tilde{a}_{dG}^\dagger
  + \tilde{a}_{dG}\tilde{m}_d^\dagger \right) \,, \label{ZdL:offshell}\\
& Z_{d_{R}} = 1  - \frac{1}{3 \pi \epsilon}\bigg[ \frac{1}{12} \alpha
    + \alpha_s \bigg]   - \frac{1}{32\pi^2\epsilon} \left(
  \tilde{c}_d^{\dagger} \tilde{c}_d \right)    \nonumber\\
& ~~~~~~~ -\frac{\tilde{e} }{16\pi^2\epsilon}
  \left(\tilde{a}_{dA}^\dagger \tilde{m}_d  + \tilde{m}_d^\dagger \tilde{a}_{dA}
  \right)+\frac{\tilde{g}_3}{4\pi^2\epsilon} \left(  \tilde{a}_{dG}^\dagger
  \tilde{m}_d + \tilde{m}_d^\dagger \tilde{a}_{dG} \right) \,,
  \label{ZdR:offshell}\\
& Z_{u_{L}} = 1  -\frac{1}{3 \pi \epsilon}\bigg[ \frac{1}{3} \alpha + \alpha_s \bigg]  - \frac{1}{32\pi^2\epsilon} \left(\tilde{c}_u \tilde{c}_u^{\dagger} \right) \nonumber  \\
& ~~~~~~~+ \frac{2 \tilde{e} }{16\pi^2\epsilon} \left( \tilde{m}_u
  \tilde{a}_{uA}^\dagger + \tilde{a}_{uA} \tilde{m}_u^\dagger
  \right)+\frac{\tilde{g}_3}{4\pi^2\epsilon} \left( \tilde{m}_u \tilde{a}_{uG}^\dagger
  + \tilde{a}_{uG}\tilde{m}_u^\dagger \right)\,,
  \label{ZuL:offshell}
  \end{align}
  \begin{align}
  & Z_{u_{R}} = 1  -\frac{1}{3 \pi \epsilon}\bigg[ \frac{1}{3} \alpha +
    \alpha_s \bigg]  - \frac{1}{32\pi^2\epsilon} \left(
  \tilde{c}_u^{\dagger}\tilde{c}_u \right)   \nonumber \\
& ~~~~~~~+ \frac{2 \tilde{e} }{16\pi^2\epsilon} \left(
  \tilde{a}_{uA}^\dagger \tilde{m}_u + \tilde{m}_u^\dagger \tilde{a}_{uA}
  \right)+\frac{\tilde{g}_3}{4\pi^2\epsilon} \left( \tilde{a}_{uG}^\dagger \tilde{m}_u
  + \tilde{m}_u^\dagger \tilde{a}_{uG} \right)\,,
  \label{ZuR:offshell}\\
& Z_{A} = 1 - \frac{\alpha}{3 \pi \epsilon} \left[ n_\ell + \frac{1}{3} n_d + \frac{4}{3} n_u \right]\nonumber \\
& ~~~~~~~ +\frac{\tilde{e}}{2\pi^2\epsilon} {\rm Tr} \left[ (\tilde{a}_{eA}^\dagger \tilde{m}_e + \tilde{m}_e^\dagger \tilde{a}_{eA}) - 2 (\tilde{a}_{uA}^\dagger \tilde{m}_u + \tilde{m}_u^\dagger \tilde{a}_{uA}) + (\tilde{a}_{dA}^\dagger \tilde{m}_d + \tilde{m}_d^\dagger \tilde{a}_{dA})\right]\,, \\
& Z_{G} = 1 + \frac{\alpha_s}{4\pi \epsilon} \left[11 - \frac{2}{3} (n_u+n_d) \right]  - \frac{\tilde{g}_3}{4\pi^2\epsilon} {\rm Tr} \left[ \tilde{a}_{dG}^\dagger \tilde{m}_d + \tilde{m}_d^\dagger \tilde{a}_{dG} + \tilde{a}_{uG}^\dagger \tilde{m}_u + \tilde{m}_u^\dagger \tilde{a}_{uG} \right]\,,  \\
& Z_s = 1 - \frac{1}{8\pi^2 \epsilon} {\rm Tr} \left[ \tilde{c}_e
    \tilde{c}_e^{\dagger} + 3 \left( \tilde{c}_d \tilde{c}_d^{\dagger}
    + \tilde{c}_u \tilde{c}_u^{\dagger} \right)\right]\,, \label{Zs:offshell}
\end{align}
where we have defined $\alpha = \tilde{e}^2/(4\pi)$ and $\alpha_s = \tilde{g_3}^2/(4\pi)$.
For the remaining couplings we only need to consider diagrams with a
single insertion of a dimension-five operator. The result is given
below, organised according to the amplitudes we have used to compute
the corresponding divergences. We provide all the relevant diagrams in Appendix~\ref{diagrams:left}.
\begin{itemize}

 \item $s(p_1)\to s(p_2)$ 
 
 The amplitude given by the diagrams in Fig.~\ref{fig:s_s} fixes the
 divergence of the mass term, 
 \begin{align}
  \tilde{m}'^2 & = -\frac{3}{4\pi^2\epsilon}  \left( {\rm Tr} \left[\tilde{m}_d^\dagger \tilde{a}_d \tilde{m}_d^\dagger \tilde{m}_d + \tilde{m}_d^\dagger \tilde{m}_d \tilde{a}_d^{\dagger} \tilde{m}_d \right]  + {\rm Tr} \left[\tilde{m}_u^\dagger \tilde{a}_u \tilde{m}_u^\dagger \tilde{m}_u + \tilde{m}_u^\dagger \tilde{m}_u \tilde{a}_u^{\dagger} \tilde{m}_u \right] \right)\nonumber \\
&  -\frac{1}{4\pi^2\epsilon}  {\rm Tr} \left[\tilde{m}_e^\dagger \tilde{a}_e \tilde{m}_e^\dagger \tilde{m}_e + \tilde{m}_e^\dagger \tilde{m}_e \tilde{a}_e^{\dagger} \tilde{m}_e \right]\,.
 \end{align}

 \item $\psi (p_1)\to\psi (p_2)$
 
 The diagrams for $\psi=e,u,d$ are shown in Figs.~\ref{fig:e_e}, \ref{fig:u_u}, \ref{fig:d_d}. They contribute to the fermion mass divergences,
 \begin{align}
  \tilde{m}_e' & = -\frac{3}{8\pi^2 \epsilon}\tilde{e} \left(\tilde{m}_e \tilde{m}_e ^\dagger \tilde{a}_{eA} + \tilde{a}_{eA} \tilde{m}_e^\dagger \tilde{m}_e\right) +\frac{1}{16\pi^2 \epsilon} \tilde{m}^2 \tilde{a}_{e}\,,\\
  \tilde{m}_u' & = \frac{2}{8\pi^2 \epsilon}\tilde{e} \left(\tilde{m}_u \tilde{m}_u ^\dagger \tilde{a}_{uA} + \tilde{a}_{uA} \tilde{m}_u^\dagger \tilde{m}_u\right) + \frac{1}{2\pi^2 \epsilon} \tilde{g}_3 \left( \tilde{m}_u \tilde{m}_u^\dagger \tilde{a}_{uG} + \tilde{a}_{uG} \tilde{m}_u^\dagger \tilde{m}_u	\right)
+\frac{1}{16\pi^2 \epsilon} \tilde{m}^2 \tilde{a}_u\,,\\
  \tilde{m}_d' & = -\frac{1}{8\pi^2 \epsilon}\tilde{e} \left(\tilde{m}_d \tilde{m}_d ^\dagger \tilde{a}_{dA} + \tilde{a}_{dA} \tilde{m}_d^\dagger \tilde{m}_d\right) + \frac{1}{2\pi^2 \epsilon} \tilde{g}_3 \left( \tilde{m}_d \tilde{m}_d^\dagger \tilde{a}_{dG} + \tilde{a}_{dG} \tilde{m}_d^\dagger \tilde{m}_d	\right)
+\frac{1}{16\pi^2 \epsilon} \tilde{m}^2 \tilde{a}_d\,,
 \end{align}
 as well as to the dimension-five contribution to the kinetic terms
 reported above.
 
 \item $s(p_1) s(p_2)\to s (p_3) s(p_4)$
 
 The corresponding amplitude, represented by the diagrams in
 Fig.~\ref{fig:ss_ss}, fixes the ALP quartic coupling, 
 \begin{align}
  \tilde{\lambda}_s' = -\frac{3}{ \pi^2 \epsilon}\bigg[{\rm Tr}^{\tilde{\lambda}}_e + 3\left({\rm Tr}^{\tilde{\lambda}}_u+{\rm Tr}^{\tilde{\lambda}}_d\right)\bigg]\,,
 \end{align}
where we have defined
\begin{equation}
{\rm Tr}^{{\tilde{\lambda}}}_\psi \equiv  {\rm Tr}\bigg[ \tilde{a}_\psi \tilde{c}_\psi^{\dagger} \tilde{c}_\psi \tilde{m}_\psi^\dagger + \tilde{a}_\psi \tilde{m}_\psi^\dagger \tilde{c}_\psi \tilde{c}_\psi^{\dagger} - \tilde{a}_\psi^{\dagger} \tilde{c}_\psi  \tilde{m}_\psi^\dagger \tilde{c}_\psi - \tilde{a}_\psi \tilde{c}_\psi^{\dagger} \tilde{m}_\psi \tilde{c}_\psi^{\dagger} + \tilde{a}_\psi^{\dagger} \tilde{c}_\psi \tilde{c}_\psi^{\dagger} \tilde{m}_\psi + \tilde{a}_\psi^{\dagger} \tilde{m}_\psi \tilde{c}_\psi^{\dagger} \tilde{c}_\psi\bigg]~.\nonumber
\end{equation}     

\item $s(p_1)\to \overline{\psi^\alpha}(p_2)\psi^\beta (p_3)$
 
 The corresponding diagrams are shown in Figs.~\ref{fig:s_ee},
 \ref{fig:s_uu}, \ref{fig:s_dd} for $\psi = e,u,d$.
  We obtain the
 following divergences for the renormalizable operators: 
 \begin{align}
  \tilde{c}_e' & =
  \frac{1}{8\pi^2 \epsilon} \left(\tilde{a}_e \tilde{m}_e^\dagger\tilde{c}_e + \tilde{c}_e \tilde{m}_e^\dagger \tilde{a}_e\right) 
 + \frac{3 \tilde{e}}{8\pi^2\epsilon} \left(\tilde{m}_e \tilde{c}_e^{\dagger} \tilde{a}_{eA} + \tilde{a}_{eA} \tilde{c}_e^{ \dagger} \tilde{m}_e- \tilde{c}_e \tilde{m}_e^\dagger \tilde{a}_{eA} - \tilde{a}_{eA} \tilde{m}_e^\dagger \tilde{c}_e\right)  \,,\\
   \tilde{c}_u'   &= 
  \frac{1}{8\pi^2 \epsilon} \left(\tilde{a}_u \tilde{m}_u^\dagger\tilde{c}_u + \tilde{c}_u \tilde{m}_u^\dagger \tilde{a}_u\right)   - \frac{2   \tilde{e}}{8\pi^2\epsilon} \left(\tilde{m}_u \tilde{c}_u^{\dagger} \tilde{a}_{uA} + \tilde{a}_{uA} \tilde{c}_u^{ \dagger} \tilde{m}_u - \tilde{c}_u \tilde{m}_u^\dagger \tilde{a}_{uA} - \tilde{a}_{uA} \tilde{m}_u^\dagger \tilde{c}_u\right) \nonumber \\
&  - \frac{\tilde{g}_3}{2\pi^2\epsilon} \left(\tilde{m}_u \tilde{c}_u^{\dagger} \tilde{a}_{uG} + \tilde{a}_{uG} \tilde{c}_u^{\dagger} \tilde{m}_u - \tilde{c}_u \tilde{m}_u^\dagger \tilde{a}_{uG} - \tilde{a}_{uG} \tilde{m}_u^\dagger \tilde{c}_u\right) 
\,,\\
  \tilde{c}_d' & =  \frac{1}{8\pi^2 \epsilon} \left(\tilde{a}_d \tilde{m}_d^\dagger\tilde{c}_d + \tilde{c}_d \tilde{m}_d^\dagger \tilde{a}_d\right)  + \frac{  \tilde{e}}{8\pi^2\epsilon} \left(\tilde{m}_d \tilde{c}_d^{\dagger} \tilde{a}_{dA} + \tilde{a}_{dA} \tilde{c}_d^{ \dagger} \tilde{m}_d - \tilde{c}_d \tilde{m}_d^\dagger \tilde{a}_{dA} - \tilde{a}_{dA} \tilde{m}_d^\dagger \tilde{c}_d\right) \nonumber \\
&  - \frac{\tilde{g}_3}{2\pi^2\epsilon} \left(\tilde{m}_d \tilde{c}_d^{\dagger}
  \tilde{a}_{dG} + \tilde{a}_{dG} \tilde{c}_d^{\dagger} \tilde{m}_d -
  \tilde{c}_d \tilde{m}_d^\dagger \tilde{a}_{dG} - \tilde{a}_{dG} \tilde{m}_d^\dagger
  \tilde{c}_d\right)  \,,
 \end{align}
 and for the non-renormalizable ones:
 \begin{align}
  \tilde{r}_{se_L}' & =  -\frac{1}{16\pi^2 \epsilon} \tilde{a}_e \tilde{c}_e^{\dagger} +  \frac{3}{2\pi \epsilon}   \alpha \tilde{a}_{s\widetilde{A}} 
-\frac{3  \tilde{e}}{16 \pi^2 \epsilon}  \tilde{c}_e \tilde{a}_{eA}^{\dagger} \,,\\
  \tilde{r}_{se_R}' & = \frac{1}{16\pi^2 \epsilon} \tilde{a}_e^\dagger \tilde{c}_e - \frac{3}{2\pi \epsilon}  \alpha \tilde{a}_{s\widetilde{A}} 
+\frac{3  \tilde{e}}{16 \pi^2 \epsilon}  \tilde{c}_e^\dagger \tilde{a}_{eA} \,,\\
  \tilde{r}_{su_L}' & =  -\frac{1}{16\pi^2 \epsilon} \tilde{a}_u \tilde{c}_u^{\dagger} +\frac{2}{\pi \epsilon} \left(\frac{1}{3} \alpha \tilde{a}_{s\widetilde{A}} + \alpha_s \tilde{a}_{s\widetilde{G}}\right)
+\frac{2   \tilde{e}}{16 \pi^2 \epsilon}  \tilde{c}_u \tilde{a}_{uA}^{\dagger} + \frac{1}{4\pi^2 \epsilon } \tilde{g}_3 \tilde{c}_u\tilde{a}_{uG}^{\dagger}\,,\\
  \tilde{r}_{su_R}' & = \frac{1}{16\pi^2 \epsilon} \tilde{a}_u^\dagger \tilde{c}_u - \frac{2}{\pi \epsilon} \left(\frac{1 }{3} \alpha \tilde{a}_{s\widetilde{A}} + \alpha_s \tilde{a}_{s\widetilde{G}}\right)
-\frac{2  \tilde{e}}{16 \pi^2 \epsilon}  \tilde{c}_u^\dagger \tilde{a}_{uA} + \frac{1}{4\pi^2 \epsilon } \tilde{g}_3 \tilde{c}_u^\dagger \tilde{a}_{uG}\,,\\
    \tilde{r}_{sd_L}' & =  -\frac{1}{16\pi^2 \epsilon} \tilde{a}_d \tilde{c}_d^{\dagger}+ \frac{2}{\pi \epsilon} \left(\frac{1 }{12} \alpha \tilde{a}_{s\widetilde{A}} + \alpha_s \tilde{a}_{s\widetilde{G}}\right)
-\frac{  \tilde{e}}{16 \pi^2 \epsilon}  \tilde{c}_d \tilde{a}_{dA}^{\dagger} + \frac{1}{4\pi^2 \epsilon } \tilde{g}_3 \tilde{c}_d\tilde{a}_{dG}^{\dagger}\,,\\
  \tilde{r}_{sd_R}' & = \frac{1}{16\pi^2 \epsilon} \tilde{a}_d^\dagger \tilde{c}_d - \frac{2}{\pi \epsilon} \left(\frac{1}{12} \alpha \tilde{a}_{s\widetilde{A}} + \alpha_s \tilde{a}_{s\widetilde{G}}\right)
+\frac{  \tilde{e}}{16 \pi^2 \epsilon}  \tilde{c}_d^\dagger \tilde{a}_{dA} + \frac{1}{4\pi^2 \epsilon } \tilde{g}_3 \tilde{c}_d^\dagger \tilde{a}_{dG}\,.
 \end{align}
 \item $A(p_1)\to \overline{\psi^\alpha}(p_2) \psi^\beta (p_3)$
 
 The corresponding diagrams are shown in Figs.~\ref{fig:a_ee},
 \ref{fig:a_uu} and \ref{fig:a_dd} for $\psi = e,u,d$,
 respectively. This process fixes the divergences 
 \begin{align}
  \tilde{a}_{eA}' &= -\frac{ \tilde{e}}{16\pi^2\epsilon}\left( \tilde{e} \tilde{a}_{eA} +  2\tilde{c}_e \tilde{a}_{s\widetilde{A}}\right)\,,\\
  \tilde{a}_{uA}' &= -\frac{   \tilde{e}}{12\pi^2\epsilon}\left(\frac{1}{3}   \tilde{e} \tilde{a}_{uA} -  \tilde{c}_u \tilde{a}_{s\widetilde{A}}\right) - \frac{  \tilde{e}}{18\pi^2 \epsilon}  \tilde{g}_3 \tilde{a}_{u G}\,,\\
  \tilde{a}_{dA}' &= -\frac{ \tilde{e}}{24\pi^2\epsilon}\left(\frac{1}{6}   \tilde{e} \tilde{a}_{dA} + \tilde{c}_d \tilde{a}_{s\widetilde{A}}\right) + \frac{ \tilde{e}}{36\pi^2 \epsilon}  \tilde{g}_3 \tilde{a}_{d G}\,,\\
  \tilde{r}_{e\square}' &= \frac{3}{8\pi^2\epsilon}   \tilde{e}
  \tilde{a}_{eA}\,,
  \label{resquare}\\
  \tilde{r}_{u\square}' &= -\frac{1}{4\pi^2\epsilon}   \tilde{e} \tilde{a}_{uA} - \frac{1}{2\pi^2 \epsilon} \tilde{g}_3 \tilde{a}_{u G} \,,\label{rusquare}\\
  \tilde{r}_{d\square}' &= \frac{1}{8\pi^2\epsilon}   \tilde{e}
  \tilde{a}_{dA} - \frac{1}{2\pi^2 \epsilon} \tilde{g}_3 \tilde{a}_{d G}
  \,, \label{rdsquare} 
 \end{align}
 as well as a contribution to the fermion kinetic term, which we
 have provided fully (\textit{i.e.} including all 
 contributions up to order $1/v$) above.

 \item $G(p_1)\to \overline{\psi^\alpha}(p_2)\psi^\beta (p_3)$
 
 The diagrams for $\psi = u,d$ are respectively shown in
 Figs.~\ref{fig:g_uu} and \ref{fig:g_dd}. Similar to the previous
 case, we obtain:
 \begin{align}
  \tilde{a}_{uG}' &= \frac{1}{8\pi^2\epsilon} \tilde{g}_3 \tilde{c}_u \tilde{a}_{s\widetilde{G}} +\frac{7}{6\pi \epsilon} \alpha_s \tilde{a}_{uG} -\frac{1}{24 \pi^2 \epsilon}\tilde{e} \tilde{g_3} \tilde{a}_{uA}
\,,\\
  \tilde{a}_{dG}' &= \frac{1}{8\pi^2\epsilon} \tilde{g}_3 \tilde{c}_d \tilde{a}_{s\widetilde{G}}  +\frac{7}{6\pi \epsilon} \alpha_s \tilde{a}_{dG} +\frac{1}{48 \pi^2 \epsilon}\tilde{e} \tilde{g_3} \tilde{a}_{dA}\,,
 \end{align}
 as well as cross-check the previous redundant operators and
 contribution to the kinetic terms for the quarks.

 \item $s(p_1) s(p_2) \to \overline{\psi^\alpha}(p_3)\psi^\beta (p_4)$
 
 The diagrams for $\psi=u,d,e$ are given in Figs.~\ref{fig:ss_ee}, \ref{fig:ss_uu}, \ref{fig:ss_dd}. We get:
 \begin{align}
  \tilde{a}_e' &=  \bigg[ \frac{1 }{ \pi \epsilon}   \alpha  - \frac{\tilde{\lambda}_s}{32\pi^2\epsilon}\bigg] \tilde{a}_u
+ \frac{1}{16 \pi^2 \epsilon} \left(\tilde{c}_e \tilde{a}_e^{\dagger} \tilde{c}_e - 2 \tilde{a}_e \tilde{c}_e^{\dagger} \tilde{c}_e - 2 \tilde{c}_e \tilde{c}_e^{\dagger} \tilde{a}_e\right) \nonumber \\
& + \frac{3  \tilde{e}}{8\pi^2\epsilon}  \left(\tilde{c}_e \tilde{c}_e^{\dagger} \tilde{a}_{eA}  + \tilde{a}_{eA} \tilde{c}_e^{\dagger} \tilde{c}_e\right) \,,\\
  \tilde{a}_u' &= \bigg[ \frac{4}{3 \pi \epsilon} \left( \frac{1}{3} \alpha + \alpha_s\right) - \frac{\tilde{\lambda}_s}{32\pi^2\epsilon}\bigg] \tilde{a}_u
+ \frac{1}{16 \pi^2 \epsilon} \left(\tilde{c}_u \tilde{a}_u^{\dagger} \tilde{c}_u - 2 \tilde{a}_u \tilde{c}_u^{\dagger} \tilde{c}_u - 2 \tilde{c}_u \tilde{c}_u^{\dagger} \tilde{a}_u\right) \nonumber \\
& - \frac{  \tilde{e}}{4\pi^2\epsilon}  \left(\tilde{c}_u \tilde{c}_u^{\dagger} \tilde{a}_{uA}  + \tilde{a}_{uA} \tilde{c}_u^{\dagger} \tilde{c}_u\right) - \frac{\tilde{g}_3}{2\pi^2\epsilon}  \left( \tilde{c}_u \tilde{c}_u^{\dagger} \tilde{a}_{uG}+ \tilde{a}_{uG} \tilde{c}_u^{\dagger} \tilde{c}_u \right) \,,\\
  \tilde{a}_d' &=\bigg[ \frac{1}{3 \pi \epsilon} \left(\frac{1}{3} \alpha + 4\alpha_s\right) - \frac{\tilde{\lambda}_s}{32\pi^2\epsilon}\bigg] \tilde{a}_d
+ \frac{1}{16 \pi^2 \epsilon} \left(\tilde{c}_d \tilde{a}_d^{\dagger} \tilde{c}_d - 2 \tilde{a}_d \tilde{c}_d^{\dagger} \tilde{c}_d - 2 \tilde{c}_d \tilde{c}_d^{\dagger} \tilde{a}_d\right) \nonumber \\
& + \frac{ \tilde{e}}{8\pi^2\epsilon}  \left(\tilde{c}_d \tilde{c}_d^{\dagger} \tilde{a}_{dA}  + \tilde{a}_{dA} \tilde{c}_d^{\dagger} \tilde{c}_d\right) - \frac{\tilde{g}_3}{2\pi^2\epsilon}  \left( \tilde{c}_d \tilde{c}_d^{\dagger} \tilde{a}_{dG}+ \tilde{a}_{dG} \tilde{c}_d^{\dagger} \tilde{c}_d \right) \,.
 \end{align}
 
 \item $s(p_1)\to V(p_2)V(p_3)$
 
 The diagrams for $V=A, G$ are given in Figs.~\ref{fig:s_aa}, \ref{fig:s_gg}. The corresponding divergences read: 
 \begin{align}
  \tilde{a}_{s\tilde{A}} &= \frac{ \tilde{e}}{8\pi^2\epsilon} {\rm Tr}\bigg[- \left(\tilde{c}_e \tilde{a}_{eA}^\dagger + \tilde{c}_e^{\dagger} \tilde{a}_{eA} \right) - \left( \tilde{c}_d \tilde{a}_{dA}^\dagger + \tilde{c}_d^{\dagger} \tilde{a}_{dA} \right) + 2 \left(\tilde{c}_u \tilde{a}_{uA}^\dagger + \tilde{c}_u^{\dagger} \tilde{a}_{uA} \right)  \bigg] \,,\\
  \tilde{a}_{s\tilde{G}} & = \frac{\tilde{g}_3} {16\pi^2\epsilon} {\rm Tr} \left[ \tilde{c}_d \tilde{a}_{dG}^\dagger  + \tilde{c}_d^{\dagger }\tilde{a}_{dG} +  \tilde{c}_u \tilde{a}_{uG}^\dagger  + \tilde{c}_u^{\dagger }\tilde{a}_{uG} \right] \,.
 \end{align}

\end{itemize}

\subsection{Eliminating redundancy}

We can now go to the on-shell basis by using the redundancy relations
in Eqs.~\eqref{red:psiBox}--\eqref{red:RspsiR}. The kinetic terms for
fermions receive an extra contribution from the coefficient of the
redundant operators,
\begin{align}
  -\delta Z_{\psi_L}
  \to& 
  -\delta Z_{\psi_L}
  -\frac{
    \tilde{r}_{\psi \Box}
    \tilde{m}_\psi^\dagger
    +\tilde{m}_\psi
    \tilde{r}_{\psi \Box}^\dagger
 }{2}\,,
  \\
  -\delta Z_{\psi_R}
  \to& 
  -\delta Z_{\psi_R}
  -\frac{
    \tilde{r}_{\psi \Box}^\dagger
    \tilde{m}_\psi
    +\tilde{m}_\psi^\dagger
    \tilde{r}_{\psi \Box}
 }{2}\,.
\end{align}
Upon replacing the values in
Eqs.~\eqref{ZeL:offshell}--\eqref{ZuR:offshell} and
Eqs.~\eqref{resquare}--\eqref{rdsquare}, we find that the contributions of
dimension-five operators to the fermion kinetic terms precisely cancel
in the on-shell basis. The resulting wave function renormalization
factors then read in the on-shell basis:
\begin{align}
& Z_{e_L}    =   1  - \frac{\alpha}{4 \pi \epsilon} -
  \frac{1}{32\pi^2\epsilon} \left(\tilde{c}_e \tilde{c}_e^{\dagger}
  \right)
  \,,
  \label{ZeL:onshell}  \\ 
& Z_{e_R}  =  1  - \frac{\alpha}{4 \pi \epsilon} -
  \frac{1}{32\pi^2\epsilon} \left(\tilde{c}_e^{\dagger} \tilde{c}_e
  \right)
    \,,
  \label{ZeR:onshell}\\ 
& Z_{d_{L}} = 1  - \frac{1}{3 \pi \epsilon}\bigg[ \frac{1}{12} \alpha
    + \alpha_s \bigg]   - \frac{1}{32\pi^2\epsilon} \left(\tilde{c}_d
  \tilde{c}_d^{\dagger} \right)
 \,, \label{ZdL:onshell}\\
& Z_{d_{R}} = 1  - \frac{1}{3 \pi \epsilon}\bigg[ \frac{1}{12} \alpha
    + \alpha_s \bigg]   - \frac{1}{32\pi^2\epsilon} \left(
 \tilde{c}_d^{\dagger} \tilde{c}_d \right)
  \,,
  \label{ZdR:onshell}\\
& Z_{u_{L}} = 1  -\frac{1}{3 \pi \epsilon}\bigg[ \frac{1}{3} \alpha +
    \alpha_s \bigg]  - \frac{1}{32\pi^2\epsilon} \left(\tilde{c}_u
  \tilde{c}_u^{\dagger} \right)
  \,,
  \label{ZuL:onshell}\\
& Z_{u_{R}} = 1  -\frac{1}{3 \pi \epsilon}\bigg[ \frac{1}{3} \alpha +
    \alpha_s \bigg]  - \frac{1}{32\pi^2\epsilon} \left(
  \tilde{c}_u^{\dagger}\tilde{c}_u \right)
  \,,
  \label{ZuR:onshell}\\
& Z_{A} = 1 - \frac{\alpha}{3 \pi \epsilon} \left[ n_\ell + \frac{1}{3} n_d + \frac{4}{3} n_u \right]\nonumber \\
& ~~~~~~~-\frac{ \tilde{e}}{2\pi^2\epsilon} {\rm Tr} \left[ - (\tilde{a}_{eA}^\dagger \tilde{m}_e + \tilde{m}_e^\dagger \tilde{a}_{eA}) + 2 (\tilde{a}_{uA}^\dagger \tilde{m}_u + \tilde{m}_u^\dagger \tilde{a}_{uA}) - (\tilde{a}_{dA}^\dagger \tilde{m}_d + \tilde{m}_d^\dagger \tilde{a}_{dA})\right]\,, \\
& Z_{G} = 1 + \frac{\alpha_s}{4\pi \epsilon} \left[11 - \frac{2}{3} (n_u+n_d) \right]  - \frac{\tilde{g}_3}{4\pi^2\epsilon} {\rm Tr} \left[ \tilde{a}_{dG}^\dagger \tilde{m}_d + \tilde{m}_d^\dagger \tilde{a}_{dG} + \tilde{a}_{uG}^\dagger \tilde{m}_u + \tilde{m}_u^\dagger \tilde{a}_{uG} \right]\,,  \\
& Z_s = 1 - \frac{1}{8\pi^2 \epsilon} {\rm Tr} \left[ \tilde{c}_e
    \tilde{c}_e^{\dagger} + 3 \left( \tilde{c}_d \tilde{c}_d^{\dagger}
    + \tilde{c}_u \tilde{c}_u^{\dagger} \right)\right]\,. \label{Zs:onshell}
\end{align}
We also have a contribution to renormalizable coefficients from
redundant ones:
\begin{equation}
  \tilde{c}_\psi \to \tilde{c}_\psi
  +
  \frac{
    \tilde{m}_\psi \tilde{r}_{\psi \Box}^\dagger \tilde{c}_\psi
   +\tilde{c}_\psi \tilde{r}_{\psi \Box}^\dagger \tilde{m}_\psi
  }{2}
    +\tilde{r}_{s\psi_L} \tilde{m}_\psi
  -\tilde{m}_{\psi} \tilde{r}_{s\psi_R}^\dagger
\,,
\end{equation}
which results in the following values:
\begin{align}
\tilde{c}_e' &= 
 \frac{3}{4\pi^2\epsilon}\tilde{e}^2 
\tilde{a}_{s\tilde{A}}  \tilde{m}_e
- \frac{1}{16 \pi^2 \epsilon}
\bigg[ \tilde{a}_e \left( \tilde{c}_e^\dagger \tilde{m}_e - 2
  \tilde{m}_e^\dagger \tilde{c}_e\right) + \left( \tilde{m}_e
  \tilde{c}_e^\dagger - 2 \tilde{c}_e \tilde{m}_e^\dagger\right)
  \tilde{a}_e\bigg] 
 \nonumber \\
 &+ \frac{3  \tilde{e}}{8\pi^2 \epsilon} \bigg[
   \tilde{m}_e\tilde{c}_e^{\dagger} \tilde{a}_{eA} + \tilde{a}_{eA}
   \tilde{c}_e^{\dagger} \tilde{m}_e - \tilde{c}_e \tilde{m}_e^\dagger
   \tilde{a}_{eA} - \tilde{a}_{eA} \tilde{m}_e^\dagger \tilde{c}_e
   \bigg] \,, \\ 
\tilde{c}_u' &=
\frac{1}{\pi^2\epsilon}\bigg[\frac{1}{3}\tilde{e}^2 
  \tilde{a}_{s\tilde{A}} + \tilde{g}_3^2
  \tilde{a}_{s\tilde{G}}\bigg] \tilde{m}_u  
-  \frac{1}{16 \pi^2 \epsilon} \bigg[ \tilde{a}_u \left(
  \tilde{c}_u^\dagger \tilde{m}_u - 2 \tilde{m}_u^\dagger
  \tilde{c}_u\right) + \left( \tilde{m}_u \tilde{c}_u^\dagger - 2
  \tilde{c}_u \tilde{m}_u^\dagger\right) \tilde{a}_u\bigg]
\nonumber \\
&- \frac{\tilde{e}}{4 \pi^2 \epsilon}
\bigg[\tilde{m}_u\tilde{c}_u^{\dagger} \tilde{a}_{uA} + \tilde{a}_{uA}
  \tilde{c}_u^{\dagger} \tilde{m}_u  - \tilde{c}_u \tilde{m}_u^\dagger
  \tilde{a}_{uA} - \tilde{a}_{uA} \tilde{m}_u^\dagger \tilde{c}_u
  \bigg] \nonumber \\ 
 &-\frac{\tilde{g}_3}{2 \pi^2 \epsilon}  \bigg[
   \tilde{m}_u\tilde{c}_u^{\dagger} \tilde{a}_{uG} + \tilde{a}_{uG}
   \tilde{c}_u^{\dagger} \tilde{m}_u - \tilde{c}_u
   \tilde{m}_u^\dagger \tilde{a}_{uG} - \tilde{a}_{uG}
   \tilde{m}_u^\dagger \tilde{c}_u  
   \bigg] \,, \\ 
\tilde{c}_d' &=
\frac{1}{4\pi^2\epsilon}\bigg[\frac{1}{3}\tilde{e}^2 
  \tilde{a}_{s\tilde{A}} + 4 \tilde{g}_3^2
  \tilde{a}_{s\tilde{G}}\bigg] \tilde{m}_d  
-  \frac{1}{16 \pi^2 \epsilon} \bigg[ \tilde{a}_d \left(
  \tilde{c}_d^\dagger \tilde{m}_d - 2 \tilde{m}_d^\dagger
  \tilde{c}_d\right) + \left( \tilde{m}_d \tilde{c}_d^\dagger - 2
  \tilde{c}_d \tilde{m}_d^\dagger\right) \tilde{a}_d\bigg]
\nonumber \\
&+ \frac{ \tilde{e}}{8 \pi^2 \epsilon}
\bigg[\tilde{m}_d\tilde{c}_d^{\dagger} \tilde{a}_{dA} + \tilde{a}_{dA}
  \tilde{c}_d^{\dagger} \tilde{m}_d  - \tilde{c}_d \tilde{m}_d^\dagger
  \tilde{a}_{dA} - \tilde{a}_{dA} \tilde{m}_d^\dagger \tilde{c}_d
  \bigg] \nonumber \\ 
 &-\frac{\tilde{g}_3}{2 \pi^2 \epsilon}  \bigg[
   \tilde{m}_d\tilde{c}_d^{\dagger} \tilde{a}_{dG} + \tilde{a}_{dG}
   \tilde{c}_d^{\dagger} \tilde{m}_d  - \tilde{c}_d
   \tilde{m}_d^\dagger \tilde{a}_{dG} - \tilde{a}_{dG}
   \tilde{m}_d^\dagger \tilde{c}_d  
\bigg]  \,.  
\end{align}
Finally, the redundancies imply the following relations for the
coefficients of non-renormalizable operators:
\begin{align}
  \tilde{a}_\psi \to&\,
  \tilde{a}_\psi
  + \tilde{c}_\psi \tilde{r}_{\psi \Box}^\dagger \tilde{c}_\psi
  +\tilde{r}_{s\psi_L} \tilde{c}_\psi
  -\tilde{c}_\psi \tilde{r}_{s\psi_R}^\dagger\,,
  \\
  \tilde{a}_{\psi A} \to&\,
  \tilde{a}_{\psi A}
  +\frac{\tilde{e} Q_\psi \tilde{r}_{\psi \Box}}{2}\,,
  \\
  \tilde{a}_{\psi G} \to&\,
  \tilde{a}_{\psi G}
  +\frac{\tilde{g}_3 \tilde{r}_{\psi \Box}}{2}\,,
\end{align}
resulting in the following on-shell non-renormalizable divergences:
\begin{align}
\tilde{a}_u' &=  \bigg[ \frac{1}{9\pi^2 \epsilon}  \tilde{e}^2  +
  \frac{1}{3\pi \epsilon}\tilde{g}_3^2 -
  \frac{\tilde{\lambda}_s}{32\pi^2\epsilon}\bigg] \tilde{a}_u
+\frac{1}{\pi^2\epsilon}\bigg[\frac{1}{3}\tilde{e}^2 
  \tilde{a}_{s\tilde{A}} +  \tilde{g}_3^2
  \tilde{a}_{s\tilde{G}}\bigg] \tilde{c}_u \nonumber \\ 
&- \frac{1}{16 \pi^2 \epsilon} \bigg[ - \tilde{c}_u
  \tilde{a}_u^{\dagger} \tilde{c}_u + 3 \tilde{a}_u
  \tilde{c}_u^{\dagger} \tilde{c}_u + 3 \tilde{c}_u
  \tilde{c}_u^{\dagger} \tilde{a}_u\bigg] 
 - \frac{1}{4\pi^2\epsilon}\tilde{e} \bigg[ \tilde{c}_u
   \tilde{c}_u^{\dagger} \tilde{a}_{uA}  + \tilde{a}_{uA}
   \tilde{c}_u^{\dagger} \tilde{c}_u\bigg]
    \nonumber \\ &
- \frac{\tilde{g}_3}{2\pi^2
   \epsilon }\bigg[ \tilde{c}_u \tilde{c}_u^{\dagger} \tilde{a}_{uG}
   + \tilde{a}_{uG} \tilde{c}_u^{\dagger} \tilde{c}_u\bigg]  
 \,,
 \\
\tilde{a}_d' &=  \bigg[ \frac{1}{36\pi^2 \epsilon}  \tilde{e}^2  +
  \frac{1}{3\pi \epsilon}\tilde{g}_3^2 -
  \frac{\tilde{\lambda}_s}{32\pi^2\epsilon}\bigg] \tilde{a}_d
+\frac{1}{4\pi^2\epsilon}\bigg[\frac{1}{3}\tilde{e}^2 
  \tilde{a}_{s\tilde{A}} + 4 \tilde{g}_3^2
  \tilde{a}_{s\tilde{G}}\bigg] \tilde{c}_d \nonumber \\ 
&- \frac{1}{16 \pi^2 \epsilon} \bigg[ - \tilde{c}_d
  \tilde{a}_d^{\dagger} \tilde{c}_d + 3 \tilde{a}_d
  \tilde{c}_d^{\dagger} \tilde{c}_d + 3 \tilde{c}_d
  \tilde{c}_d^{\dagger} \tilde{a}_d\bigg] 
 + \frac{1}{8\pi^2\epsilon} \tilde{e} \bigg[ \tilde{c}_d
   \tilde{c}_d^{\dagger} \tilde{a}_{dA}  + \tilde{a}_{dA}
   \tilde{c}_d^{\dagger} \tilde{c}_d\bigg]
    \nonumber \\ &
- \frac{\tilde{g}_3}{2\pi^2
   \epsilon }\bigg[ \tilde{c}_d \tilde{c}_d^{\dagger} \tilde{a}_{dG}
   + \tilde{a}_{dG} \tilde{c}_d^{\dagger} \tilde{c}_d\bigg]  
 \,,
 \\
 \tilde{a}_e' &= \bigg[ \frac{1}{4\pi^2 \epsilon}  \tilde{e}^2  -
  \frac{\tilde{\lambda}_s}{32\pi^2\epsilon}\bigg] \tilde{a}_e
+\frac{3}{4\pi^2\epsilon}\tilde{e}^2  \tilde{a}_{s\tilde{A}}
\tilde{c}_e\nonumber	\\ 
 &- \frac{1}{16 \pi^2 \epsilon} \bigg[ - \tilde{c}_e
   \tilde{a}_e^{\dagger} \tilde{c}_e + 3 \tilde{a}_e
   \tilde{c}_e^{\dagger} \tilde{c}_e + 3 \tilde{c}_e
   \tilde{c}_e^{\dagger} \tilde{a}_e\bigg]  
 + \frac{3}{8\pi^2\epsilon} \tilde{e} \bigg[ \tilde{c}_e
   \tilde{c}_e^{\dagger} \tilde{a}_{eA}  + \tilde{a}_{eA}
   \tilde{c}_e^{\dagger} \tilde{c}_e\bigg]  \,,
 \\%
\tilde{a}_{eA}' &=  -\frac{  \tilde{e}}{8\pi^2\epsilon}\left[ 2 
  \tilde{e} \tilde{a}_{eA} +  \tilde{c}_e
  \tilde{a}_{s\tilde{A}}\right] \,,  \\ 
\tilde{a}_{uA}' &=  -\frac{  \tilde{e}}{12\pi^2\epsilon}\left[\frac{4}{3}
  \tilde{e} \tilde{a}_{uA} - \tilde{c}_u
  \tilde{a}_{s\tilde{A}}\right] - \frac{2 \tilde{e} }{9\pi^2
  \epsilon}  \tilde{g}_3 \tilde{a}_{u\tilde{G}} \,,  \\ 
\tilde{a}_{dA}' &=  -\frac{  \tilde{e}}{24\pi^2\epsilon}\left[\frac{2}{3}
  \tilde{e} \tilde{a}_{dA} + \tilde{c}_d
  \tilde{a}_{s\tilde{A}}\right] + \frac{ \tilde{e} }{9\pi^2
  \epsilon}  \tilde{g}_3 \tilde{a}_{d\tilde{G}} \,,  \\ 
\tilde{a}_{uG}' &=  -\frac{\tilde{g}_3}{8\pi^2\epsilon}\left[
  \frac{4}{3}\tilde{e} \tilde{a}_{uA} -\tilde{c}_u
  \tilde{a}_{s\tilde{G}} \right] + \frac{1}{24\pi^2\epsilon}
\tilde{g}_3^2 \tilde{a}_{uG} \,,   \\
\tilde{a}_{dG}' &=  \frac{\tilde{g}_3}{8\pi^2\epsilon}\left[
  \frac{2}{3}\tilde{e} \tilde{a}_{dA} +\tilde{c}_d
  \tilde{a}_{s\tilde{G}} \right] + \frac{1}{24\pi^2\epsilon}
\tilde{g}_3^2 \tilde{a}_{dG} \,.
\end{align}

\subsection{Anomalous dimensions}

Once we have parametrised all the relevant divergences in the
on-shell basis, we can obtain the beta functions
of the different parameters following the standard procedure outlined
in Section~\ref{sec:rges}. We start reporting the beta functions for
the renormalizable couplings, which read:
\begin{align}\label{eq:dim4runLEFT}
 \beta_{\tilde{e}} &= \frac{4}{3}\left[ n_\ell + \frac{1}{3} n_d + \frac{4}{3} n_u \right] \tilde{e}^3  \\
& \hd{ +8 \tilde{e}^2 {\rm Tr} \bigg[- (\tilde{a}_{eA}^\dagger \tilde{m}_e + \tilde{m}_e^\dagger \tilde{a}_{eA}) + 2 (\tilde{a}_{uA}^\dagger \tilde{m}_u + \tilde{m}_u^\dagger \tilde{a}_{uA}) - (\tilde{a}_{dA}^\dagger \tilde{m}_d + \tilde{m}_d^\dagger \tilde{a}_{dA})\bigg]}\nonumber\,,
\label{eq:erunLEFT}
\end{align}
\begin{align}
 \beta_{\tilde{g}_3} &= [-11+\frac{2}{3}(n_u+n_d)]\tilde{g}_3^3 \hd{+ 4 \tilde{g}_3^2 {\rm Tr}\left[\tilde{a}_{dG}^\dagger \tilde{m}_d + \tilde{m}_d^\dagger \tilde{a}_{dG} + \tilde{a}_{uG}^\dagger \tilde{m}_u + \tilde{m}_u^\dagger \tilde{a}_{uG}\right]}\,,
\end{align}
\begin{align}\nonumber
  \beta_{\tilde{m}_e} = &
  -6\tilde{e}^2 \tilde{m}_e
  + \frac{1}{2}(\tilde{m}_e \tilde{c}_e^{\dagger}\tilde{c}_e
  + \tilde{c}_e \tilde{c}_e^{\dagger} \tilde{m}_e
  - 4 \tilde{c}_e\tilde{m}_e^\dagger\tilde{c}_e)\\
%
& \hd{+12 \tilde{e} \left(\tilde{m}_e \tilde{m}_e ^\dagger \tilde{a}_{eA} + \tilde{a}_{eA} \tilde{m}_e^\dagger \tilde{m}_e\right)  - 2 \tilde{m}^2 \tilde{a}_e} \,,
\end{align}
\begin{align}\nonumber
  \beta_{\tilde{m}_u} = &
  -8 \tilde{g}_3^2 \tilde{m}_u - \frac{8}{3}\tilde{e}^2\tilde{m}_u
  + \frac{1}{2}(\tilde{m}_u\tilde{c}_u^{\dagger}\tilde{c}_u
  +\tilde{c}_u\tilde{c}_u^{\dagger}\tilde{m}_u
  - 4\tilde{c}_u\tilde{m}_u^\dagger\tilde{c}_u)\\
 %
&\hd{ -8 \tilde{e} \left(\tilde{m}_u \tilde{m}_u ^\dagger \tilde{a}_{uA} + \tilde{a}_{uA} \tilde{m}_u^\dagger \tilde{m}_u\right) -16 \tilde{g}_3 \left( \tilde{m}_u \tilde{m}_u^\dagger \tilde{a}_{uG} + \tilde{a}_{uG} \tilde{m}_u^\dagger \tilde{m}_u	\right) - 2 \tilde{m}^2 \tilde{a}_u } \,,
\end{align}
\begin{align}\nonumber
  \beta_{\tilde{m}_d} = &
  -8\tilde{g}_3^2\tilde{m}_d-\frac{2}{3}\tilde{e}^2\tilde{m}_d
  + \frac{1}{2}(\tilde{m}_d\tilde{c}_d^{\dagger}\tilde{c}_d
  +\tilde{c}_d\tilde{c}_d^{\dagger}\tilde{m}_d
  - 4\tilde{c}_d\tilde{m}_d^\dagger\tilde{c}_d)\\
&\hd{ 4 \tilde{e} \left(\tilde{m}_d \tilde{m}_d ^\dagger \tilde{a}_{dA} + \tilde{a}_{dA} \tilde{m}_d^\dagger \tilde{m}_d\right) -16 \tilde{g}_3 \left( \tilde{m}_d \tilde{m}_d^\dagger \tilde{a}_{dG} + \tilde{a}_{dG} \tilde{m}_d^\dagger \tilde{m}_d	\right) - 2 \tilde{m}^2 \tilde{a}_d }\,,
\end{align}
\begin{align}\nonumber
  \beta_{\tilde{m}^2} &= \tilde{\lambda}_s \tilde{m}^2
  +4 \tilde{m}^2 \text{Tr}(\tilde{c}_e \tilde{c}_e^{\dagger})
  + 12\tilde{m}^2\text{Tr}(\tilde{c}_d\tilde{c}_d^{\dagger})
  + 12\tilde{m}^2\text{Tr}(\tilde{c}_u\tilde{c}_u^{\dagger})\\\nonumber
  &-24\text{Tr}(\tilde{c}_u\tilde{c}_u^{\dagger}\tilde{m}_u\tilde{m}_u^\dagger
  +\tilde{c}_d\tilde{c}_d^{\dagger}\tilde{m}_d\tilde{m}_d^\dagger)
  -
  18\text{Tr}(\tilde{c}_u^{\dagger}\tilde{c}_u\tilde{m}_u^\dagger\tilde{m}_u
  +\tilde{c}_d^{\dagger}\tilde{c}_d\tilde{m}_d^\dagger\tilde{m}_d)\\\nonumber
  &+12\text{Tr}(\tilde{c}_u\tilde{m}_u^\dagger\tilde{c}_u\tilde{m}_u^\dagger
  + \tilde{c}_u^{\dagger}\tilde{m}_u\tilde{c}_u^{\dagger}\tilde{m}_u
  + \tilde{c}_d\tilde{m}_d^\dagger\tilde{c}_d\tilde{m}_d^\dagger
  + \tilde{c}_d^{\dagger}\tilde{m}_d\tilde{c}_d^{\dagger}\tilde{m}_d)\\\nonumber
  &-6\text{Tr}(\tilde{c}_u\tilde{m}_u^\dagger\tilde{m}_u\tilde{c}_u^{\dagger}
  + \tilde{m}_d\tilde{c}_d^{\dagger}\tilde{c}_d\tilde{m}_d^\dagger
  + \tilde{c}_e^{\dagger}\tilde{c}_e\tilde{m}_e^\dagger\tilde{m}_e)\\
  &-2\text{Tr}(4\tilde{c}_e\tilde{c}_e^{\dagger}\tilde{m}_e\tilde{m}_e^\dagger
  -2\tilde{c}_e\tilde{m}_e^\dagger\tilde{c}_e\tilde{m}_e^\dagger
  -2\tilde{c}_e^\dagger\tilde{m}_e\tilde{c}_e^\dagger\tilde{m}_e+\tilde{c}_e\tilde{m}_e^\dagger\tilde{m}_e\tilde{c}_e^\dagger) \nonumber \\
&+ \hd{  8 \bigg[3  {\rm Tr} \left(\tilde{m}_d^\dagger \tilde{a}_d \tilde{m}_d^\dagger \tilde{m}_d + \tilde{m}_d^\dagger \tilde{m}_d \tilde{a}_d^{\dagger} \tilde{m}_d \right)  + 3 {\rm Tr} \left(\tilde{m}_u^\dagger \tilde{a}_u \tilde{m}_u^\dagger \tilde{m}_u + \tilde{m}_u^\dagger \tilde{m}_u \tilde{a}_u^{\dagger} \tilde{m}_u \right) }   \nonumber \\
&  \hd{ +  {\rm Tr} \left(\tilde{m}_e^\dagger \tilde{a}_e \tilde{m}_e^\dagger \tilde{m}_e + \tilde{m}_e^\dagger \tilde{m}_e \tilde{a}_e^{\dagger} \tilde{m}_e \right)\bigg]}
\,,
\end{align}
\begin{align}\nonumber
  \beta_{\tilde{\lambda}_s} &=
  3\tilde{\lambda}_s^2
  -144\text{Tr}(\tilde{c}_d\tilde{c}_d^{\dagger}
  \tilde{c}_d\tilde{c}_d^{\dagger})
  -144\text{Tr}(\tilde{c}_u\tilde{c}_u^{\dagger}
  \tilde{c}_u\tilde{c}_u^{\dagger})
  -48\text{Tr}(\tilde{c}_e\tilde{c}_e^{\dagger}
  \tilde{c}_e\tilde{c}_e^{\dagger})\\
  &+24\tilde{\lambda}_s \text{Tr}(\tilde{c}_d\tilde{c}_d^{\dagger})
  + 24\tilde{\lambda}_s \text{Tr}(\tilde{c}_u\tilde{c}_u^{\dagger})
  +8\tilde{\lambda}_s \text{Tr}(\tilde{c}_e\tilde{c}_e^{\dagger})
+ \hd{96 \bigg[{\rm Tr}^{\tilde{\lambda}}_e + 3\left({\rm Tr}^{\tilde{\lambda}}_u+{\rm Tr}^{\tilde{\lambda}}_d\right)\bigg] } \,,
\end{align}
\begin{align}
  \beta_{\tilde{c}_u} &=
  -\frac{24}{9}(\tilde{e}^2 + 3\tilde{g}_3^2) \tilde{c}_u
  +3\tilde{c}_u\tilde{c}_u^{\dagger}\tilde{c}_u
  + 2\bigg[\text{Tr}(\tilde{c}_e\tilde{c}_e^{\dagger})
    + 3\text{Tr}(\tilde{c}_d\tilde{c}_d^{\dagger})
    +3\text{Tr}(\tilde{c}_u\tilde{c}_u^{\dagger})\bigg]\tilde{c}_u+\nonumber\\ 
&\hd{- 32\bigg[\frac{1}{3}\tilde{e}^2  \tilde{a}_{s\tilde{A}} + \tilde{g}_3^2 \tilde{a}_{s\tilde{G}}\bigg] \tilde{m}_u  + 2\bigg[ \tilde{a}_u \left( \tilde{c}_u^\dagger \tilde{m}_u - 2 \tilde{m}_u^\dagger \tilde{c}_u\right) + \left( \tilde{m}_u \tilde{c}_u^\dagger - 2 \tilde{c}_u \tilde{m}_u^\dagger\right) \tilde{a}_u\bigg]} \nonumber\\
& \hd{+ 8 \tilde{e} \bigg[\tilde{m}_u\tilde{c}_u^{\dagger} \tilde{a}_{uA} + \tilde{a}_{uA} \tilde{c}_u^{\dagger} \tilde{m}_u  - \tilde{c}_u \tilde{m}_u^\dagger \tilde{a}_{uA} - \tilde{a}_{uA} \tilde{m}_u^\dagger \tilde{c}_u  \bigg] }\nonumber \\
 & \hd{+16 \tilde{g}_3\bigg[ \tilde{m}_u\tilde{c}_u^{\dagger} \tilde{a}_{uG} + \tilde{a}_{uG} \tilde{c}_u^{\dagger} \tilde{m}_u  - \tilde{c}_u \tilde{m}_u^\dagger \tilde{a}_{uG} - \tilde{a}_{uG} \tilde{m}_u^\dagger \tilde{c}_u \bigg] } \,,
\end{align}
\begin{align}
  \beta_{\tilde{c}_d} &= -\frac{2}{3}(\tilde{e}^2
  + 12\tilde{g}_3^2) \tilde{c}_d
  +3\tilde{c}_d\tilde{c}_d^{\dagger}\tilde{c}_d
  + 2\bigg[\text{Tr}(\tilde{c}_e\tilde{c}_e^{\dagger})
    + 3\text{Tr}(\tilde{c}_d\tilde{c}_d^{\dagger})
    +3\text{Tr}(\tilde{c}_u\tilde{c}_u^{\dagger})\bigg]\tilde{c}_d\nonumber\\ 
&\hd{- 8\bigg[\frac{1}{3}\tilde{e}^2  \tilde{a}_{s\tilde{A}} + 4 \tilde{g}_3^2 \tilde{a}_{s\tilde{G}}\bigg] \tilde{m}_d  + 2\bigg[ \tilde{a}_d \left( \tilde{c}_d^\dagger \tilde{m}_d - 2 \tilde{m}_d^\dagger \tilde{c}_d\right) + \left( \tilde{m}_d \tilde{c}_d^\dagger - 2 \tilde{c}_d \tilde{m}_d^\dagger\right) \tilde{a}_d\bigg]} \nonumber\\
& \hd{-4 \tilde{e} \bigg[\tilde{m}_d\tilde{c}_d^{\dagger} \tilde{a}_{dA} + \tilde{a}_{dA} \tilde{c}_d^{\dagger} \tilde{m}_d  - \tilde{c}_d \tilde{m}_d^\dagger \tilde{a}_{dA} - \tilde{a}_{dA} \tilde{m}_d^\dagger \tilde{c}_d  \bigg] }\nonumber \\
 & \hd{+16 \tilde{g}_3\bigg[ \tilde{m}_d\tilde{c}_d^{\dagger} \tilde{a}_{dG} + \tilde{a}_{dG} \tilde{c}_d^{\dagger} \tilde{m}_d - \tilde{c}_d \tilde{m}_d^\dagger \tilde{a}_{dG} - \tilde{a}_{dG} \tilde{m}_d^\dagger \tilde{c}_d \bigg] } \,,
\end{align}
\begin{align}
  \beta_{\tilde{c}_e} &= -6\tilde{e}^2
  \tilde{c}_e
  +3\tilde{c}_e\tilde{c}_e^{\dagger}\tilde{c}_e
  + 2\bigg[\text{Tr}(\tilde{c}_e\tilde{c}_e^{\dagger})
    + 3\text{Tr}(\tilde{c}_d\tilde{c}_d^{\dagger})
    +3\text{Tr}(\tilde{c}_u\tilde{c}_u^{\dagger})\bigg]\tilde{c}_e\nonumber\\ 
&\hd{- 8\bigg[3\tilde{e}^2  \tilde{a}_{s\tilde{A}} \bigg] \tilde{m}_e  + 2\bigg[ \tilde{a}_e \left( \tilde{c}_e^\dagger \tilde{m}_e - 2 \tilde{m}_e^\dagger \tilde{c}_e\right) + \left( \tilde{m}_e \tilde{c}_e^\dagger - 2 \tilde{c}_e \tilde{m}_e^\dagger\right) \tilde{a}_e\bigg]} \nonumber\\
& \hd{-12  \tilde{e} \bigg[\tilde{m}_e\tilde{c}_e^{\dagger} \tilde{a}_{eA} + \tilde{a}_{eA} \tilde{c}_e^{\dagger} \tilde{m}_e  - \tilde{c}_e \tilde{m}_e^\dagger \tilde{a}_{eA} - \tilde{a}_{eA} \tilde{m}_e^\dagger \tilde{c}_e  \bigg] } \,;
\label{eq:cerunLEFT}
\end{align}
where the contributions from the effective operators are apparent from
the presence of the corresponding Wilson coefficients, and we have used
\texttt{Pyr@te}~\cite{Lyonnet:2016xiz} with manual cross-checks to compute the part of the beta functions that depend only on
renormalizable couplings.

As a byproduct of this work, we have
reproduced the 
anomalous dimensions of purely SM operators to dimension five given in
Ref.~\cite{Jenkins:2017dyc}. One can also trivially reproduce the $\log{(m_W/m_\psi)}$ piece of the ALP-fermion-fermion couplings induced by ALP-vector-vector ones in Eqs. 3.15  and 3.20 of Ref.~\cite{Bauer:2017ris}.

In the case of the non-renormalizable Wilson coefficients, the beta
functions read:
\begin{align}
\beta_{\tilde{a}_u} &=\bigg[-\frac{8}{3} \tilde{e}^2  - 8 \tilde{g}_3^2 +  \tilde{\lambda}_s\bigg] \tilde{a}_u  -32 \bigg[\frac{1}{3}\tilde{e}^2  \tilde{a}_{s\tilde{A}} + \tilde{g}_3^2 \tilde{a}_{s\tilde{G}}\bigg] \tilde{c}_u \nonumber \\
& + 2 \bigg[ - \tilde{c}_u \tilde{a}_u^{\dagger} \tilde{c}_u + \frac{13}{4} \tilde{a}_u \tilde{c}_u^{\dagger} \tilde{c}_u + \frac{13}{4} \tilde{c}_u \tilde{c}_u^{\dagger} \tilde{a}_u\bigg] +  4 {\rm Tr} \left[ \tilde{c}_e \tilde{c}_e^{\dagger} + 3 \left( \tilde{c}_d \tilde{c}_d^{\dagger} + \tilde{c}_u \tilde{c}_u^{\dagger} \right)\right] \tilde{a}_u\nonumber \\
& + 8\tilde{e}\bigg[ \tilde{c}_u \tilde{c}_u^{\dagger} \tilde{a}_{uA}  + \tilde{a}_{uA} \tilde{c}_u^{\dagger} \tilde{c}_u\bigg] + 16 \tilde{g}_3\bigg[ \tilde{c}_u \tilde{c}_u^{\dagger} \tilde{a}_{uG}  + \tilde{a}_{uG} \tilde{c}_u^{\dagger} \tilde{c}_u\bigg] ~,\\
\beta_{\tilde{a}_d} &=\bigg[-\frac{2}{3} \tilde{e}^2  - 8 \tilde{g}_3^2 +  \tilde{\lambda}_s\bigg] \tilde{a}_d  -32 \bigg[\frac{1}{12}\tilde{e}^2  \tilde{a}_{s\tilde{A}} + \tilde{g}_3^2 \tilde{a}_{s\tilde{G}}\bigg] \tilde{c}_d \nonumber \\
& + 2 \bigg[ - \tilde{c}_d \tilde{a}_d^{\dagger} \tilde{c}_d + \frac{13}{4} \tilde{a}_d \tilde{c}_d^{\dagger} \tilde{c}_d + \frac{13}{4} \tilde{c}_d \tilde{c}_d^{\dagger} \tilde{a}_d\bigg] +  4 {\rm Tr} \left[ \tilde{c}_e \tilde{c}_e^{\dagger} + 3 \left( \tilde{c}_d \tilde{c}_d^{\dagger} + \tilde{c}_u \tilde{c}_u^{\dagger} \right)\right] \tilde{a}_d \nonumber \\
& -4 \tilde{e}\bigg[ \tilde{c}_d \tilde{c}_d^{\dagger} \tilde{a}_{dA}  + \tilde{a}_{dA} \tilde{c}_d^{\dagger} \tilde{c}_d\bigg] + 16 \tilde{g}_3\bigg[ \tilde{c}_d\tilde{c}_d^{\dagger} \tilde{a}_{dG}  + \tilde{a}_{dG} \tilde{c}_d^{\dagger} \tilde{c}_d\bigg] ~,\\
\beta_{\tilde{a}_e} &=\bigg[-6 \tilde{e}^2  +  \tilde{\lambda}_s\bigg] \tilde{a}_e -24\tilde{e}^2  \tilde{a}_{s\tilde{A}} \tilde{c}_e - 12  \tilde{e} \bigg[ \tilde{c}_e \tilde{c}_e^{\dagger} \tilde{a}_{eA}  + \tilde{a}_{eA} \tilde{c}_e^{\dagger} \tilde{c}_e\bigg] \nonumber \\
& + 2 \bigg[ - \tilde{c}_e \tilde{a}_e^{\dagger} \tilde{c}_e + \frac{13}{4} \tilde{a}_e \tilde{c}_e^{\dagger} \tilde{c}_e + \frac{13}{4} \tilde{c}_e \tilde{c}_e^{\dagger} \tilde{a}_e\bigg] +  4 {\rm Tr} \left[ \tilde{c}_e \tilde{c}_e^{\dagger} + 3 \left( \tilde{c}_d \tilde{c}_d^{\dagger} + \tilde{c}_u \tilde{c}_u^{\dagger} \right)\right] \tilde{a}_e \,,\\ \nonumber
\end{align}
\begin{align}
\beta_{\tilde{a}_{s\widetilde{A}}} &=4\tilde{e} {\rm Tr}\bigg[ \left(\tilde{c}_e \tilde{a}_{eA}^\dagger + \tilde{c}_e^{\dagger} \tilde{a}_{eA} \right) + \left( \tilde{c}_d \tilde{a}_{dA}^\dagger + \tilde{c}_d^{\dagger} \tilde{a}_{dA} \right) - 2 \left(\tilde{c}_u \tilde{a}_{uA}^\dagger + \tilde{c}_u^{\dagger} \tilde{a}_{uA} \right)  \bigg] \nonumber \\
& + 2 {\rm Tr} \left[ \tilde{c}_e \tilde{c}_e^{\dagger} + 3 \left( \tilde{c}_d \tilde{c}_d^{\dagger} + \tilde{c}_u \tilde{c}_u^{\dagger} \right)\right] \tilde{a}_{s\tilde{A}} + \frac{8}{3}\tilde{e}^2 \bigg[  n_\ell + \frac{1}{3} n_d  + \frac{4}{3} n_u \bigg]\tilde{a}_{s\tilde{A}} \,,\\
 \beta_{\tilde{a}_{s\widetilde{G}}} &=- 2\tilde{g}_3{\rm Tr} \left[ \tilde{c}_d \tilde{a}_{dG}^\dagger  + \tilde{c}_d^{\dagger }\tilde{a}_{dG} +  \tilde{c}_u \tilde{a}_{uG}^\dagger  + \tilde{c}_u^{\dagger }\tilde{a}_{uG} \right] +  2 {\rm Tr} \left[ \tilde{c}_e \tilde{c}_e^{\dagger} + 3 \left( \tilde{c}_d \tilde{c}_d^{\dagger} + \tilde{c}_u \tilde{c}_u^{\dagger} \right)\right] \tilde{a}_{s\tilde{G}} \nonumber \\
& + 2 \tilde{g}_3^2 \left[\frac{2}{3} (n_u+n_d) -11\right] \tilde{a}_{s\tilde{G}} \,,\\
 \beta_{\tilde{a}_{eA}} &= 10 \tilde{e}^2 \tilde{a}_{eA} + 4  \tilde{e} \tilde{c}_e \tilde{a}_{s\tilde{A}}+  \frac{1}{2} \left( \tilde{c}_e \tilde{c}_e^{\dagger}\tilde{a}_{eA} + \tilde{a}_{eA}  \tilde{c}_e^{\dagger} \tilde{c}_e \right) \nonumber \\
&+  \frac{4}{3}\tilde{e}^2 \bigg[ n_\ell + \frac{1}{3} n_d  + \frac{4}{3} n_u\bigg]  \tilde{a}_{eA}\,, \\
 \beta_{\tilde{a}_{uA}} &= \frac{40}{9} \tilde{e}^2 \tilde{a}_{uA} -   \frac{8}{3}\tilde{e} \tilde{c}_u \tilde{a}_{s\tilde{A}} +  \frac{64  }{9} \tilde{e} \tilde{g}_3 \tilde{a}_{u G}  + \frac{8}{3} \tilde{g}_3^2 \tilde{a}_{uA} + \frac{1}{2} \left( \tilde{c}_u \tilde{c}_u^{\dagger}\tilde{a}_{uA} + \tilde{a}_{uA}  \tilde{c}_u^{\dagger} \tilde{c}_u \right)  \nonumber \\
&+\frac{4}{3}\tilde{e}^2 \bigg[ n_\ell + \frac{1}{3} n_d  + \frac{4}{3} n_u\bigg] \tilde{a}_{uA}\,, \\
 \beta_{\tilde{a}_{dA}} &= \frac{10}{9}\tilde{e}^2 \tilde{a}_{dA} +   \frac{4}{3} \tilde{e} \tilde{c}_d \tilde{a}_{s\tilde{A}} -  \frac{32  \tilde{e} }{9}  \tilde{g}_3 \tilde{a}_{d G}  + \frac{8}{3} \tilde{g}_3^2 \tilde{a}_{dA} + \frac{1}{2} \left( \tilde{c}_d \tilde{c}_d^{\dagger}\tilde{a}_{dA} + \tilde{a}_{dA}  \tilde{c}_d^{\dagger} \tilde{c}_d \right)  \nonumber \\
& +\frac{4}{3}\tilde{e}^2 \bigg[ n_\ell + \frac{1}{3} n_d  + \frac{4}{3} n_u\bigg]\tilde{a}_{dA}\,, \\
 \beta_{\tilde{a}_{uG}} &= \frac{16}{3} \tilde{g}_3  \tilde{e} \tilde{a}_{uA} -  4\tilde{g}_3\tilde{c}_u \tilde{a}_{s\tilde{G}} + \frac{8}{9} \tilde{e}^2 \tilde{a}_{uG} + \frac{1}{2} \left( \tilde{c}_u \tilde{c}_u^{\dagger}\tilde{a}_{uG} + \tilde{a}_{uG}  \tilde{c}_u^{\dagger} \tilde{c}_u \right) \nonumber \\
&+\frac{1}{3}\bigg[ 2 \left(n_u + n_d\right) - 29  \bigg] \tilde{g}_3^ 2\tilde{a}_{uG}\,, \\
 \beta_{\tilde{a}_{dG}} &= -\frac{8}{3} \tilde{g}_3 \tilde{e} \tilde{a}_{dA} -  4\tilde{g}_3\tilde{c}_d^1 \tilde{a}_{s\tilde{G}} + \frac{2}{9} \tilde{e}^2 \tilde{a}_{dG} + \frac{1}{2} \left( \tilde{c}_d \tilde{c}_d^{\dagger}\tilde{a}_{dG} + \tilde{a}_{dG}  \tilde{c}_d^{\dagger} \tilde{c}_d \right) \nonumber \\
& +\frac{1}{3}\bigg[ 2 \left(n_u + n_d\right) - 29  \bigg] \tilde{g}_3^ 2\tilde{a}_{dG}\,;
\end{align}
where $n_u$, $n_d$ and $n_e$ are the number of dynamical up-type
quarks, down-type quarks and charged leptons, respectively, for the
EFT we are considering. The appearence of fermionic dipole moments induced by ALP couplings~\cite{Marciano:2016yhf} is manifest.

The equations above are fully generic, meaning that they hold
irrespective of whether the EFT in the UV is the one we have assumed
in Section~\ref{sec:eft} and that leads to the matching conditions in
Eqs.~\eqref{eq:matching}--\eqref{eq:matching2}, or rather any other EFT containing different
degrees of freedom such as for example a second scalar doublet with
arbitrary Yukawa couplings which would lead to non vanishing
$\tilde{a}_{u,d,e}$. In the former case, though, one should take
into account that $\tilde{c}_{u,d,e}$ are already $1/\Lambda$
suppressed and therefore terms with more than one appearance of these
Wilson coefficients should be neglected for consistency. 

Within our EFT(s), the parameters $\tilde{a}_u$, $\tilde{a}_d$ and
$\tilde{a}_e$ vanish at all scales, and the renormalizable fermion
and ALP masses do not get contributions from dimension-five
operators. In general this is not the case and we have, for example,
\begin{equation}
  \delta\tilde{m}^2\sim \tilde{m}_e^3\tilde{a}_e\,,
  \quad \delta\tilde{m}_e\sim \tilde{m}^2\tilde{a}_e\,,
\end{equation}
and likewise for quarks.

The running of the operators involving coloured particles should not be taken at face value at energies close or below $\Lambda_{QCD}\sim 200$ MeV, where QCD becomes strongly coupled. Otherwise, the equations above, together with Eqs.~\eqref{eq:beta10}--\eqref{eq:beta18}, can be used to make predictions within the ALP EFT to leading-log accuracy across all energy scales.

\section{Some phenomenological applications}
\label{sec:pheno}
The mixing of different operators can have a significant
impact in the understanding of extensions of the SM with ALPs. To
exemplify this point, we consider in this section the following simple
Lagrangian, defined at the scale $\Lambda=10$ TeV:
\begin{equation}
  \mathcal{L}
  = \mathcal{L}_{\mathrm{SM}}+ \frac{1}{2}\partial_\mu s\partial^\mu s + \frac{1}{2}\tilde{m}^2 s^2 
 + \frac{a_{s\widetilde{Z}}}{c_\omega^2 - s_\omega^2} s \left( c_\omega^2 W_{\mu\nu} \widetilde{W}^{\mu\nu} - s_\omega^2 B_{\mu\nu} \widetilde{B}^{\mu\nu} \right)
\,,
\end{equation}
where $\mathcal{L}_{\mathrm{SM}}$ stands for the SM Lagrangian. In
this Lagrangian, the ALP couples to pairs of $Z$ bosons but not to
pairs of photons. Despite 
its simplicity, this structure arises for example in the 
next-to-minimal composite Higgs model based on $SO(6)/SO(5)$ as a
result of quantum anomalies~\cite{Gripaios:2016mmi}. Within this
framework, the photophobic condition is stable, namely
$a_{s\tilde{A}}$ remains vanishing at all scales. (The ALP can still
couple to photons proportionally to $\tilde{m}$ via loops of heavy gauge
bosons~\cite{Craig:2018kne}, for instance.) However, below the EW scale, the
ALP coupling to photons could be induced by (purely SM) dipole
operators even at dimension five if the high-energy theory is not just
the ALP EFT, but it rather involves other states near the EW scale. 

Let us assume that the physical ALP mass is $\mathcal{O}(\text{KeV})$.
While $a_{s\widetilde{Z}}$ can be directly constrained at colliders,
\textit{e.g.} in $pp\to Zs$, the corresponding bounds are very
weak. For example, values of $a_{s\widetilde{Z}}$ larger
than $0.2\,\mathrm{TeV}^{-1}$
can be bound, depending on the value of $\tilde{m}$, from
LHC Run II data~\cite{Brivio:2017ije}.  This
bound can be extended to values above $0.04\, \mathrm{TeV}^{-1}$ for the High-Luminosity
phase of the LHC~\cite{Brivio:2017ije}. 
However, $a_{s\widetilde{Z}}$ does generate, through mixing, other operators with
  non-vanishing Wilson coefficients, in particular
  $a_{se\phi}$, which is 
  very constrained experimentally.
Indeed, the most stringent bound for ALPs with masses $\sim \mathrm{KeV}$
comes from the modification of Red Giants cooling due to ALP
radiation. This sets a bound on the ALP coupling to electrons
$\tilde{c}_{e} \lesssim 3\times10^{-13}$, for a typical  core
temperature of $T \approx 10^{8}$ K~\cite{Raffelt:2006cw}. Since we
are assuming that there are no degrees of freedom other than the ALP
and the SM ones below $\Lambda$, $\tilde{c}_e$ runs 
proportional to itself.  Resumming 
Eqs.~\eqref{eq:erunLEFT} and \eqref{eq:cerunLEFT}, including
$\mathcal{O}(1/\Lambda)$ effects,  
we obtain at the EW scale:
\begin{equation}
\tilde{c}_e(v)\lesssim 2.8\times 10^{-13}, \label{cev}
\end{equation}
which translates into
\begin{equation}
a_{se\phi} (v) \lesssim 1.6 \times 10^{-12} \; \mathrm{TeV}^{-1}\,.
\end{equation}
Solving numerically
Eqs.~\eqref{eq:beta10}-\eqref{eq:beta14},~\eqref{eq:betag1}-\eqref{eq:betag3}
and \eqref{eq:betayu}-\eqref{eq:betaye} for $\lambda_{s\phi} = 0$, we
can compute the maximum allowed value for $a_{s\widetilde{Z}}(10\,\text{TeV})$:
\begin{equation}
  a_{s\widetilde{Z}}(10\,\text{TeV}) \lesssim
  4.8 \times 10^{-6} \; \mathrm{TeV}^{-1}\,.
\end{equation}
Despite the electron Yukawa suppression,
  this is four orders of magnitude stronger than prospects from direct
searches. Such analysis of the photophobic ALP was previously
performed in Ref.~\cite{Craig:2018kne} by considering solely the
running of the gauge operators. As shown in Eq. (\ref{cev}) the
running from the EW scale down to the KeV amounts to a $\sim 6\%$
effect, which
can be taken as a systematic theory error when using only
the EFT before EWSB. See also
Ref.~\cite{Terol-Calvo:2019vck} for an analysis of the RGE effects in bounds on neutrino
interactions resulting in similar numbers. 

As another example, we consider the case of a top-philic ALP at $\Lambda=10$ TeV,
with Lagrangian 
\begin{equation}
  \mathcal{L}
  =
  \mathcal{L}_{\mathrm{SM}}+\frac{1}{2}\partial_\mu s\partial^\mu s + \frac{1}{2}\tilde{m}^2 s^2 
  + a_t s[i \overline{q}_L\tilde{\phi}t_R + \text{h.c.}] \,,
\end{equation}
where $q_L$ stands for the third generation quark doublet and, in our
notation, $a_t=(a_{su\phi})^{33}$.
As before, $a_t$ generates, via renormalization mixing, a
non-vanishing $a_{se\phi}$.
Proceeding in the same way as above, we obtain
$a_t(10\,\text{TeV}) \lesssim 4.3 \times 10^{-6}\; \mathrm{TeV}^{-1}$
from the bound $\tilde{c}_e(\mu\sim\mathrm{KeV}) \lesssim 3\times
10^{-13}$.
Direct bounds on this coupling could in principle be obtained from
$pp\to t\overline{t} s$, but they are likely to be very weak due to
the challenging final state. Other indirect constraints on $a_t$ have
been studied in~Ref.\cite{Ebadi:2019gij} but they are again much weaker
than the one we have obtained.

Other interesting phenomenological implications, like the
possible impact of non-SM degrees of freedom using the generic
RGEs, and in particular the mixing  between operators of different
dimensions and different ALP content, are left for future work.

\section{Conclusions}
\label{sec:conclusions}

In this article we have investigated the EFT for
ALPs up to order $\mathcal{O}(1/\Lambda)$ in the
cutoff scale $\Lambda$. We have worked in a complete basis of EFT
  operators to dimension five, including both shift-preserving and
  shift-breaking interactions.  
Assuming that CP is conserved in the UV,
we have computed, at one loop, the
evolution of the CP-even effective operators under renormalization group
running from arbitrarily high energies down to
the EW scale.
In the ALP LEFT, relevant at smaller energies, in which the heavy top
  quark and the Higgs, $Z$ and $W$ bosons are no longer
  dynamical, we have also computed the renormalization of all,
  relevant and marginal, parameters. We have found that, in general,  effective
interactions can renormalize dimension-four ones, and
  operators 
involving the ALP mix with pure SM operators; although the latter effect
vanishes if the theory above the EW scale involves only the ALP and SM
degrees of freedom. For this latter case we have also provided the
matching conditions between the EFTs above and below the EW
threshold. Interestingly, we have shown that in the presence of SM
dimension-five interactions below the EW scale, the ALP coupling to
photons no longer renormalizes proportionally to itself. 

To make our work more useful, we have not only given the full list
of beta functions in a 
minimal basis but we have also explicitly written the
 corresponding counterterms of all independent (off-shell) Green
 functions. This is 
 important for two reasons: first because without this information,
 the RGEs of extensions of our EFT, \textit{e.g} by adding
 right-handed neutrinos, can not be built on our
 results, as redundancies are different in different EFTs;
 second because while
 in analytical one-loop computations within $\overline{\text{MS}}$
 the counterterms might be in principle ignored by just dropping the
 $1/\epsilon$ poles, their precise value is of fundamental importance
 in numerical Monte Carlo simulations. 

For the ALP EFT in the unbroken phase, 
several RGEs have been obtained previously in the literature. These assumed however shift invariance, and were therefore presented in a different set of operators, in which the ALP comes always in derivatives.
We have discussed some redundancies that appear in this set of
operators, and obtained conditions on the corresponding coefficients
under which they actually form a basis of independent
interactions. Upon relating this basis to ours, we have obtained the
RGEs in the former. 
In this way, we have not only completely solved the
full dimension-five ALP EFT to leading-log order, but also cross-checked several of the partial results which
were somewhat spread over different references~\cite{Bauer:2016lbe,Choi:2017gpf,Bauer:2017ris}.

Finally, as an example of the utilization of our results, we have
explored the possibility of indirectly probing ALP-$Z$ as well as
ALP-top interactions through their contribution to the ALP-electron
coupling, which is bounded by very low-energy measurements. (We leave the interesting possibility of testing ALP interactions through its mixing into pure SM ones, and vice-versa, to future work.)
This shows
the 
potential of our results to study the ALP EFT phenomenology to
leading-log accuracy across all energy scales. 

\section*{Acknowledgments}
We would like to thank Adri\'an Carmona for useful feedback and for
suggesting the title. MC thanks also Sara Khatibi for useful discussions.
We would also like to thank
Supratim Das Bakshi and Jonathan Machado for pointing out typos in previous versions of
the manuscript.
MC and JS are partially supported 
by the Ministry of Science and Innovation
under grant number FPA2016-78220-C3-1/3-P (FEDER), SRA under grant 
PID2019-106087GB-C21/C22 (10.13039/501100011033), and 
by the Junta de Andaluc\'ia grants FQM 101, and A-FQM-211-UGR18 
and P18-FR-4314 (FEDER).
MC is also supported by the Spanish MINECO under the Ram\'on y Cajal programme.
GG and MR acknowledge support by LIP 
(FCT, COMPETE2020-Portugal2020, FEDER, POCI-01-0145-FEDER-007334) as well as by FCT under project CERN/FIS-PAR/0024/2019.
GG is also supported by FCT under the grant SFRH/BD/144244/2019.
MR is also supported by Funda\c{c}\~ao para a Ci\^encia e Tecnologia (FCT) under the grant PD/BD/142773/2018.

\appendix

\section{One-loop diagrams in the ALP SMEFT}\label{app:HEdiags}



\section{Shift symmetry and relations between different bases}
\label{app:bases}
It is often interesting to assume that the ALP EFT is strictly
invariant under the shift $s\to s+\sigma$, with $\sigma$ being
constant. 
In this case, the following Lagrangian is more commonly used
in the literature:
\begin{equation}\label{eq:newbasis}
\mathcal{L}_{dim-5} = \sum_{\Psi}\left(\partial_\mu s\right)
\overline{\Psi} C_\Psi \gamma^\mu \Psi + C_B g_1^2  sB_{\mu\nu}\widetilde{B}^{\mu\nu} + C_W g_2^2
sW_{\mu\nu}^a \widetilde{W}_{a}^{\mu\nu}+ C_G g_3^2  sG_{\mu\nu}^A
\widetilde{G}_{A}^{\mu\nu}\,,
\end{equation}
where $\Psi$ runs over $q_L, l_L, u_R, d_R, e_R$ and $C_\Psi$ are hermitian
matrices in flavour space.
The advantage of this parametrisation is that the shift
symmetry is explicit. However, it must be noted that not all terms in
the Lagrangian are independent. Indeed, focusing for concreteness on
the leptons, we note that there are $9+9=18$ independent real
parameters from $C_e$ and $C_l$. This is the same number of parameters
that we have in the lepton sector of the (non-redundant) Lagrangian
used in Section~\ref{sec:eft}, but the latter involves all
possible terms, including those that are not shift-invariant. 

One could be tempted to think that the LH operators can be traded,
on-shell, by the RH ones. This is however not true. In general, the
following relation holds on shell: 
\begin{equation}
 \partial_\mu s\overline{l_L} C_l \gamma^\mu l_L = \partial_\mu s
 \overline{e_R}H\gamma^\mu e_R +
 s\overline{e_R} (\ii A)
 \ii\overleftrightarrow{\slashed{D}}e_R
 \,, \label{dmuslvsdmusr}
\end{equation}
where $\overleftrightarrow{D_\mu}\equiv D_\mu - \overleftarrow{D_\mu}$
and $H$ and $A$ are hermitian and anti-hermitian matrices,
respectively, given by  
\begin{align}
  A &= \frac{1}{2} \bigg[ (y^e)^{-1} C_l y^e - y^{e\dagger}C_l
    (y^e)^{-1\dagger} \bigg]\,,\label{Amat}\\ 
 H &= -\frac{1}{2} \bigg[(y^e)^{-1} C_l y^e + y^{e\dagger}C_l
   (y^e)^{-1\dagger}\bigg]\,. \label{Hmat}
\end{align}
So the LH and RH operators are only equivalent if $A$ vanishes. A
sufficient condition for this to happen is that the Yukawa coupling
commutes with $C_l$, which
happens in particular if the ALP couples to a
single lepton family.

The on-shell relation in Eq.~\eqref{dmuslvsdmusr} might seem
counter-intuitive at first, since the left-hand side is explicitly
shift-invariant but the term in the right-hand side proportional to
$A$ does not seem to be. However a closer inspection shows that this
term is in fact shift-invariant as it should. To prove it, let us consider the
operator
\begin{equation}
s\overline{e_R} C
 \ii\overleftrightarrow{\slashed{D}}e_R\,,
\end{equation}
with $C$ being an arbitrary hermitian matrix of order $1/\Lambda$.
In order to see under which conditions this operator is shift
invariant we perform a shift,
$s\to s+\sigma$, 
\begin{equation}
 s\overline{e_R} C
 \ii\overleftrightarrow{\slashed{D}}e_R
 \to
  (s+\sigma)\overline{e_R} C
 \ii\overleftrightarrow{\slashed{D}}e_R\,,
\end{equation}
and we get an extra contribution to the $e$ kinetic term, which
can be canonically normalized via the following field redefinition,
\begin{equation}
  e_R \to (1-\sigma C) e_R\,.
\end{equation}
Since $C$ is of order $1/\Lambda$, the only effect of this field
redefinition happens in the renormalizable Lagrangian and therefore
only in the Yukawa term:
\begin{equation}
  \overline{l_L} \phi y^e e \to
  \overline{l_L} \phi (y^e -\sigma y^e C) e_R.
\end{equation}
We can now perform arbitrary chiral rotations which, to the order we
are considering, read
\begin{equation}
  l_L \to (1+A_L) l_L, \qquad
  e_R \to (1+A_R) e_R\,, \label{chiral:rotations}
\end{equation}
where $A_{L,R}$ are anti-hermitian matrices of order $\sigma/\Lambda$. Under
these rotations, the Yukawa Lagrangian receives the following correction:
\begin{equation}
  \overline{l_L} \phi y^e e_R \to
  \overline{l_L} \phi (y^e - \sigma y^eC-A_L y^e + y^e A_R) e_R\,,
\end{equation}
which is independent of $\sigma$ (and therefore shift invariant) if
\begin{equation}
  y^e C= \frac{y^e A_R-A_L y^e}{\sigma}\,.
\end{equation}
Using the hermiticity of $C$ we can eliminate $A_R$ and get the
following sufficient condition for shift invariance:
\begin{equation}
  C^{\mathrm{shift-inv}}=y^{e\,\dagger} A_L  (y^{e\,\dagger})^{-1} -
  (y^{e})^{-1} A_L y^e,
\end{equation}
with $A_L$ being an arbitrary anti-hermitian matrix of order $1/\Lambda$ (we
have absorbed in its 
definition a factor of $1/2\sigma$). We see from
Eq.~\eqref{Amat} that $\ii A$ has precisely that form with the
identification $A_L= -\ii C_l/2$ and therefore Eq.~\eqref{dmuslvsdmusr}
is, as 
anticipated, shift invariant.

Eq.~\eqref{dmuslvsdmusr} shows that the Lagrangian in
Eq.~\eqref{eq:newbasis} is in general redundant but also that, if $A$
vanishes, a minimal basis is given by eliminating the
operators with LH fields in this equation. In this case the Wilson
coefficients are univocally related in both bases and we can translate
the anomalous dimensions to the new one. The relations between the
Wilson coefficients are:
\begin{align}
a_{sB} &= C_B g_1^2\,,\\
a_{sW} &= C_W g_2^2\,,\\
a_{sG} &= C_G g_3^2\,,\\
a_{s\psi \phi} &=   -\mathrm{Re}(y^\psi C_{\psi_R})
\,,\\
a_{\widetilde{s \psi \phi}} &=  \mathrm{Im}(y^\psi C_{\psi_R})
\,,
\end{align}
where $\psi$ stands for $u$, $d$ or $e$.
Using the results in Section~\ref{sec:rges} and the SM RGEs in
Appendix~\ref{app:4dim}, we obtain the following RGEs for in the new
basis, valid in the limit of vanishing complex phases for the
  Yukawa couplings:
%
\begin{align} 
  \beta_{C_{u}}  =& \,y^{u\,\dagger} y^u C_{u}  +2 C_u y^{u\,\dagger} y^u
+2 \lambda_{s\phi} C_u + (y^u)^{-1} y^d C_d y^{d\,\dagger} y^u
\nonumber \\
&
+   \frac{17}{3} g_1^4 C_B +9
g_2^4 C_W + 32 C_G g_3^4 C_G
- 2 \gamma'
\,, 
\\
  \beta_{C_{d}}  =& \,y^{d\,\dagger} y^d C_{d}  +2 C_d y^{d\,\dagger} y^d
+2 \lambda_{s\phi} C_d + (y^d)^{-1} y^u C_u y^{u\,\dagger} y^d
\nonumber \\
&
+   \frac{5}{3}g_1^4 C_B +9 g_2^4 C_W + 32  g_3^4 C_G
+ 2 \gamma'
\,, 
\\
  \beta_{C_{e}}  =& \,y^{e\,\dagger} y^e C_{e}  +2 C_e y^{e\,\dagger} y^e
  +2 \lambda_{s\phi} C_e
%
+ 15 g_1^4 C_B +9 g_2^4 C_W 
+ 2 \gamma'
\,, 
\end{align}
%
with
\begin{equation}
  \gamma' \equiv  {\rm Tr}\left[
    y^e C_e y^{e\,\dagger} 
   +3y^d C_d y^{d\,\dagger} 
   -3y^u C_u y^{u\,\dagger} 
\right]\,.
\end{equation}

Note that the ALP-fermion-fermion operators in the new basis are not renormalized by gauge interactions. The gauge contributions to the $s\overline{\Psi} \phi \Psi$ operators is exactly cancelled by the running of the Yukawa couplings.
This result can be ultimately rooted to the fact that
$\overline{\psi}\gamma_\mu \psi$ is a conserved current of $U(1)$ and
therefore does not renormalize to itself
off-shell~\cite{Collins:2005nj}, and by noticing that the non-abelian
structure of the EW and color gauge interactions does not manifest in
the renormalization of $\partial_\mu s \overline{\psi}\gamma_\mu \psi$
at one loop.

Given the discussion so far, we find worth discussing shift invariance
in the basis of operators introduced in Section~\ref{sec:eft}, which
contrary to Eq.~\eqref{eq:newbasis} is minimal.
We find the following sufficient conditions on the Wilson coefficients
for a shift-invariant theory:
\begin{align}
  a_{s u \phi}^{\mathrm{shift-inv}}
  &= \text{Re}(H_{q_L}y^u + y^u H_{u_R})\,,
   \label{asuf:shiftinv}
\\
 a_{\widetilde{s u \phi}}^{\mathrm{shift-inv}} &= -\text{Im}(H_{q_L}y^u + y^u H_{u_R}) \,,\\
  a_{s d \phi}^{\mathrm{shift-inv}} &= \text{Re}(H_{q_L}y^d + y^d H_{d_R})\,,\\
 a_{\widetilde{s d \phi}}^{\mathrm{shift-inv}} &= -\text{Im}(H_{q_L}y^d + y^d H_{d_R}) \,,\\
  a_{s e \phi}^{\mathrm{shift-inv}} &= \text{Re}(H_{l_L}y^e + y^e H_{e_R})\,,\\
 a_{\widetilde{s e \phi}}^{\mathrm{shift-inv}} &= -\text{Im}(H_{l_L}y^e + y^e H_{e_R}) \,,
 \label{aseft:shiftinv}
\end{align}
with $H_{q_L,l_L,u_R,d_R,e_R}$ being arbitrary hermitian matrices.

Let us see that this is indeed the case. We show it explicitly for the
case of leptons, quarks being a straightforward generalization.
Let us consider the following Lagrangian:
\begin{equation}
  \mathcal{L}= -\overline{l_L} ( y^e - \ii s \alpha_e)
  \phi e_R + \text{h.c.}+\ldots\,,
\end{equation}
where $\alpha_e$ is an
arbitrary matrix in flavour space of order $1/\Lambda$ and the dots
stand for other interactions that are not relevant for the present
discussion. 
A shift $s\to s + \sigma$ induces
the following change in the Lagrangian:
\begin{equation}
  \mathcal{L}\to \mathcal{L} -\left[\overline{l_L} ( - \ii \sigma
    \alpha_e) 
  \phi e_R + \text{h.c.}\right]\,.
\end{equation}
Performing again arbitrary chiral rotations as in
Eq.~\eqref{chiral:rotations} we get, to the order we 
are considering,
\begin{equation}
  \mathcal{L}\to  \mathcal{L}-\left[\overline{l_L} (y^e A_{e_R} -
    A_{l_L}  y^e - \ii \sigma \alpha_e)
  \phi e_R + \text{h.c.}\right]\,. \label{extrarot}
\end{equation}
We therefore see that a sufficient condition for $\alpha_e$ to 
keep shift invariance is that it has the form
\begin{equation}
  \alpha_e^{\mathrm{shift-inv}}=H_{l_L}  y^e+ y^e H_{e_R}\,,
\end{equation}
with $H_{l_L,e_R}$ being arbitrary hermitian matrices, corresponding to $\pm
\ii A_{l_L,e_R}/\sigma$ in Eq.~\eqref{extrarot}, respectively.
Taking now into account the
relation between $\alpha_e$ and the Wilson coefficients in our
basis, namely,
\begin{equation}
  a_{s e \phi}=\text{Re}\alpha_e,
  \qquad
  a_{\widetilde{s e \phi}}=-\text{Im}\alpha_e\,,
\end{equation}
we obtain the condition in Eqs.~\eqref{asuf:shiftinv}-\ref{aseft:shiftinv}.

\section{Running of renormalizable parameters above the electroweak scale}
\label{app:4dim}
The renormalizable parameters of the theory above EWSB evolve following:
\begin{align}
\label{eq:betag1}
 \beta_{g_1} &= \frac{41}{6}g_1^3\,,\\
 \beta_{g_2} &= -\frac{19}{6}g_2^3\,,\\
\label{eq:betag3}
 \beta_{g_3} &= -7 g_3^3\,,\\
 %
%
 %
 \beta_{m^2} &= 4 \lambda_{s\phi} \mu^2 + \lambda_s m^2\,,\\
 \beta_{\mu^2} &= \lambda_{s\phi} m^2 + \left[2 \text{Tr}(y^e {y^e}^\dagger) + 6\text{Tr}(y^u {y^u}^\dagger) + 6\text{Tr}(y^d {y^d}^\dagger) -\frac{3}{2}g_1^2-\frac{9}{2}g_2^2 + 12\lambda\right] \mu^2\,,\\
 %
%
 %
 \beta_{\lambda_s} &= 3\lambda_s^2 + 12 \lambda_{s\phi}^2\,,\\
 \beta_{\lambda_{s\phi}} &= \left[\lambda_s + 4\lambda_{s\phi} + 2\text{Tr}(y^e {y^e}^\dagger)+6\text{Tr}(y^u {y^u}^\dagger)+6\text{Tr}(y^d {y^d}^\dagger)+12\lambda-\frac{3}{2}g_1^2-\frac{9}{2}g_2^2\right]\lambda_{s\phi}\,,\\\nonumber
 %
 %
 \beta_{\lambda} &= \frac{1}{2}\lambda_{s\phi}^2 -6\text{Tr}(y^u {y^u}^\dagger y^u {y^u}^\dagger)-6\text{Tr}(y^d {y^d}^\dagger y^d {y^d}^\dagger)-2\text{Tr}(y^e {y^e}^\dagger y^e {y^e}^\dagger)+\frac{3}{8}g_1^4+\frac{9}{8}g_2^4+\frac{3}{4}g_1^2g_2^2\\
 &+\left[24\lambda -3g_1^2-9g_2^2+ 4\text{Tr}(y^e {y^e}^\dagger)+12\text{Tr}(y^u {y^u}^\dagger)+12\text{Tr}(y^d {y^d}^\dagger)\right]\lambda\,,\\
 %
%
 %
\label{eq:betayu}
 \beta_{y^u} &= \left\lbrace\frac{3}{2} y^u {y^u}^\dagger - \frac{3}{2} y^d {y^d}^\dagger + 3\left[\text{Tr}(y^u {y^u}^\dagger) + \text{Tr}(y^d {y^d}^\dagger)\right]+\text{Tr}(y^e {y^e}^\dagger)-\frac{17}{12}g_1^2-\frac{9}{4}g_2^2-8g_3^2\right\rbrace y^u\,,\\
 \beta_{y^d} &= \left\lbrace\frac{3}{2} y^d {y^d}^\dagger - \frac{3}{2} y^u {y^u}^\dagger + 3\left[\text{Tr}(y^u {y^u}^\dagger) + \text{Tr}(y^d {y^d}^\dagger)\right]+\text{Tr}(y^e {y^e}^\dagger)-\frac{5}{12}g_1^2-\frac{9}{4}g_2^2-8g_3^2\right\rbrace y^d\,,\\
\label{eq:betaye}
 \beta_{y^e} &= \left\lbrace\frac{3}{2} y^e {y^e}^\dagger + 3\left[\text{Tr}(y^u {y^u}^\dagger) + \text{Tr}(y^d {y^d}^\dagger)\right]+\text{Tr}(y^e {y^e}^\dagger)-\frac{15}{4}g_1^2-\frac{9}{4}g_2^2\right\rbrace y^e\,.
\end{align}

\section{One-loop diagrams in the ALP LEFT}
\label{diagrams:left}

\begin{figure}[H]
 \centering
 \includegraphics[width=0.20\columnwidth]{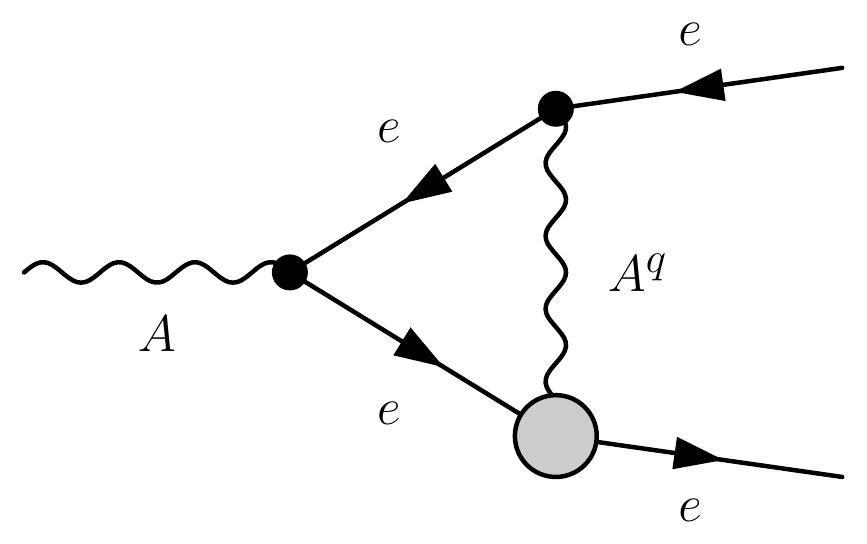}
 \includegraphics[width=0.20\columnwidth]{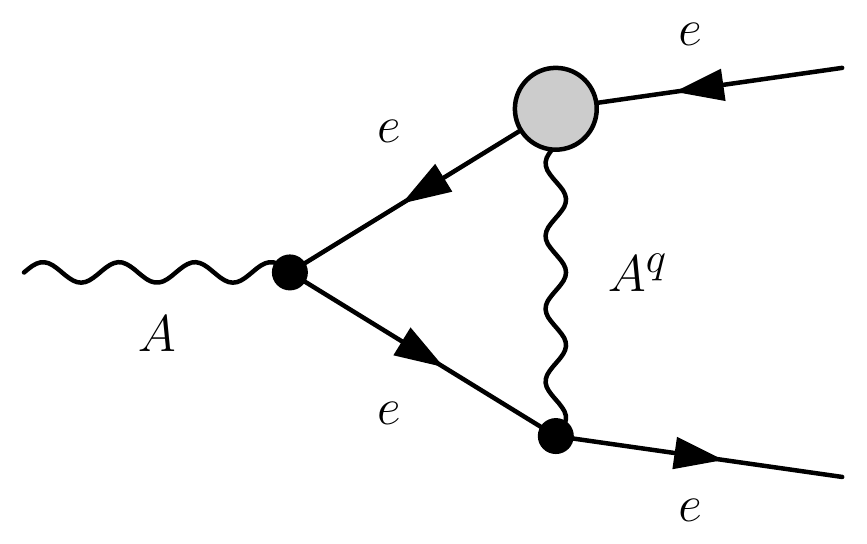}
 \includegraphics[width=0.20\columnwidth]{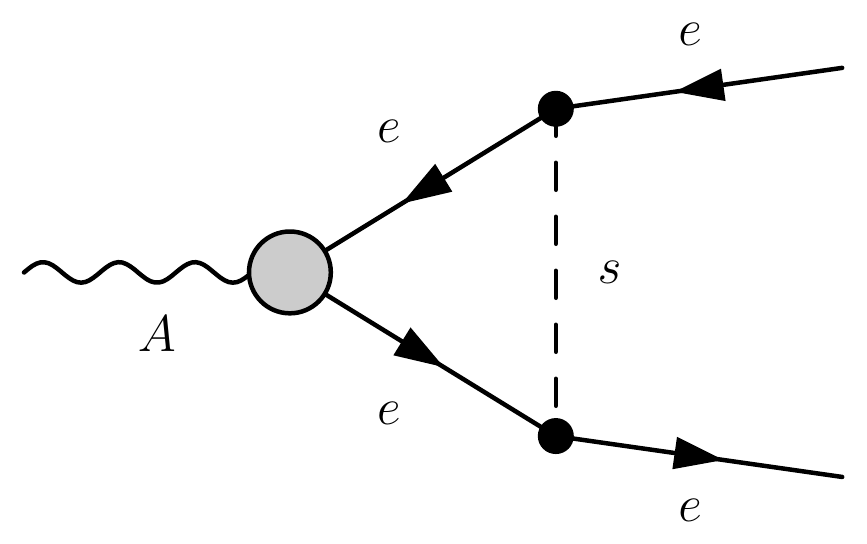}
 \includegraphics[width=0.20\columnwidth]{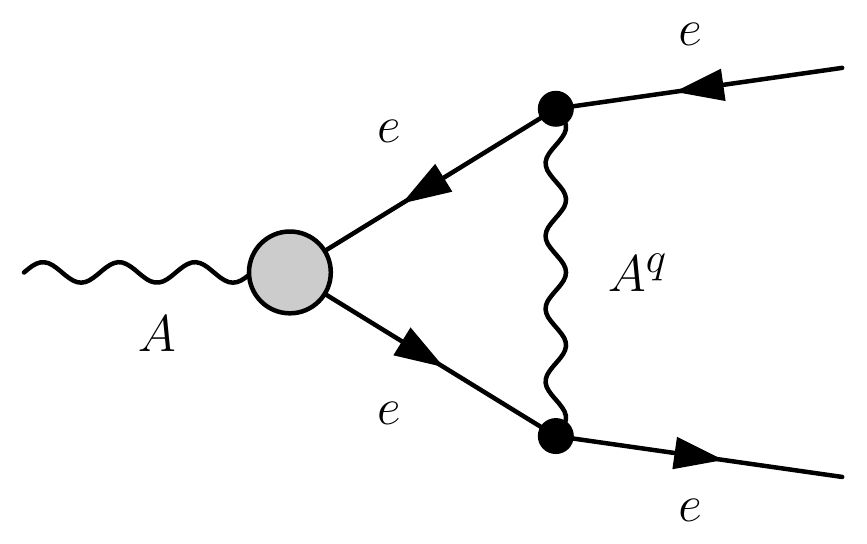}
 \includegraphics[width=0.20\columnwidth]{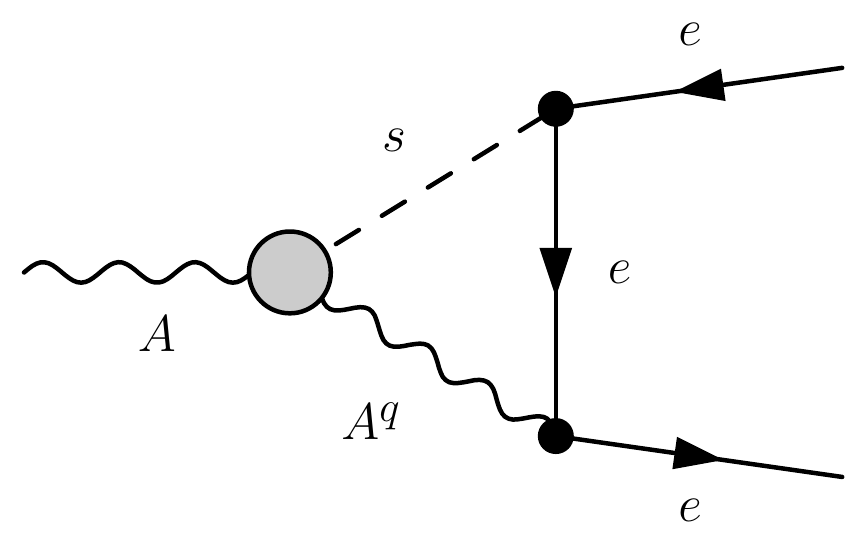}
 \includegraphics[width=0.20\columnwidth]{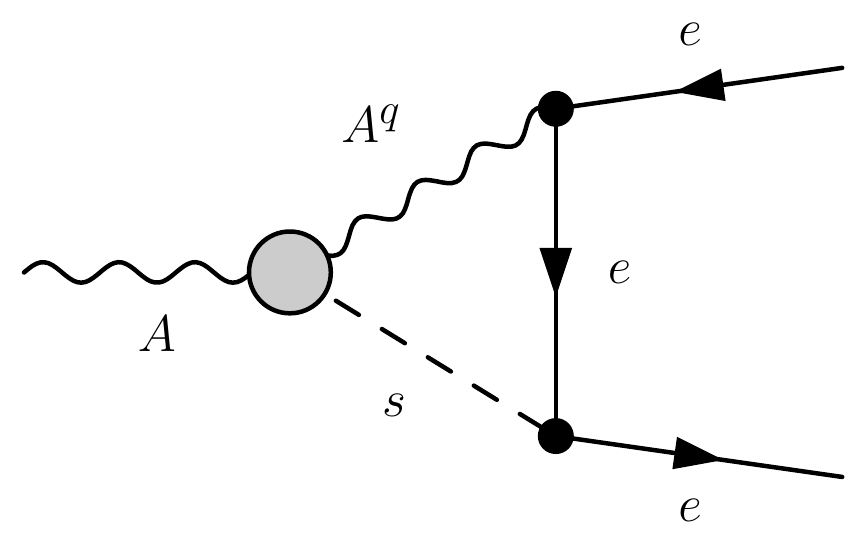}
 \caption{\it Feynman diagrams for $A(p_1)\to\overline{e}(p_2)
   e(p_3)$.
 }
 \label{fig:a_ee}  
\end{figure}

\begin{figure}[H]
 \centering\vspace{1cm}
 \includegraphics[width=0.20\columnwidth]{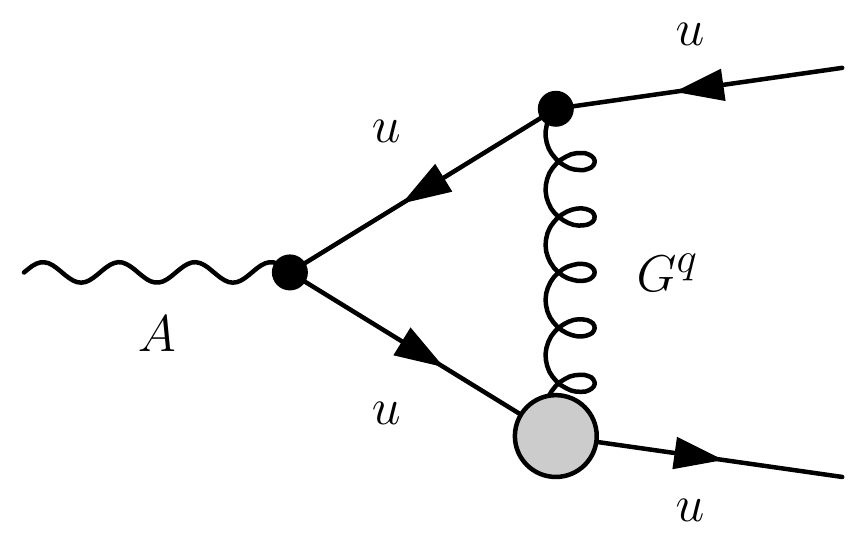}
 \includegraphics[width=0.20\columnwidth]{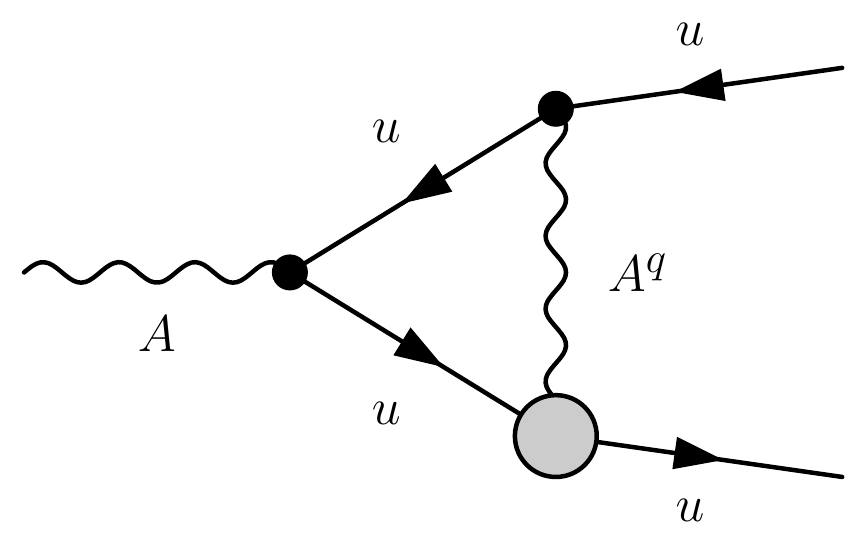}
 \includegraphics[width=0.20\columnwidth]{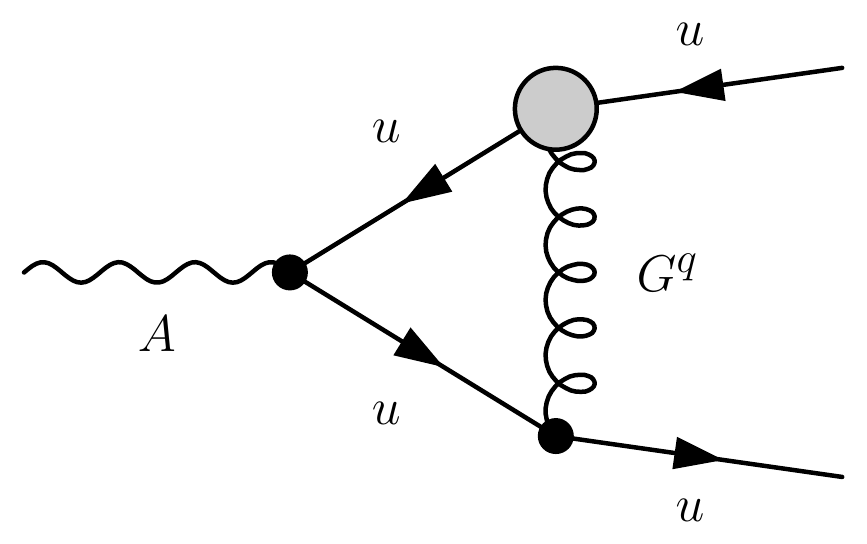}\\[0.5cm]
 \includegraphics[width=0.20\columnwidth]{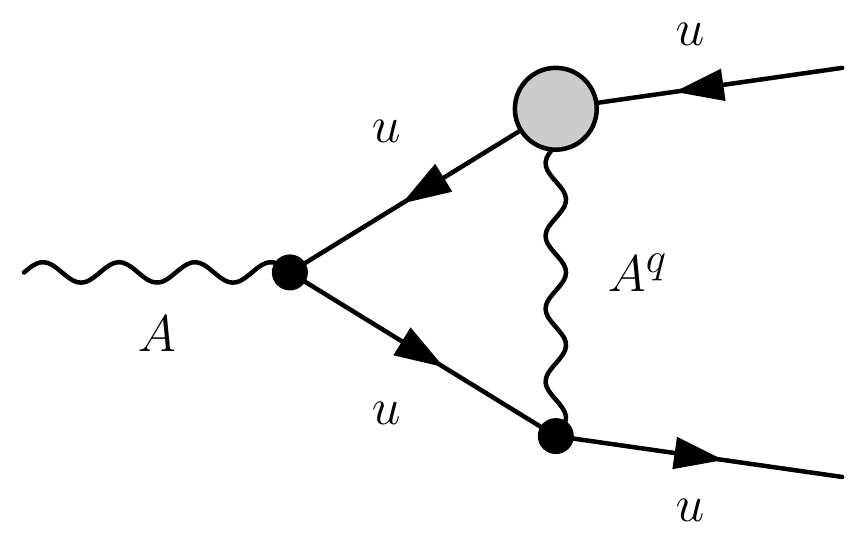}
 \includegraphics[width=0.20\columnwidth]{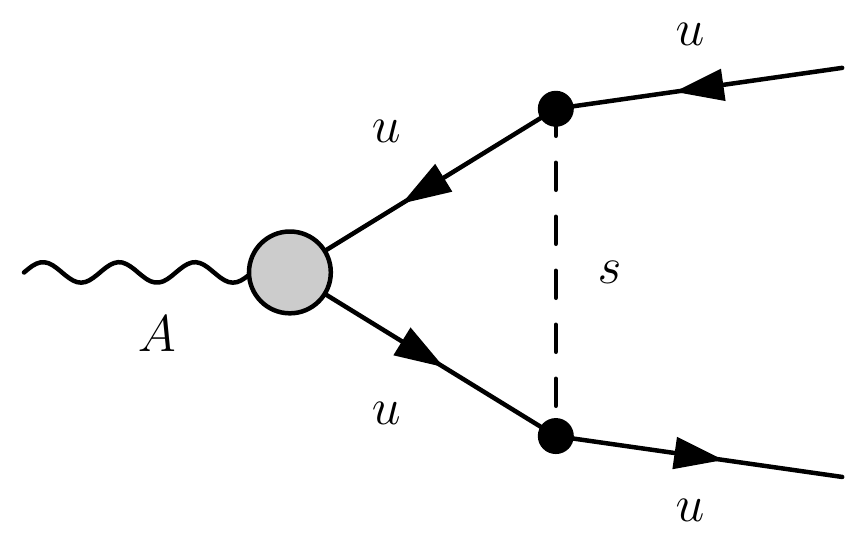}
 \includegraphics[width=0.20\columnwidth]{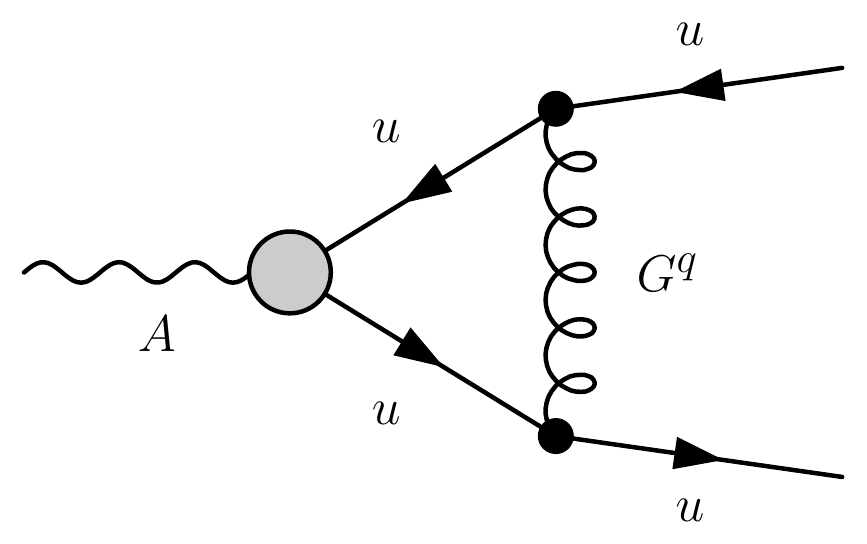}\\[0.5cm]
 \includegraphics[width=0.20\columnwidth]{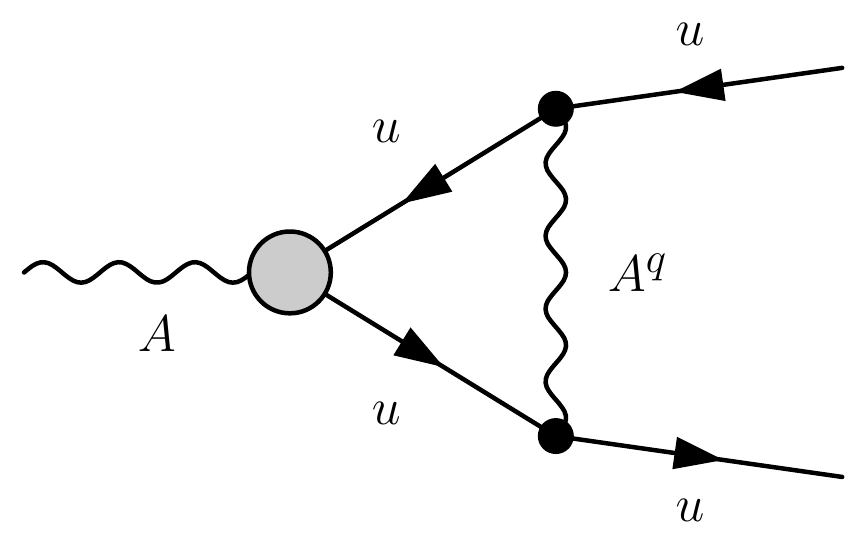}
 \includegraphics[width=0.20\columnwidth]{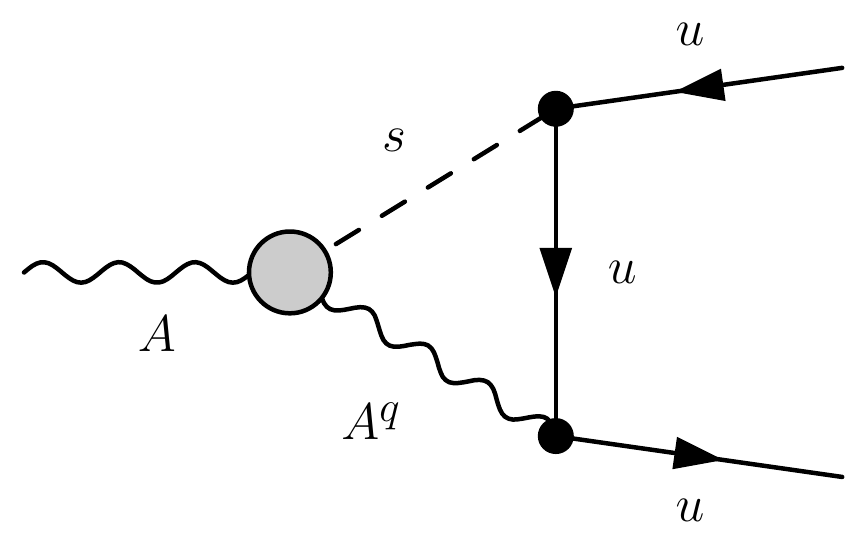}
 \includegraphics[width=0.20\columnwidth]{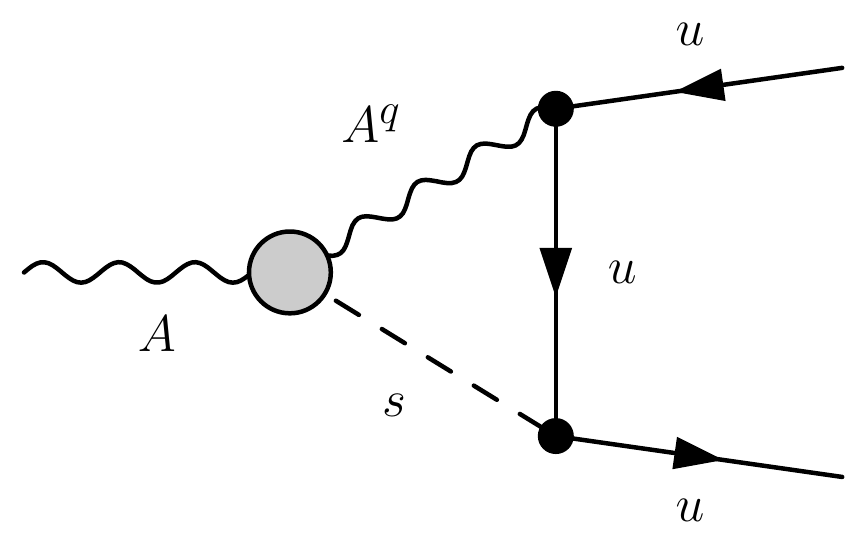}
 \caption{\it Feynman diagrams for $A(p_1)\to\overline{u}(p_2) u(p_3)$.}\label{fig:a_uu}
\end{figure}

\vspace{1cm}

\begin{figure}[H]
 \centering
 \includegraphics[width=0.20\columnwidth]{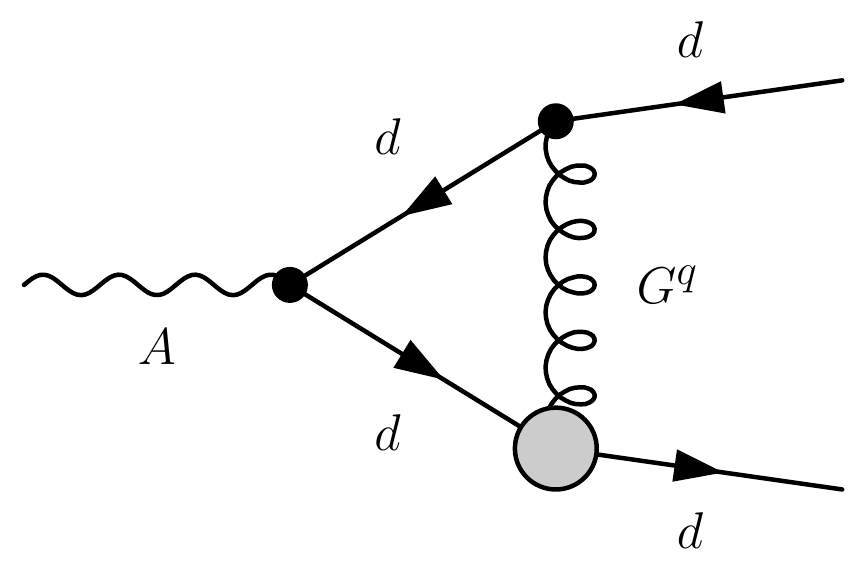}
 \includegraphics[width=0.20\columnwidth]{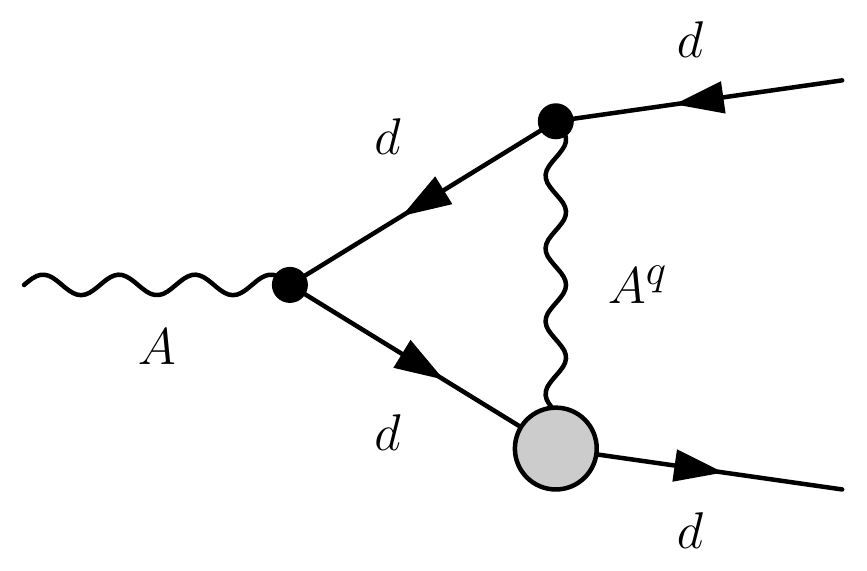}
 \includegraphics[width=0.20\columnwidth]{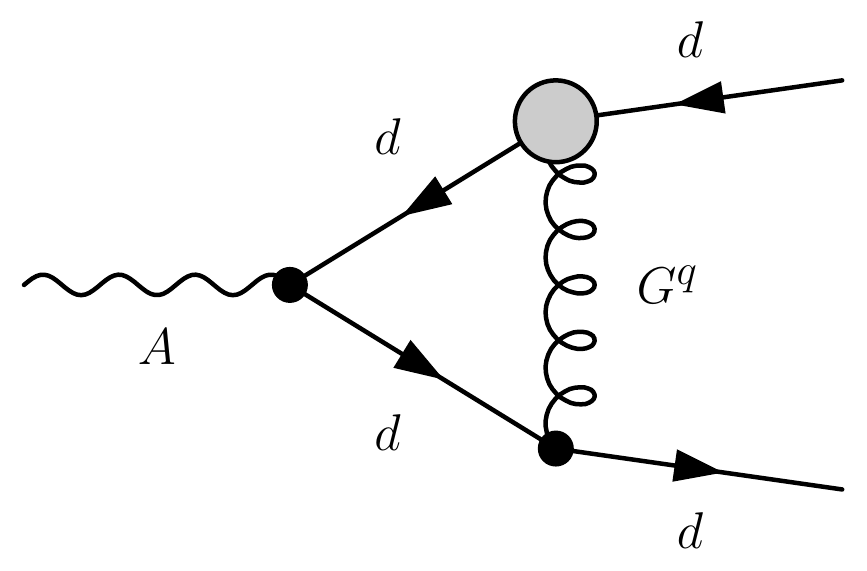}\\[0.5cm]
 \includegraphics[width=0.20\columnwidth]{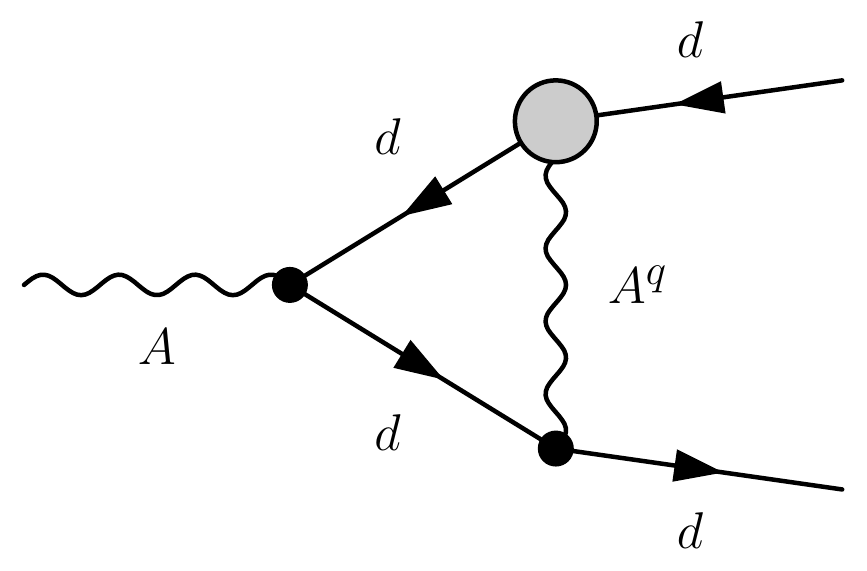}
 \includegraphics[width=0.20\columnwidth]{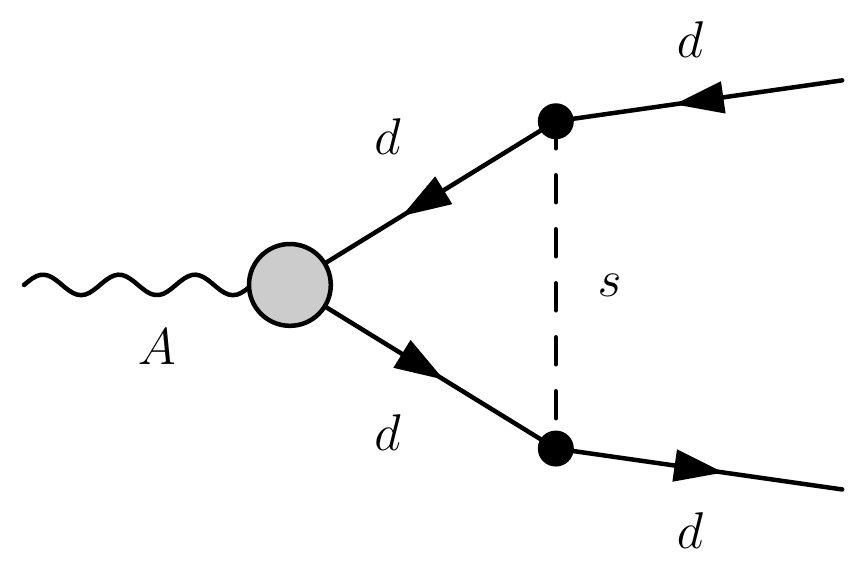}
 \includegraphics[width=0.20\columnwidth]{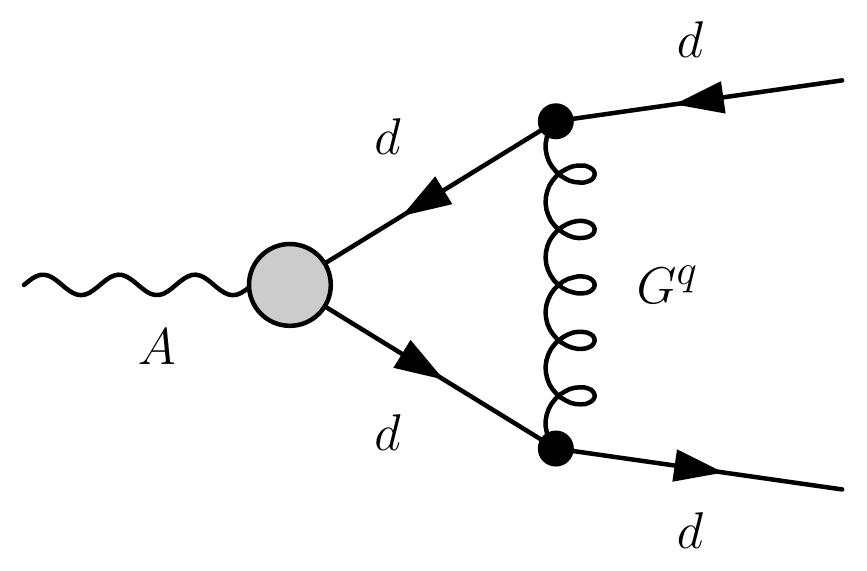}\\[0.5cm]
 \includegraphics[width=0.20\columnwidth]{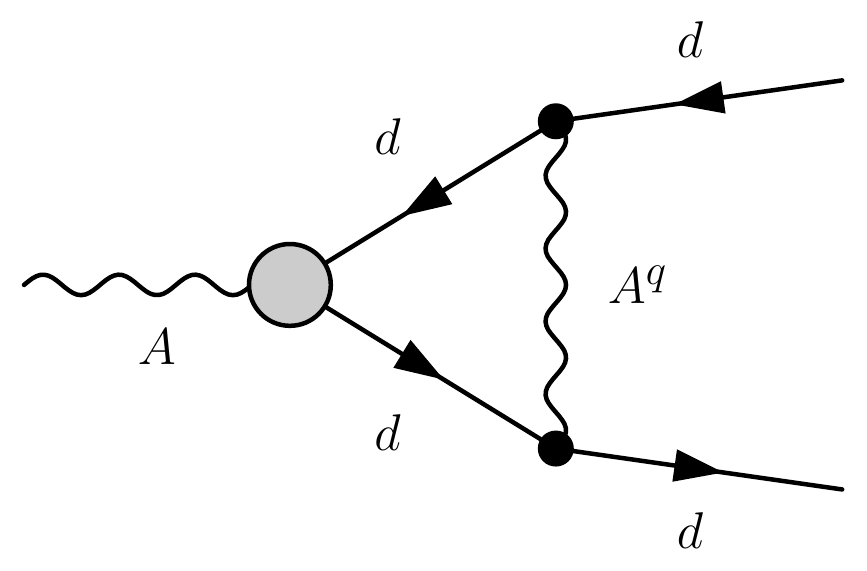}
 \includegraphics[width=0.20\columnwidth]{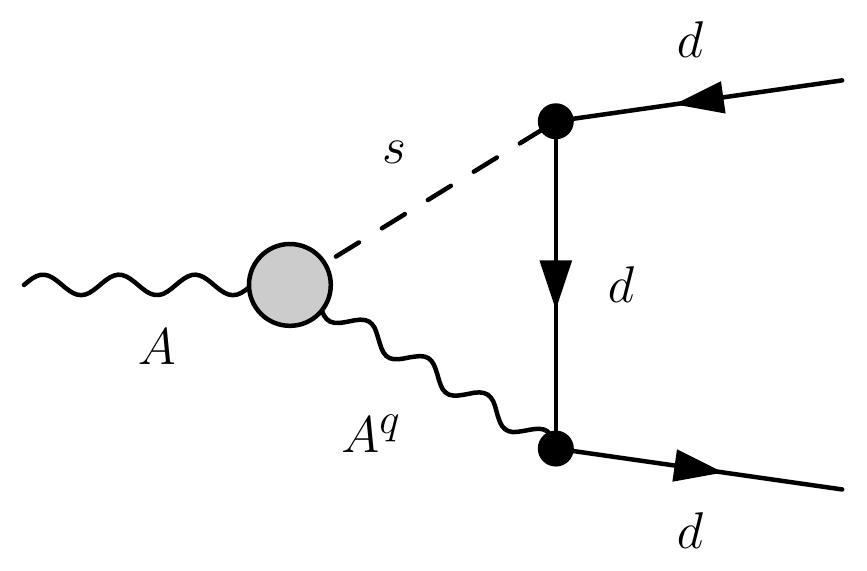}
 \includegraphics[width=0.20\columnwidth]{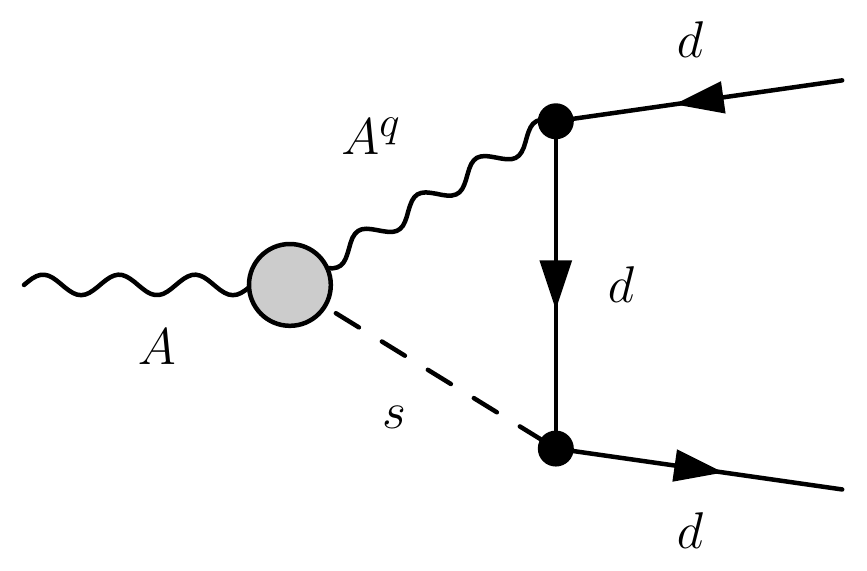}
 \caption{\it Feynman diagrams for $A(p_1)\to\overline{d}(p_2) d(p_3)$.}\label{fig:a_dd}
\end{figure}

\begin{figure}[H]
 \centering
 \includegraphics[width=0.20\columnwidth]{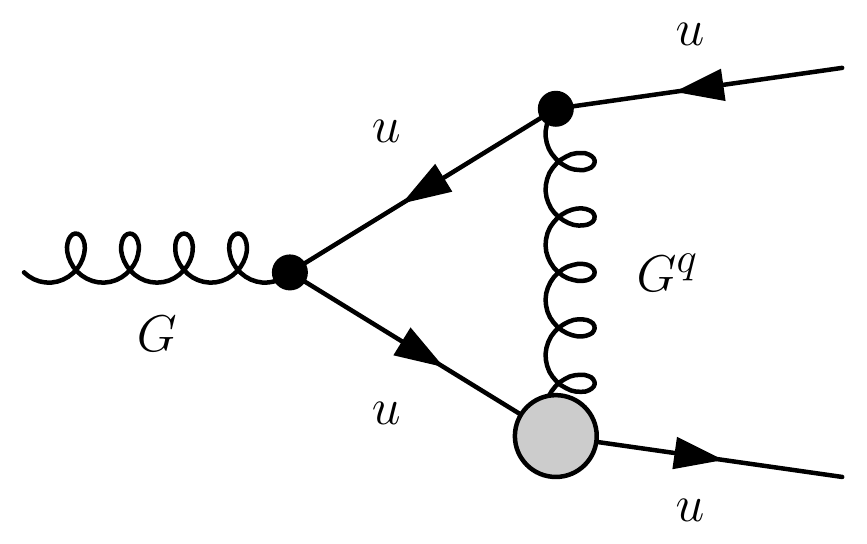}
 \includegraphics[width=0.20\columnwidth]{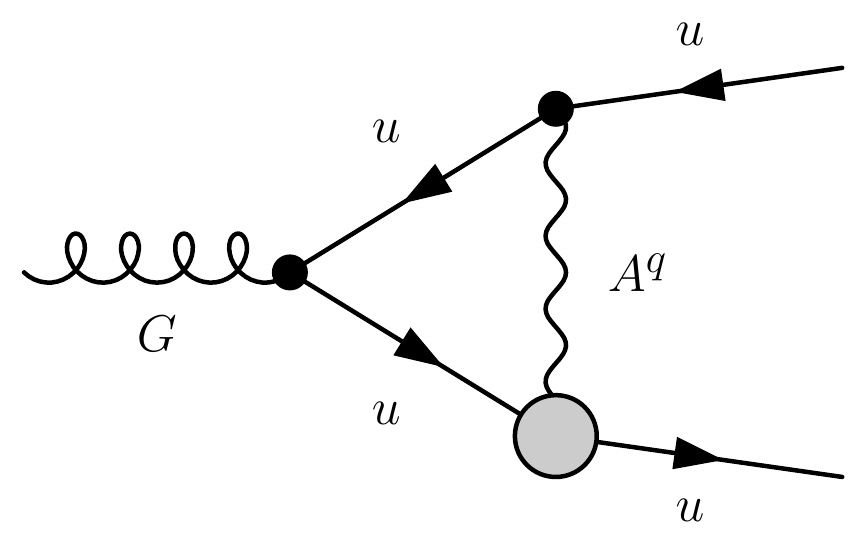}
 \includegraphics[width=0.20\columnwidth]{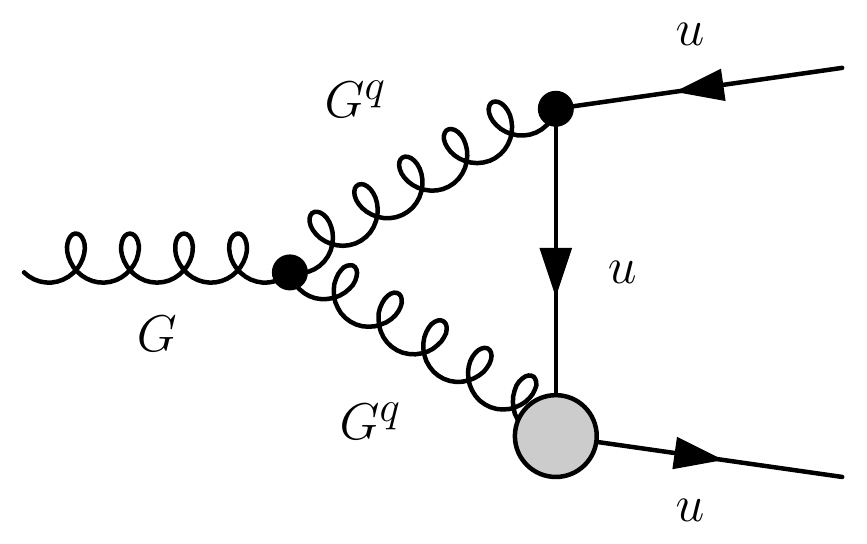}
 \includegraphics[width=0.20\columnwidth]{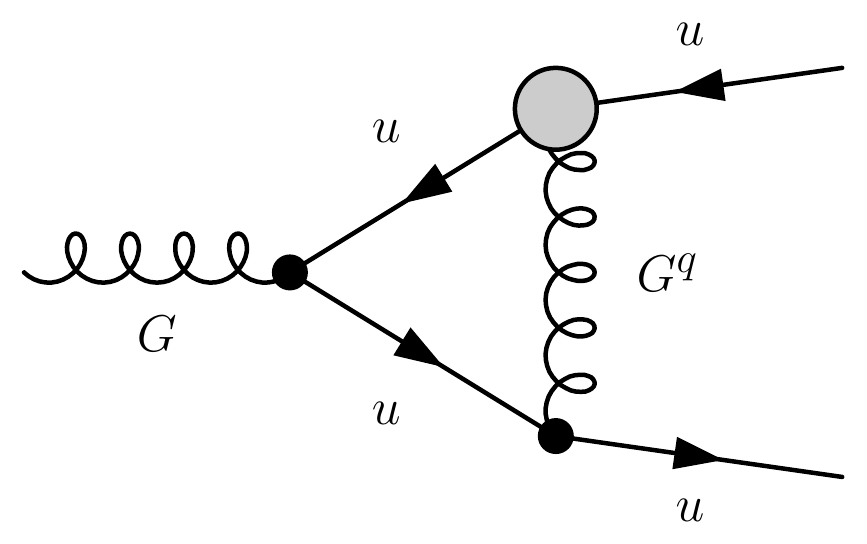}
 \includegraphics[width=0.20\columnwidth]{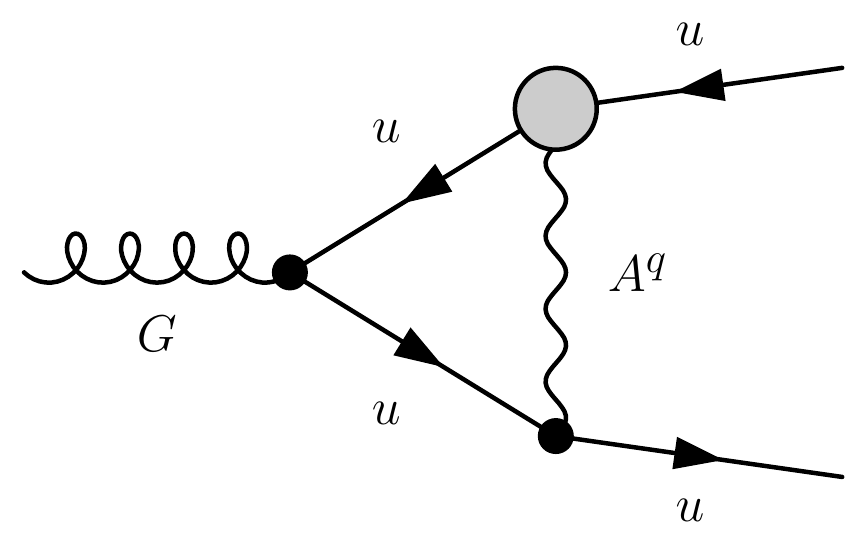}
 \includegraphics[width=0.20\columnwidth]{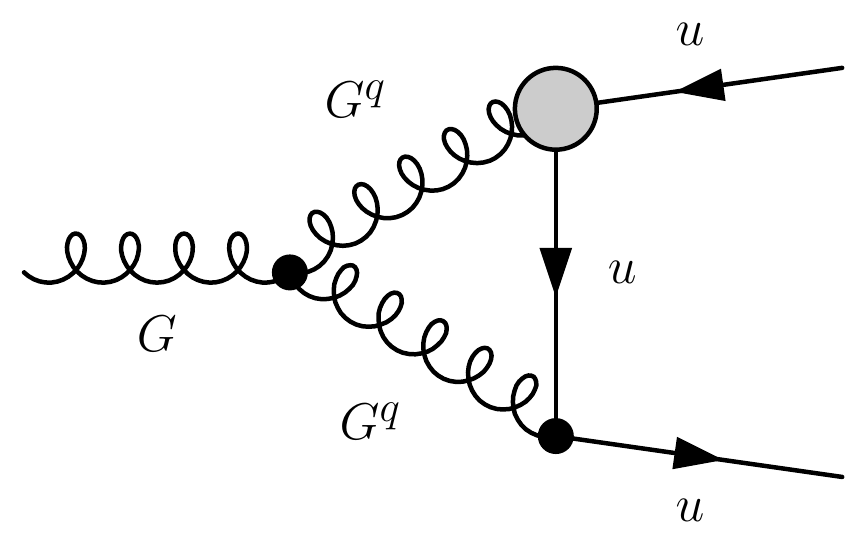}
 \includegraphics[width=0.20\columnwidth]{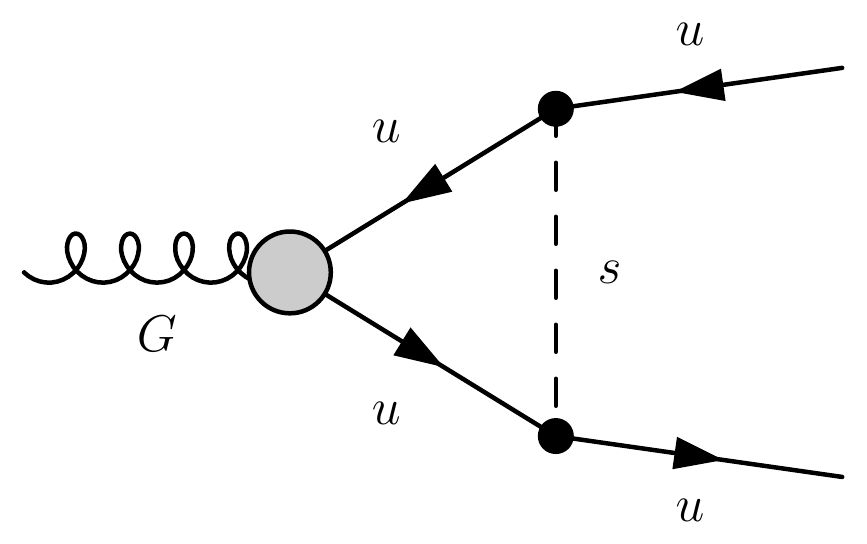}
 \includegraphics[width=0.20\columnwidth]{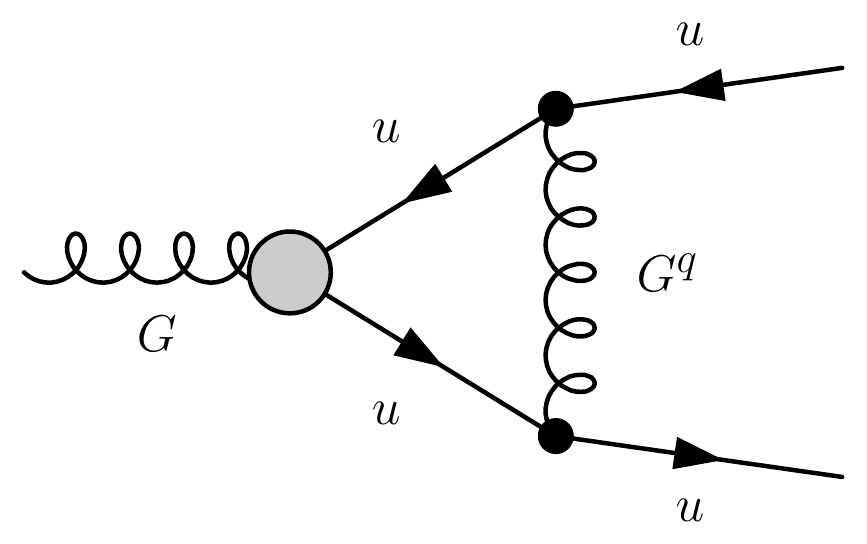}
 \includegraphics[width=0.20\columnwidth]{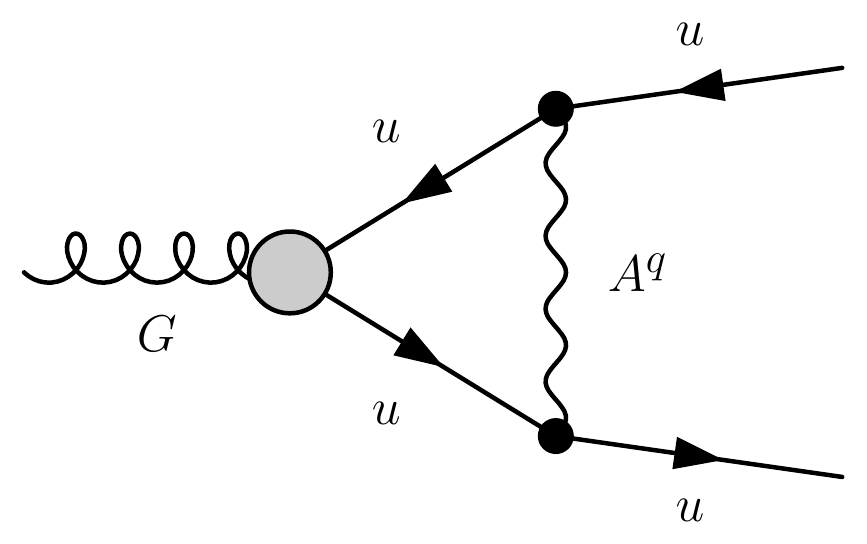}
 \includegraphics[width=0.20\columnwidth]{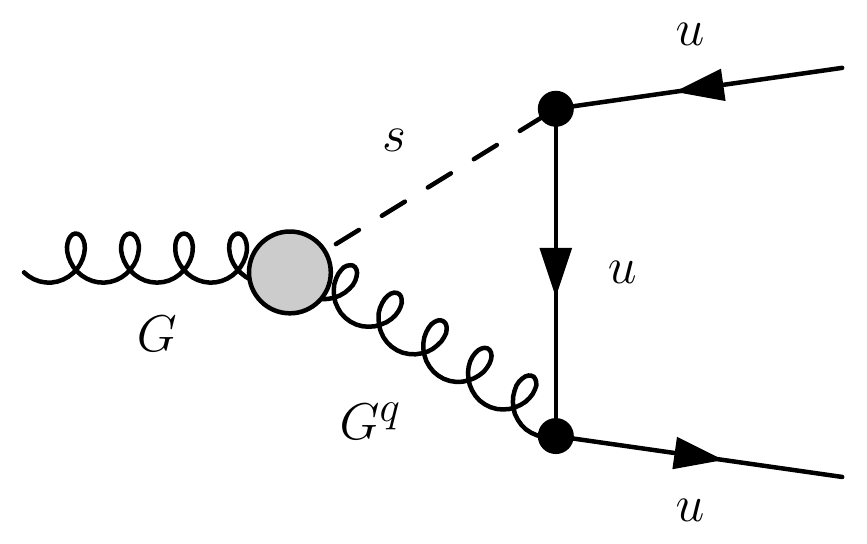}
 \includegraphics[width=0.20\columnwidth]{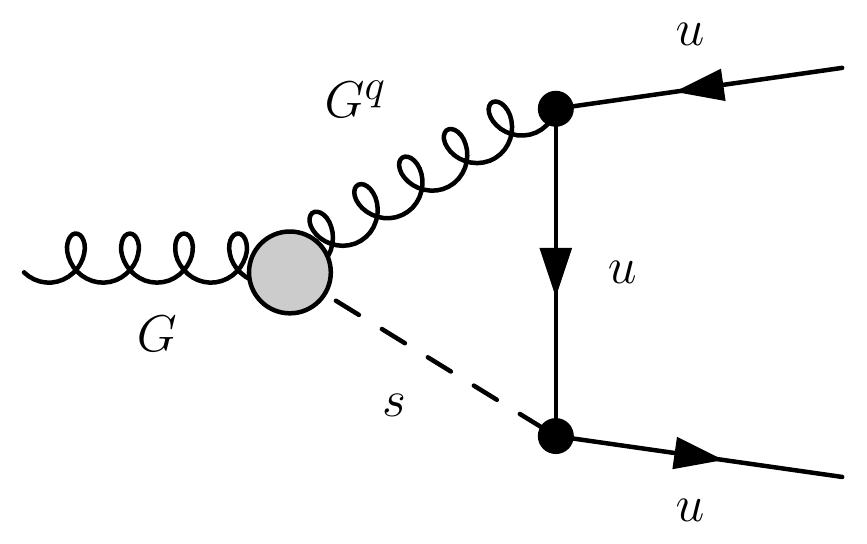}
 \includegraphics[width=0.20\columnwidth]{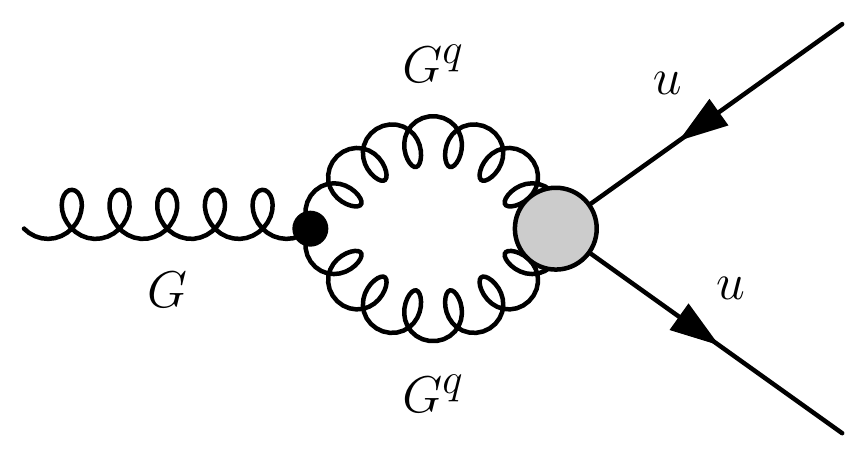}
 \includegraphics[width=0.20\columnwidth]{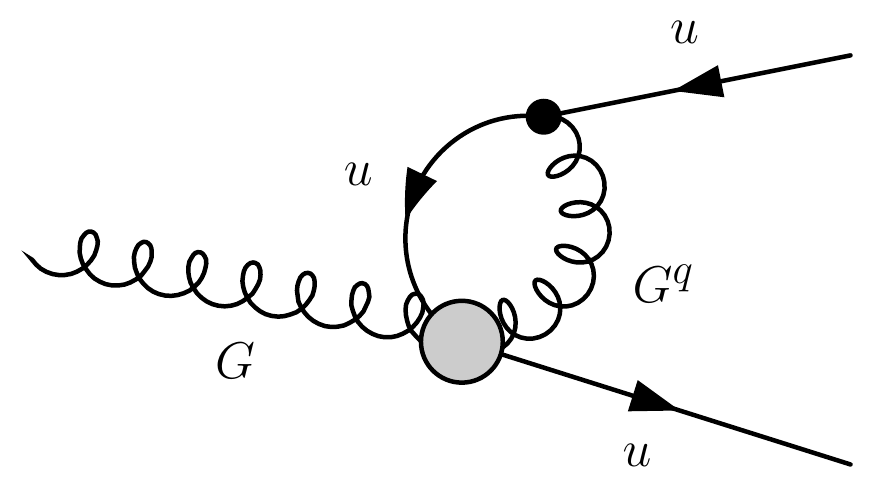}
 \includegraphics[width=0.20\columnwidth]{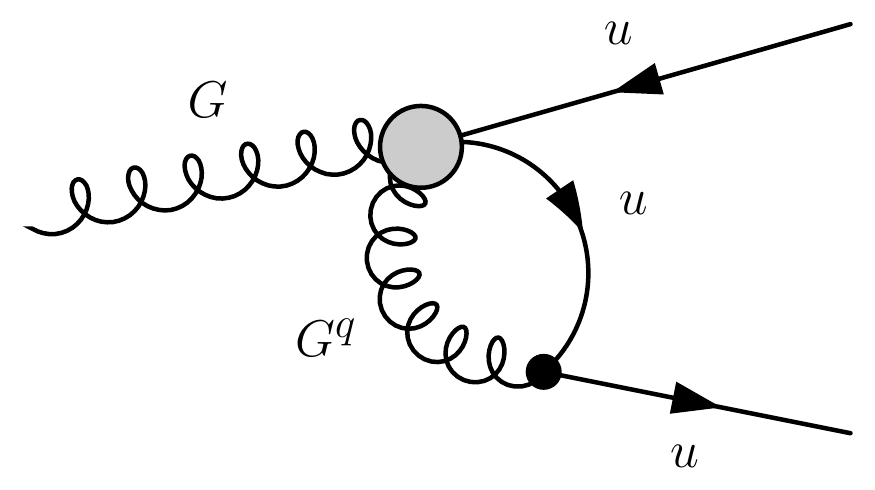}
 \caption{\it Feynman diagrams for $G(p_1)\to\overline{u}(p_2) u(p_3)$.}\label{fig:g_uu}
\end{figure}

\begin{figure}[H]
 \centering
 \includegraphics[width=0.20\columnwidth]{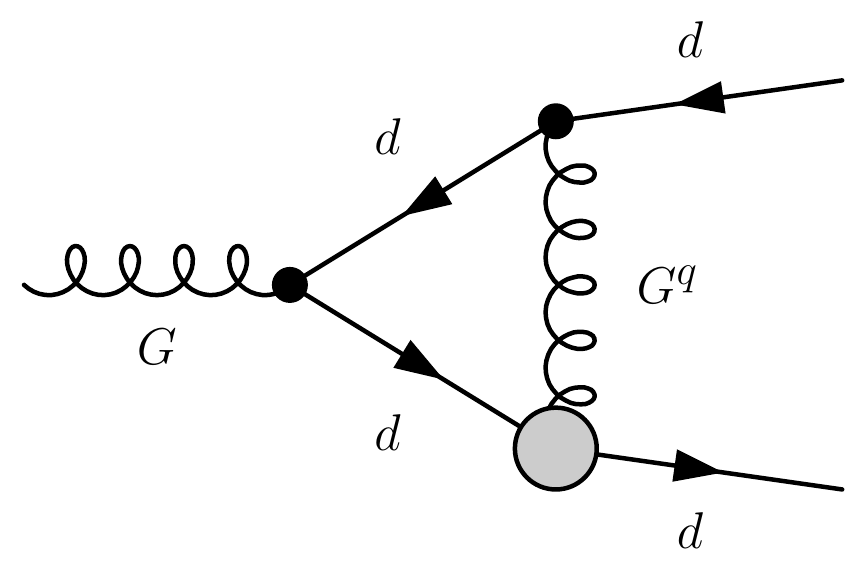}
 \includegraphics[width=0.20\columnwidth]{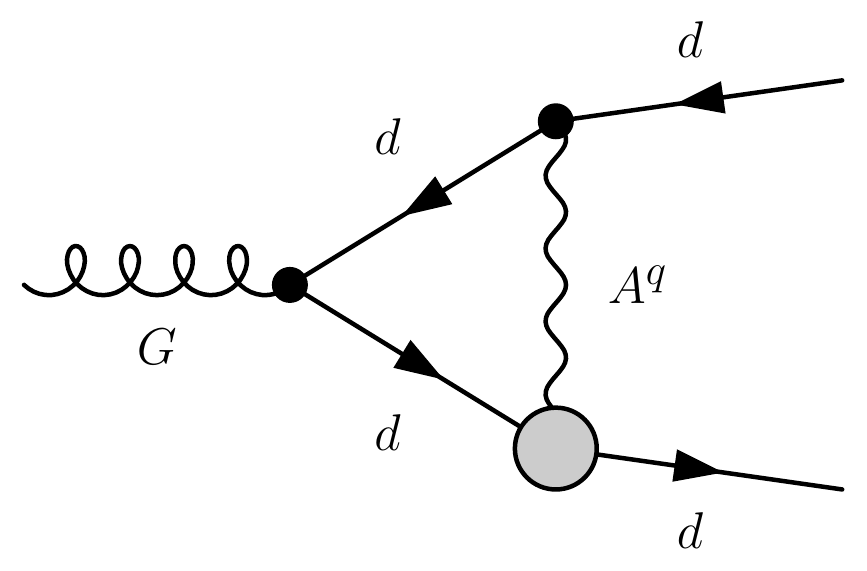}
 \includegraphics[width=0.20\columnwidth]{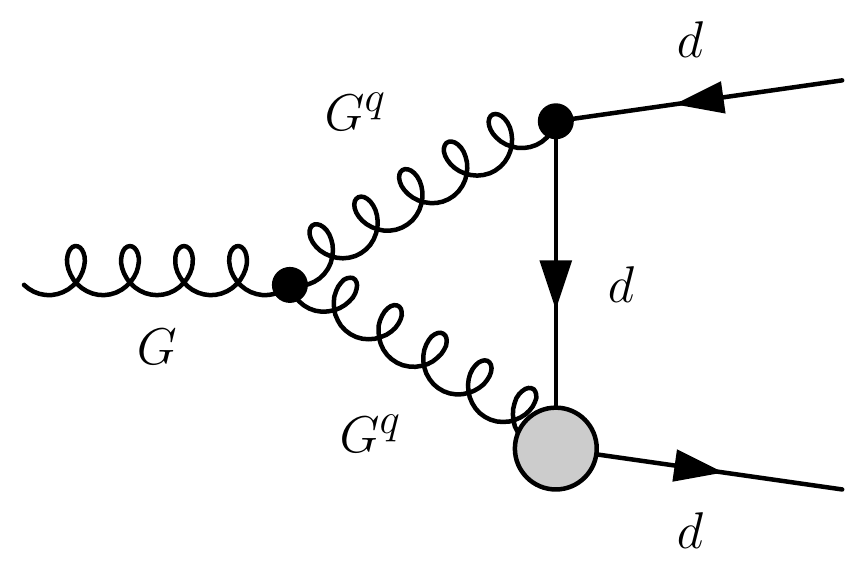}
 \includegraphics[width=0.20\columnwidth]{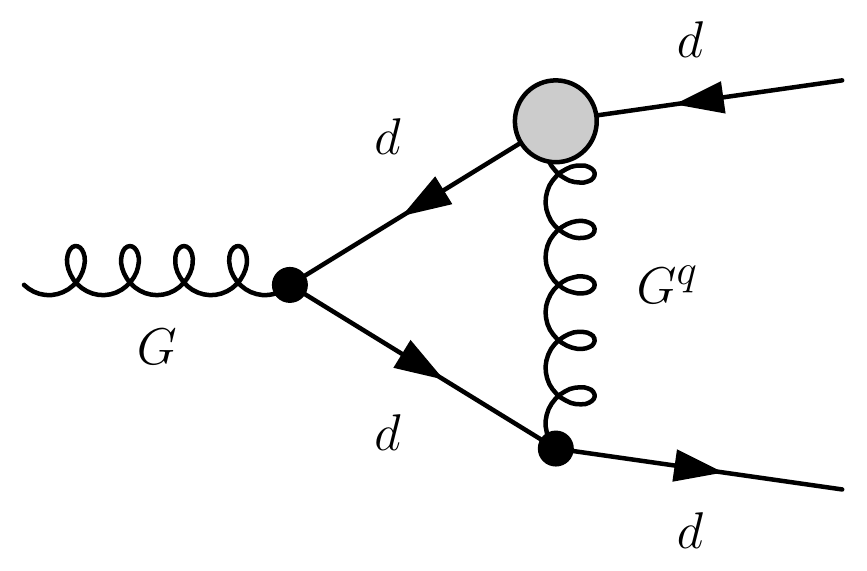}
 \includegraphics[width=0.20\columnwidth]{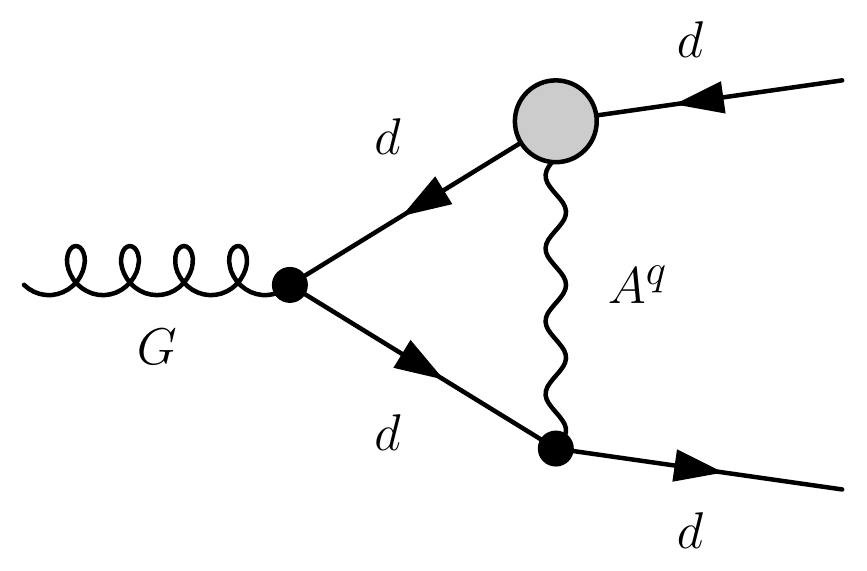}
 \includegraphics[width=0.20\columnwidth]{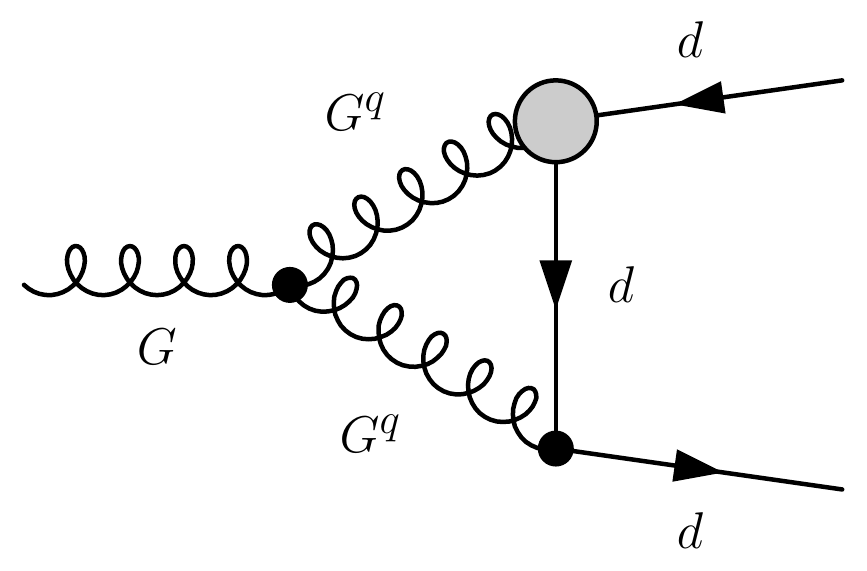}
 \includegraphics[width=0.20\columnwidth]{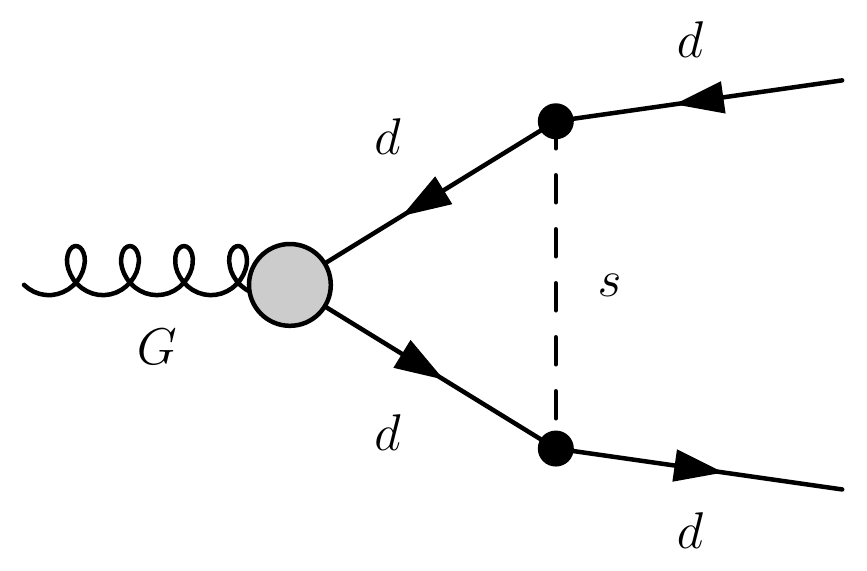}
 \includegraphics[width=0.20\columnwidth]{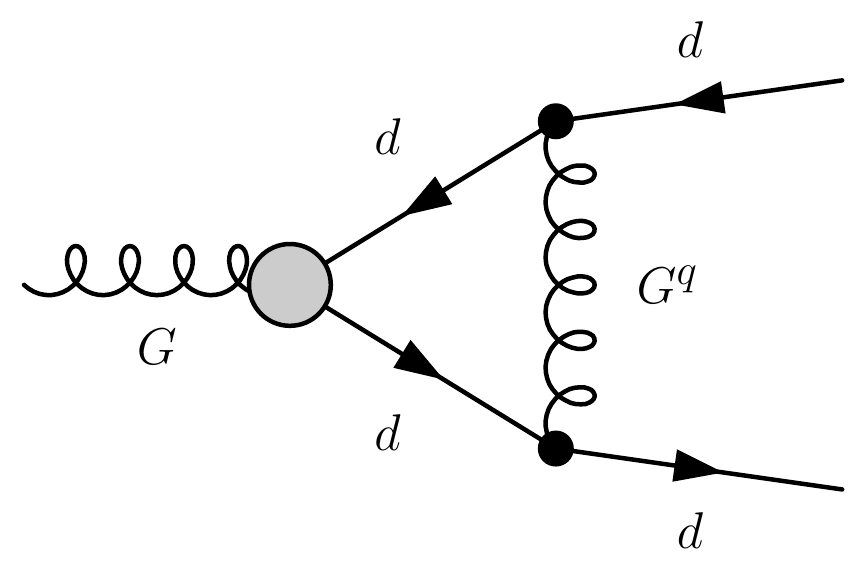}
 \includegraphics[width=0.20\columnwidth]{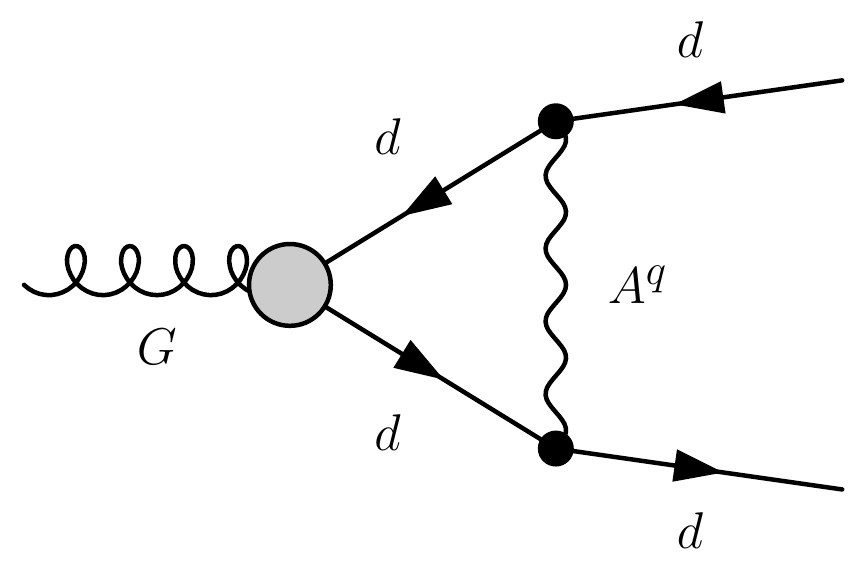}
 \includegraphics[width=0.20\columnwidth]{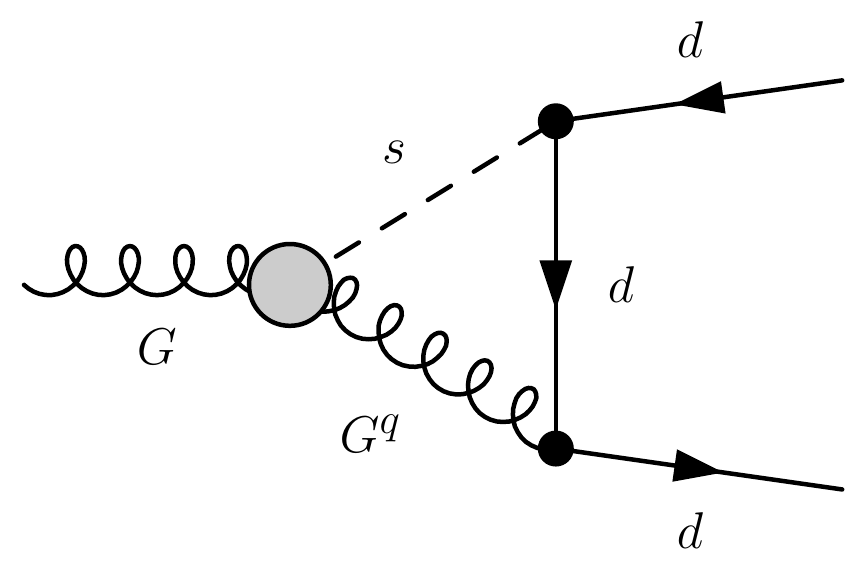}
 \includegraphics[width=0.20\columnwidth]{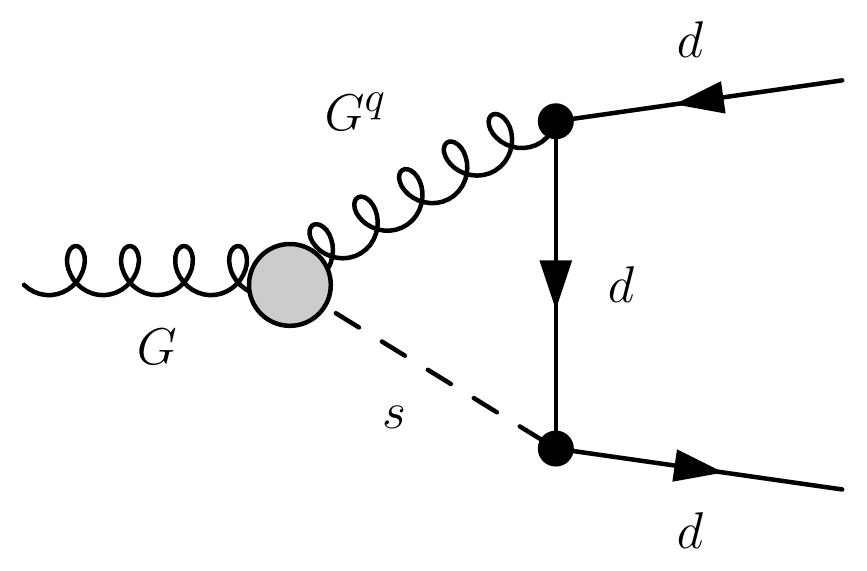}
 \includegraphics[width=0.20\columnwidth]{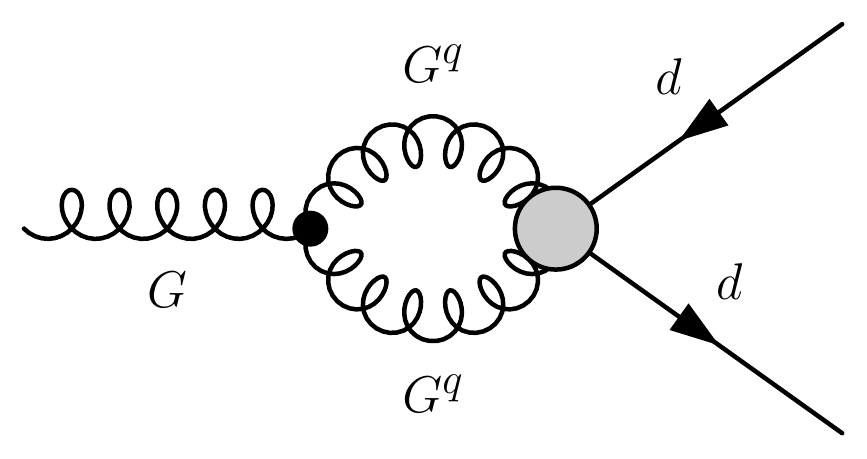}
 \includegraphics[width=0.20\columnwidth]{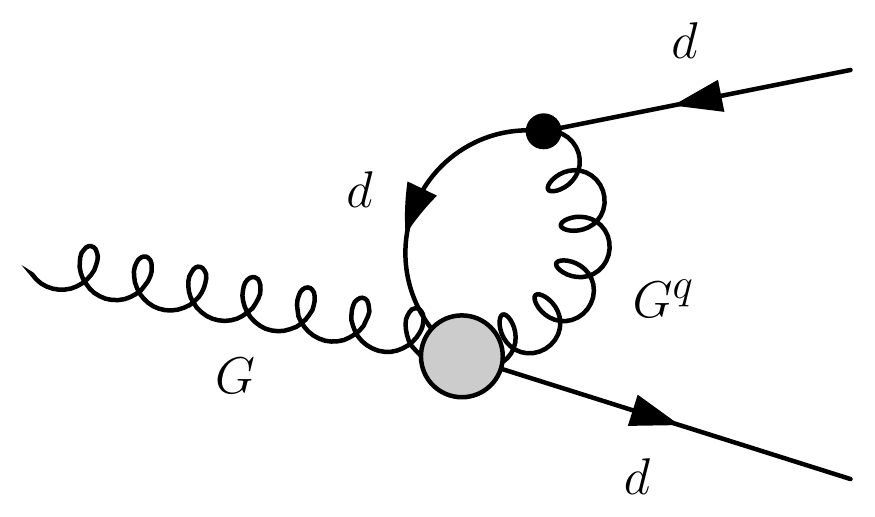}
 \includegraphics[width=0.20\columnwidth]{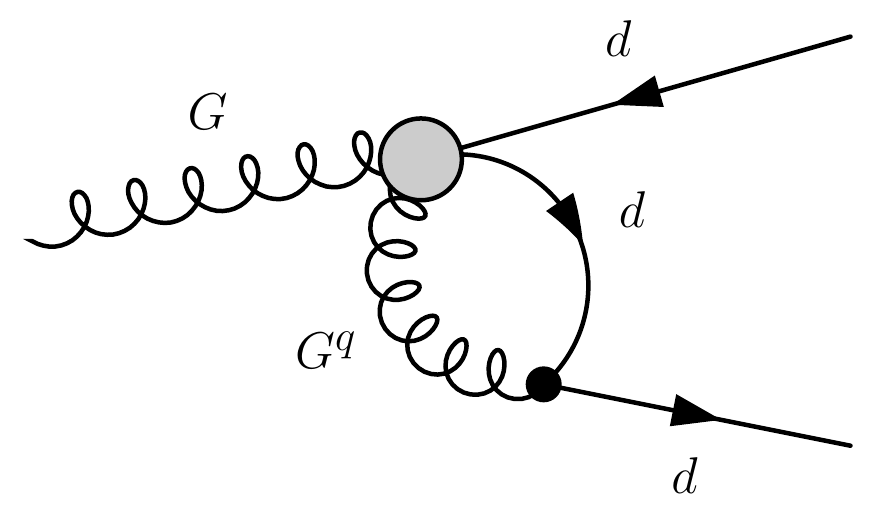}
 \caption{\it Feynman diagrams for $G(p_1)\to\overline{d}(p_2) d(p_3)$.}\label{fig:g_dd}
\end{figure}

\begin{figure}[H]
 \centering
 \includegraphics[width=0.20\columnwidth]{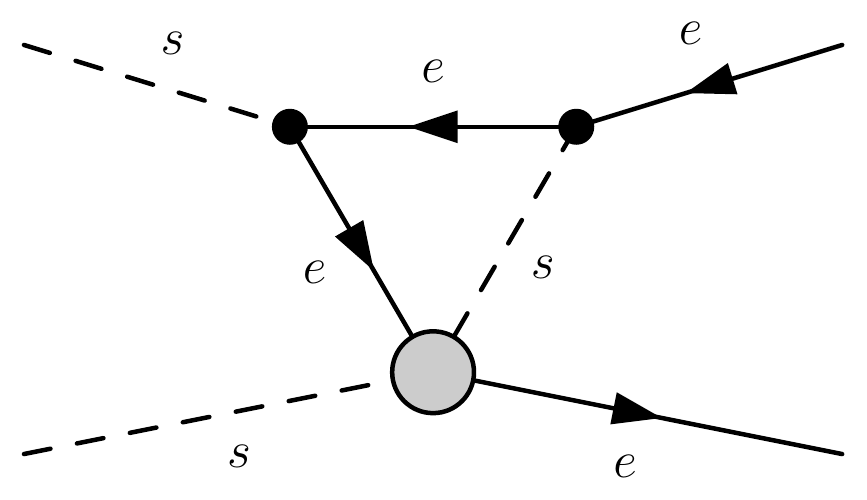}
 \includegraphics[width=0.20\columnwidth]{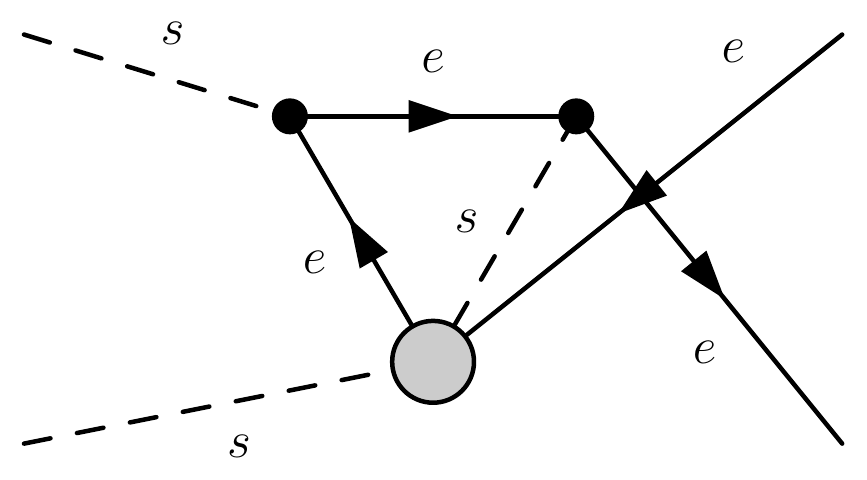}
 \includegraphics[width=0.20\columnwidth]{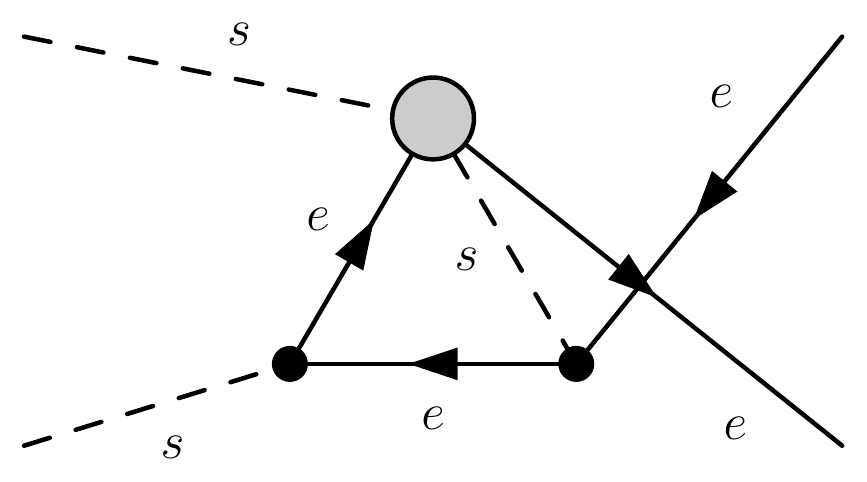}
 \includegraphics[width=0.20\columnwidth]{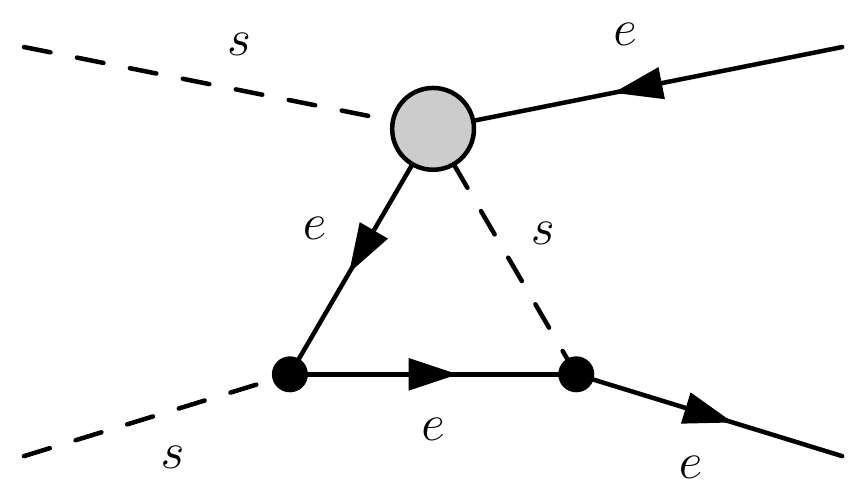}
 \includegraphics[width=0.20\columnwidth]{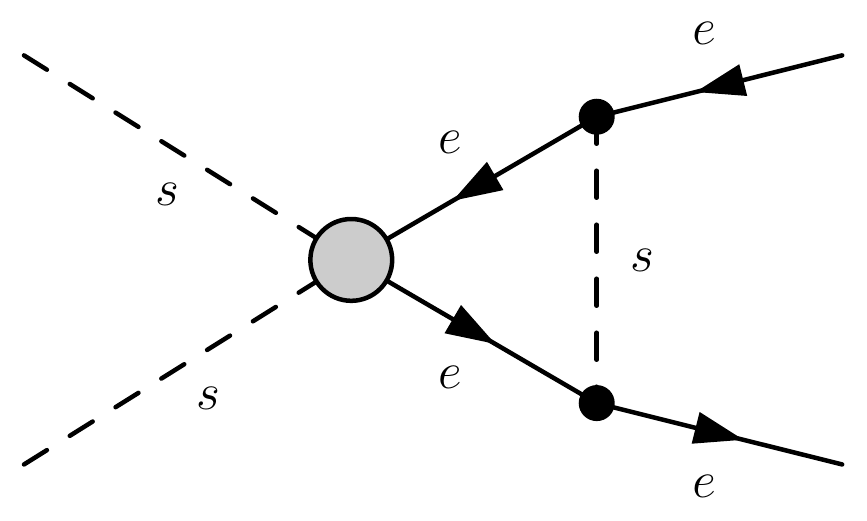}
 \includegraphics[width=0.20\columnwidth]{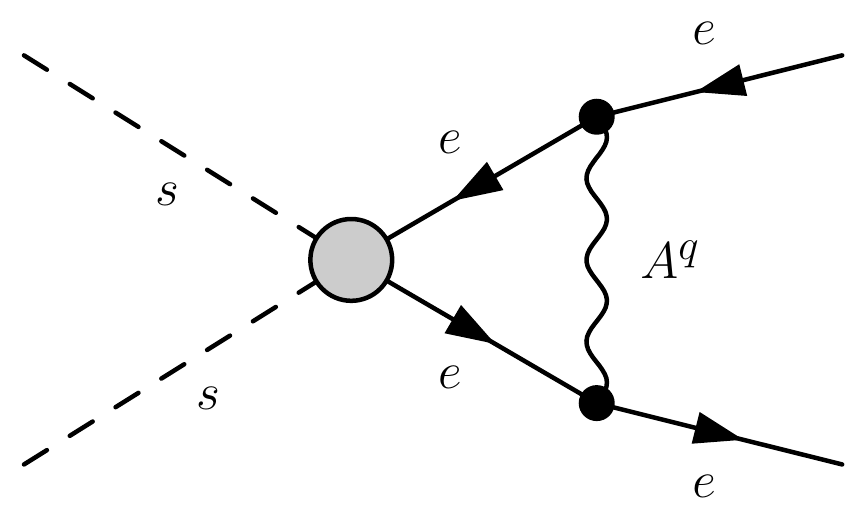}
 \includegraphics[width=0.20\columnwidth]{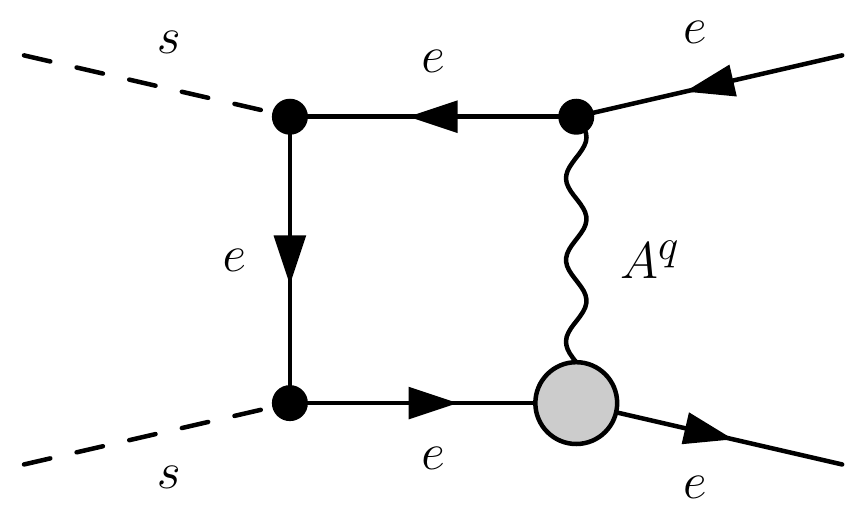}
 \includegraphics[width=0.20\columnwidth]{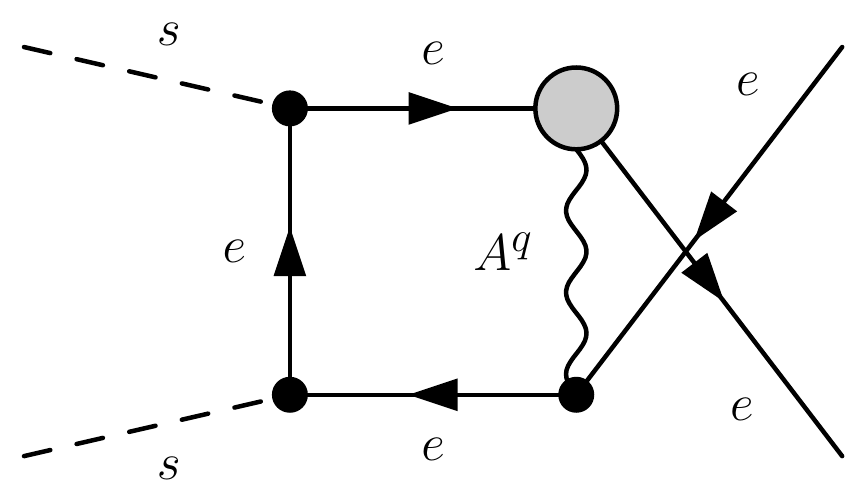}
 \includegraphics[width=0.20\columnwidth]{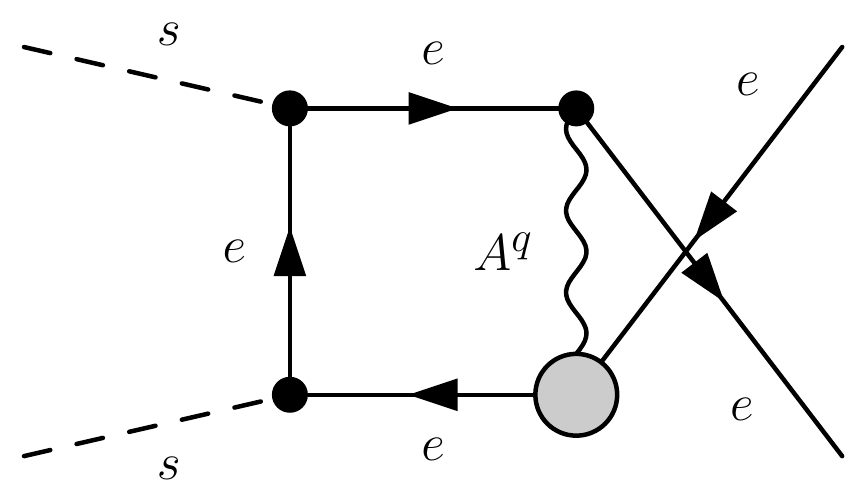}
 \includegraphics[width=0.20\columnwidth]{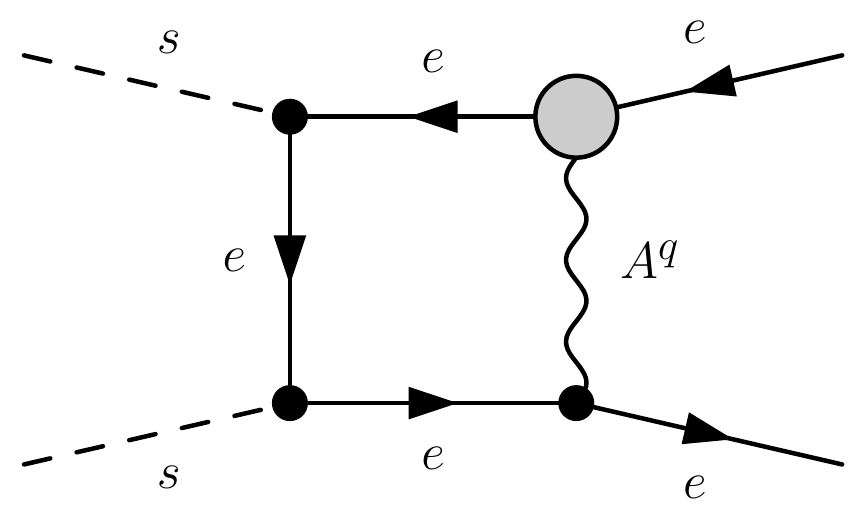}
 \includegraphics[width=0.20\columnwidth]{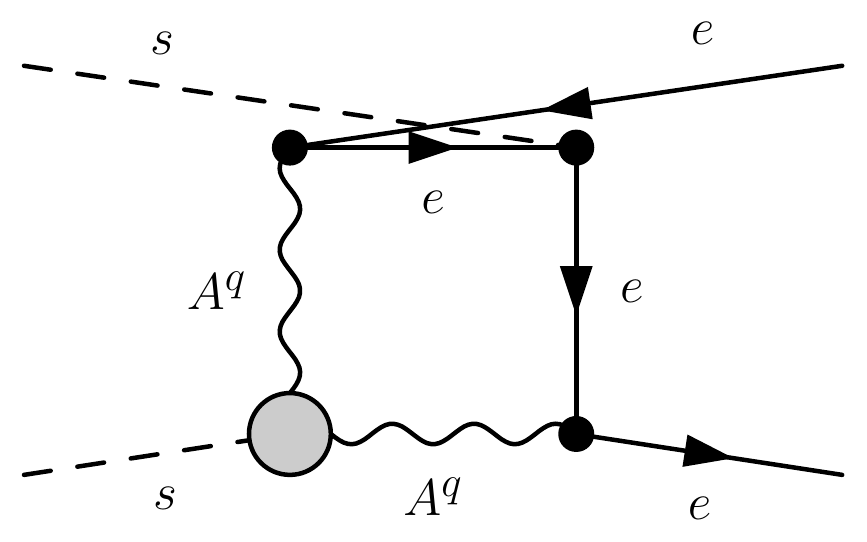}
 \includegraphics[width=0.20\columnwidth]{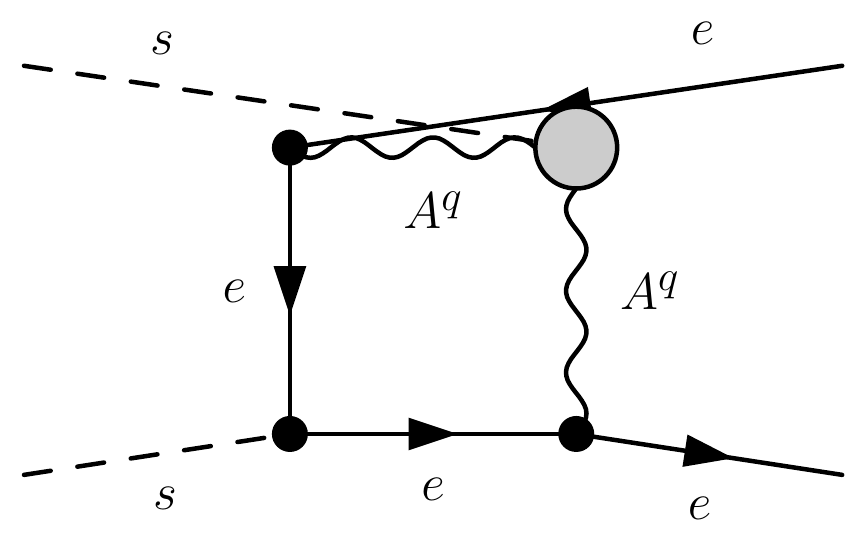}
 \includegraphics[width=0.20\columnwidth]{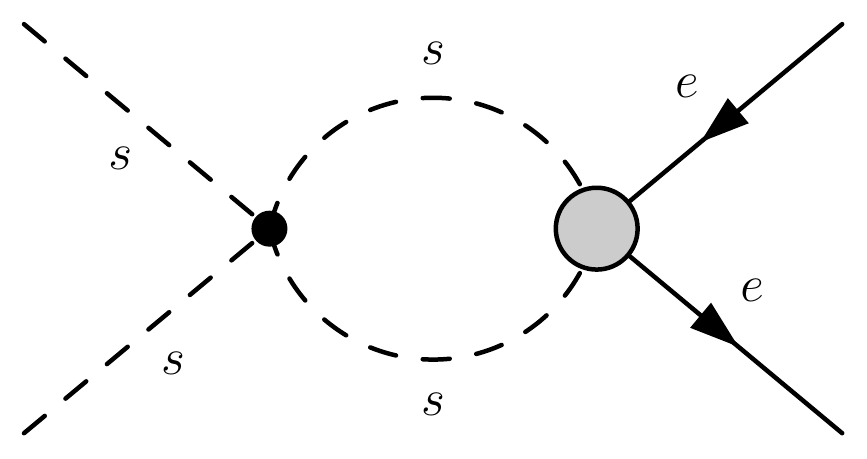}
 \caption{\it Feynman diagrams for $s(p_1) s(p_2)\to\overline{e}(p_3) e(p_4)$.}\label{fig:ss_ee}
\end{figure}

\begin{figure}[H]
 \centering
 \includegraphics[width=0.20\columnwidth]{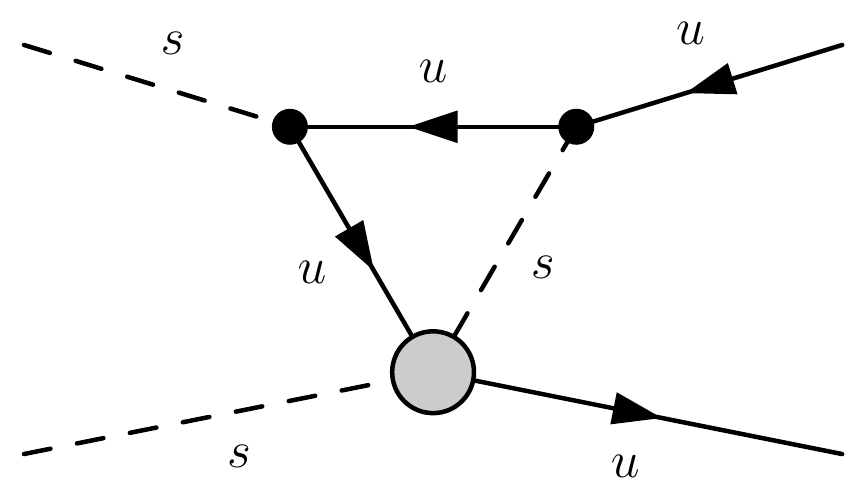}
 \includegraphics[width=0.20\columnwidth]{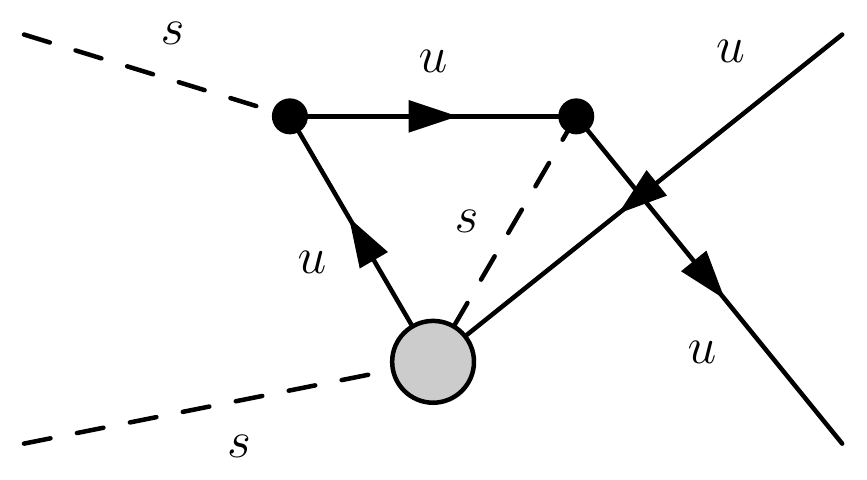}
 \includegraphics[width=0.20\columnwidth]{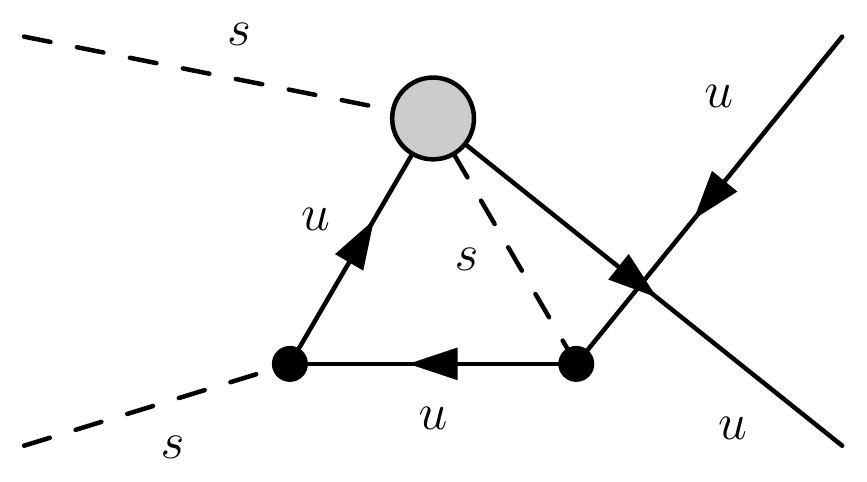}
 \includegraphics[width=0.20\columnwidth]{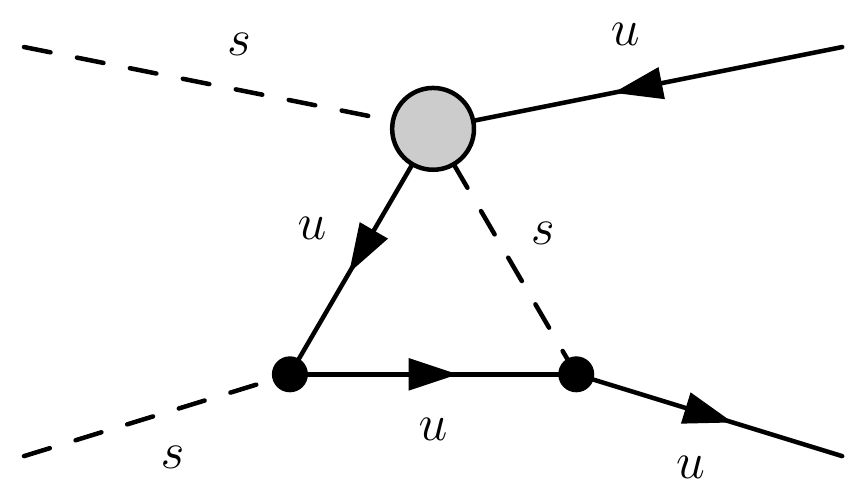}
 \includegraphics[width=0.20\columnwidth]{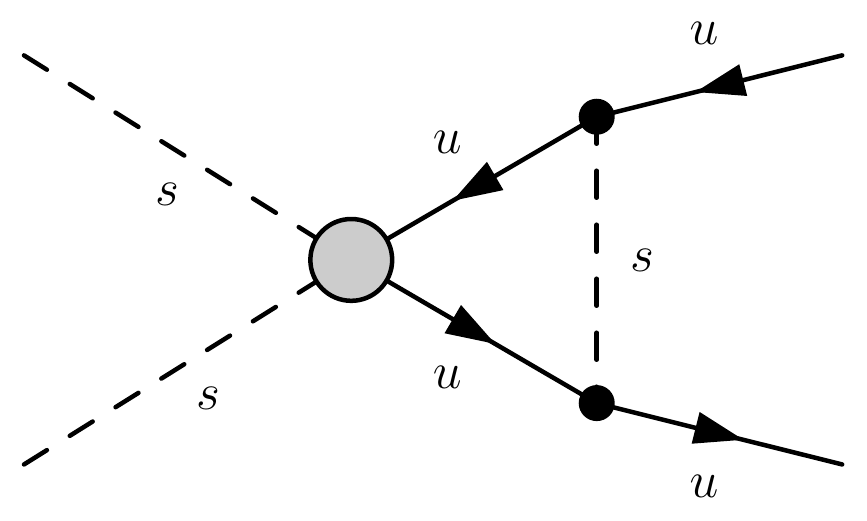}
 \includegraphics[width=0.20\columnwidth]{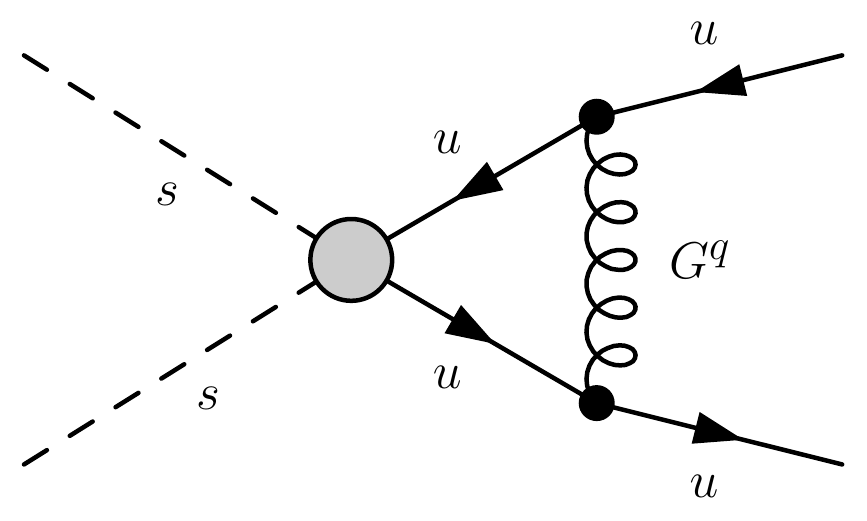}
 \includegraphics[width=0.20\columnwidth]{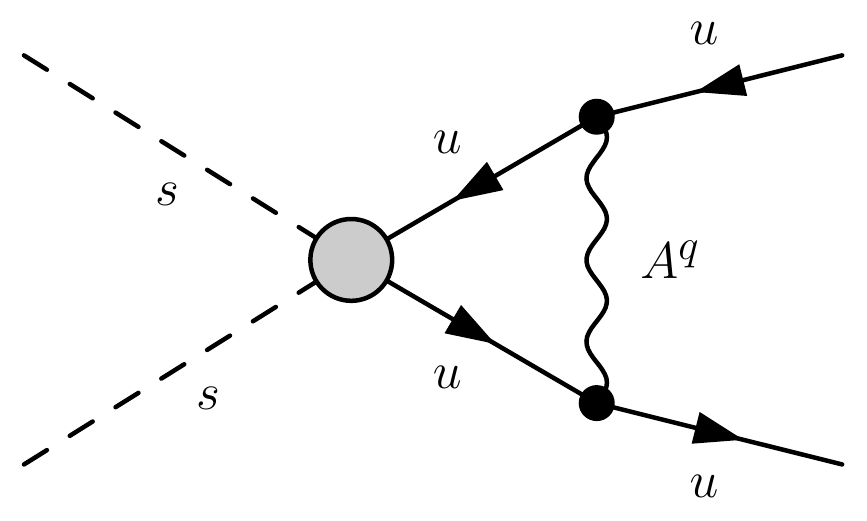}
 \includegraphics[width=0.20\columnwidth]{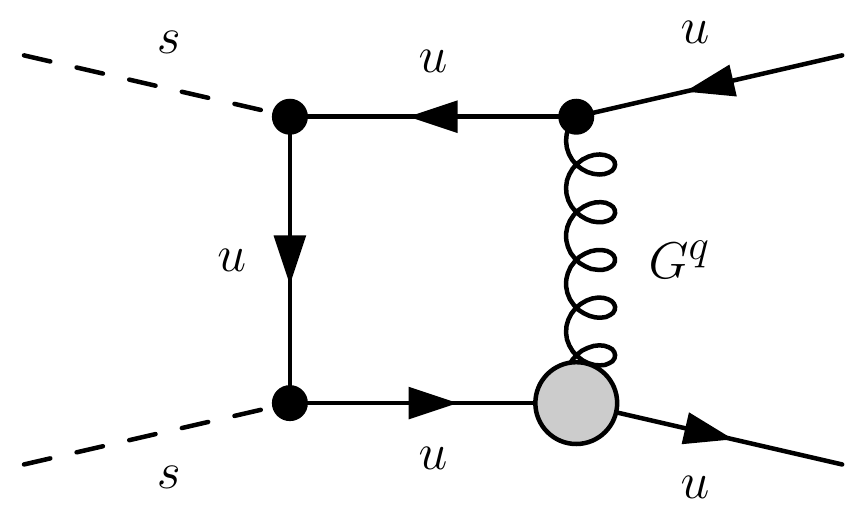}
 \includegraphics[width=0.20\columnwidth]{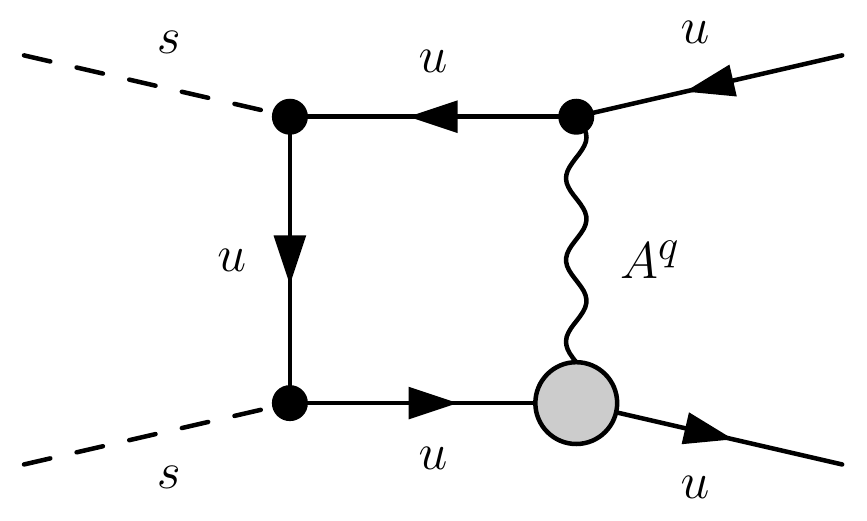}
 \includegraphics[width=0.20\columnwidth]{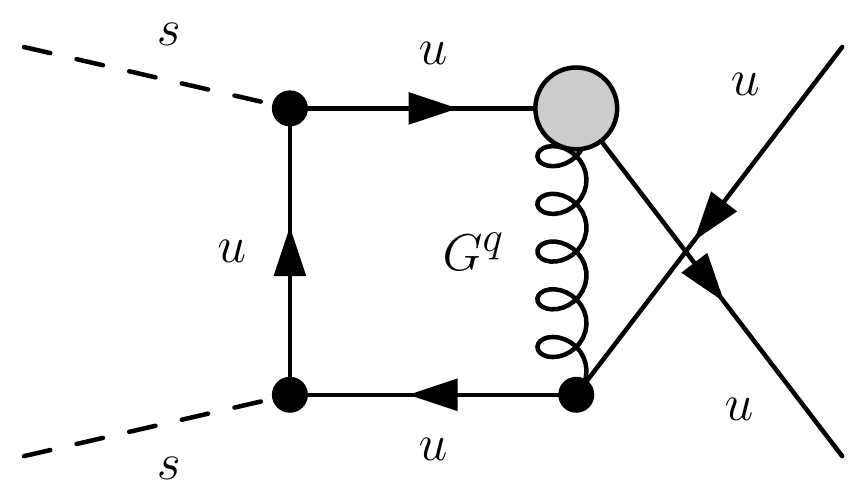}
 \includegraphics[width=0.20\columnwidth]{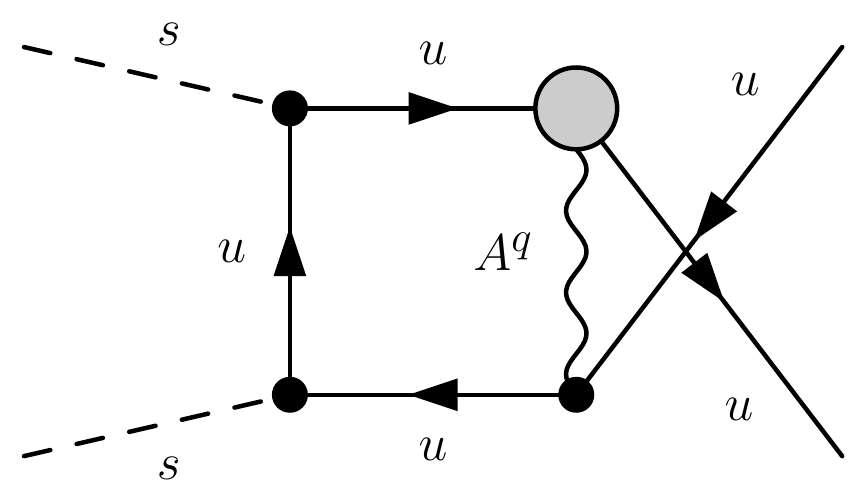}
 \includegraphics[width=0.20\columnwidth]{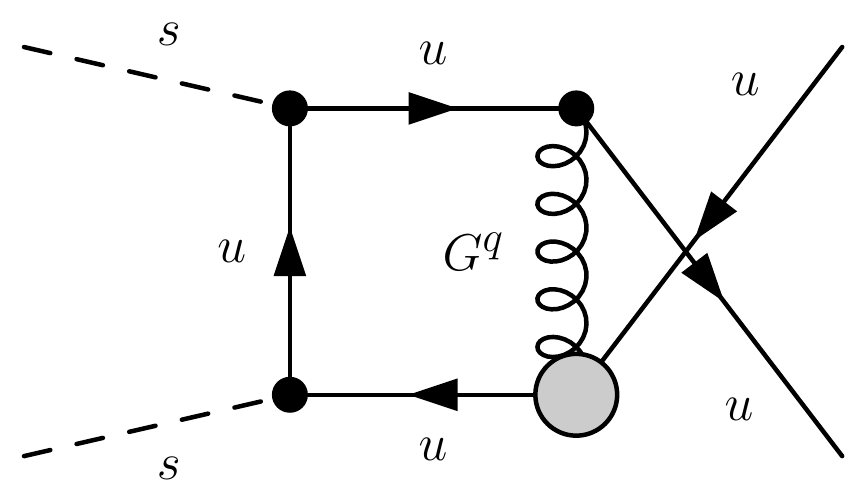}
 \includegraphics[width=0.20\columnwidth]{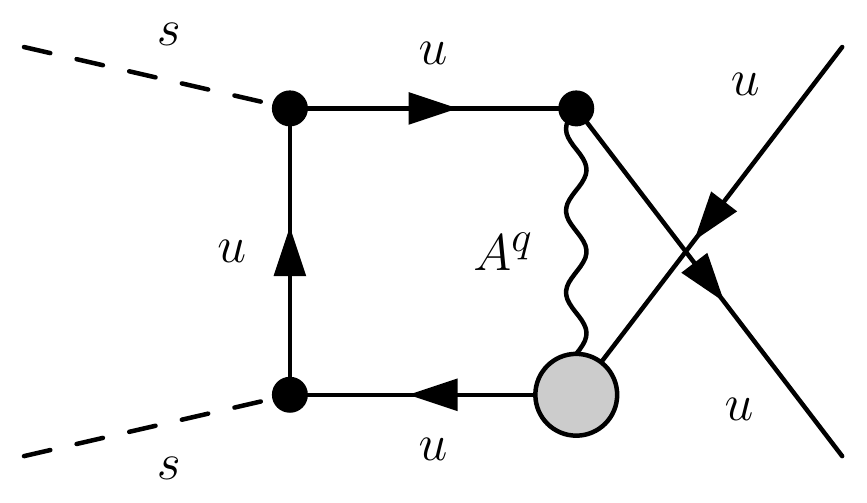}
 \includegraphics[width=0.20\columnwidth]{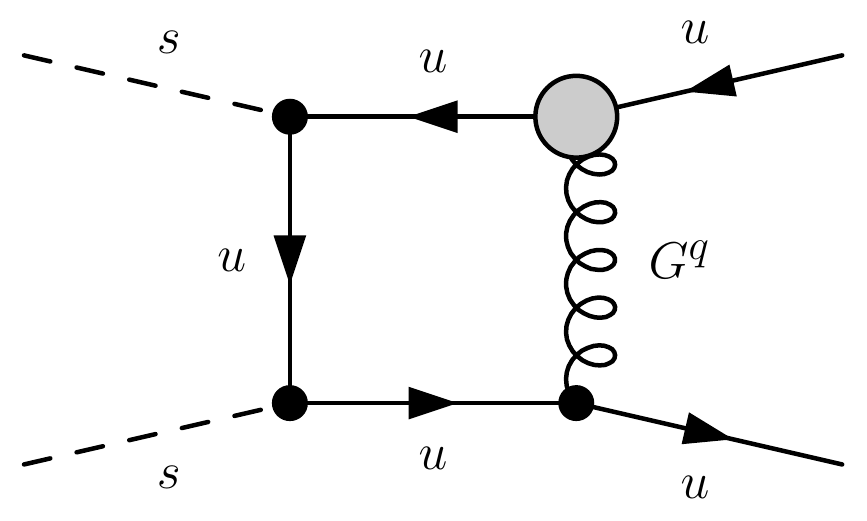}
 \includegraphics[width=0.20\columnwidth]{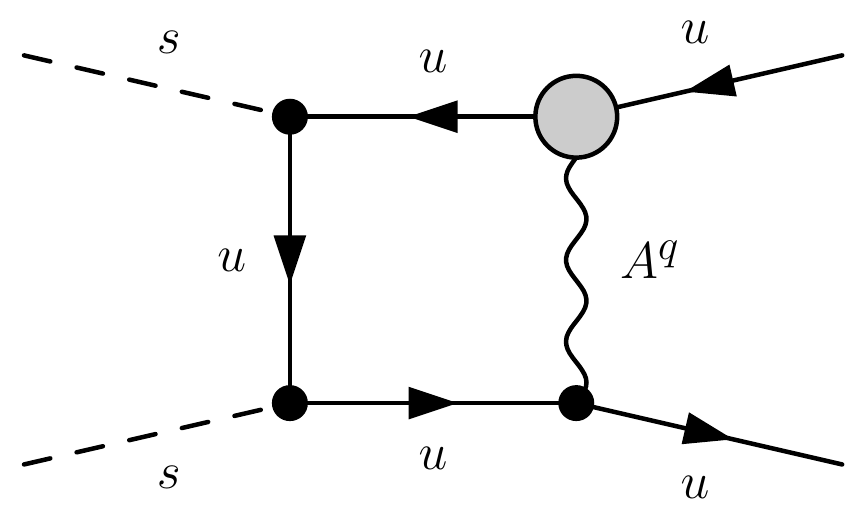}
 \includegraphics[width=0.20\columnwidth]{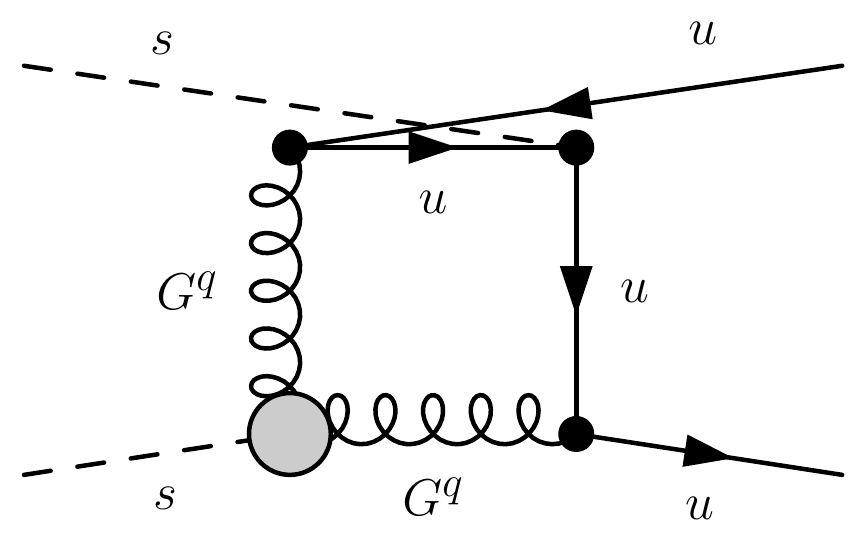}
 \includegraphics[width=0.20\columnwidth]{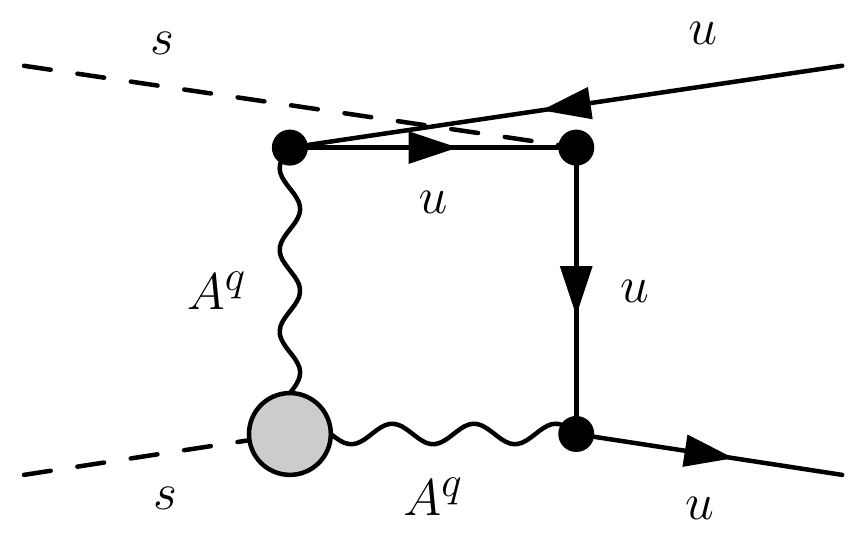}
 \includegraphics[width=0.20\columnwidth]{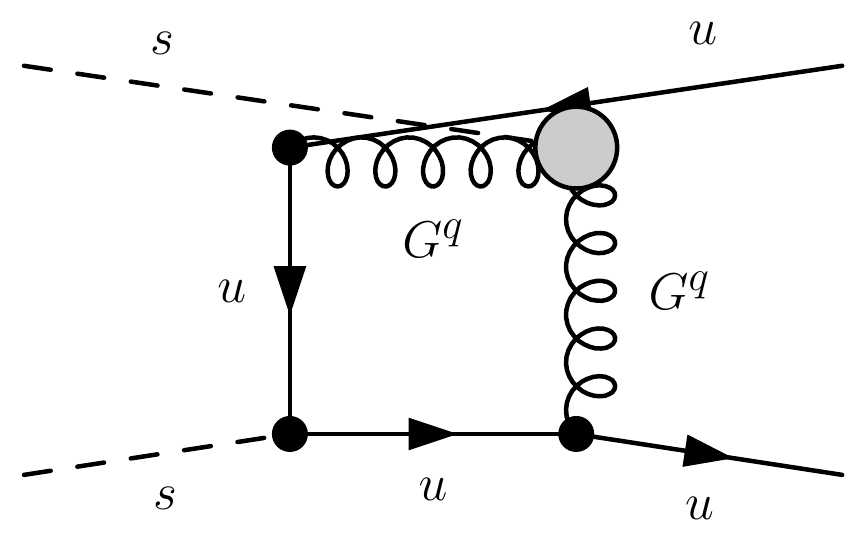}
 \includegraphics[width=0.20\columnwidth]{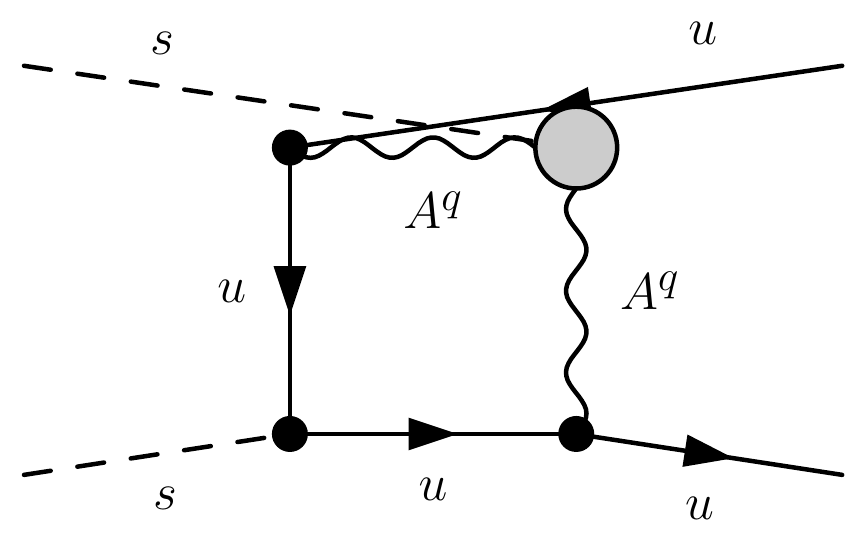}
 \includegraphics[width=0.20\columnwidth]{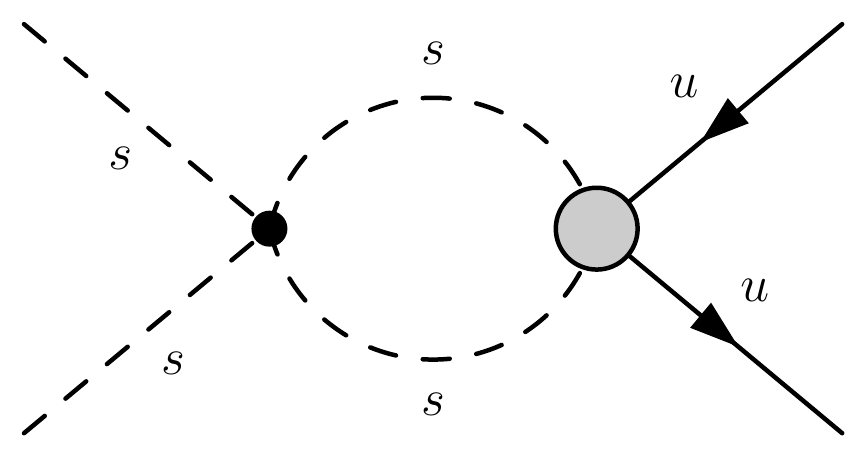}
 \caption{\it Feynman diagrams for $s(p_1)s(p_2)\to\overline{u}(p_3) u(p_4)$.}\label{fig:ss_uu}
\end{figure}

\begin{figure}[H]
 \centering
 \includegraphics[width=0.20\columnwidth]{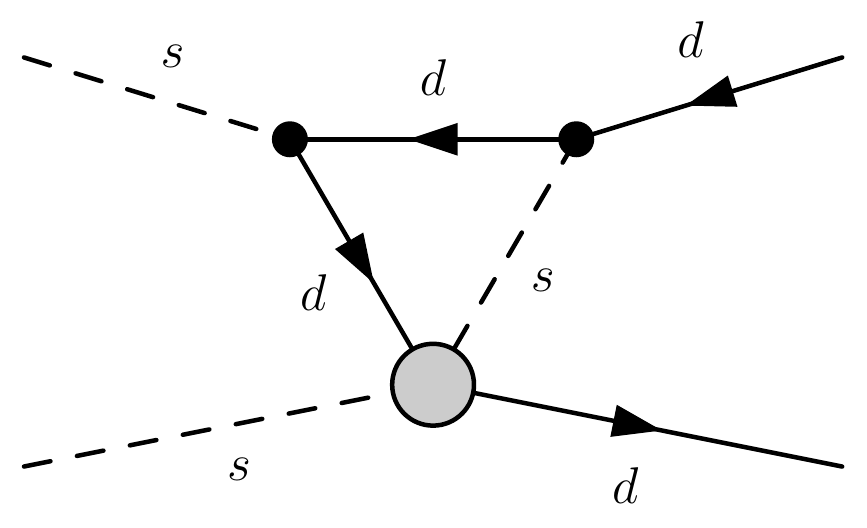}
 \includegraphics[width=0.20\columnwidth]{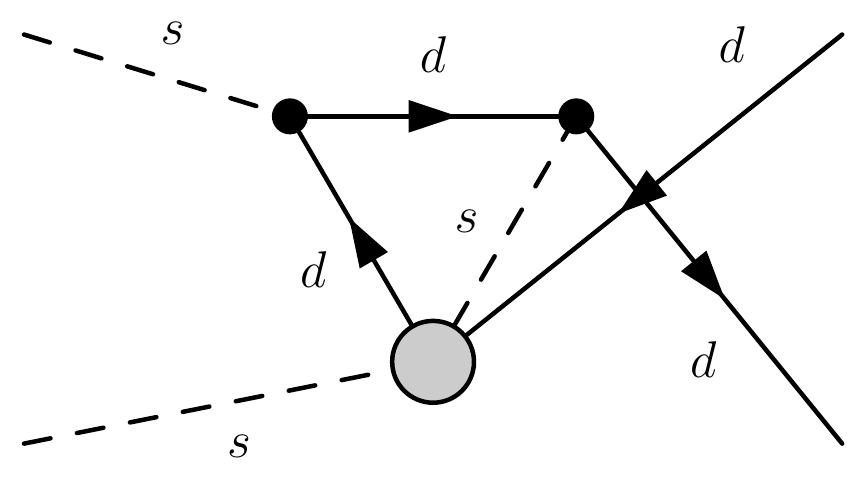}
 \includegraphics[width=0.20\columnwidth]{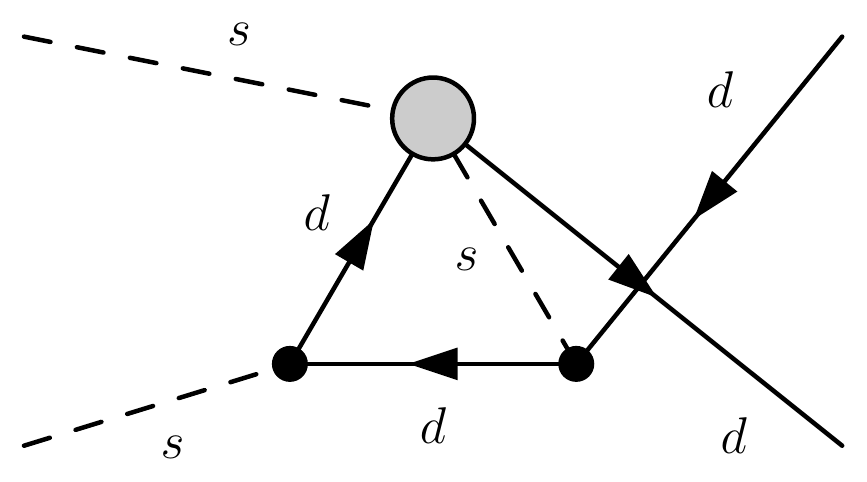}
 \includegraphics[width=0.20\columnwidth]{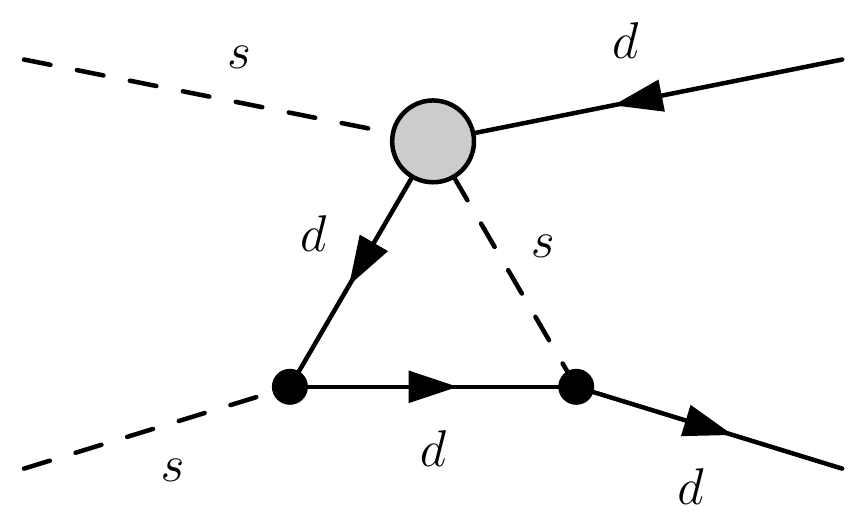}
 \includegraphics[width=0.20\columnwidth]{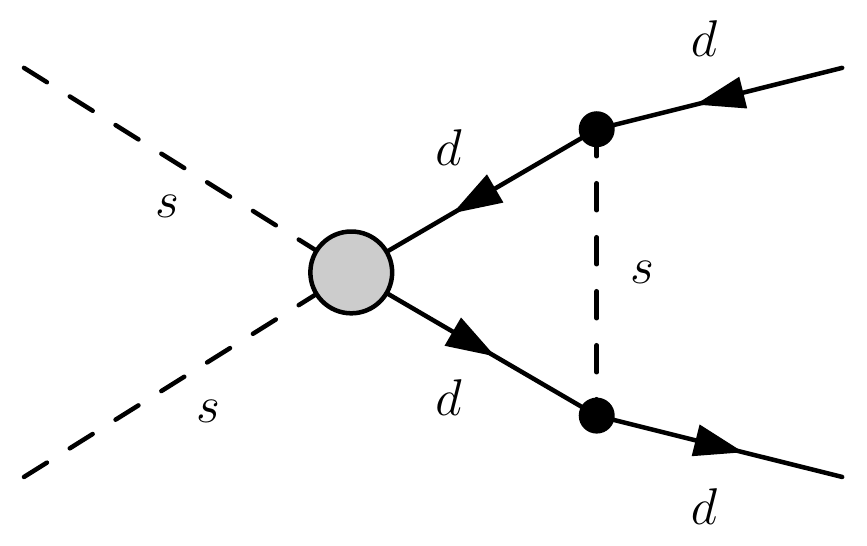}
 \includegraphics[width=0.20\columnwidth]{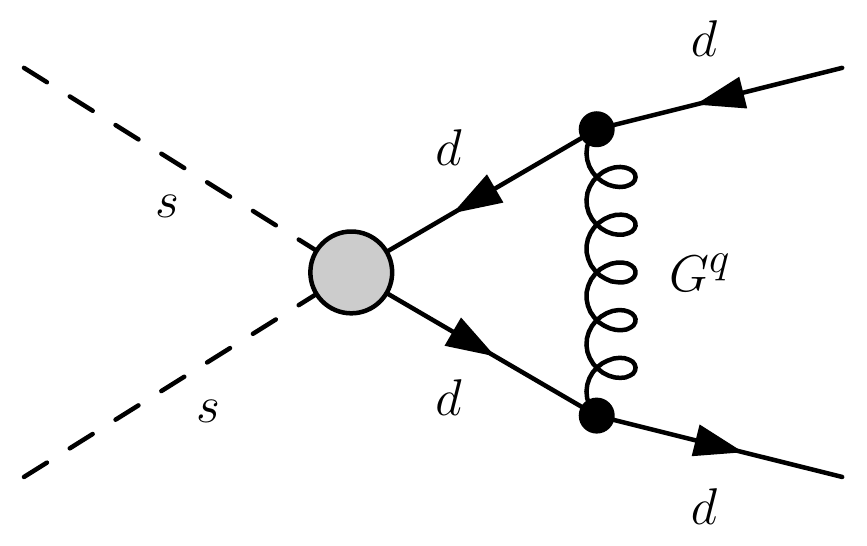}
 \includegraphics[width=0.20\columnwidth]{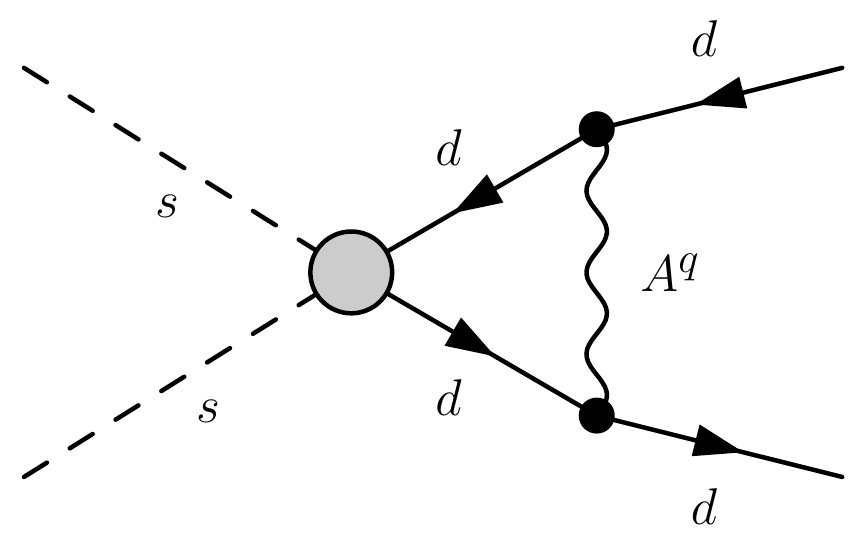}
 \includegraphics[width=0.20\columnwidth]{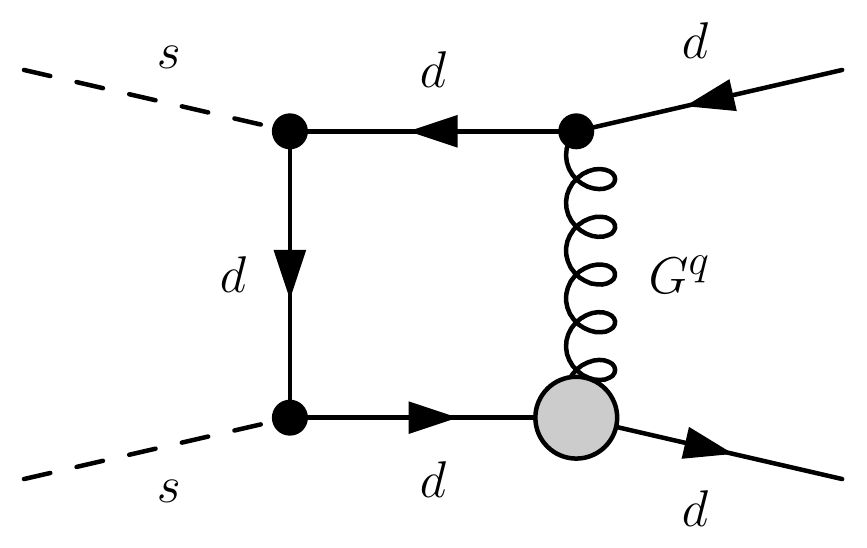}
 \includegraphics[width=0.20\columnwidth]{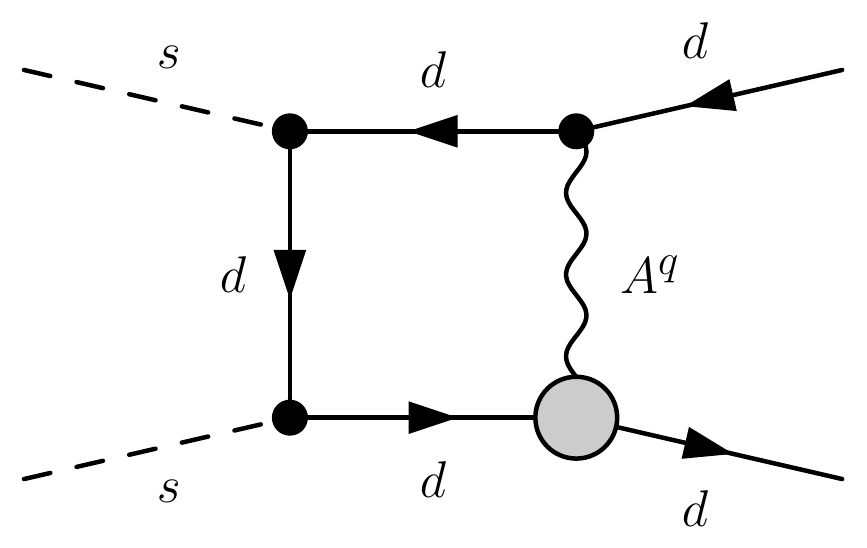}
 \includegraphics[width=0.20\columnwidth]{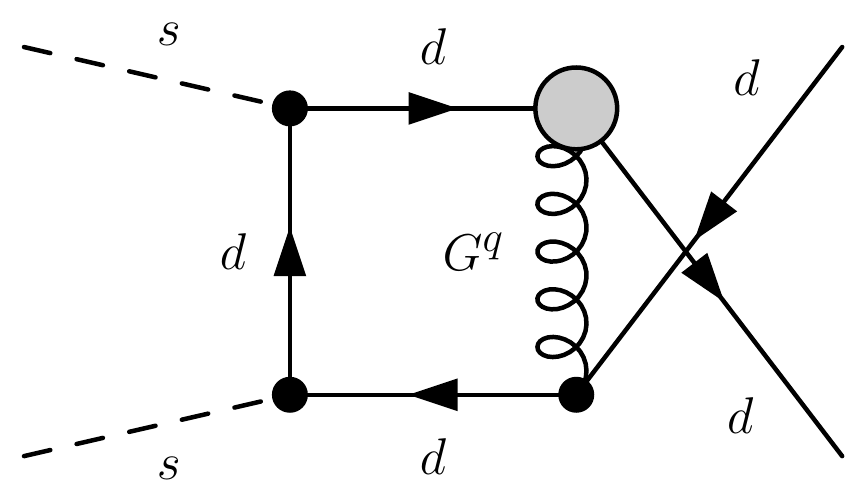}
 \includegraphics[width=0.20\columnwidth]{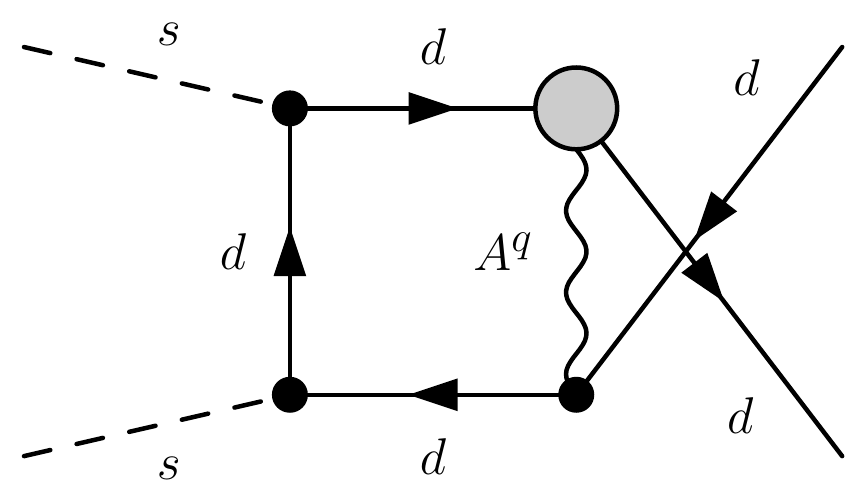}
 \includegraphics[width=0.20\columnwidth]{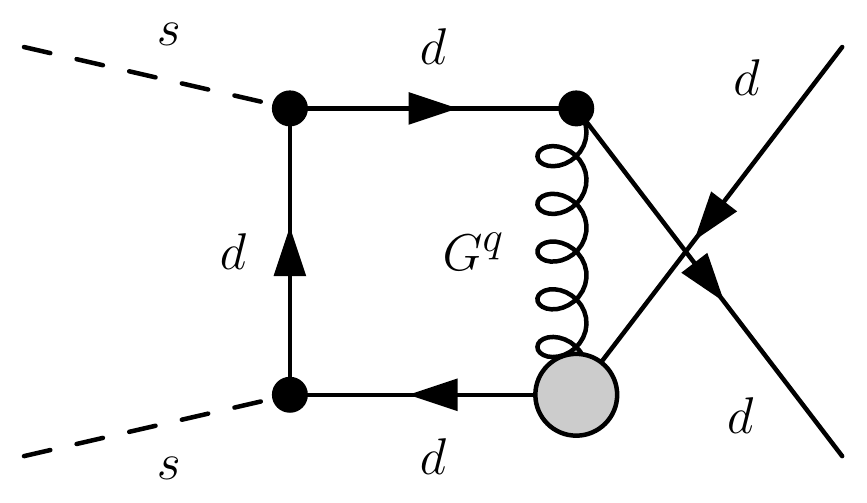}
 \includegraphics[width=0.20\columnwidth]{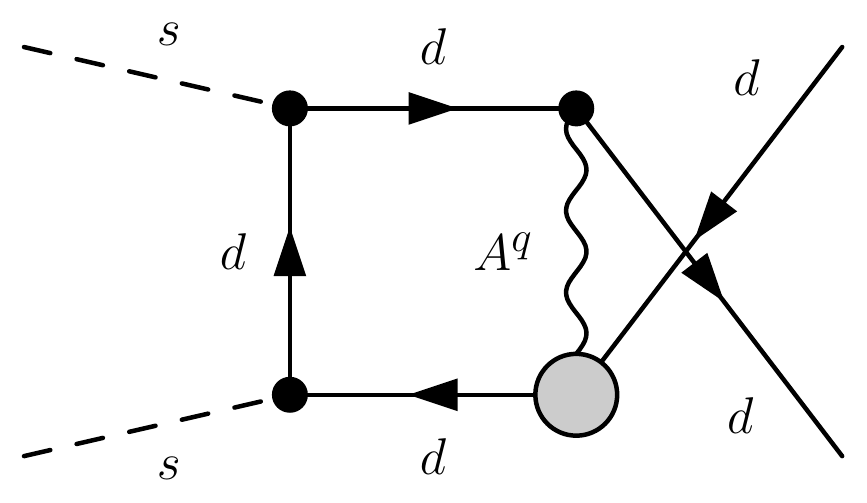}
 \includegraphics[width=0.20\columnwidth]{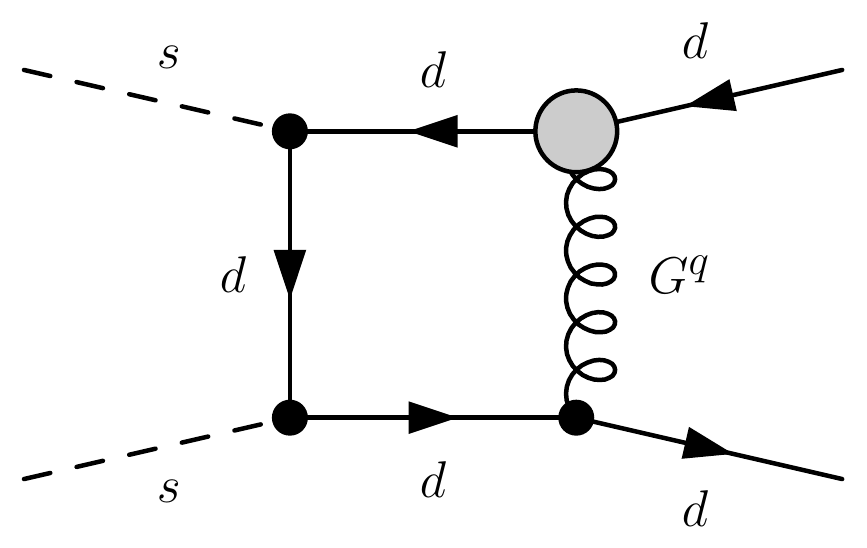}
 \includegraphics[width=0.20\columnwidth]{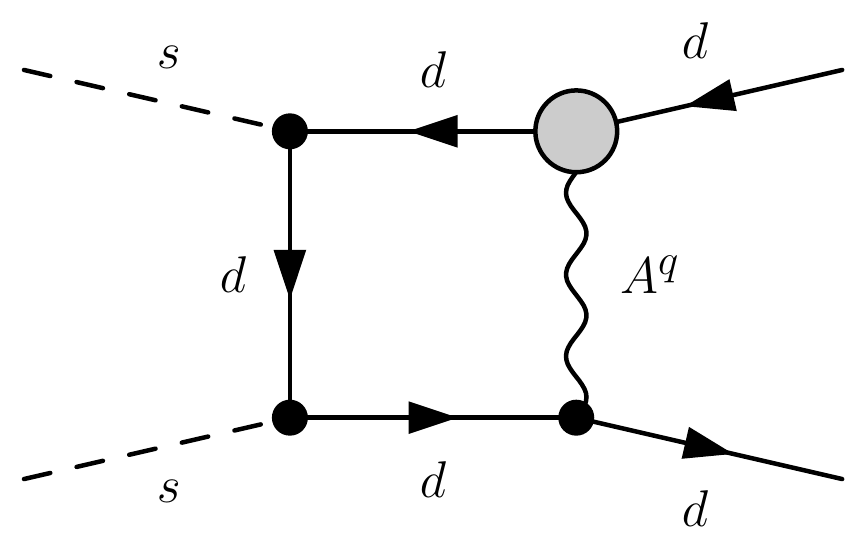}
 \includegraphics[width=0.20\columnwidth]{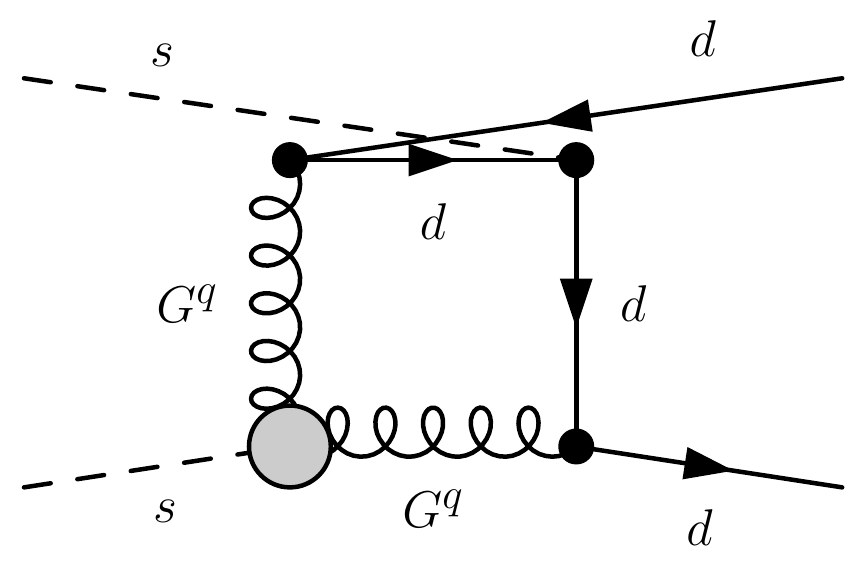}
 \includegraphics[width=0.20\columnwidth]{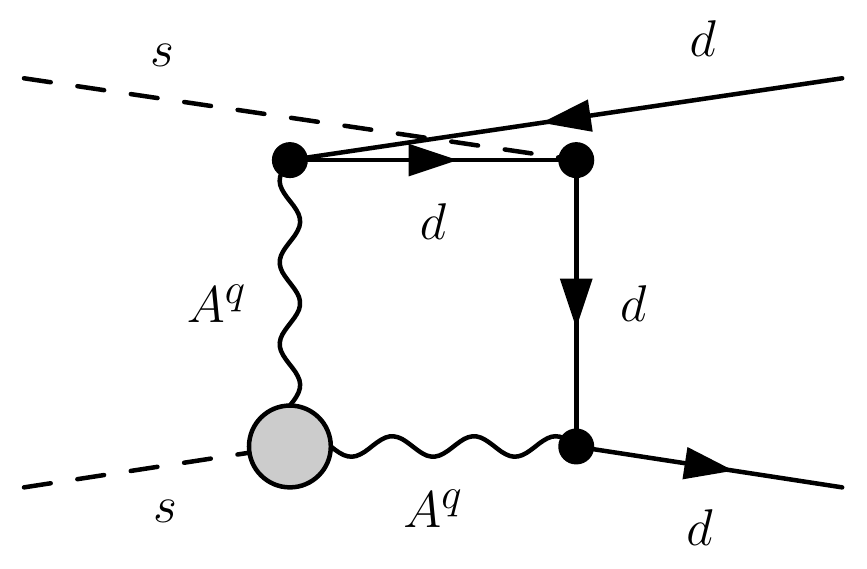}
 \includegraphics[width=0.20\columnwidth]{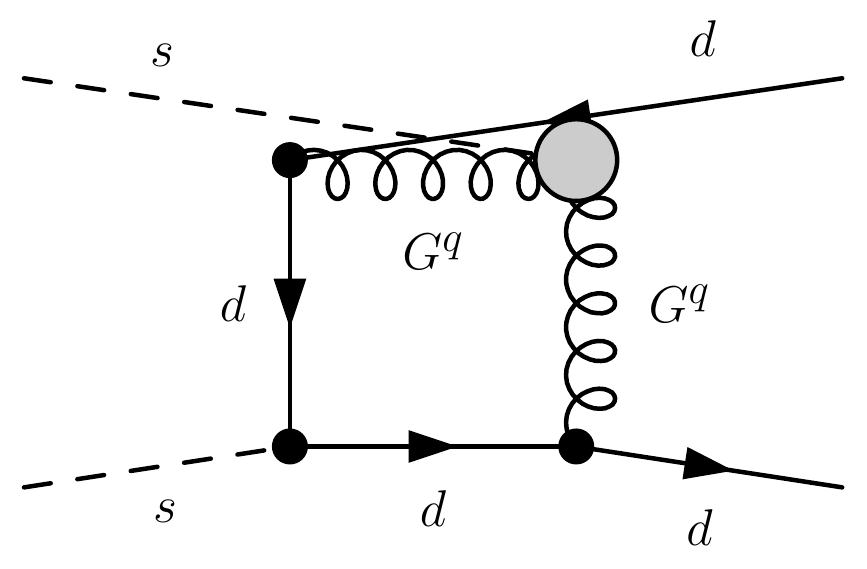}
 \includegraphics[width=0.20\columnwidth]{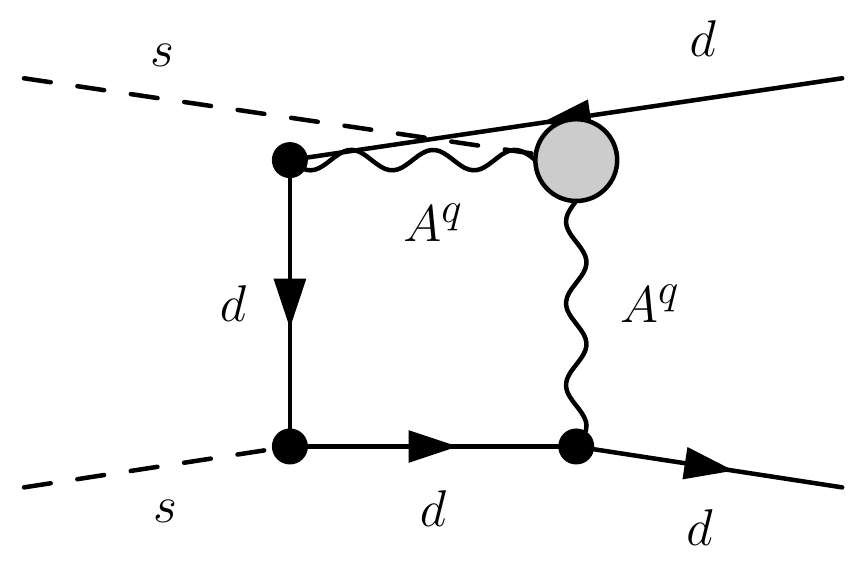}
 \includegraphics[width=0.20\columnwidth]{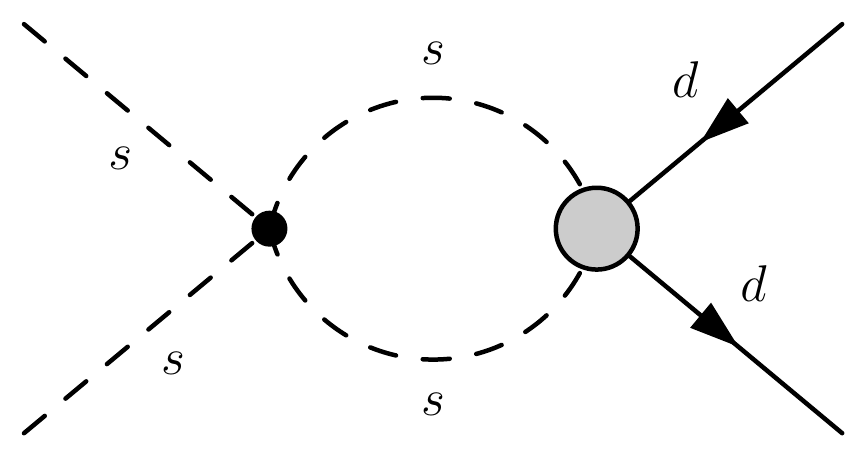}
 \caption{\it Feynman diagrams for $s(p_1)s(p_2)\to\overline{d}(p_3) d(p_4)$.}\label{fig:ss_dd}
\end{figure}

\begin{figure}[H]
 \centering
 \includegraphics[width=0.20\columnwidth]{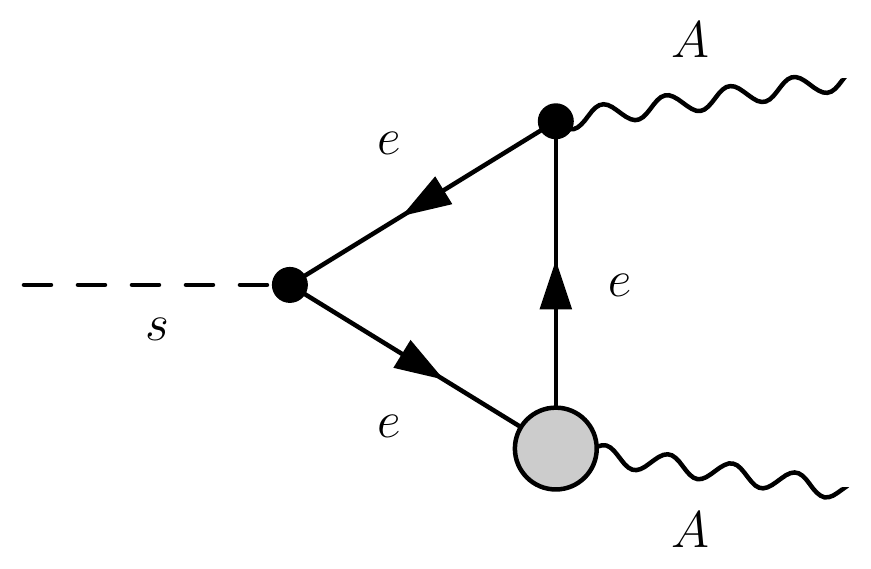}
 \includegraphics[width=0.20\columnwidth]{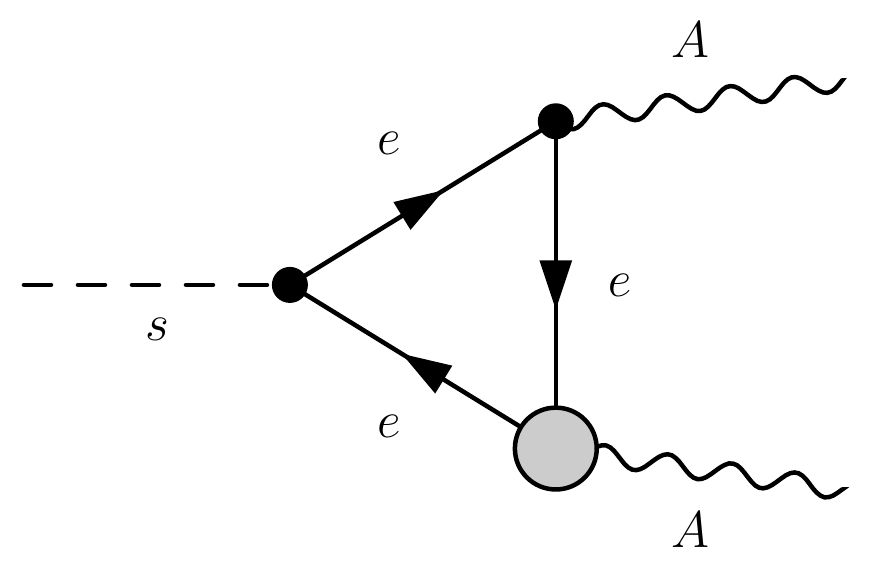}
 \includegraphics[width=0.20\columnwidth]{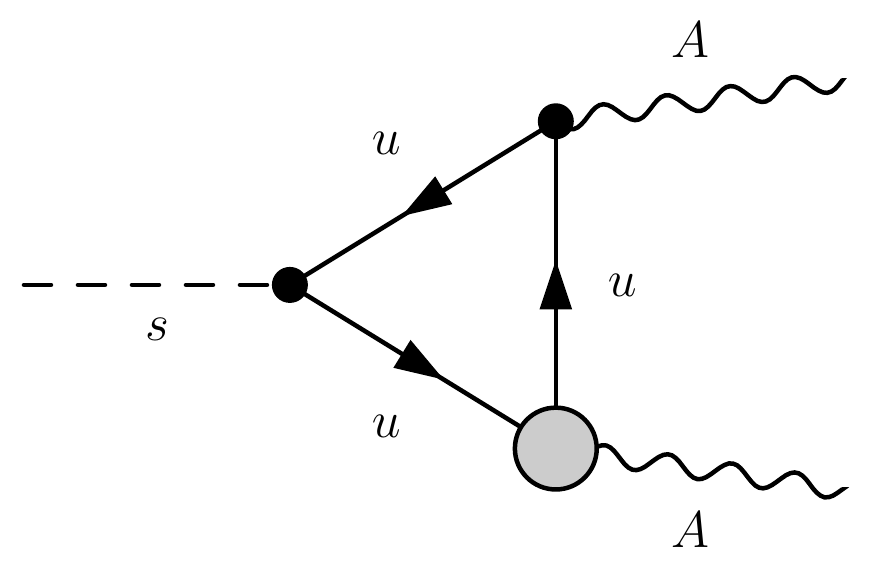}
 \includegraphics[width=0.20\columnwidth]{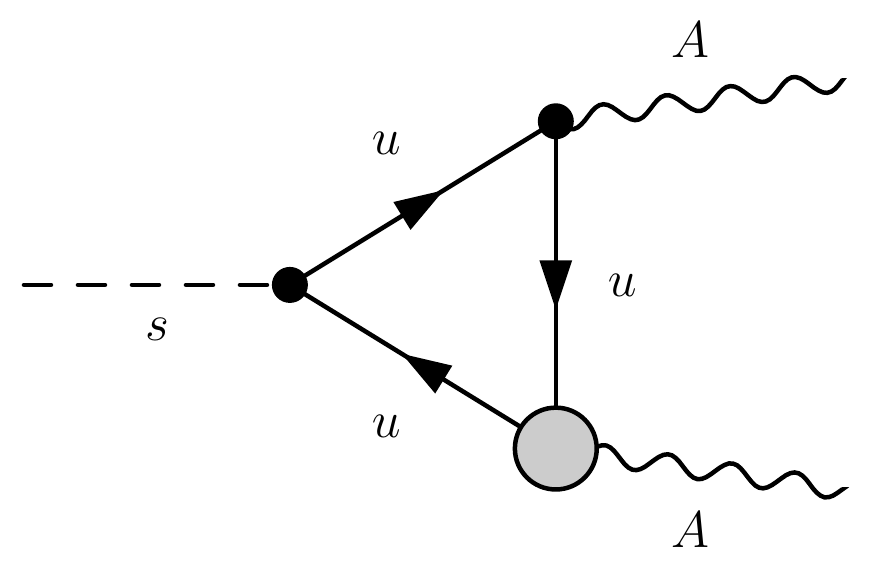}
 \includegraphics[width=0.20\columnwidth]{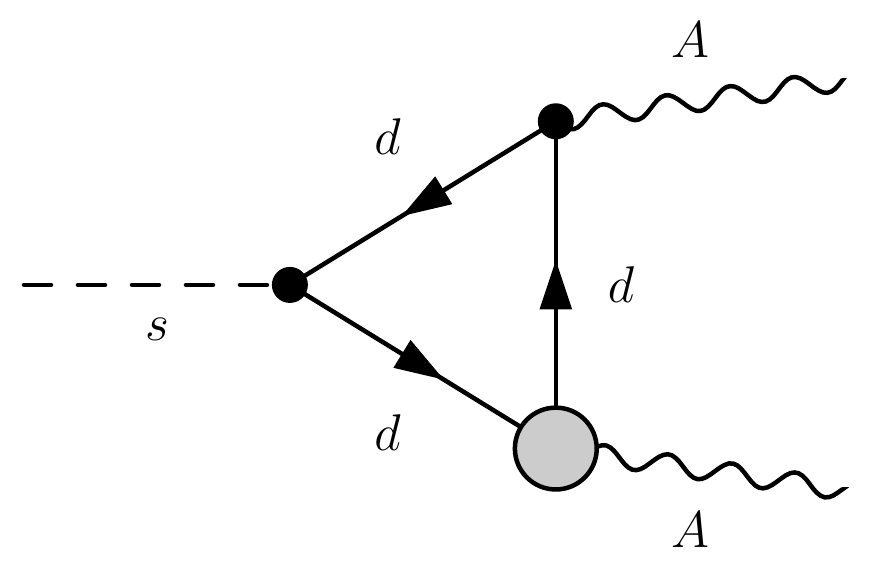}
 \includegraphics[width=0.20\columnwidth]{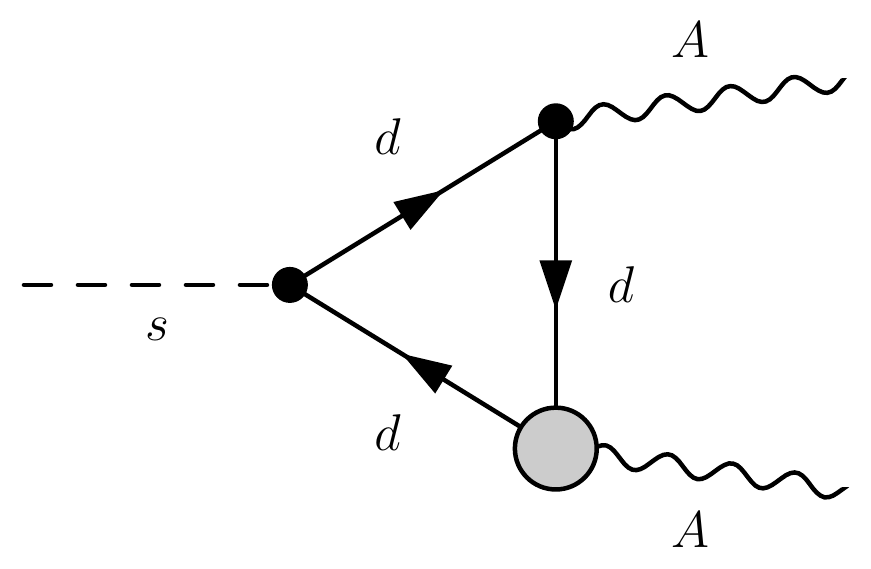}
 \includegraphics[width=0.20\columnwidth]{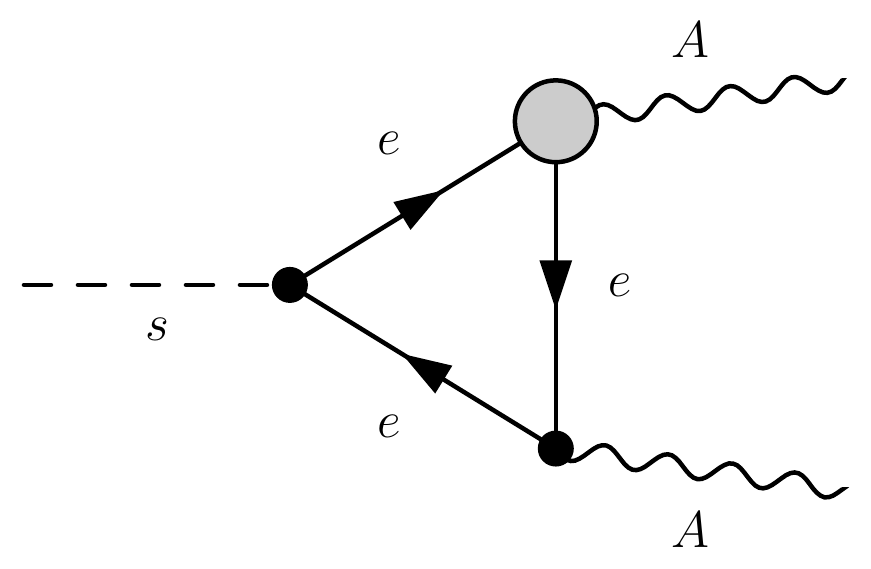}
 \includegraphics[width=0.20\columnwidth]{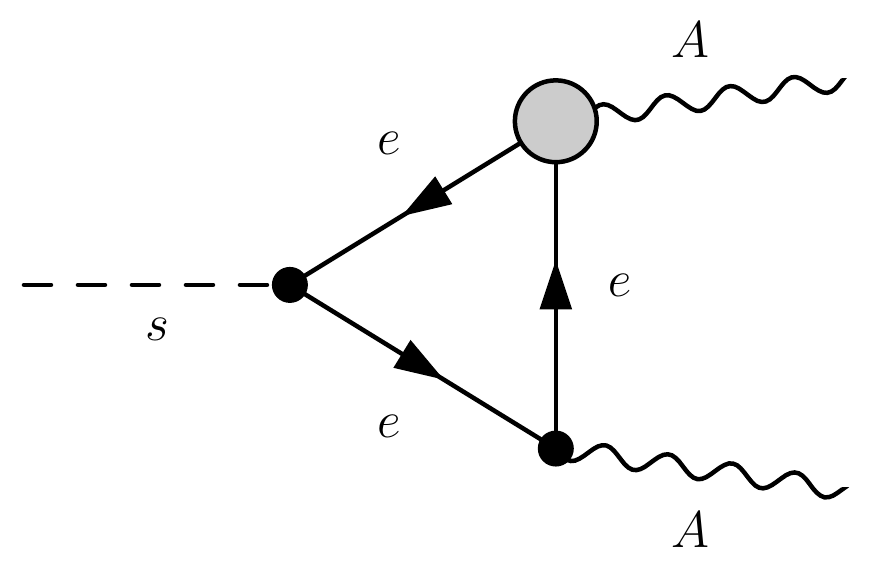}
 \includegraphics[width=0.20\columnwidth]{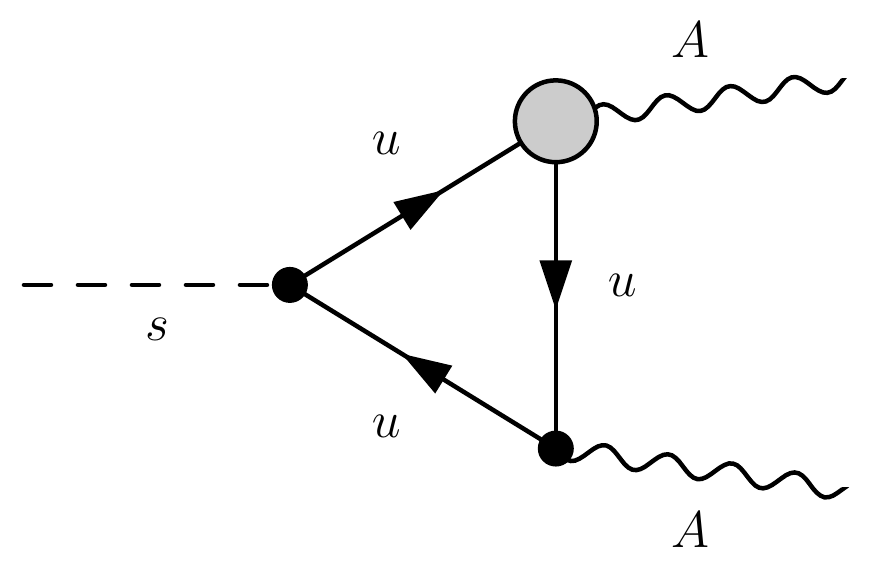}
 \includegraphics[width=0.20\columnwidth]{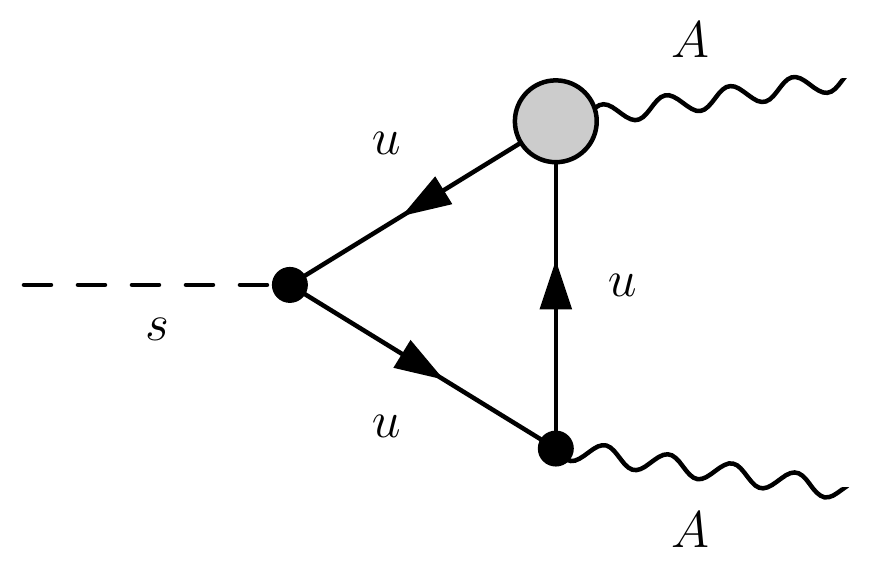}
 \includegraphics[width=0.20\columnwidth]{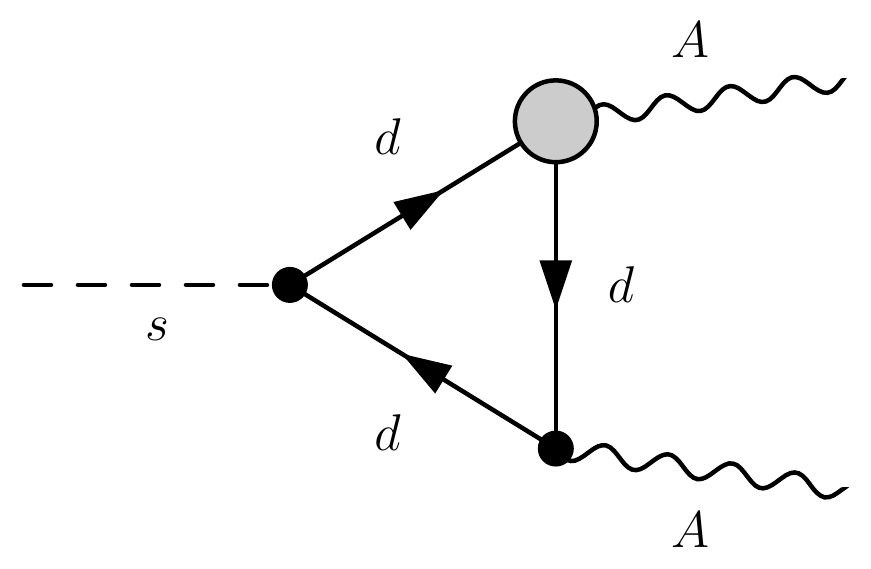}
 \includegraphics[width=0.20\columnwidth]{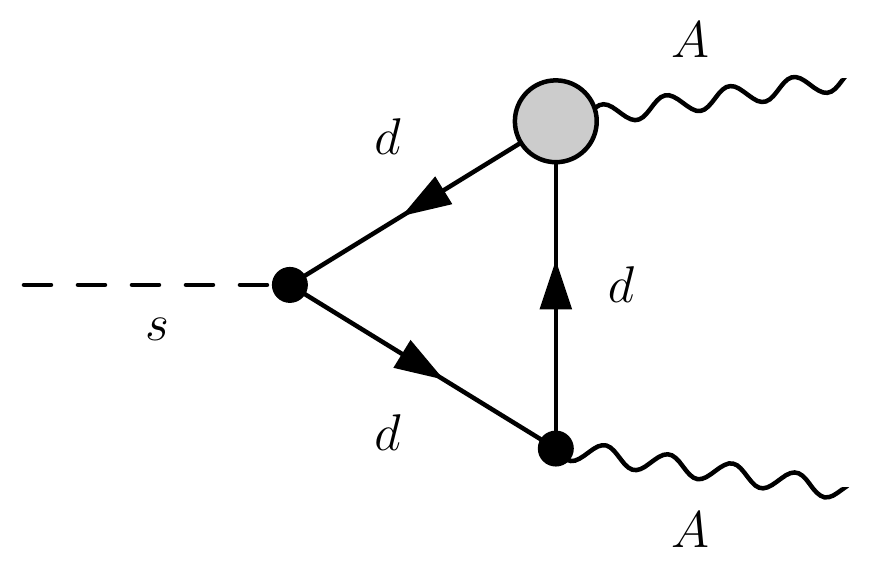}
 \caption{\it Feynman diagrams for $s(p_1)\to A(p_2) A(p_3)$.}\label{fig:s_aa}
\end{figure}

\begin{figure}[H]
 \centering
 \includegraphics[width=0.20\columnwidth]{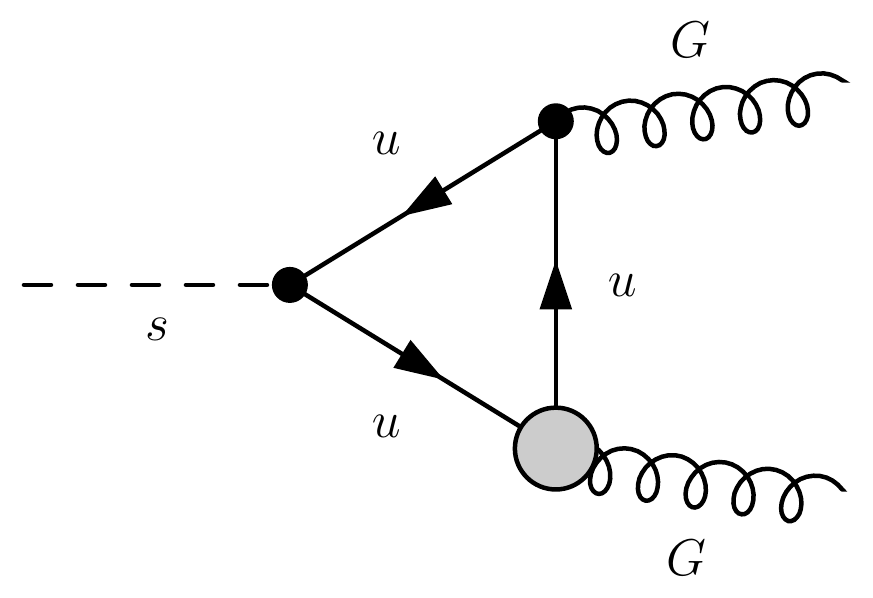}
 \includegraphics[width=0.20\columnwidth]{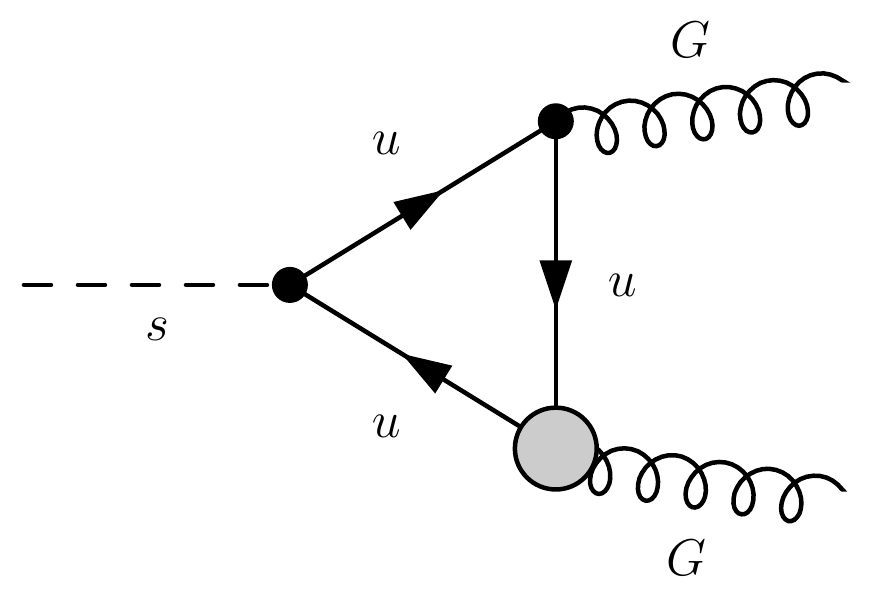}
 \includegraphics[width=0.20\columnwidth]{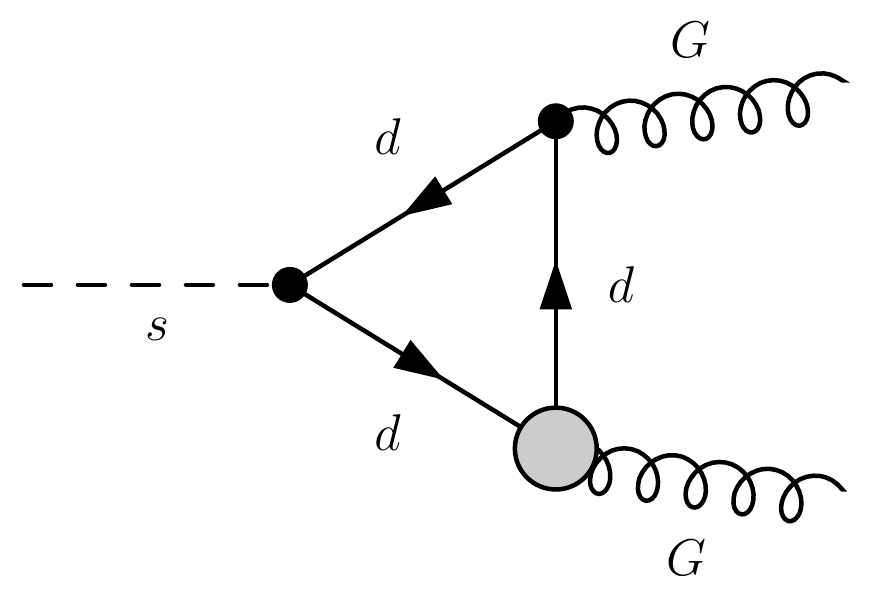}
 \includegraphics[width=0.20\columnwidth]{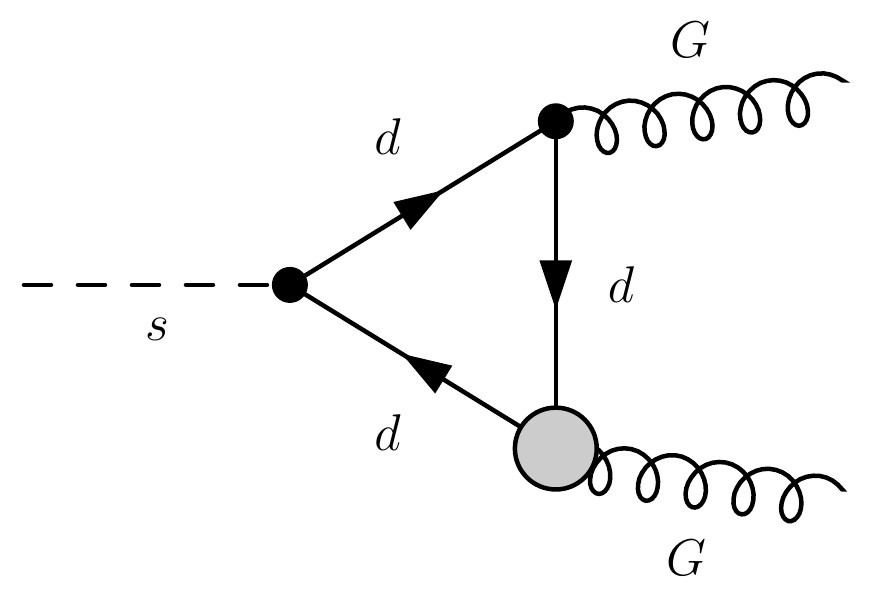}
 \includegraphics[width=0.20\columnwidth]{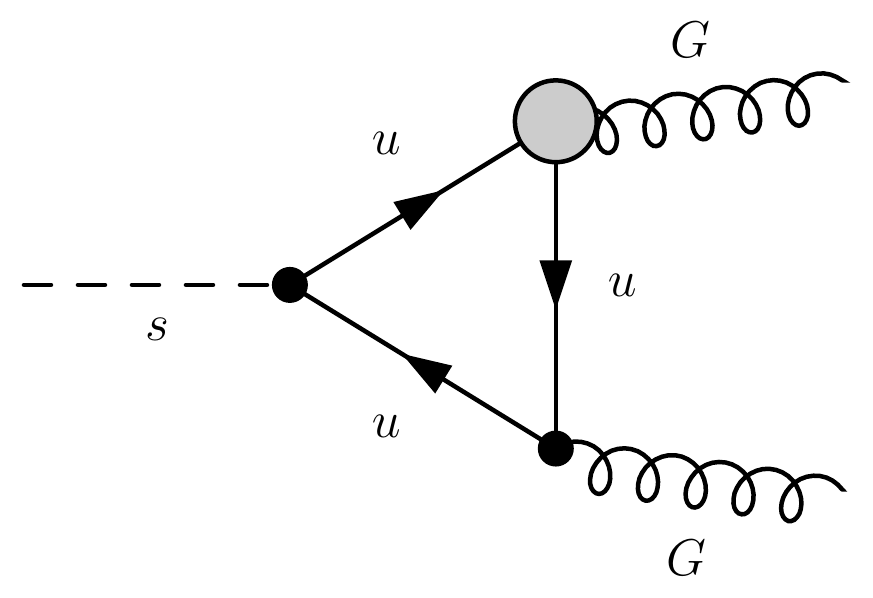}
 \includegraphics[width=0.20\columnwidth]{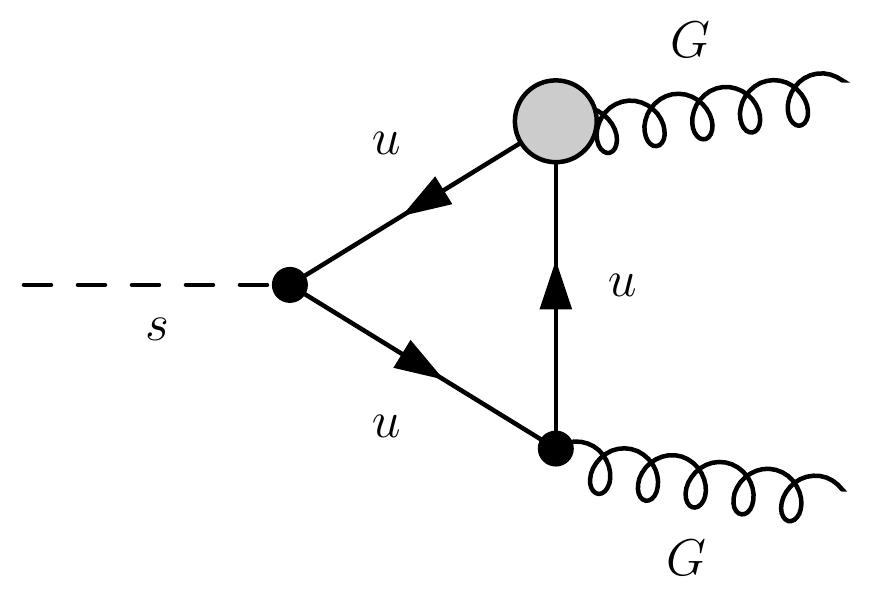}
 \includegraphics[width=0.20\columnwidth]{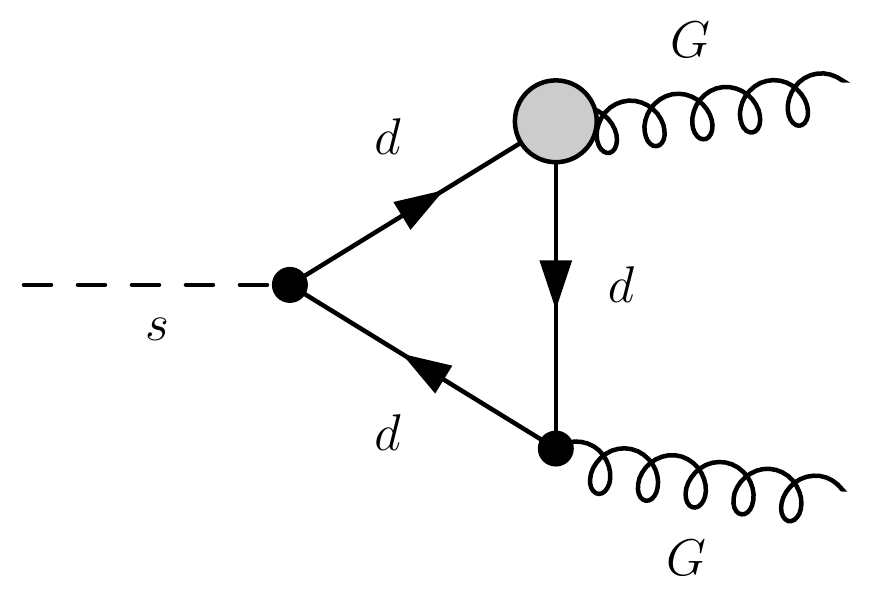}
 \includegraphics[width=0.20\columnwidth]{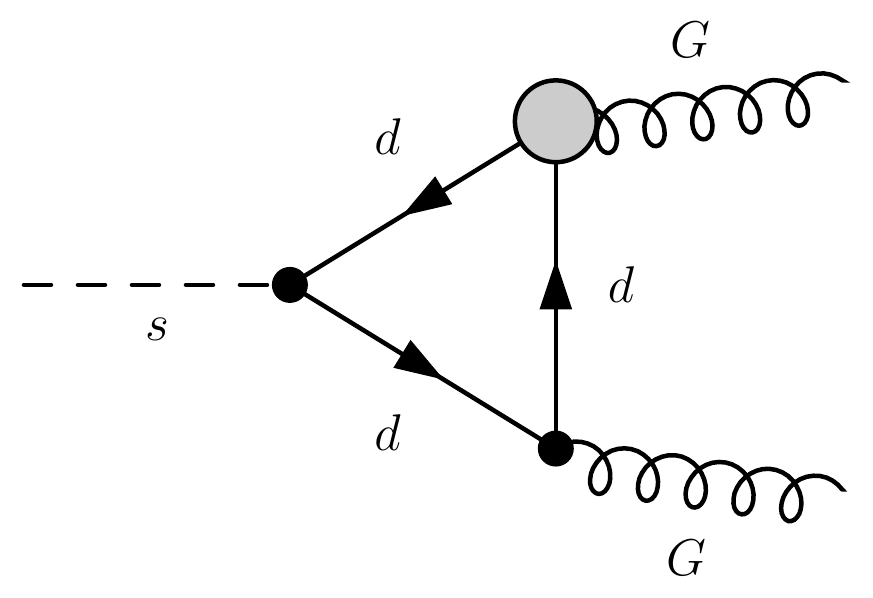}
 \includegraphics[width=0.20\columnwidth]{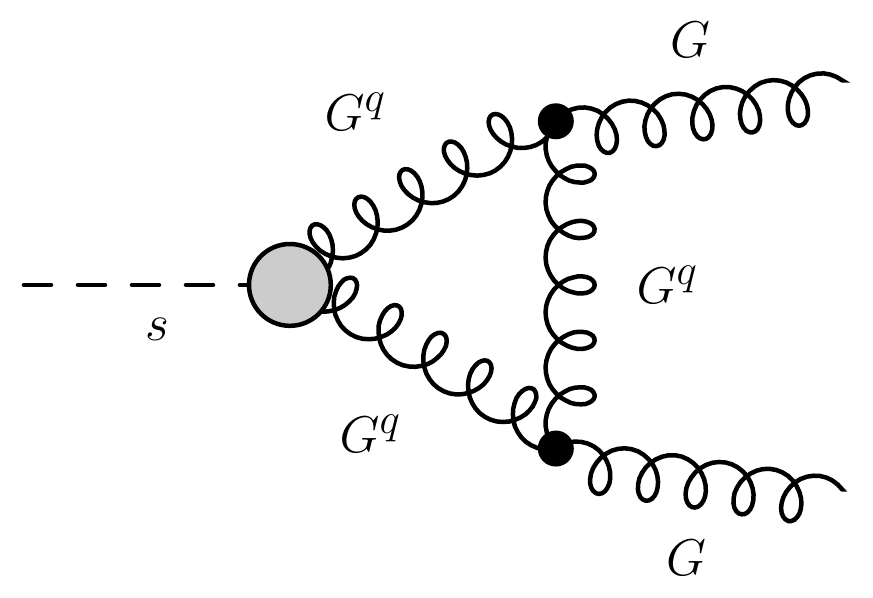}
 \includegraphics[width=0.20\columnwidth]{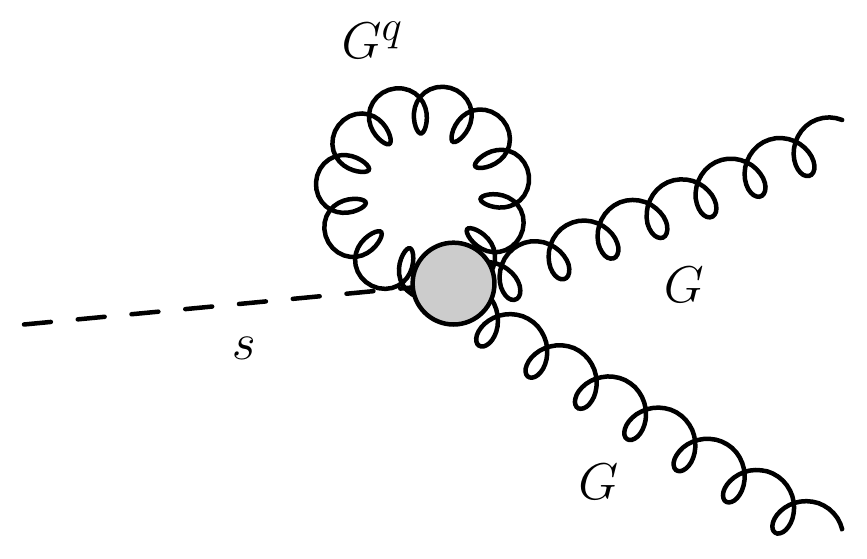}
 \includegraphics[width=0.20\columnwidth]{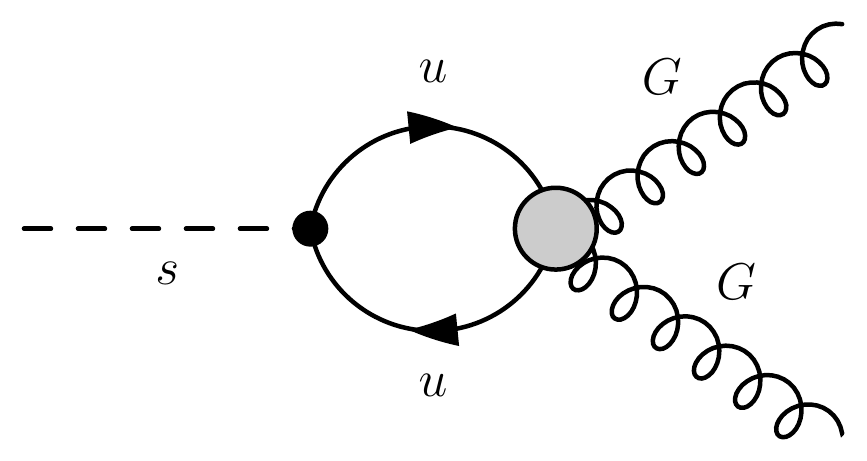}
 \includegraphics[width=0.20\columnwidth]{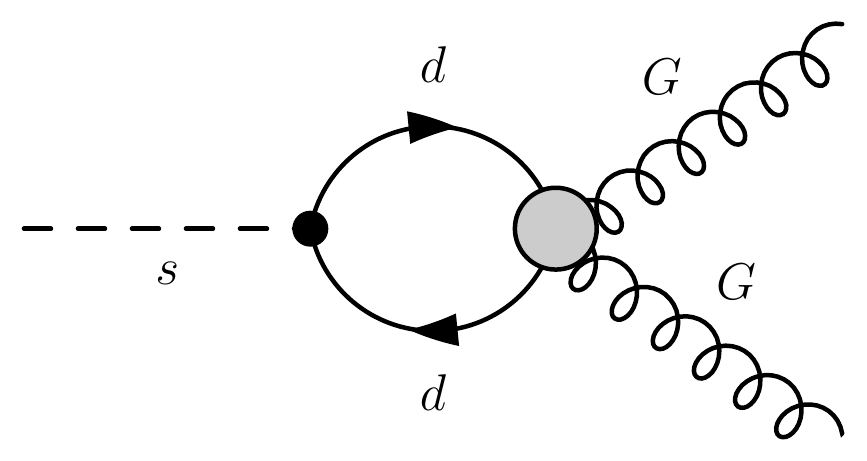}
 \includegraphics[width=0.20\columnwidth]{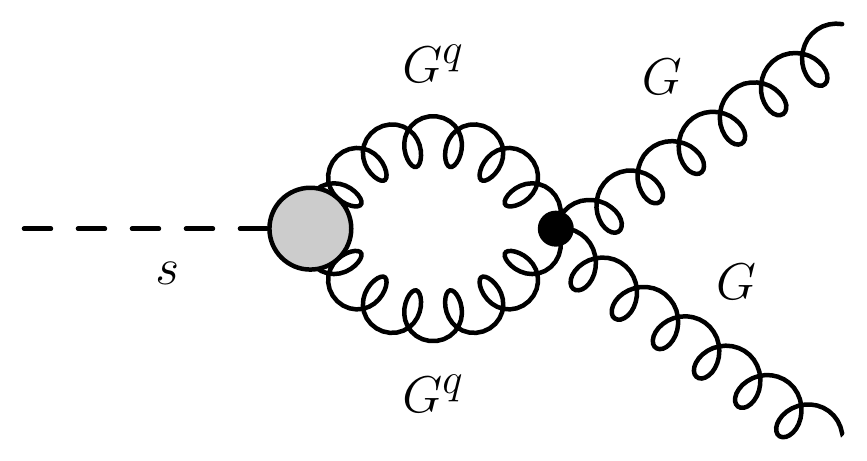}
 \includegraphics[width=0.20\columnwidth]{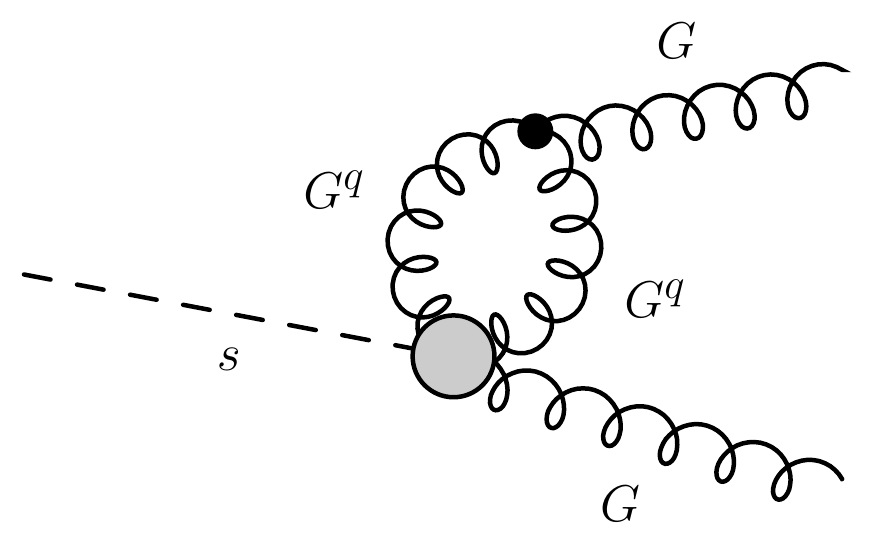}
 \includegraphics[width=0.20\columnwidth]{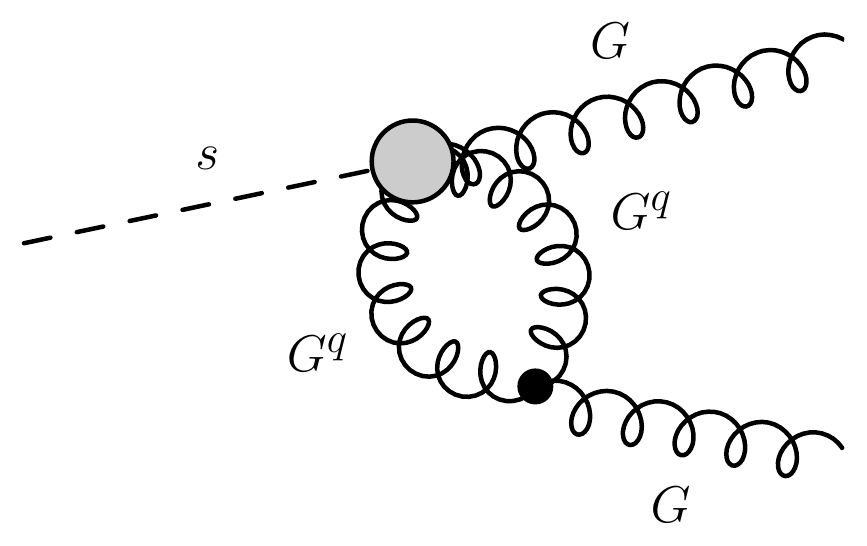}
 \caption{\it Feynman diagrams for $s(p_1)\to G(p_2) G(p_3)$.}\label{fig:s_gg}
\end{figure}

\bibliographystyle{JHEP}
\bibliography{notes}

\providecommand{\href}[2]{#2}\begingroup\raggedright\begin{thebibliography}{10}

\bibitem{Peccei:1977hh}
R.~Peccei and H.~R. Quinn, \emph{{CP Conservation in the Presence of
  Instantons}}, \href{https://doi.org/10.1103/PhysRevLett.38.1440}{\emph{Phys.
  Rev. Lett.} {\bfseries 38} (1977) 1440}.

\bibitem{Peccei:1977ur}
R.~Peccei and H.~R. Quinn, \emph{{Constraints Imposed by CP Conservation in the
  Presence of Instantons}},
  \href{https://doi.org/10.1103/PhysRevD.16.1791}{\emph{Phys. Rev. D}
  {\bfseries 16} (1977) 1791}.

\bibitem{Weinberg:1977ma}
S.~Weinberg, \emph{{A New Light Boson?}},
  \href{https://doi.org/10.1103/PhysRevLett.40.223}{\emph{Phys. Rev. Lett.}
  {\bfseries 40} (1978) 223}.

\bibitem{Wilczek:1977pj}
F.~Wilczek, \emph{{Problem of Strong $P$ and $T$ Invariance in the Presence of
  Instantons}}, \href{https://doi.org/10.1103/PhysRevLett.40.279}{\emph{Phys.
  Rev. Lett.} {\bfseries 40} (1978) 279}.

\bibitem{Gripaios:2009pe}
B.~Gripaios, A.~Pomarol, F.~Riva and J.~Serra, \emph{{Beyond the Minimal
  Composite Higgs Model}},
  \href{https://doi.org/10.1088/1126-6708/2009/04/070}{\emph{JHEP} {\bfseries
  04} (2009) 070} [\href{https://arxiv.org/abs/0902.1483}{{\ttfamily
  0902.1483}}].

\bibitem{Gripaios:2016mmi}
B.~Gripaios, M.~Nardecchia and T.~You, \emph{{On the Structure of Anomalous
  Composite Higgs Models}},
  \href{https://doi.org/10.1140/epjc/s10052-017-4603-5}{\emph{Eur. Phys. J. C}
  {\bfseries 77} (2017) 28} [\href{https://arxiv.org/abs/1605.09647}{{\ttfamily
  1605.09647}}].

\bibitem{Chala:2017sjk}
M.~Chala, G.~Durieux, C.~Grojean, L.~de~Lima and O.~Matsedonskyi,
  \emph{{Minimally extended SILH}},
  \href{https://doi.org/10.1007/JHEP06(2017)088}{\emph{JHEP} {\bfseries 06}
  (2017) 088} [\href{https://arxiv.org/abs/1703.10624}{{\ttfamily
  1703.10624}}].

\bibitem{Wilczek:1982rv}
F.~Wilczek, \emph{{Axions and Family Symmetry Breaking}},
  \href{https://doi.org/10.1103/PhysRevLett.49.1549}{\emph{Phys. Rev. Lett.}
  {\bfseries 49} (1982) 1549}.

\bibitem{Chikashige:1980ui}
Y.~Chikashige, R.~N. Mohapatra and R.~Peccei, \emph{{Are There Real Goldstone
  Bosons Associated with Broken Lepton Number?}},
  \href{https://doi.org/10.1016/0370-2693(81)90011-3}{\emph{Phys. Lett. B}
  {\bfseries 98} (1981) 265}.

\bibitem{Preskill:1982cy}
J.~Preskill, M.~B. Wise and F.~Wilczek, \emph{{Cosmology of the Invisible
  Axion}}, \href{https://doi.org/10.1016/0370-2693(83)90637-8}{\emph{Phys.
  Lett. B} {\bfseries 120} (1983) 127}.

\bibitem{Abbott:1982af}
L.~Abbott and P.~Sikivie, \emph{{A Cosmological Bound on the Invisible Axion}},
  \href{https://doi.org/10.1016/0370-2693(83)90638-X}{\emph{Phys. Lett. B}
  {\bfseries 120} (1983) 133}.

\bibitem{Dine:1982ah}
M.~Dine and W.~Fischler, \emph{{The Not So Harmless Axion}},
  \href{https://doi.org/10.1016/0370-2693(83)90639-1}{\emph{Phys. Lett. B}
  {\bfseries 120} (1983) 137}.

\bibitem{Davidson:1981zd}
A.~Davidson and K.~C. Wali, \emph{{Minimal flavour unification via
  multigenerational Peccei-Quinn symmetry}},
  \href{https://doi.org/10.1103/PhysRevLett.48.11}{\emph{Phys. Rev. Lett.}
  {\bfseries 48} (1982) 11}.

\bibitem{Ema:2016ops}
Y.~Ema, K.~Hamaguchi, T.~Moroi and K.~Nakayama, \emph{{Flaxion: a minimal
  extension to solve puzzles in the standard model}},
  \href{https://doi.org/10.1007/JHEP01(2017)096}{\emph{JHEP} {\bfseries 01}
  (2017) 096} [\href{https://arxiv.org/abs/1612.05492}{{\ttfamily
  1612.05492}}].

\bibitem{Calibbi:2016hwq}
L.~Calibbi, F.~Goertz, D.~Redigolo, R.~Ziegler and J.~Zupan, \emph{{Minimal
  axion model from flavor}},
  \href{https://doi.org/10.1103/PhysRevD.95.095009}{\emph{Phys. Rev. D}
  {\bfseries 95} (2017) 095009}
  [\href{https://arxiv.org/abs/1612.08040}{{\ttfamily 1612.08040}}].

\bibitem{Graham:2015cka}
P.~W. Graham, D.~E. Kaplan and S.~Rajendran, \emph{{Cosmological Relaxation of
  the Electroweak Scale}},
  \href{https://doi.org/10.1103/PhysRevLett.115.221801}{\emph{Phys. Rev. Lett.}
  {\bfseries 115} (2015) 221801}
  [\href{https://arxiv.org/abs/1504.07551}{{\ttfamily 1504.07551}}].

\bibitem{Povey:2010hs}
R.~Povey, J.~Hartnett and M.~Tobar, \emph{{Microwave cavity light shining
  through a wall optimization and experiment}},
  \href{https://doi.org/10.1103/PhysRevD.82.052003}{\emph{Phys. Rev. D}
  {\bfseries 82} (2010) 052003}
  [\href{https://arxiv.org/abs/1003.0964}{{\ttfamily 1003.0964}}].

\bibitem{Wagner:2010mi}
{\scshape ADMX} collaboration, \emph{{A Search for Hidden Sector Photons with
  ADMX}}, \href{https://doi.org/10.1103/PhysRevLett.105.171801}{\emph{Phys.
  Rev. Lett.} {\bfseries 105} (2010) 171801}
  [\href{https://arxiv.org/abs/1007.3766}{{\ttfamily 1007.3766}}].

\bibitem{Essig:2013lka}
R.~Essig et~al., \emph{{Working Group Report: New Light Weakly Coupled
  Particles}},  in \emph{{Community Summer Study 2013}: {Snowmass on the
  Mississippi}}, 10, 2013, \href{https://arxiv.org/abs/1311.0029}{{\ttfamily
  1311.0029}}.

\bibitem{Betz:2013dza}
M.~Betz, F.~Caspers, M.~Gasior, M.~Thumm and S.~Rieger, \emph{{First results of
  the CERN Resonant Weakly Interacting sub-eV Particle Search (CROWS)}},
  \href{https://doi.org/10.1103/PhysRevD.88.075014}{\emph{Phys. Rev. D}
  {\bfseries 88} (2013) 075014}
  [\href{https://arxiv.org/abs/1310.8098}{{\ttfamily 1310.8098}}].

\bibitem{Bjorken:2009mm}
J.~D. Bjorken, R.~Essig, P.~Schuster and N.~Toro, \emph{{New Fixed-Target
  Experiments to Search for Dark Gauge Forces}},
  \href{https://doi.org/10.1103/PhysRevD.80.075018}{\emph{Phys. Rev. D}
  {\bfseries 80} (2009) 075018}
  [\href{https://arxiv.org/abs/0906.0580}{{\ttfamily 0906.0580}}].

\bibitem{Andreas:2012mt}
S.~Andreas, C.~Niebuhr and A.~Ringwald, \emph{{New Limits on Hidden Photons
  from Past Electron Beam Dumps}},
  \href{https://doi.org/10.1103/PhysRevD.86.095019}{\emph{Phys. Rev. D}
  {\bfseries 86} (2012) 095019}
  [\href{https://arxiv.org/abs/1209.6083}{{\ttfamily 1209.6083}}].

\bibitem{Jaeckel:2015jla}
J.~Jaeckel and M.~Spannowsky, \emph{{Probing MeV to 90 GeV axion-like particles
  with LEP and LHC}},
  \href{https://doi.org/10.1016/j.physletb.2015.12.037}{\emph{Phys. Lett. B}
  {\bfseries 753} (2016) 482}
  [\href{https://arxiv.org/abs/1509.00476}{{\ttfamily 1509.00476}}].

\bibitem{Bauer:2017ris}
M.~Bauer, M.~Neubert and A.~Thamm, \emph{{Collider Probes of Axion-Like
  Particles}}, \href{https://doi.org/10.1007/JHEP12(2017)044}{\emph{JHEP}
  {\bfseries 12} (2017) 044}
  [\href{https://arxiv.org/abs/1708.00443}{{\ttfamily 1708.00443}}].

\bibitem{Craig:2018kne}
N.~Craig, A.~Hook and S.~Kasko, \emph{{The Photophobic ALP}},
  \href{https://doi.org/10.1007/JHEP09(2018)028}{\emph{JHEP} {\bfseries 09}
  (2018) 028} [\href{https://arxiv.org/abs/1805.06538}{{\ttfamily
  1805.06538}}].

\bibitem{Knapen:2016moh}
S.~Knapen, T.~Lin, H.~K. Lou and T.~Melia, \emph{{Searching for Axionlike
  Particles with Ultraperipheral Heavy-Ion Collisions}},
  \href{https://doi.org/10.1103/PhysRevLett.118.171801}{\emph{Phys. Rev. Lett.}
  {\bfseries 118} (2017) 171801}
  [\href{https://arxiv.org/abs/1607.06083}{{\ttfamily 1607.06083}}].

\bibitem{Brivio:2017ije}
I.~Brivio, M.~Gavela, L.~Merlo, K.~Mimasu, J.~No, R.~del Rey et~al.,
  \emph{{ALPs Effective Field Theory and Collider Signatures}},
  \href{https://doi.org/10.1140/epjc/s10052-017-5111-3}{\emph{Eur. Phys. J. C}
  {\bfseries 77} (2017) 572}
  [\href{https://arxiv.org/abs/1701.05379}{{\ttfamily 1701.05379}}].

\bibitem{Bauer:2017nlg}
M.~Bauer, M.~Neubert and A.~Thamm, \emph{{LHC as an Axion Factory: Probing an
  Axion Explanation for $(g-2)_\mu$ with Exotic Higgs Decays}},
  \href{https://doi.org/10.1103/PhysRevLett.119.031802}{\emph{Phys. Rev. Lett.}
  {\bfseries 119} (2017) 031802}
  [\href{https://arxiv.org/abs/1704.08207}{{\ttfamily 1704.08207}}].

\bibitem{Alonso-Alvarez:2018irt}
G.~Alonso-\'Alvarez, M.~Gavela and P.~Quilez, \emph{{Axion couplings to
  electroweak gauge bosons}},
  \href{https://doi.org/10.1140/epjc/s10052-019-6732-5}{\emph{Eur. Phys. J. C}
  {\bfseries 79} (2019) 223}
  [\href{https://arxiv.org/abs/1811.05466}{{\ttfamily 1811.05466}}].

\bibitem{Ebadi:2019gij}
J.~Ebadi, S.~Khatibi and M.~Mohammadi~Najafabadi, \emph{{New probes for
  axionlike particles at hadron colliders}},
  \href{https://doi.org/10.1103/PhysRevD.100.015016}{\emph{Phys. Rev. D}
  {\bfseries 100} (2019) 015016}
  [\href{https://arxiv.org/abs/1901.03061}{{\ttfamily 1901.03061}}].

\bibitem{Gavela:2019cmq}
M.~Gavela, J.~No, V.~Sanz and J.~de~Troc\'oniz, \emph{{Nonresonant Searches for
  Axionlike Particles at the LHC}},
  \href{https://doi.org/10.1103/PhysRevLett.124.051802}{\emph{Phys. Rev. Lett.}
  {\bfseries 124} (2020) 051802}
  [\href{https://arxiv.org/abs/1905.12953}{{\ttfamily 1905.12953}}].

\bibitem{Coelho:2020saz}
R.~Coelho, V.~Goncalves, D.~Martins and M.~Rangel, \emph{{Production of
  axionlike particles in $PbPb$ collisions at the LHC, HE\textendash{}LHC and
  FCC: A phenomenological analysis}},
  \href{https://doi.org/10.1016/j.physletb.2020.135512}{\emph{Phys. Lett. B}
  {\bfseries 806} (2020) 135512}
  [\href{https://arxiv.org/abs/2002.06027}{{\ttfamily 2002.06027}}].

\bibitem{Haghighat:2020nuh}
G.~Haghighat, D.~Haji~Raissi and M.~Mohammadi~Najafabadi, \emph{{New collider
  searches for axionlike particles coupling to gluons}},
  \href{https://doi.org/10.1103/PhysRevD.102.115010}{\emph{Phys. Rev. D}
  {\bfseries 102} (2020) 115010}
  [\href{https://arxiv.org/abs/2006.05302}{{\ttfamily 2006.05302}}].

\bibitem{Goncalves:2020bqi}
V.~P. Goncalves and W.~K. Sauter, \emph{{Exclusive axionlike particle
  production by gluon -- induced interactions in hadronic collisions}},
  \href{https://doi.org/10.1016/j.physletb.2020.135981}{\emph{Phys. Lett. B}
  {\bfseries 811} (2020) 135981}
  [\href{https://arxiv.org/abs/2006.16716}{{\ttfamily 2006.16716}}].

\bibitem{Bauer:2018uxu}
M.~Bauer, M.~Heiles, M.~Neubert and A.~Thamm, \emph{{Axion-Like Particles at
  Future Colliders}},
  \href{https://doi.org/10.1140/epjc/s10052-019-6587-9}{\emph{Eur. Phys. J. C}
  {\bfseries 79} (2019) 74} [\href{https://arxiv.org/abs/1808.10323}{{\ttfamily
  1808.10323}}].

\bibitem{Yue:2019gbh}
C.-X. Yue, M.-Z. Liu and Y.-C. Guo, \emph{{Searching for axionlike particles at
  future $ep$ colliders}},
  \href{https://doi.org/10.1103/PhysRevD.100.015020}{\emph{Phys. Rev. D}
  {\bfseries 100} (2019) 015020}
  [\href{https://arxiv.org/abs/1904.10657}{{\ttfamily 1904.10657}}].

\bibitem{Inan:2020aal}
S.~\.Inan and A.~Kisselev, \emph{{A search for axion-like particles in
  light-by-light scattering at the CLIC}},
  \href{https://doi.org/10.1007/JHEP06(2020)183}{\emph{JHEP} {\bfseries 06}
  (2020) 183} [\href{https://arxiv.org/abs/2003.01978}{{\ttfamily
  2003.01978}}].

\bibitem{Feng:2018pew}
J.~L. Feng, I.~Galon, F.~Kling and S.~Trojanowski, \emph{{Axionlike particles
  at FASER: The LHC as a photon beam dump}},
  \href{https://doi.org/10.1103/PhysRevD.98.055021}{\emph{Phys. Rev. D}
  {\bfseries 98} (2018) 055021}
  [\href{https://arxiv.org/abs/1806.02348}{{\ttfamily 1806.02348}}].

\bibitem{Ariga:2018uku}
{\scshape FASER} collaboration, \emph{{FASER\textquoteright{}s physics reach
  for long-lived particles}},
  \href{https://doi.org/10.1103/PhysRevD.99.095011}{\emph{Phys. Rev. D}
  {\bfseries 99} (2019) 095011}
  [\href{https://arxiv.org/abs/1811.12522}{{\ttfamily 1811.12522}}].

\bibitem{Kling:2020mch}
F.~Kling and S.~Trojanowski, \emph{{Looking forward to test the KOTO anomaly
  with FASER}}, \href{https://doi.org/10.1103/PhysRevD.102.015032}{\emph{Phys.
  Rev. D} {\bfseries 102} (2020) 015032}
  [\href{https://arxiv.org/abs/2006.10630}{{\ttfamily 2006.10630}}].

\bibitem{Arik:2011rx}
{\scshape CAST} collaboration, \emph{{CAST search for sub-eV mass solar axions
  with 3He buffer gas}},
  \href{https://doi.org/10.1103/PhysRevLett.107.261302}{\emph{Phys. Rev. Lett.}
  {\bfseries 107} (2011) 261302}
  [\href{https://arxiv.org/abs/1106.3919}{{\ttfamily 1106.3919}}].

\bibitem{Raffelt:2006cw}
G.~G. Raffelt, \emph{{Astrophysical axion bounds}},
  \href{https://doi.org/10.1007/978-3-540-73518-2_3}{\emph{Lect. Notes Phys.}
  {\bfseries 741} (2008) 51}
  [\href{https://arxiv.org/abs/hep-ph/0611350}{{\ttfamily hep-ph/0611350}}].

\bibitem{Lee:2018lcj}
J.~S. Lee, \emph{{Revisiting Supernova 1987A Limits on Axion-Like-Particles}},
  \href{https://arxiv.org/abs/1808.10136}{{\ttfamily 1808.10136}}.

\bibitem{Chang:2018rso}
J.~H. Chang, R.~Essig and S.~D. McDermott, \emph{{Supernova 1987A Constraints
  on Sub-GeV Dark Sectors, Millicharged Particles, the QCD Axion, and an
  Axion-like Particle}},
  \href{https://doi.org/10.1007/JHEP09(2018)051}{\emph{JHEP} {\bfseries 09}
  (2018) 051} [\href{https://arxiv.org/abs/1803.00993}{{\ttfamily
  1803.00993}}].

\bibitem{Jaeckel:2019xpa}
J.~Jaeckel and L.~J. Thormaehlen, \emph{{Axions as a probe of solar metals}},
  \href{https://doi.org/10.1103/PhysRevD.100.123020}{\emph{Phys. Rev. D}
  {\bfseries 100} (2019) 123020}
  [\href{https://arxiv.org/abs/1908.10878}{{\ttfamily 1908.10878}}].

\bibitem{Carenza:2019pxu}
P.~Carenza, T.~Fischer, M.~Giannotti, G.~Guo, G.~Mart\'\i{}nez-Pinedo and
  A.~Mirizzi, \emph{{Improved axion emissivity from a supernova via
  nucleon-nucleon bremsstrahlung}},
  \href{https://doi.org/10.1088/1475-7516/2019/10/016}{\emph{JCAP} {\bfseries
  10} (2019) 016} [\href{https://arxiv.org/abs/1906.11844}{{\ttfamily
  1906.11844}}].

\bibitem{Ertas:2020xcc}
F.~Ertas and F.~Kahlhoefer, \emph{{On the interplay between astrophysical and
  laboratory probes of MeV-scale axion-like particles}},
  \href{https://doi.org/10.1007/JHEP07(2020)050}{\emph{JHEP} {\bfseries 07}
  (2020) 050} [\href{https://arxiv.org/abs/2004.01193}{{\ttfamily
  2004.01193}}].

\bibitem{Lucente:2020whw}
G.~Lucente, P.~Carenza, T.~Fischer, M.~Giannotti and A.~Mirizzi, \emph{{Heavy
  axion-like particles and core-collapse supernovae: constraints and impact on
  the explosion mechanism}},
  \href{https://doi.org/10.1088/1475-7516/2020/12/008}{\emph{JCAP} {\bfseries
  12} (2020) 008} [\href{https://arxiv.org/abs/2008.04918}{{\ttfamily
  2008.04918}}].

\bibitem{Carenza:2020cis}
P.~Carenza, B.~Fore, M.~Giannotti, A.~Mirizzi and S.~Reddy, \emph{{Enhanced
  Supernova Axion Emission and its Implications}},
  \href{https://arxiv.org/abs/2010.02943}{{\ttfamily 2010.02943}}.

\bibitem{Marciano:2016yhf}
W.~Marciano, A.~Masiero, P.~Paradisi and M.~Passera, \emph{{Contributions of
  axionlike particles to lepton dipole moments}},
  \href{https://doi.org/10.1103/PhysRevD.94.115033}{\emph{Phys. Rev. D}
  {\bfseries 94} (2016) 115033}
  [\href{https://arxiv.org/abs/1607.01022}{{\ttfamily 1607.01022}}].

\bibitem{Gavela:2019wzg}
M.~Gavela, R.~Houtz, P.~Quilez, R.~Del~Rey and O.~Sumensari, \emph{{Flavor
  constraints on electroweak ALP couplings}},
  \href{https://doi.org/10.1140/epjc/s10052-019-6889-y}{\emph{Eur. Phys. J. C}
  {\bfseries 79} (2019) 369}
  [\href{https://arxiv.org/abs/1901.02031}{{\ttfamily 1901.02031}}].

\bibitem{Bauer:2019gfk}
M.~Bauer, M.~Neubert, S.~Renner, M.~Schnubel and A.~Thamm, \emph{{Axionlike
  Particles, Lepton-Flavor Violation, and a New Explanation of $a_\mu$ and
  $a_e$}}, \href{https://doi.org/10.1103/PhysRevLett.124.211803}{\emph{Phys.
  Rev. Lett.} {\bfseries 124} (2020) 211803}
  [\href{https://arxiv.org/abs/1908.00008}{{\ttfamily 1908.00008}}].

\bibitem{Cornella:2019uxs}
C.~Cornella, P.~Paradisi and O.~Sumensari, \emph{{Hunting for ALPs with Lepton
  Flavor Violation}},
  \href{https://doi.org/10.1007/JHEP01(2020)158}{\emph{JHEP} {\bfseries 01}
  (2020) 158} [\href{https://arxiv.org/abs/1911.06279}{{\ttfamily
  1911.06279}}].

\bibitem{Calibbi:2020jvd}
L.~Calibbi, D.~Redigolo, R.~Ziegler and J.~Zupan, \emph{{Looking forward to
  Lepton-flavor-violating ALPs}},
  \href{https://arxiv.org/abs/2006.04795}{{\ttfamily 2006.04795}}.

\bibitem{MartinCamalich:2020dfe}
J.~Martin~Camalich, M.~Pospelov, P.~N.~H. Vuong, R.~Ziegler and J.~Zupan,
  \emph{{Quark Flavor Phenomenology of the QCD Axion}},
  \href{https://doi.org/10.1103/PhysRevD.102.015023}{\emph{Phys. Rev. D}
  {\bfseries 102} (2020) 015023}
  [\href{https://arxiv.org/abs/2002.04623}{{\ttfamily 2002.04623}}].

\bibitem{DiLuzio:2020oah}
L.~Di~Luzio, R.~Grober and P.~Paradisi, \emph{{Hunting for the CP violating
  ALP}},  \href{https://arxiv.org/abs/2010.13760}{{\ttfamily 2010.13760}}.

\bibitem{DiLuzio:2020wdo}
L.~Di~Luzio, M.~Giannotti, E.~Nardi and L.~Visinelli, \emph{{The landscape of
  QCD axion models}},
  \href{https://doi.org/10.1016/j.physrep.2020.06.002}{\emph{Phys. Rept.}
  {\bfseries 870} (2020) 1} [\href{https://arxiv.org/abs/2003.01100}{{\ttfamily
  2003.01100}}].

\bibitem{Choi:2020rgn}
K.~Choi, S.~H. Im and C.~S. Shin, \emph{{Recent progresses in physics of axions
  or axion-like particles}},
  \href{https://arxiv.org/abs/2012.05029}{{\ttfamily 2012.05029}}.

\bibitem{Bauer:2016lbe}
M.~Bauer, C.~H\"orner and M.~Neubert, \emph{{Diphoton Resonance from a Warped
  Extra Dimension}}, \href{https://doi.org/10.1007/JHEP07(2016)094}{\emph{JHEP}
  {\bfseries 07} (2016) 094}
  [\href{https://arxiv.org/abs/1603.05978}{{\ttfamily 1603.05978}}].

\bibitem{Choi:2017gpf}
K.~Choi, S.~H. Im, C.~B. Park and S.~Yun, \emph{{Minimal Flavor Violation with
  Axion-like Particles}},
  \href{https://doi.org/10.1007/JHEP11(2017)070}{\emph{JHEP} {\bfseries 11}
  (2017) 070} [\href{https://arxiv.org/abs/1708.00021}{{\ttfamily
  1708.00021}}].

\bibitem{Grzadkowski:2010es}
B.~Grzadkowski, M.~Iskrzynski, M.~Misiak and J.~Rosiek, \emph{{Dimension-Six
  Terms in the Standard Model Lagrangian}},
  \href{https://doi.org/10.1007/JHEP10(2010)085}{\emph{JHEP} {\bfseries 10}
  (2010) 085} [\href{https://arxiv.org/abs/1008.4884}{{\ttfamily 1008.4884}}].

\bibitem{Criado:2019ugp}
J.~C. Criado, \emph{{BasisGen: automatic generation of operator bases}},
  \href{https://doi.org/10.1140/epjc/s10052-019-6769-5}{\emph{Eur. Phys. J. C}
  {\bfseries 79} (2019) 256}
  [\href{https://arxiv.org/abs/1901.03501}{{\ttfamily 1901.03501}}].

\bibitem{Georgi:1986df}
H.~Georgi, D.~B. Kaplan and L.~Randall, \emph{{Manifesting the Invisible Axion
  at Low-energies}},
  \href{https://doi.org/10.1016/0370-2693(86)90688-X}{\emph{Phys. Lett. B}
  {\bfseries 169} (1986) 73}.

\bibitem{Alloul:2013bka}
A.~Alloul, N.~D. Christensen, C.~Degrande, C.~Duhr and B.~Fuks,
  \emph{{FeynRules 2.0 - A complete toolbox for tree-level phenomenology}},
  \href{https://doi.org/10.1016/j.cpc.2014.04.012}{\emph{Comput. Phys. Commun.}
  {\bfseries 185} (2014) 2250}
  [\href{https://arxiv.org/abs/1310.1921}{{\ttfamily 1310.1921}}].

\bibitem{Hahn:2000kx}
T.~Hahn, \emph{{Generating Feynman diagrams and amplitudes with FeynArts 3}},
  \href{https://doi.org/10.1016/S0010-4655(01)00290-9}{\emph{Comput. Phys.
  Commun.} {\bfseries 140} (2001) 418}
  [\href{https://arxiv.org/abs/hep-ph/0012260}{{\ttfamily hep-ph/0012260}}].

\bibitem{Hahn:1998yk}
T.~Hahn and M.~Perez-Victoria, \emph{{Automatized one loop calculations in
  four-dimensions and D-dimensions}},
  \href{https://doi.org/10.1016/S0010-4655(98)00173-8}{\emph{Comput. Phys.
  Commun.} {\bfseries 118} (1999) 153}
  [\href{https://arxiv.org/abs/hep-ph/9807565}{{\ttfamily hep-ph/9807565}}].

\bibitem{Nogueira:1991ex}
P.~Nogueira, \emph{{Automatic Feynman graph generation}},
  \href{https://doi.org/10.1006/jcph.1993.1074}{\emph{J. Comput. Phys.}
  {\bfseries 105} (1993) 279}.

\bibitem{Cheung:2015aba}
C.~Cheung and C.-H. Shen, \emph{{Nonrenormalization Theorems without
  Supersymmetry}},
  \href{https://doi.org/10.1103/PhysRevLett.115.071601}{\emph{Phys. Rev. Lett.}
  {\bfseries 115} (2015) 071601}
  [\href{https://arxiv.org/abs/1505.01844}{{\ttfamily 1505.01844}}].

\bibitem{Bern:2019wie}
Z.~Bern, J.~Parra-Martinez and E.~Sawyer, \emph{{Nonrenormalization and
  Operator Mixing via On-Shell Methods}},
  \href{https://doi.org/10.1103/PhysRevLett.124.051601}{\emph{Phys. Rev. Lett.}
  {\bfseries 124} (2020) 051601}
  [\href{https://arxiv.org/abs/1910.05831}{{\ttfamily 1910.05831}}].

\bibitem{Buchalla:2019wsc}
G.~Buchalla, A.~Celis, C.~Krause and J.-N. Toelstede, \emph{{Master Formula for
  One-Loop Renormalization of Bosonic SMEFT Operators}},
  \href{https://arxiv.org/abs/1904.07840}{{\ttfamily 1904.07840}}.

\bibitem{Chala:2020pbn}
M.~Chala and A.~Titov, \emph{{One-loop running of dimension-six Higgs-neutrino
  operators and implications of a large neutrino dipole moment}},
  \href{https://doi.org/10.1007/JHEP09(2020)188}{\emph{JHEP} {\bfseries 09}
  (2020) 188} [\href{https://arxiv.org/abs/2006.14596}{{\ttfamily
  2006.14596}}].

\bibitem{Lyonnet:2016xiz}
F.~Lyonnet and I.~Schienbein, \emph{{PyR@TE 2: A Python tool for computing RGEs
  at two-loop}}, \href{https://doi.org/10.1016/j.cpc.2016.12.003}{\emph{Comput.
  Phys. Commun.} {\bfseries 213} (2017) 181}
  [\href{https://arxiv.org/abs/1608.07274}{{\ttfamily 1608.07274}}].

\bibitem{Jenkins:2017dyc}
E.~E. Jenkins, A.~V. Manohar and P.~Stoffer, \emph{{Low-Energy Effective Field
  Theory below the Electroweak Scale: Anomalous Dimensions}},
  \href{https://doi.org/10.1007/JHEP01(2018)084}{\emph{JHEP} {\bfseries 01}
  (2018) 084} [\href{https://arxiv.org/abs/1711.05270}{{\ttfamily
  1711.05270}}].

\bibitem{Jenkins:2017jig}
E.~E. Jenkins, A.~V. Manohar and P.~Stoffer, \emph{{Low-Energy Effective Field
  Theory below the Electroweak Scale: Operators and Matching}},
  \href{https://doi.org/10.1007/JHEP03(2018)016}{\emph{JHEP} {\bfseries 03}
  (2018) 016} [\href{https://arxiv.org/abs/1709.04486}{{\ttfamily
  1709.04486}}].

\bibitem{Terol-Calvo:2019vck}
J.~Terol-Calvo, M.~T\'ortola and A.~Vicente, \emph{{High-energy constraints
  from low-energy neutrino nonstandard interactions}},
  \href{https://doi.org/10.1103/PhysRevD.101.095010}{\emph{Phys. Rev. D}
  {\bfseries 101} (2020) 095010}
  [\href{https://arxiv.org/abs/1912.09131}{{\ttfamily 1912.09131}}].

\bibitem{Collins:2005nj}
J.~C. Collins, A.~V. Manohar and M.~B. Wise, \emph{{Renormalization of the
  vector current in QED}},
  \href{https://doi.org/10.1103/PhysRevD.73.105019}{\emph{Phys. Rev. D}
  {\bfseries 73} (2006) 105019}
  [\href{https://arxiv.org/abs/hep-th/0512187}{{\ttfamily hep-th/0512187}}].

\end{thebibliography}\endgroup

\end{document}